\documentstyle [12pt,epsf]{article}
\begin{document}
\setcounter{page}{1}
\def\theequation{\arabic{section}.\arabic{equation}}
\def\theequation{\thesection.\arabic{equation}}
\setcounter{section}{0}

\title{Solar neutrino processes \\ in the relativistic field theory model of the deuteron}

\author{A. N. Ivanov~\thanks{E--mail: ivanov@kph.tuwien.ac.at,
Tel.: +43--1--58801--14261, Fax: +43--1--5864203}~${\textstyle ^\ddagger}$, H. Oberhummer~\thanks{E--mail: ohu@kph.tuwien.ac.at, Tel.: +43--1--58801--14251, Fax: +43--1--5864203} ,  
N. I. Troitskaya~\thanks{Permanent Address:
State Technical University, Department of Nuclear
Physics, 195251 St. Petersburg, Russian Federation} ,
M. Faber~\thanks{E--mail: faber@kph.tuwien.ac.at,
Tel.: +43--1--58801--14261, Fax: +43--1--5864203}}
\date{}

\maketitle

\vspace{-0.5in}
\begin{center}
{\it Institut f\"ur Kernphysik, Technische Universit\"at Wien,\\ 
Wiedner Hauptstr. 8-10, A-1040 Vienna, Austria}
\end{center}

\begin{center}
\begin{abstract}
The generalized version of the relativistic field theory model of the deuteron (RFMD) is applied to the
description of processes of astrophysical interest and low--energy elastic NN scattering. The value of the astrophysical factor $S_{\rm pp}(0) = 5.52\,\times 10^{-25} {\rm MeV\,\rm b}$ for the solar proton burning  p + p $\to$ D + e$^+$ + $\nu_{\rm e}$ is found to be enhanced by a factor of $1.42$ with respect to the classical value  $S^*_{\rm pp}(0) = 3.89\,\times 10^{-25}\,{\rm MeV\,\rm b}$ obtained by Kamionkowski and Bahcall in the potential model approach (PMA). The astrophysical aspects of this enhancement are discussed. The cross sections for the disintegration of the deuteron by (anti--)neutrinos $\nu_{\rm e}$ + D $\to$ e$^-$ + p + p, $\bar{\nu}_{\rm e}$ + D $\to$ ${\rm e}^+$ + n + n and $\nu_{\rm e}(\bar{\nu}_{\rm e})$ + D $\to$ $\nu_{\rm e}(\bar{\nu}_{\rm e})$ + n + p  are calculated for the energies of $\nu_{\rm e}(\bar{\nu}_{\rm e})$ ranging from thresholds up to  10$\,{\rm MeV}$. The results are discussed in comparison  with the PMA data.  The cross sections for $\bar{\nu}_{\rm e}$ + D $\to$ ${\rm e}^+$ + n + n and $\bar{\nu}_{\rm e}$ + D $\to$ $\bar{\nu}_{\rm e}$ + n + p  averaged over the reactor anti--neutrino energy spectrum agree well with experimental data. The astrophysical factor $S_{\rm pep}(0)$ for the process p + e$^-$ + p $\to$ $\nu_{\rm e}$ + D (or pep--process) is calculated relative to $S_{\rm pp}(0)$ in complete agreement with the result obtained by Bahcall and May. The reaction rate for the neutron--proton radiative capture is calculated in agreement with the PMA result obtained for pure M1 transition. It is shown that in the RFMD one can describe low--energy elastic NN scattering in complete agreement with low--energy nuclear phenomenology.
\end{abstract}
\end{center}
\begin{center}
PACS: 11.10.Ef, 13.75.Cs, 14.20.Dh, 21.30.Fe, 25.40.Lw, 26.65.+t\\
\noindent Keywords: relativistic field theory, deuteron, proton--proton fusion,  Coulomb repulsion, neutrino disintegration, nucleon nucleon scattering
\end{center}

\newpage

\section{Introduction}
\setcounter{equation}{0}

The relativistic field theory model of the deuteron (RFMD) suggested in Refs.~[1--4] gives a new approach to the description of strong low--energy interactions of hadrons and light nuclei. The basis of the model is the one--nucleon loop origin of a physical deuteron produced by low--energy fluctuations of the proton and the neutron. This imposes the constraint for the deuteron to be coupled to itself and other particles through the one--nucleon loop exchanges only. In terms of one--nucleon loop exchanges we describe in the RFMD a non--trivial wave function of the relative movement of the nucleons inside the  deuteron.

For the description of the low--energy processes of the deuteron coupled to the nucleon--nucleon (NN) system in the ${^1}{\rm S}_0$--state we have postulated the effective local four--nucleon interaction [2,4]
\begin{eqnarray}\label{label1.1}
&&{\cal L}^{\rm NN \to NN}_{\rm eff}(x)= G_{\rm \pi
NN}\,\int d^3\rho\,\delta^{(3)}(\vec{\rho}\,)\nonumber\\
&&\times \{[\bar{n}(t,\vec{x} + \frac{1}{2}\vec{\rho}\,)\gamma_{\mu}
\gamma^5 p^c(t,\vec{x} - \frac{1}{2}\vec{\rho}\,)] [\bar{p^c}(t,\vec{x}
+ \frac{1}{2}\vec{\rho}\,)\gamma^{\mu}\gamma^5 n(t,\vec{x} -
\frac{1}{2}\vec{\rho}\,)]\nonumber\\
&&+\frac{1}{2}\,
[\bar{n}(t,\vec{x} + \frac{1}{2}\vec{\rho}\,)\gamma_{\mu} \gamma^5 n^c(t,\vec{x}
- \frac{1}{2}\vec{\rho}\,)] [\bar{n^c}(t,\vec{x} + \frac{1}{2}\vec{\rho}\,)
\gamma^{\mu}\gamma^5
n(t,\vec{x} - \frac{1}{2}\vec{\rho}\,)]\nonumber\\
&&+ \frac{1}{2}\,[\bar{p}(t,\vec{x} + \frac{1}{2}\vec{\rho}\,)\gamma_{\mu}
\gamma^5 p^c(t,\vec{x} - \frac{1}{2}\vec{\rho}\,)]
[\bar{p^c}(t,\vec{x} + \frac{1}{2}\vec{\rho}\,)\gamma^{\mu}\gamma^5 p(t,\vec{x}
- \frac{1}{2}\vec{\rho}\,)] \nonumber\\
&&+ (\gamma_{\mu}\gamma^5 \otimes \gamma^{\mu}\gamma^5 \to \gamma^5 \otimes
\gamma^5)\},
\end{eqnarray}
where $n(t,\vec{x} \pm\frac{1}{2}\vec{\rho}\,)$ and $p(t,\vec{x} \pm
\frac{1}{2}\vec{\rho}\,)$ are the operators of the neutron and the proton interpolating fields, $\vec{\rho}$ is a radius--vector of a relative movement, $n^c(t,\vec{x} \pm\frac{1}{2}\vec{\rho}\,) = C \bar{n}^T(t,\vec{x} \pm\frac{1}{2}\vec{\rho}\,)$, etc. The effective coupling constant $G_{\rm \pi NN}$ is defined by
\begin{eqnarray}\label{label1.2}
G_{\rm \pi NN} = \frac{g^2_{\rm \pi NN}}{4M^2_{\pi}} - \frac{2\pi a_{\rm np}}{M_{\rm N}} = 3.27\times 10^{-3}\,{\rm MeV}^{-2},
\end{eqnarray}
where $g_{\rm \pi NN}= 13.4$ [5] is the coupling constant of the ${\rm \pi NN}$ interaction, $M_{\pi}=135\,{\rm MeV}$ is the pion mass, $M_{\rm p} = M_{\rm n} = M_{\rm N} = 940\,{\rm MeV}$ is the mass of the proton and the neutron neglecting the electromagnetic mass difference and $a_{\rm np} = (-23.748\pm 0.010)\,{\rm fm}$ is the scattering length of the np scattering in the ${^1}{\rm S}_0$--state [5]. 

The first term in the effective coupling constant $G_{\rm \pi NN}$ is caused by the one--pion exchange, whereas the second is a phenomenological one representing the collective contribution of heavy meson exchanges [4]. The effective interaction Eq.~(\ref{label1.1}) is written in the isotopically invariant form, and the coupling constant $G_{\rm \pi NN}$ can be never equal zero at $a_{\rm np} \neq 0$ due to negative value of $a_{\rm np}$ imposed by nuclear forces [6].

In Refs.~[2,4] the RFMD supplemented by the effective local four--nucleon interaction Eq.~(\ref{label1.1}) has been applied to calculation of the cross sections for the neutron--proton radiative capture n + p $\to$ D + $\gamma$ for thermal neutrons and the photomagnetic disintegration of the deuteron $\gamma$ + D $\to$ n + p, being the inverse process for the neutron--proton radiative capture, the astrophysical factor for the solar proton burning p + p $\to$ D + e$^+$ + $\nu_{\rm e}$ and the cross section for the disintegration of the deuteron by anti--neutrinos $\bar{\nu}_{\rm e}$ + D $\to$ e$^+$ + n + n. The former process due to charge independence of weak interaction strength is valued as a terrestrial equivalent of the solar proton burning. The obtained results have been found in agreement with the potential model approach (PMA) within an accuracy better than 10$\%$. For example, for the astrophysical factor of the solar proton burning we have found the value $S_{\rm pp}(0) = 4.02 \times 10^{-25}\,{\rm MeV\,b}$\footnote{In order to get this numerical value of the astrophysical factor $S_{\rm pp}(0)$ we have to multiply the astrophysical factor $\delta S_{\rm pp}(0)$ calculated in [2] (Erratum) by a factor 2 caused by the symmetrization of the wave function of the protons in the initial state. For the detailed calculations of the amplitude of the solar proton burning we relegate readers to Appendix C of this paper and Appendix A of Ref.~[4].} which agrees well with the classical value $S^*_{\rm pp}(0) = 3.89\,\times 10^{-25}\,{\rm MeV\,b}$ obtained by Kamionkowski and Bahcall [7] in the PMA (see also Refs.~[8,9]) and the value $S^*_{\rm pp}(0) = 4.05\,\times 10^{-25}\,{\rm MeV\,b}$ calculated in the Effective Field Theory (EFT) approach [10--12] by Park {\it et al.} [13].

The main problem which has been encountered for the calculation of the astrophysical factor $S_{\rm pp}(0)$ through the local four--nucleon interaction Eq.~(\ref{label1.1}) lays in the impossibility to describe the Coulomb repulsion between the protons. The Coulomb repulsion has been taken into account in terms of the Gamow penetration factor only, and, apart from the weak interactions, the obtained value of the astrophysical factor $S_{\rm pp}(0) = 4.02 \times 10^{-25}\,{\rm MeV\,b}$ is caused by strong low--energy nuclear forces.  However, as has been stated by Kamionkowski and Bahcall [7] the  Coulomb repulsion between the protons should give an important contribution to the amplitude of the process p + p $\to$ D + e$^+$ + $\nu_{\rm e}$. Indeed, more than 60$\%$ of the value $S^*_{\rm pp}(0) = 3.89\,\times 10^{-25}\,{\rm MeV\,b}$ are defined by the Coulomb repulsion between the protons [7]. Therefore, if only due to strong low--energy nuclear forces the RFMD predicts the value of the astrophysical factor comparable with that obtained by Kamionkowski and Bahcall [7], where more than 60$\%$ of the magnitude are caused by the Coulomb repulsion between the protons, so accounting for the Coulomb repulsion between the protons one can expect to get an enhancement with respect to the values of the astrophysical factor calculated in the PMA and the EFT approach [7--9,13].

In this paper we develop a generalized version of the RFMD [1--4] admitting the description of the Coulomb repulsion of between the protons for the solar proton burning in terms of the explicit Coulomb wave function. The modification of the model is connected with the replacement of the $\delta$--potential $\delta^{(3)}(\vec{\rho}\,)$ in the effective four--nucleon interaction by a smeared one. The reason of this smearing is in the following. A relative movement of the protons in the process p + p $\to$ D +  e$^+$ + $\nu_{\rm e}$ should be described in terms of the exact Coulomb wave function. This wave function contains a regular and  an irregular solution of Schr\"odinger equation [7--9]. Since, as we show below, the main contribution comes from the irregular solution, a $\delta$--function interaction should lead to unphysical singularities and should be smeared:
\begin{eqnarray}\label{label1.3}
\delta^{(3)}(\vec{\rho}\,) \to U(\rho),
\end{eqnarray}
For the description of the NN system strongly coupled in the ${^1}{\rm S}_0$--states we propose to use the effective Yukawa potential defined by one--pion exchange, i.e.,
\begin{eqnarray}\label{label1.4}
U(\rho) = {\displaystyle
\frac{M^2_{\pi}}{4\pi}\,\frac{e^{\displaystyle - M_{\pi}\rho}}{\rho}}.
\end{eqnarray}
Making a change $\delta^{(3)}(\vec{\rho}\,) \to U(\rho)$ in Eq.~(\ref{label1.1}) we arrive at the interaction
\begin{eqnarray}\label{label1.5}
&&{\cal L}^{\rm NN \to NN}_{\rm eff}(x) = G_{\rm \pi
NN}\,\int d^3\rho\,U(\rho)\nonumber\\
&&\times \{[\bar{n}(t,\vec{x} + \frac{1}{2}\vec{\rho}\,)\gamma_{\mu}
\gamma^5 p^c(t,\vec{x} - \frac{1}{2}\vec{\rho}\,)] [\bar{p^c}(t,\vec{x}
+ \frac{1}{2}\vec{\rho}\,)\gamma^{\mu}\gamma^5 n(t,\vec{x} -
\frac{1}{2}\vec{\rho}\,)]\nonumber\\
&&+\frac{1}{2}\,
[\bar{n}(t,\vec{x} + \frac{1}{2}\vec{\rho}\,)\gamma_{\mu} \gamma^5 n^c(t,\vec{x}
- \frac{1}{2}\vec{\rho}\,)] [\bar{n^c}(t,\vec{x} + \frac{1}{2}\vec{\rho}\,)
\gamma^{\mu}\gamma^5
n(t,\vec{x} - \frac{1}{2}\vec{\rho}\,)]\nonumber\\
&&+ \frac{1}{2}\,[\bar{p}(t,\vec{x} + \frac{1}{2}\vec{\rho}\,)\gamma_{\mu}
\gamma^5 p^c(t,\vec{x} - \frac{1}{2}\vec{\rho}\,)]
[\bar{p^c}(t,\vec{x} + \frac{1}{2}\vec{\rho}\,)\gamma^{\mu}\gamma^5 p(t,\vec{x}
- \frac{1}{2}\vec{\rho}\,)] \nonumber\\
&&+ (\gamma_{\mu}\gamma^5 \otimes \gamma^{\mu}\gamma^5 \to \gamma^5 \otimes
\gamma^5)\},
\end{eqnarray}
Such a modification means that for the description of strong low--energy NN interactions we take into account the one--pion exchange in the form of the Yukawa potential and assume that the contribution of heavy meson exchanges has the range and the shape of the Yukawa potential defined by the one--pion exchange. This is not very strong assumption if to take into account that the spatial region of the proton and the neutron fluctuations forming the physical deuteron coincides with the range of the Yukawa potential defined by the one--pion exchange (see Appendix A).

Then, we would like to implicate more nuclear phenomenology for the definition of the effective coupling constants caused by heavy meson exchange contributions. We have expressed these contributions in terms of the S--wave scattering length $a_{\rm np}$ of the np scattering in the ${^1}{\rm S}_0$--state. However, from low--energy elastic NN scattering phenomenology it is known that the S--wave scattering lengths of the np, nn and pp scattering in the ${^1}{\rm S}_0$--states caused by nuclear forces differ each other. The experimental values of them amount to [5]:
\begin{eqnarray}\label{label1.6}
a_{\rm np} &=& (-23.748\pm 0.010)\,{\rm fm},\nonumber\\
a_{\rm nn} &=& (- 16.40 \pm 0.09)\,{\rm fm}\nonumber\\
a_{\rm pp} &=& (-17.10\pm 0.20)\,{\rm fm}.
\end{eqnarray}
In order to employ these data in the RFMD we suggest to use the values of the scattering lengths Eq.~(\ref{label1.6}) for the definition of numerical values of heavy meson exchange contributions to different channels of NN scattering. By virtue of this the resultant effective four--nucleon interaction should read
\begin{eqnarray}\label{label1.7}
&&{\cal L}^{\rm NN \to NN}_{\rm eff}(x) = \,\int d^3\rho\,U(\rho)\nonumber\\
&&\times \{G_{\rm \pi np}\,[\bar{n}(t,\vec{x} + \frac{1}{2}\vec{\rho}\,)\gamma_{\mu}
\gamma^5 p^c(t,\vec{x} - \frac{1}{2}\vec{\rho}\,)] [\bar{p^c}(t,\vec{x}
+ \frac{1}{2}\vec{\rho}\,)\gamma^{\mu}\gamma^5 n(t,\vec{x} -
\frac{1}{2}\vec{\rho}\,)]\nonumber\\
&&+\frac{1}{2}\,G_{\rm \pi nn}\,
[\bar{n}(t,\vec{x} + \frac{1}{2}\vec{\rho}\,)\gamma_{\mu} \gamma^5 n^c(t,\vec{x}
- \frac{1}{2}\vec{\rho}\,)] [\bar{n^c}(t,\vec{x} + \frac{1}{2}\vec{\rho}\,)
\gamma^{\mu}\gamma^5
n(t,\vec{x} - \frac{1}{2}\vec{\rho}\,)]\nonumber\\
&&+ \frac{1}{2}\,G_{\rm \pi pp}\, [\bar{p}(t,\vec{x} + \frac{1}{2}\vec{\rho}\,)\gamma_{\mu}
\gamma^5 p^c(t,\vec{x} - \frac{1}{2}\vec{\rho}\,)]
[\bar{p^c}(t,\vec{x} + \frac{1}{2}\vec{\rho}\,)\gamma^{\mu}\gamma^5 p(t,\vec{x}
- \frac{1}{2}\vec{\rho}\,)] \nonumber\\
&&+ (\gamma_{\mu}\gamma^5 \otimes \gamma^{\mu}\gamma^5 \to \gamma^5 \otimes
\gamma^5)\},
\end{eqnarray}
where the effective coupling constants $G_{\rm \pi np}$, $G_{\rm \pi nn}$ and $G_{\rm \pi pp}$ are determined as
\begin{eqnarray}\label{label1.8}
G_{\rm \pi np} &=&\frac{g^2_{\rm \pi NN}}{4M^2_{\pi}} - \frac{2\pi a_{\rm np}}{M_{\rm N}} = 3.27\times 10^{-3}\,{\rm MeV}^{-2},\nonumber\\
G_{\rm \pi nn} &=&\frac{g^2_{\rm \pi NN}}{4M^2_{\pi}} - \frac{2\pi a_{\rm nn}}{M_{\rm N}} = 3.02\times 10^{-3}\,{\rm MeV}^{-2},\nonumber\\
G_{\rm \pi pp} &=&\frac{g^2_{\rm \pi NN}}{4M^2_{\pi}} - \frac{2\pi a_{\rm pp}}{M_{\rm N}} = 3.04\times 10^{-3}\,{\rm MeV}^{-2},
\end{eqnarray}
calculated for the S--wave scattering lengths given by Eq.~(\ref{label1.6}).  Thus, for the description of the NN system in the ${^1}{\rm S}_0$--state coupled at low energies to the deuteron we will use below the effective four--nucleon interaction given by Eq.~(\ref{label1.7}) with the effective coupling constants defined by Eq.~(\ref{label1.8}).

The process of the solar proton burning p + p $\to$ D + e$^+$ + $\nu_{\rm e}$ is closely related to processes of disintegration of the deuteron by neutrinos and anti--neutrinos $\nu_{\rm e}$ + D $\to$ e$^-$ + p + p, $\bar{\nu}_{\rm e}$ + D $\to$ ${\rm e}^+$ + n + n and $\nu_{\rm e}(\bar{\nu}_{\rm e})$ + D $\to$ $\nu_{\rm e}(\bar{\nu}_{\rm e})$ + n + p. Since these reactions are governed by the same dynamics of strong low--energy nuclear forces, we apply the generalized RFMD\footnote{Below we retain the abbreviation the RFMD for the generalized version of the RFMD.} 
to the computation of the cross sections for the reactions $\nu_{\rm e}$ + D $\to$ e$^-$ + p + p, $\bar{\nu}_{\rm e}$ + D $\to$ ${\rm e}^+$ + n + n and $\nu_{\rm e}(\bar{\nu}_{\rm e})$ + D $\to$ $\nu_{\rm e}(\bar{\nu}_{\rm e})$ + n + p  for energies of neutrinos and anti--neutrinos ranging from thresholds up to  10$\,{\rm MeV}$. 

The paper is organized as follows. In Sect.~2 we discuss the RFMD in outline and draw similarity between the RFMD and effective quark models motivated by QCD applied to the derivation of Effective Chiral Lagrangians with chiral $U(3)\times U(3)$ symmetry. In Sect.~3 we discuss the wave function of the relative movement of two protons accounting for the Coulomb repulsion. In Sect.~4 we compute the astrophysical factor for the solar proton burning. We give the value $S_{\rm pp}(0) = 5.52\,\times 10^{-25} {\rm MeV\,\rm b}$ which is enhanced by a factor of $1.42$ with respect to the classical value  $S^*_{\rm pp}(0) = 3.89\,\times 10^{-25}\,{\rm MeV\,\rm b}$ obtained 
by Kamionkowski and Bahcall in the PMA. In Sect.~5 we discuss astrophysical consequences of such an enhancement and 
estimate the solar neutrino fluxes. In Sect.~6 we calculate the cross section for the process $\nu_{\rm e}$ + D $\to$ e$^-$ + p + p. In Sect.~7 we derive the astrophysical factor $S_{\rm pep}(0)$ for the process p + e$^-$ + p $\to$ $\nu_{\rm e}$ + D (or pep--process) by using the cross section for the process of $\nu_{\rm e}$ + D $\to$ e$^-$ + p + p calculated in Sect.~6 and the astrophysical factor $S_{\rm pp}(0)$ obtained in Sect.~4. We give  the ratio $S_{\rm pep}(0)/S_{\rm pp}(0)$ in complete agreement with the result obtained by Bahcall and May. 
In Sects.~8 and 9 we calculate the cross sections for the processes $\bar{\nu}_{\rm e}$ + D $\to$ e$^+$ + n + n and $\nu_{\rm e}(\bar{\nu}_{\rm e})$ + D $\to$  $\nu_{\rm e}(\bar{\nu}_{\rm e})$ + n + p caused by the charged and neutral weak currents, respectively. Our results for the cross sections for the processes of the disintegration of the deuteron by neutrinos and anti--neutrinos can be applied to the analysis of solar neutrino experiments at Sudbury Neutrino Observatory (SNO) [14]. In Sect.~10 we calculate the reaction rate for the neutron--proton radiative capture n + p $\to$ D + $\gamma$ for thermal neutrons caused by pure M1 transition. We find a complete agreement with our former result obtained in Refs.~[2,4] and the PMA.  In Sect.~11 we show that in the RFMD one can describe  low--energy elastic NN scattering in terms of the S--wave scattering length and the effective range in complete agreement with low--energy nuclear phenomenology. This confutes the critique by Bahcall and Kamionkowski [15] that the RFMD is unable to describe low--energy elastic NN scattering with non--zero effective range. In Conclusion we discuss the obtained results and outline the perspectives and further applications of the RFMD. 
In Appendix A we calculate the binding energy of the deuteron in one-- and two--nucleon loop approximation and reestimate the theoretical uncertainty of the model. We find now that the theoretical uncertainty of the RFMD is not more than 9.5$\%$ for amplitudes and, correspondingly, 19$\%$ for  cross sections.  
In Appendix B we derive the effective four--nucleon 
interaction of Eq.~(\ref{label1.1}). In Appendix C, D and E we give detailed calculations of the matrix elements of the  solar proton burning, the process $\nu_{\rm e}$ + D $\to$ e$^-$ + p + p and the process $\nu_{\rm e}$ + D $\to$ $\nu_{\rm e}$ + n + p, respectively. The computation of the matrix element of the process $\bar{\nu}_{\rm e}$ + D $\to$ e$^+$ + n + n is analogous to the process  $\nu_{\rm e}$ + D $\to$ e$^-$ + p + p. In Appendix F we calculate in details the amplitude of the neutron--proton radiative capture. In Appendix G we show how to calculate in the quantum field theory approach like the RFMD the cross sections for the low--energy elastic pp and np scattering caused by strong local four--nucleon interaction. In Appendix H we calculate 
the amplitude and the cross section for the process $\bar{\nu}_{\rm e}$ + D $\to$ e$^+$ + n + n near threshold by applying the 
local four--nucleon interaction Eq.~(\ref{label1.1}) [2,4].

\section{Outline of the RFMD}
\setcounter{equation}{0}

The RFMD describing strong low--energy nuclear interactions of the deuteron coupled to nucleons and other particles through one--nucleon loop exchanges suggests dynamics of strong low--energy nuclear forces completely different to the PMA and the EFT approach but very similar to dynamics of effective quark models motivated by QCD like the extended Nambu--Jona--Lasinio (ENJL) model with chiral $U(3)\times U(3)$ symmetry [16--19] applied to the derivation of Effective Chiral Lagrangians [20--22]. 

In the RFMD the deuteron is represented by a local field operator $D_{\mu}(x)$ (or $D^{\dagger}_{\mu}(x)$), the action of which on a vacuum state annihilates (or creates) the deuteron.  All low--energy interactions come through the one--nucleon loop exchanges. The nucleon--deuteron vertices in the one--nucleon loop diagrams are point--like and defined by a phenomenological {\it local} conserving nucleon current $J^{\mu}(x) = - i g_{\rm V} [\bar{p}(x) \gamma^{\mu} n^c(x) - \bar{n}(x) \gamma^{\mu} p^c(x)]$, i.e., $\partial_{\mu}J^{\mu}(x) = 0$, accounting for spinorial and isotopical properties of the deuteron, and $g_{\rm V}$ is a dimensionless phenomenological coupling constant. The virtual nucleons are described by Green functions of free nucleons and anti--nucleons with a constant mass $M_{\rm N}$.

In order to couple to the deuteron through the one--nucleon loop exchange the nucleons should pass through intermediate interactions providing low--energy transitions N + N $\to$ N + N. For the description of the NN system coupled in the ${^1}{\rm S}_0$--state to the deuteron we apply a low--energy four--nucleon interaction given by Eq.~(\ref{label1.1}) and Eq.~(\ref{label1.7}), where the one--pion exchange plays a dominant role [2,4]. The former distinguishes the RFMD from the PMA and the EFT approach, where for the correct description of low--energy nuclear forces there should be introduced a phenomenological NN potential, for instance, the Argonne $v_{18}$ [23], and the one--pion exchange contribution is taken only as a perturbation. Indeed, in power counting [9--11] the interaction induced by the one--pion exchange is of order $O(k^2)$, where $k$ is a relative momentum of the NN system. This behaviour is caused by the $\gamma^5$ matrix due to which large components of Dirac bispinors of wave functions of interacting nucleons become suppressed and only small components are material. The phenomenological part of the effective four--nucleon interactions Eq.~(\ref{label1.1}) and Eq.~(\ref{label1.7}) is expressed in terms of the S--wave scattering lengths of low--energy elastic NN scattering in the form accepted in the EFT approach [9--11]. The appearance of this part is motivated by heavy meson exchange contributions [4].This phenomenological part of the effective four--nucleon interaction makes up less than 33$\%$ of the one--pion exchange contribution. Such a dominant role of the one--pion exchange is completely a peculiarity of dynamics of the one--nucleon loop exchanges related to one--fermion loop anomalies [2,4, 24--26].

Indeed, the effective Lagrangian Eq.~(\ref{label1.1}) and, correspondingly, Eq.~(\ref{label1.7}) taken in the low--energy limit vanishes due to the reduction $[\bar{N}(x)\gamma_{\mu}\gamma^5 N^c(y)][\bar{N^c}(y) \gamma^{\mu}\gamma^5 N(x)] \to  - [\bar{N}(x) \gamma^5 N^c(y)][\bar{N^c}(y) \gamma^5 N(x)]$ or shortly $\gamma_{\mu}\gamma^5 \otimes \gamma^{\mu}\gamma^5 \to - \gamma^5 \otimes \gamma^5$ (see Eq.~({\rm C}.41) and Eq.~({\rm D}.22)). This agrees with the vanishing of the one--pion exchange contribution to the NN potential in the low--energy limit. Hence, the effective four--nucleon interactions Eq.~(\ref{label1.1}) and  Eq.~(\ref{label1.7}) applied in the 
tree--approximation to the description of the NN system coupled to the deuteron would scarcely give a significant contribution compared with the PMA and the EFT approach. However, 
due to the one--nucleon loop approach the contributions of the interactions $\gamma_{\mu}\gamma^5 \otimes \gamma^{\mu}\gamma^5$ and $\gamma^5 \otimes \gamma^5$ to amplitudes of nuclear processes are different and do not cancel each other in the 
low--energy limit. For instance, in the case of the neutron--proton radiative capture and the photomagnetic disintegration of the deuteron the amplitudes of the processes are defined by the triangle one--nucleon loop diagrams with  AVV (axial--vector--vector) and PVV (pseudoscalar--vector--vector) vertices [2,4] caused by $\gamma_{\mu}\gamma^5 \otimes \gamma^{\mu}\gamma^5$ and $\gamma^5 \otimes \gamma^5$ interactions, respectively. These diagrams are well--known in particle physics in connection with  the Adler--Bell--Jackiw axial anomaly [24] which plays a dominant role for the processes of the decays $\pi^0 \to \gamma \gamma$, $\omega \to \pi^0 \gamma$ and so on [24--26]. Since 
the results of the calculation of these diagrams differ each other, they give different contributions to the amplitudes of the processes and do not cancel themselves in the low--energy limit. Then, the amplitudes of the solar proton burning and the anti--neutrino disintegration of the deuteron are defined by the one--nucleon loop diagrams with AAV and APV vertices caused by $\gamma_{\mu}\gamma^5 \otimes \gamma^{\mu}\gamma^5$ and $\gamma^5 \otimes \gamma^5$ interactions, respectively [2,4]. The contribution of the 
diagrams with APV vertices, calculated for the local four--nucleon interactions Eq.~(\ref{label1.1}), turns out to be divergent and, therefore, negligibly small in comparison with the contribution of the diagrams with AAV vertices [2,4], which contains a non--trivial convergent part related to one--fermion loop anomalies [2,4,25]. In the case of the four--nucleon interactions given by Eq.~(\ref{label1.7}) the contribution of the diagrams with PAV vertices is convergent and cancels partly the contribution of the diagrams with AAV vertices (see Appendix C).

The main problem which we encounter for the practical
realization of the derivation of effective Lagrangians of 
low--energy interactions of the deuteron coupled to nucleons and other particles through the one--nucleon loop exchanges lies in the necessity to satisfy requirement of {\it locality} of these interactions  related to {\it the condition of microscopic causality} in a quantum field theory approach [27]. 
Since in the RFMD one--nucleon loop diagrams are defined by the point--like vertices and the Green functions of free virtual nucleons with constant masses, there is only a naive way to satisfy to requirement of {\it locality} of effective interactions through a formal application of a long--wavelength approximation to the computation of one--nucleon loop diagrams [2,4]. 
This approximation implies the expansion of one--nucleon loop diagrams in powers of external momenta by keeping only leading terms of the expansion. Of course, the application of such an approximation to the computation of one--nucleon loop diagrams, when on--mass shell the energy of the deuteron exceeds twice the masses of 
virtual nucleons, can seem rather unjustified.

However, in this connection we would like 
to recall that the analogous problem encounters itself for 
the derivation of Effective Chiral Lagrangians [20,21] within effective quark models motivated by QCD like 
the ENJL model with chiral $U(3)\times U(3)$ symmetry [16--19]. Indeed, all phenomenological low--energy interactions predicted by Effective Chiral Lagrangians [20,21] for the nonet of vector mesons ($\rho(770)$, $\omega(780)$ and so  on) can be derived within the ENJL model by calculating one--constituent quark loop diagrams at leading order in the long--wavelength approximation. As has turned out the long--wavelength approximation works very good in spite of the fact that the constituent quark loop diagrams are 
defined by point--like vertices of quark--meson interactions and Green functions of free constituent quarks with constant masses $M_q \sim 330\,{\rm MeV}$, and, moreover, the masses of vector 
mesons exceed twice the constituent quark mass. A formal justification of the validity of the long--wavelength approximation can be given by attracting the Vector Dominance (VD) hypothesis [20,28] 
due to which {\it the effective vertices of low--energy interactions of vector mesons should be smooth functions of squared 4--momenta of interacting mesons varying from on--mass shell to zero values}.  Due to this hypothesis one can calculate the vertices of low--energy interactions of vector mesons keeping them off--mass shell around zero values 
of their squared momenta [28], and then, having had kept leading terms of the long--wavelength expansion, continue the resultant expression on--mass shell. Such a procedure describes perfectly well [16--19] all phenomenological vertices of low--energy 
interactions of vector mesons predicted by Effective Chiral Lagrangians [20,21]. 

One cannot say exactly, whether we really have in the RFMD some 
kind of the VD hypothesis, i.e., smooth dependence of effective low--energy interactions of the deuteron coupled to other particles on squared 4--momenta of interacting external particles including the deuteron. However, the application of the long--wavelength approximation to the computation of one--nucleon loop diagrams leads eventually to effective local Lagrangians  describing reasonably well dynamics of strong low--energy nuclear interactions. 
The static parameters of the deuteron and amplitudes of strong low--energy interactions of the deuteron coupled to nucleons and other particles can be described in the RFMD in complete agreement with the philosophy and technique of the derivation of Effective Chiral Lagrangians within effective quark models motivated by QCD. 

The agreement between the reaction rates for the neutron--proton radiative capture, caused by pure M1 transition, calculated 
in the RFMD and the PMA is not surprising [2,4]. Indeed, it is known from particle physics that the radiative decays of pseudoscalar and vector mesons like $\pi^0 \to \gamma\gamma$, $\omega \to \pi^0 \gamma$ and so on, caused by M1 transitions, can be computed both in the non--relativistic quark model [29], which is some kind of 
the PMA, and in the Effective Chiral Lagrangian approach [21]. In the non--relativistic quark model the matrix elements of these decays are given in terms of magnetic moments of constituent quarks proportional to $1/M_q$, whereas in the Effective Chiral Lagrangian approach they are defined by the axial anomaly and proportional to $1/F_{\pi}$, the inverse power of the PCAC constant $F_{\pi}= 92.4\,{\rm MeV}$ [24,30]. Equating the matrix elements of these decays calculated in the non--relativistic quark model and in the Effective Chiral Lagrangian approach  one can express a constituent quark mass in terms of the PCAC constant $F_{\pi}$ [30]. The estimated value of the constituent quark mass  $M_q \simeq 400\,{\rm MeV}$ [30] is comparable with the values $M_q = 330\div 380\;{\rm MeV}$ accepted in the literature [28]. This testifies that both the non--relativistic quark model and the Effective Chiral Lagrangian approach describe equally well dynamics of strong low--energy interactions of low--lying mesons even if for the decays caused by  M1 transitions. Referring to this example the agreement between the reaction rates for the neutron--proton radiative capture calculated in the RFMD and in the PMA, respectively, is understandable. Our prediction for the astrophysical factor value $S_{\rm pp}(0) = 4.02\times 10^{-25}\,{\rm MeV\,b}$ for the solar proton burning [2,4] is rather promising. Indeed, it agrees very good with the classical result $S_{\rm pp}(0) = 3.89\times 10^{-25}\,{\rm MeV\,b}$ obtained by Kamionkowski and Bahcall [7] in spite of the Coulomb repulsion has been taken into account in the form of the Gamow penetration factor only. However, as has been claimed by Kamionkowski and Bahcall [7] the Coulomb repulsion defines more than 60$\%$ of the value of the astrophysical factor. Hence, one can expect that the implication of the Coulomb repulsion between the protons in terms of the explicit Coulomb wave function that we intend to do in the generalized RFMD should provide an enhancement of the astrophysical factor value compared with the classical result.

\section{Wave function of relative movement of two protons for solar proton burning}
\setcounter{equation}{0}

For the description of the relative movement of the protons for the process of the solar proton burning we need
the wave
functions $\psi_{\rm pp}(\vec{\rho}\,)_{\rm in}$
and $\psi_{\rm pp}(\vec{\rho}\,)_{\rm out}$ of two protons in the initial
and the
final (inside the one--nucleon loop) scattering states, respectively. Following Ref.~[8] these wave
functions are
normalized per unit density at infinity. In the RFMD the amplitude of the solar proton burning contains the vertex of the transition p + p $\to$ p + p defined by the integral
\begin{eqnarray}\label{label3.1}
\int d^3\rho\,\psi^*_{\rm pp}(\vec{\rho}\,)_{\rm out}\,{\displaystyle
\frac{M^2_{\pi}}{4\pi}\,\frac{e^{\displaystyle - M_{\pi}\rho}}{\rho}}\,
\psi_{\rm pp}(\vec{\rho}\,)_{\rm in}. 
\end{eqnarray}
For the construction of the wave function $\psi_{\rm pp}(\vec{\rho}\,)_{\rm in}$
we have to take into account the Coulomb repulsion between the protons. Since
in the
initial state the protons couple at low energies the wave
function $\psi_{\rm pp}(\vec{\rho}\,)_{\rm in}$ can be written in the form
[7--9]
\begin{eqnarray}\label{label3.2}
\psi_{\rm pp}(\vec{\rho}\,)_{\rm in} = C(\eta)\,\frac{- a^{\rm e}_{\rm
pp}}{\rho}\,\Phi(\rho),
\end{eqnarray}
where $C(\eta)$ is the Gamow penetration factor defined as
\begin{eqnarray}\label{label3.3}
 C(\eta) = \sqrt{\frac{\displaystyle 2\pi\eta }{\displaystyle e^{\displaystyle 2\pi\eta} - 1}} \to \sqrt{2\pi\eta}\,e^{\displaystyle - \pi\eta}
\end{eqnarray}
depending on the relative velocity of the protons $v$ as $\eta =\alpha/v$ at $v\to 0$ and  $\alpha=1/137$, the fine structure constant, then $a^{\rm e}_{\rm pp}= (-7.828\pm 0.008)\,{\rm fm}$ is the S--wave scattering length of the pp scattering in the ${^1}{\rm S}_0$--state accounting for the Coulomb repulsion [5]. The function $\Phi(\rho)$ is given by [7--9,31]
\begin{eqnarray}\label{label3.4}
\Phi(\rho) = 2\sqrt{x}\,K_1(2\sqrt{x}) - \frac{r_{\rm C}}{a^{\rm e}_{\rm pp}}\,
\sqrt{x}\,I_1(2\sqrt{x}),
\end{eqnarray}
where $x=\rho/r_{\rm C}$, $r_{\rm C} = 1/M_{\rm N}\alpha = 28.82\,{\rm fm}$,
and $K_1(2\sqrt{x})$ and $I_1(2\sqrt{x})$ are Modified Bessel functions related to the irregular and regular solutions of Schr\"odinger equation, respectively. 

Since, as it is shown below, in the one--nucleon loop diagram the main
contribution comes from high virtual momenta, the Coulomb repulsion between protons can be included perturbatively. In leading order the wave function $\psi_{\rm pp}(\vec{\rho}\,)_{\rm out}$ can be taken in the form of a plane wave, i.e.,  $\psi_{\rm pp}(\vec{\rho}\,)_{\rm out} = \exp(i\vec{k}\cdot \vec{\rho})$, where $\vec{k}$ is a relative momentum of the protons.

\section{Astrophysical factor for solar proton burning}
\setcounter{equation}{0}

In the RFMD the amplitude of the solar proton burning is defined by one--nucleon loop diagrams and reads [2,4] (see Appendix C):
\begin{eqnarray}\label{label4.1}
&&i{\cal M}({\rm p} + {\rm p} \to {\rm D} + {\rm e}^+ + \nu_{e}) = - \,C(\eta)
\,g_{\rm A} M_{\rm N} \frac{G_{\rm V}}{\sqrt{2}}\,\frac{3g_{\rm V}}{4\pi^2}\,
G_{\rm \pi pp}\,{\cal F}^{\rm e}_{\rm pp}\nonumber\\
&&\hspace{1in}\times\,e^*_{\mu}(k_{\rm D})\,[\bar{u}(k_{\nu_{\rm e}})\gamma^{\mu} (1-\gamma^5) v(k_{\rm e^+})]\,[\bar{u^c}(p_2) \gamma^5 u(p_1)],
\end{eqnarray}
where $G_{\rm V} = G_{\rm F}\cos \vartheta_{\rm C}$ with $G_{\rm F}=1.166\times
10^{-11}\,{\rm MeV}^{-2}$ and $\vartheta_{\rm C}$ are is the Fermi weak coupling constant and the Cabibbo angle $\cos \vartheta_{\rm C} = 0.975$. Then $g_{\rm A}=1.260\pm 0.012$ describes the renormalization of the weak axial hadron current by strong interactions [5], $g_{\rm V}$ is the effective coupling constant of the RFMD related to the electric quadrupole moment of the deuteron: $g^2_{\rm V} = 2\pi^2 Q_{\rm D} M^2_{\rm N}$ [2] with $Q_{\rm D} = 0.286\,{\rm fm}^2$ [5], $e^*_{\mu}(k_{\rm D})$ is a 4--vector of a polarization of the deuteron and $\bar{u}(k_{\nu_{\rm e}})$, $v(k_{\rm e^+})$, $\bar{u^c}(p_2)$ and $u(p_1)$ are the Dirac bispinors of the neutrino, the positron, and the protons, respectively. For the binding energy of the deuteron we use the value $\varepsilon_{\rm D}=2.225\,{\rm MeV}$ [5].

The detailed computation of the amplitude of the solar proton burning Eq.~(\ref{label4.1}) and the factor ${\cal F}^{\rm e}_{\rm pp}$ is given  in Appendix C. It is found that the factor ${\cal F}^{\rm e}_{\rm pp}$ (see Eq.~({\rm C}.62)) amounts to
\begin{eqnarray}\label{label4.2}
{\cal F}^{\rm e}_{\rm pp} &=&-\sqrt{2}\,a^{\rm e}_{\rm pp}\,\frac{28}{27}\,\Bigg[\frac{M^2_{\pi}}{\sqrt{M^2_{\rm N} - M^2_{\pi}}}\,{\rm arctg}\frac{\sqrt{M^2_{\rm N} - M^2_{\pi}}}{M_{\pi}} + \frac{8}{7}\,\frac{M^3_{\pi}}{M^2_{\rm N} - M^2_{\pi}}\nonumber\\
&&- \frac{8}{7}\,\frac{M^4_{\pi}}{(M^2_{\rm N} - M^2_{\pi})^{3/2}}\,{\rm arctg}\frac{\sqrt{M^2_{\rm N} - M^2_{\pi}}}{M_{\pi}}\Bigg] = 1.78.
\end{eqnarray}
The cross section for the low--energy p + p $\to$ D + ${\rm e}^+$ + $\nu_{\rm e}$ reaction is defined
\begin{eqnarray}\label{label4.3}
\hspace{-0.5in}&&\sigma({\rm pp} \to {\rm D e^+ \nu_{\rm e}}) = \frac{1}{v}\,\frac{1}{4E_1E_2}\int\,\overline{|{\cal M}({\rm p} + {\rm p} \to {\rm D} + {\rm e}^+ + \nu_{\rm e})|^2}\,\nonumber\\
\hspace{-0.5in}&&\times (2\pi)^4\,\delta^{(4)}(k_{\rm D} + k_{\rm e^+ } + k_{\nu_{\rm e}} - p_1 - p_2)\,\frac{d^3k_{\rm D}}{(2\pi)^3 2E_{\rm D}}\frac{d^3k_{\rm e^+}}{(2\pi)^3 2E_{\rm e^+}}\frac{d^3k_{\nu_{\rm e}}}{(2\pi)^3 2E_{\nu_{\rm e}}}\,,
\end{eqnarray}
where $v$ is a relative velocity of the protons and $E_i\,(i=1,2)$ are the energies of the protons in the center of mass frame. 

Then, $\overline{|{\cal M}({\rm p} + {\rm p} \to {\rm D} + {\rm e}^+ +\nu_{e})|^2}$ is the squared amplitude averaged over polarizations of protons and summed over polarizations of final particles:
\begin{eqnarray}\label{label4.4}
\hspace{-0.5in}&&\overline{|{\cal M}({\rm p} + {\rm p} \to {\rm D} + {\rm e}^+ + \nu_{e})|^2}= C^2(\eta)\,g^2_{\rm A}M^4_{\rm N}G^2_{\rm V}\frac{9Q_{\rm D}}{16\pi^2}\,G^2_{\rm \pi pp}\,|{\cal F}^{\rm e}_{\rm pp}|^2\,\times\,\Bigg(- g^{\alpha\beta}+\frac{k^{\alpha}_{\rm D}k^{\beta}_{\rm D}}{M^2_{\rm D}}\Bigg)\nonumber\\
\hspace{-0.5in}&&\times {\rm tr}\{(- m_{\rm e} + \hat{k}_{\rm e^+})\gamma_{\alpha}(1-\gamma^5) \hat{k}_{\nu_{\rm e}}\gamma_{\beta}(1-\gamma^5)\}\times \frac{1}{4}\times {\rm tr}\{(M_{\rm N} - \hat{p}_2) \gamma^5 (M_{\rm N} + \hat{p}_1) \gamma^5\},
\end{eqnarray}
where $m_{\rm e}=0.511\;{\rm MeV}$ is the mass of positron, and we have used the relation $g^2_{\rm V}/\pi^2 = 2\,Q_{\rm D}\,M^2_{\rm N}$.

The computation of the traces yields:
\begin{eqnarray}\label{label4.5}
\hspace{-0.5in}&&\Bigg(- g^{\alpha\beta}+\frac{k^{\alpha}_{\rm D}k^{\beta}_{\rm D}}{M^2_{\rm D}}\Bigg)\,
\times\,{\rm tr}
 \{( - m_{\rm e} + \hat{k}_{\rm e^+})
 \gamma_{\alpha}(1-\gamma^5) \hat{k}_{\nu_{\rm e}}\gamma_{\beta}(1-\gamma^5)\}= \nonumber\\
\hspace{-0.5in}&& = 24\,\Bigg( E_{\rm e^+} E_{\nu_{\rm e}} - \frac{1}{3}\vec{k}_{\rm e^+}\cdot \vec{k}_{\nu_{\rm e}}\,\Bigg) ,\nonumber\\
\hspace{-0.5in}&&\frac{1}{4}\,\times\,{\rm tr}\{(M_{\rm N} - \hat{p}_2) \gamma^5 (M_{\rm N} + \hat{p}_1) \gamma^5\} = 2\,M^2_{\rm N},
\end{eqnarray}
where we have neglected the relative kinetic energy of the protons with respect to the mass of the proton.

Substituting Eq.~(\ref{label4.5}) in Eq.~(\ref{label4.4}) we get
\begin{eqnarray}\label{label4.6}
\hspace{-0.5in}\overline{|{\cal M}({\rm p} + {\rm p} \to {\rm D} + {\rm e}^+ + \nu_{e})|^2} &=& C^2(\eta) g^2_{\rm A} M^6_{\rm N}\,G^2_{\rm V}\,G^2_{\rm \pi pp}\,|{\cal F}^{\rm e}_{\rm pp}|^2\,\frac{27 Q_{\rm D}}{\pi^2}\nonumber\\
\hspace{-0.5in}&&\times\,\Bigg( E_{\rm e^+} E_{\nu_{\rm e}} - \frac{1}{3}\vec{k}_{\rm e^+}\cdot \vec{k}_{\nu_{\rm e}}\Bigg).
\end{eqnarray}
The integration over the phase volume of the final ${\rm D}{\rm e}^+ \nu_{\rm e}$--state we perform in the non--relativistic limit
\begin{eqnarray}\label{label4.7}
\hspace{-0.5in}&&\int\frac{d^3k_{\rm D}}{(2\pi)^3 2E_{\rm D}}\frac{d^3k_{\rm e^+}}{(2\pi)^3 2E_{\rm e^+}}\frac{d^3k_{\nu_{\rm e}}}{(2\pi)^3 2 E_{\nu_{\rm e}}}\,(2\pi)^4\,\delta^{(4)}(k_{\rm D} + k_{\ell} - p_1 - p_2)\,\Bigg( E_{\rm e^+} E_{\nu_{\rm e}} - \frac{1}{3}\vec{k}_{\rm e^+}\cdot \vec{k}_{\nu_{\rm e}}\,\Bigg)\nonumber\\
\hspace{-0.5in}&&= \frac{1}{32\pi^3 M_{\rm N}}\,\int^{W + T_{\rm pp}}_{m_{\rm e}}\sqrt{E^2_{\rm e^+}-m^2_{\rm e}}\,E_{\rm e^+}(W + T_{\rm pp} - E_{\rm e^+})^2\,d E_{\rm e^+} = \frac{(W + T_{\rm pp})^5}{960\pi^3 M_{\rm N}}\,f(\xi),
\end{eqnarray}
where $W = \varepsilon_{\rm D} - (M_{\rm n} - M_{\rm p}) = (2.225 -1.293)\,{\rm MeV} = 0.932\,{\rm MeV}$, $T_{\rm pp} = M_{\rm N}\,v^2/4$ is the kinetic energy of the relative movement of the protons, and  $\xi=m_{\rm e}/(W + T_{\rm pp})$. The function $f(\xi)$ is defined by the integral
\begin{eqnarray}\label{label4.8}
\hspace{-0.5in}f(\xi)&=&30\,\int^1_{\xi}\sqrt{x^2 -\xi^2}\,x\,(1-x)^2 dx=(1 - \frac{9}{2}\,\xi^2 - 4\,\xi^4)\,\sqrt{1-\xi^2}\nonumber\\
\hspace{-0.5in}&&+ \frac{15}{2}\,\xi^4\,{\ell n}\Bigg(\frac{1+\sqrt{1-\xi^2}}{\xi}\Bigg)\Bigg|_{T_{\rm pp} = 0} =  0.222
\end{eqnarray}
and normalized to unity at $\xi=0$.

Thus, the cross section for the solar proton burning is given by
\begin{eqnarray}\label{label4.9}
\sigma({\rm pp} \to {\rm D e^+\nu_{\rm e}}) &=& \frac{e^{\displaystyle - 2\pi\eta}}{v^2}\,\alpha\,\frac{9g^2_{\rm A} G^2_{\rm V} Q_{\rm D} M^3_{\rm N}}{640\,\pi^4}\,G^2_{\rm \pi pp}\,|{\cal F}^{\rm e}_{\rm pp}|^2\,(W + T_{\rm pp})^5\,f\Bigg(\frac{m_{\rm e}}{W + T_{\rm pp}}\Bigg)=\nonumber\\
&=& \frac{S_{\rm pp}(T_{\rm pp})}{T_{\rm pp}}\,e^{\displaystyle - 2\pi\eta}.
\end{eqnarray}
For the astrophysical factor $S_{\rm pp}(T_{\rm pp})$ we obtain the following expression [2,4]:
\begin{eqnarray}\label{label4.10}
S_{\rm pp}(T_{\rm pp}) = \alpha\,\frac{9g^2_{\rm A}G^2_{\rm V}Q_{\rm
D}M^4_{\rm N}}
{2560\pi^4}\,G^2_{\rm \pi pp}\,|{\cal F}^{\rm e}_{\rm pp}|^2\,(W + T_{\rm pp})^5\,f\Bigg(\frac{m_{\rm
e}}{W + T_{\rm pp}}
\Bigg), 
\end{eqnarray}
At zero kinetic energy of the protons $T_{\rm pp} =0$ the astrophysical factor $S_{\rm pp}(0)$ reads
\begin{eqnarray}\label{label4.11}
S_{\rm pp}(0) &=&\alpha\,\frac{9g^2_{\rm A}G^2_{\rm V}Q_{\rm
D}M^4_{\rm N}}{2560\pi^4}\,G^2_{\rm \pi pp}\,|{\cal
F}^{\rm e}_{\rm pp}|^2\,W^5\,f\Bigg(\frac{m_{\rm
e}}{W}\Bigg) = \nonumber\\
&=& 5.52\,\times 10^{-25}\,{\rm MeV\,b}.
\end{eqnarray}
The value $S_{\rm pp}(0) = 5.52\times 10^{-25}\,{\rm MeV\,\rm b}$ is enhanced by a factor of $1.42$ with respect to the classical value $S^*_{\rm pp}(0) = 3.89\,(1 \pm 0.011)\times 10^{-25}\,{\rm MeV\,\rm b}$ obtained by Kamionkowski and Bahcall in the PMA [7].

\section{Astrophysical consequences}
\setcounter{equation}{0}

The solar luminosity $L_{\odot} = (3.846\pm 0.008)\times 10^{26}\,{\rm W}$ is normalized in the Standard Solar Model (SSM) [32] to the reaction rate for the solar proton burning calculated in the PMA [7]. This defines the temperature in the solar core equal to  $T_{\rm c} = 15.6 \times 10^6$\,K [32].  By virtue of the enhancement factor $1.42\,(1\pm 19\%)$\footnote{Here 19$\%$ is the assumed theoretical uncertainty of the RFMD (see Appendix A). The real theoretical uncertainty of the approach can turn out to be much less.} the temperature in the solar core has to be reduced from the standard value of $T_{\rm c} = 15.6 \times 10^6$\,K to $T_{\rm c} = 15.1^{-0.25}_{+0.44} \times 10^6$\,K. This leads to the decrease of the solar neutrino fluxes relative to the solar neutrino fluxes calculated in the SSM [33]. For example, the solar ${^8}{\rm B}$ neutrino flux becomes equal [34]: $\Phi_{\rm B}[10^6\;{\rm cm}^{-2}{\rm s}^{-1}] = 2.00^{-0.75}_{+1.53}$ which agrees reasonably well with the experimental data by (SUPER)KAMIOKANDE [35]: ${\rm S}_{\rm KAM}(E_{\nu}\ge 7\;{\rm MeV}) = \Phi_{\rm B}[10^6\;{\rm cm}^{-2}{\rm s}^{-1}]=2.44\pm 0.05^{+ 0.09}_{-0.07} \pm 0.18$. The estimates of the solar neutrino fluxes observed by the Gallium [36] and the  Chlorine detectors [37] are adduced in Table~1.
\vspace{0.2in}

\noindent Table 1.  Contributions from the main components of the neutrino fluxes to the signals (SNU) in the Gallium [36] and Chlorine [37] experiments according to the SSM [33] and our approach. The errors are due to the assumed 30\,\% uncertainty of the reaction rate for p + p $\to$ D + $e^{\,+}$ + $\nu_{\rm e}$.
The power--law parameters $\alpha_i$ have been taken from Table X of Ref.~[34].
\vspace{0.2in}

\begin{tabular}{|c||c||c|c||c||c|c||c|  }\hline
 & \multicolumn{3}{c||}{S$_{\rm Ga}(E_{\nu}\ge 0.233\;{\rm MeV})$}& \multicolumn{3}{c||}{S$_{\rm Cl}(E_{\nu}\ge 0.814\;{\rm MeV})$}& \\ \cline{2-7}
& SSM & RFMD& experiment & SSM & RFMD & experiment & \raisebox{1.5ex}[-1.5ex]{$\alpha_i$}  \\ \hline
pp & 69.6 & 71.3$^{+}_{-1.0}$ &  & 0.00 & 0.00 & &0.07 \\
pep & 2.8 & 2.9$^{+0.0}_{-0.1}$ &  & 0.20 & 0.21$^{+0.00}_{-0.01}$ & & 0.07 \\
$^7{\rm Be}$ & 34.4 & 23.4$^{-4.1}_{+6.1}$ &  & 1.15 & 0.78$^{-0.13}_{+0.21}$ & &$-1.1$ \\
$^{8}{\rm B}$ & 12.4 & 4.8$^{-1.8}_{+3.7}$ &  & 5.90 &  2.29$^{-0.86}_{+1.76}$ & &$-2.7$ \\ 
$^{13}{\rm N}$ & 3.7 & 1.7$^{-0.5}_{+1.0}$ &  & 0.10 & 0.05$^{-0.02}_{+0.02}$ & &$-2.2$ \\
$^{15}{\rm O}$ & 6.0 & 2.8$^{-0.9}_{+1.6}$ &  & 0.40 & 0.19$^{-0.06}_{+0.10}$ & &$-2.2$ \\ \hline
 & $129^{+8}_{-6}$ & 107$^{-9}_{+13}$ & $77.5 \pm 6.2$ & $7.7^{+1.2}_{-1.0}$ & 3.52$^{-1.07}_{+2.08}$ & $2.56\pm 0.16$ &\\ \hline
\end{tabular}\\
\vspace{0.2in}

Thus, one can note that by using the mean value of the astrophysical factor $S_{\rm pp}(0)$ obtained in the RFMD, $S_{\rm pp}(0)/S^*_{\rm pp}(0) = 1.42$, the discrepancies of the Solar Neutrino Problem [38] can be relaxed considerably. 

However, an enhancement of the astrophysical factor $S_{\rm pp}(0)$ may be restricted by the data on helioseismology [39], which presently admit  deviations from the classical value $S^*_{\rm pp}(0) = 3.89\times 10^{-25}\,{\rm MeV\,\rm b}$ [7] not more than 20$\%$ of the magnitude.

\section{Cross section for $\nu_{\rm e}$ + D $\to$ ${\rm e}^-$ + p + p}
\setcounter{equation}{0}

The calculation of the amplitude of the neutrino disintegration of the
deuteron $\nu_{\rm e}$ + D $\to$ ${\rm e}^-$ + p + p can be performed by
analogy with the amplitude of the solar proton burning. The details of the calculation  are given in Appendix D. To the description of the low--energy transition p + p $\to$ p + p we apply the effective four--proton interaction Eq.~(\ref{label1.7})
\begin{eqnarray}\label{label6.1}
&&{\cal L}^{\rm pp \to pp}_{\rm eff}(x) = \frac{1}{2}\,G_{\rm \pi
pp }\,\int d^3\rho\,\frac{M^2_{\pi}}{4\pi}\,{\displaystyle
\frac{\displaystyle e^{\displaystyle - M_{\pi}\rho}}{\rho}}\nonumber\\
&&\times \{[\bar{p}(t,\vec{x} - \frac{1}{2}\vec{\rho}\,)\gamma_{\mu}
\gamma^5 p^c(t,\vec{x} + \frac{1}{2}\vec{\rho}\,)]
[\bar{p^c}(t,\vec{x} - \frac{1}{2}\vec{\rho}\,)\gamma^{\mu}\gamma^5
p(t,\vec{x} + \frac{1}{2}\vec{\rho}\,)] \nonumber\\
&&+ (\gamma_{\mu}\gamma^5 \otimes \gamma^{\mu}\gamma^5 \to \gamma^5 \otimes
\gamma^5)\}.
\end{eqnarray}
The amplitude of transition p + p $\to$ p + p enters into the amplitude of the process $\nu_{\rm e}$ + D $\to$ ${\rm e}^-$ + p + p in the form
\begin{eqnarray}\label{label6.2}
\int d^3\rho\,\psi^*_{\rm pp}(\vec{\rho}\,)_{\rm out}\,{\displaystyle
\frac{M^2_{\pi}}{4\pi}\,\frac{e^{\displaystyle -
M_{\pi}\rho}}{\rho}}\,\psi_{\rm pp}(\vec{\rho}\,)_{\rm in}\quad.
\end{eqnarray}
The wave function of the relative movement of the protons inside the nucleon loop $\psi_{\rm pp}(\vec{\rho}\,)_{\rm in}$ can be taken in the form of the plane wave. This is due to that the main contribution to the momentum integrals defining the one--nucleon loops comes from high virtual momenta and, inside the one--nucleon loop, the Coulomb repulsion between protons can be described perturbatively. 

The wave function $\psi_{\rm pp}(\vec{\rho}\,)_{\rm out}$ describes the
relative movement of the protons in the final state. Since it is the ${^1}{\rm S}_0$--state, the wave function $\psi_{\rm
pp}(\vec{\rho}\,)_{\rm out}$ should read [7--9,31]
\begin{eqnarray}\label{label6.3}
\psi_{\rm pp}(\vec{\rho}\,)_{\rm out} = {\displaystyle e^{\displaystyle
i\delta^{\rm e}_{\rm pp}(k)}\cos\delta^{\rm e}_{\rm pp}(k)\,[F_0(k\rho) +
{\rm tg}\delta^{\rm e}_{\rm pp}(k)\,G_0(k\rho)]\,\frac{1}{k\rho}},
\end{eqnarray}
where $k = \sqrt{M_{\rm N}T_{\rm pp}}$ is the relative momentum and $T_{\rm pp}$ is the kinetic energy of the protons, the Coulomb wave functions $F_0(k\rho)$ and $G_0(k\rho)$ are regular and irregular at $\rho\to 0$ [7--9,31]. At $k \to 0$ the wave function Eq.~(\ref{label6.3}) takes the form of Eq.~(\ref{label3.2}).

The wave function $\psi_{\rm pp}(\vec{\rho}\,)_{\rm out}$ describes
properly in terms of the Coulomb wave functions $F_0(k\rho)$ and $G_0(k\rho)$ the relative movement of the protons at distances $\rho > r_{\rm NF}$, where $r_{\rm NF}\sim 1/M_{\pi}$ is a radius of nuclear forces [6,40]. The contribution of distances $0 \le \rho \le r_{\rm NF}$ is represented in terms of the  phase shift $\delta^{\rm e}_{\rm pp}(k)$ depending on the phenomenological parameters such as the S--wave scattering length $a^{\rm e}_{\rm pp}= (- 7.828\pm 0.008)\,{\rm fm}$ and the effective range $r^{\rm e}_{\rm pp}= (2.80\pm 0.02)\,{\rm fm}$ of the low--energy elastic pp scattering including a Coulomb repulsion [5]. Since strong low--energy interactions of the protons coupled to the deuteron are described in the RFMD through the one--nucleon loop exchanges taking into account the contribution of distances $\rho \le r_{\rm NF}$, one does not need to get more detailed information about short--distance behaviour of the wave  function of the protons than that included in the phase shift $\delta^{\rm e}_{\rm pp}(k)$.

Up to the energies $T_{\rm pp} \le 10\,{\rm MeV}$ the phase shift
$\delta^{\rm e}_{\rm pp}(k)$ satisfies the relation [40]
\begin{eqnarray}\label{label6.4}
{\rm ctg}\delta^{\rm e}_{\rm pp}(k) = \frac{1}{\displaystyle
C^2(k)\,k}\,\Bigg[ - \frac{1}{a^{\rm e}_{\rm pp}} + \frac{1}{2}\,r^{\rm
e}_{\rm pp}k^2 + \frac{1}{r_{\rm C}}\,h(2 k r_{\rm C})\Bigg],
\end{eqnarray}
where $r_{\rm C}= 1/M_{\rm N}\alpha=28.82\,{\rm fm}$ with a fine structure constant $\alpha = 1/137$. The Gamow penetration factor $C(k)$ is given by [31]
\begin{eqnarray}\label{label6.5}
C^2(k) = {\displaystyle \frac{\pi}{\displaystyle k r_{\rm
C}}\frac{1}{\displaystyle e^{\displaystyle \pi/k r_{\rm C}}-1}}
\end{eqnarray}
and
\begin{eqnarray}\label{label6.6}
h(2 k r_{\rm C}) = \sum^{\infty}_{n=1}\frac{1}{n(1 + 4 n^2 k^2 r^2_{\rm C})} - \gamma + {\ell n}(2 k r_{\rm C}),
\end{eqnarray}
where $\gamma=0.5772\ldots$ is Euler$^{\prime}$s constant.

As has been shown in the Appendix C the resultant integrals over $\rho$, obtained after the calculation of the corresponding momentum integrals defining the one--nucleon loop diagrams, are concentrated in the region $0\le \rho \le 1/M_{\rm N}$. Therefore, the wave functions $F_0(k\rho)$ and $G_0(k\rho)$ can be taken at $\rho=0$. This gives $F_0(0) = 0$ and $G_0(0) = C^{-1}(k)$, respectively [31]. Thus, the wave function $\psi_{\rm pp}(\vec{\rho}\,)_{\rm out}$ reduces to the form
\begin{eqnarray}\label{label6.7}
\psi_{\rm pp}(\vec{\rho}\,)_{\rm out} = {\displaystyle e^{\displaystyle
i\delta^{\rm e}_{\rm pp}(k)}\frac{\displaystyle \sin\delta^{\rm e}_{\rm
pp}(k)}{\displaystyle C(k) k}\,\frac{1}{\rho}}.
\end{eqnarray}
The amplitude of the process $\nu_{\rm e}$ + D $\to$ ${\rm e}^-$ + p + p is calculated in Appendix D. Here we adduce the result
\begin{eqnarray}\label{label6.8}
&&i{\cal M}(\nu_{\rm e} + {\rm D} \to {\rm
e}^- + {\rm p} + {\rm p}) = -\,g_{\rm A} M_{\rm N} \frac{G_{\rm
V}}{\sqrt{2}}\,\frac{3g_{\rm V}}{2\pi^2}\,
G_{\rm \pi pp}\,{\cal F}^{\rm e}_{\rm ppe^-}\nonumber\\
&&\times e_{\mu}(k_{\rm D})\,[\bar{u}(k_{\rm e^-})\gamma^{\mu}(1-\gamma^5)
u(k_{\nu_{\rm e}})]\,[\bar{u}(p_1) \gamma^5 u^c(p_2)]\,{\displaystyle
e^{\displaystyle i\delta^{\rm e}_{\rm pp}(k)}\frac{\displaystyle
\sin\delta^{\rm e}_{\rm pp}(k)}{\displaystyle a^{\rm e}_{\rm pp} k C(k)}},
\end{eqnarray}
where $\bar{u}(k_{\rm e^-})$, $u(k_{\nu_{\rm e}})$, $\bar{u}(p_1)$
and $u^c(p_2)$ are the Dirac bispinors of the electron, the neutrino and the protons, $e_{\mu}(k_{\rm D})$ is the 4--vector of the polarization of the deuteron. We have taken into account that $\bar{u}(p_2) \gamma^5 u^c(p_1) = -\bar{u}(p_1) \gamma^5 u^c(p_2)$. The factor ${\cal F}^{\rm e}_{\rm ppe^-} =1.70$ is defined by Eq.~({\rm D}.34).

The amplitude Eq.~(\ref{label6.8}) squared, averaged over polarizations of the deuteron and summed over polarizations of the
final particles reads
\begin{eqnarray}\label{label6.9}
&&\overline{|{\cal M}(\nu_{\rm e} + {\rm D} \to {\rm e}^- + {\rm p} + {\rm
p})|^2} = g^2_{\rm A}M^6_{\rm N}\frac{144 G^2_{\rm V}Q_{\rm
D}}{\pi^2}\,G^2_{\rm \pi pp}\,|{\cal F}^{\rm e}_{\rm ppe^-}|^2\,\nonumber\\
&&\times {\displaystyle \frac{\displaystyle C^2(k)}{\displaystyle (a^{\rm
e}_{\rm pp})^2k^2 C^4(k) + \Bigg[1 - \frac{1}{2}a^{\rm e}_{\rm pp} r^{\rm
e}_{\rm pp} k^2 + \frac{a^{\rm e}_{\rm pp}}{r_{\rm C}}\,h(2kr_{\rm
C})\Bigg]^2}}
\Bigg( E_{{\rm e^-}}E_{{\nu}_{\rm e}} - \frac{1}{3}\vec{k}_{{\rm e^-}}\cdot
\vec{k}_{{\nu}_{\rm e}}\Bigg).
\end{eqnarray}
Following the prescription [4] the amplitude Eq.~(\ref{label6.9}) should be extrapolated for the neutrino energies far from threshold. According to the procedure suggested in [4] we obtain
\begin{eqnarray}\label{label6.10}
&&\overline{|{\cal M}(\nu_{\rm e} + {\rm D} \to {\rm e}^- + {\rm p} + {\rm
p})|^2} = g^2_{\rm A}M^6_{\rm N}\frac{144 G^2_{\rm V}Q_{\rm
D}}{\pi^2}\,G^2_{\rm \pi pp}\,|{\cal F}^{\rm e}_{\rm ppe^-}|^2\,F_{\rm D}(k^2)\,F(Z, E_{\rm
e^-})\nonumber\\
&&\times {\displaystyle \frac{\displaystyle C^2(k)}{\displaystyle (a^{\rm
e}_{\rm pp})^2k^2 C^4(k) + \Bigg[1 - \frac{1}{2}a^{\rm e}_{\rm pp} r^{\rm
e}_{\rm pp} k^2 + \frac{a^{\rm e}_{\rm pp}}{r_{\rm C}}\,h(2kr_{\rm
C})\Bigg]^2}}
\Bigg( E_{{\rm e^-}}E_{{\nu}_{\rm e}} - \frac{1}{3}\vec{k}_{{\rm e^-}}\cdot
\vec{k}_{{\nu}_{\rm e}}\Bigg).
\end{eqnarray}
The form factor $F_{\rm D}(k^2)$, describing a spatial smearing of the deuteron, is defined as [4]
\begin{eqnarray}\label{label6.11}
F_{\rm D}(k^2) = \frac{1}{1 + r^2_{\rm D}k^2}\,,
\end{eqnarray}
where $r_{\rm D}= 1/\sqrt{M_{\rm N}\varepsilon_{\rm D}} = 4.315\,{\rm fm}$ is the radius of the deuteron [5]. We do need to include a form factor for a spatial smearing of the pp system [4], since such a smearing has been included in terms of the wave function Eq.~(\ref{label6.7}). Then, $F(Z, E_{\rm e^-})$ is well--known Fermi function [41] describing the Coulomb interaction of the electron with the nuclear system having the charge $Z$. In the case of the reaction $\nu_{\rm e}$ + D $\to$ e$^-$ + p + p we have $Z=2$. At $\alpha^2 Z^2 \ll 1$ the Fermi function $F(Z, E_{\rm e^-})$ reads [41]:
\begin{eqnarray}\label{label6.12}
F(Z, E_{\rm e^-}) = {\displaystyle \frac{2\pi\eta_{\rm e^-}}{\displaystyle 1 -
e^{\displaystyle -2\pi\eta_{\rm e^-}}}},
\end{eqnarray}
where $\eta_{\rm e^-}=Z\alpha/v_{\rm e^-} = Z\alpha\,E_{\rm e}/\sqrt{E^2_{\rm e^-} - m^2_{\rm e}}$ and $v_{\rm e^-} = \sqrt{E^2_{\rm e^-} - m^2_{\rm e}}/E_{\rm e^-}$ is a velocity of the electron.

The r.h.s. of Eq.~(\ref{label6.10}) can be expressed in terms of the
astrophysical factor $S_{\rm pp}(0)$ for the solar proton burning defined as
\begin{eqnarray}\label{label6.13}
\hspace{-0.5in}&&\overline{|{\cal M}(\nu_{\rm e} + {\rm D} \to {\rm e}^- + {\rm p} + {\rm
p})|^2} = S_{\rm pp}(0)\,\frac{2^{13}5\pi^2}{\Omega_{\rm D e^+ \nu_{\rm
e}}}\,\frac{r_{\rm C}M^3_{\rm N}}{m^5_{\rm e}}\,\frac{|{\cal F}^{\rm e}_{\rm ppe^-}|^2}{|{\cal F}^{\rm e}_{\rm pp}|^2}\,F_{\rm D}(k^2)\,F(Z, E_{\rm e^-})\nonumber\\
\hspace{-0.5in}&&\times {\displaystyle \frac{\displaystyle C^2(k)}{\displaystyle (a^{\rm
e}_{\rm pp})^2k^2 C^4(k) + \Bigg[1 - \frac{1}{2}a^{\rm e}_{\rm pp} r^{\rm e}_{\rm pp} k^2 + \frac{a^{\rm e}_{\rm pp}}{r_{\rm C}}\,h(2kr_{\rm C})\Bigg]^2}}
\Bigg( E_{{\rm e^-}}E_{{\nu}_{\rm e}} - \frac{1}{3}\vec{k}_{{\rm e^-}} \cdot
\vec{k}_{{\nu}_{\rm e}}\Bigg). 
\end{eqnarray}
We have used here the expression for the astrophysical factor 
\begin{eqnarray}\label{label6.14}
S_{\rm pp}(0) = |{\cal F}^{\rm e}_{\rm pp}|^2\,\frac{9g^2_{\rm A}G^2_{\rm V}Q_{\rm
D}M^3_{\rm N}}{2560\pi^4r_{\rm C}}\,G^2_{\rm \pi pp}\,m^5_{\rm
e}\,\Omega_{\rm D e^+ \nu_{\rm e}},
\end{eqnarray}
where $m_{\rm e}=0.511\,{\rm MeV}$ is the electron mass, and $\Omega_{\rm D e^+ \nu_{\rm e}} = (W/m_{\rm e})^5 f(m_{\rm e}/W) = 4.481$ at $W=0.932\,{\rm MeV}$ [2]. The function $f(m_{\rm e}/W)$ is defined by Eq.~(\ref{label4.8}). 

In the rest frame of the deuteron the cross section for the process $\nu_{\rm e}$ + D $\to$ ${\rm e}^-$ + p + p is defined as
\begin{eqnarray}\label{label6.15}
&&\sigma^{\nu_{\rm e} D}_{\rm cc}(E_{\nu_{\rm e}}) =
\frac{1}{4M_{\rm D}E_{\nu_{\rm e}}}\int\,\overline{|{\cal
M}(\nu_{\rm e} +
{\rm D} \to {\rm e}^- + {\rm p} + {\rm p})|^2}\nonumber\\
&&\frac{1}{2}\,(2\pi)^4\,\delta^{(4)}(k_{\rm D} + k_{\nu_{\rm e}} - p_1 -
p_2 - k_{\rm e^-})\,
\frac{d^3p_1}{(2\pi)^3 2E_1}\frac{d^3 p_2}{(2\pi)^3 2E_2}\frac{d^3k_{{\rm e^-}}}{(2\pi)^3 2E_{{\rm e^-}}},
\end{eqnarray}
where $E_{\nu_{\rm e}}$, $E_1$, $E_2$  and $E_{{\rm
e^-}}$ are the energies of the neutrino, the protons and the electron. The abbreviation (cc) means the charged current. The integration over the phase volume of the (${\rm p p e^-}$)--state we perform in the non--relativistic limit and in the rest frame of
the deuteron,
\begin{eqnarray}\label{label6.16}
&&\frac{1}{2}\,\int\frac{d^3p_1}{(2\pi)^3 2E_1}\frac{d^3p_2}{(2\pi)^3 2E_2}
\frac{d^3k_{\rm e}}{(2\pi)^3 2E_{\rm e^-}}(2\pi)^4\,\delta^{(4)}(k_{\rm D} +
k_{\nu_{\rm e}} - p_1 - p_2 - k_{\rm e^-})\,\nonumber\\
&&{\displaystyle \frac{\displaystyle C^2(\sqrt{M_{\rm N}T_{\rm pp}})\,F(Z,
E_{\rm e^-})}{\displaystyle (a^{\rm e}_{\rm pp})^2 M_{\rm N}T_{\rm pp}
C^4(\sqrt{M_{\rm N}T_{\rm pp}}) + \Bigg[1 - \frac{1}{2}a^{\rm e}_{\rm pp}
r^{\rm e}_{\rm pp}M_{\rm N}T_{\rm pp} + \frac{a^{\rm e}_{\rm pp}}{r_{\rm C}}\, h(2 r_{\rm C}\sqrt{M_{\rm N}T_{\rm pp}})\Bigg]^2}}\nonumber\\
&&\Bigg( E_{\rm e^-} E_{\nu_{\rm e}} - \frac{1}{3} \vec{k}_{\rm e^-} \cdot \vec{k}_{\nu_{\rm e}}\Bigg)\,F_{\rm D}(M_{\rm N}T_{\rm pp})
=\frac{E_{\bar{\nu}_{\rm e}}M^3_{\rm N}}{128\pi^3}\,\Bigg(\frac{E_{\rm
th}}{M_{\rm N}}
\Bigg)^{\!\!7/2}\Bigg(\frac{2 m_{\rm e}}{E_{\rm
th}}\Bigg)^{\!\!3/2}\frac{1}{E^2_{\rm th}}\nonumber\\
&&\int\!\!\!\int dT_{\rm e^-} dT_{\rm pp}\delta(E_{\nu_{\rm e}}- E_{\rm th} -
T_{\rm e^-} - T_{\rm pp}) \sqrt{T_{\rm e^-}T_{\rm pp}}\Bigg(1 + \frac{T_{\rm e^-}}{m_{\rm e}}\Bigg)\,{\displaystyle \sqrt{1 + \frac{T_{\rm e^-}}{2 m_{\rm e}}}}\nonumber\\
&&{\displaystyle \frac{\displaystyle C^2(\sqrt{M_{\rm N}T_{\rm pp}})
\,F_{\rm D}(M_{\rm N}T_{\rm pp})\,F(Z, E_{\rm e^-})}{\displaystyle (a^{\rm
e}_{\rm pp})^2 M_{\rm N}T_{\rm pp} C^4(\sqrt{M_{\rm N}T_{\rm pp}}) +
\Bigg[1 - \frac{1}{2}a^{\rm e}_{\rm pp} r^{\rm e}_{\rm pp}M_{\rm N}T_{\rm
pp} + \frac{a^{\rm e}_{\rm pp}}{r_{\rm C}}\,h(2 r_{\rm C}\sqrt{M_{\rm
N}T_{\rm pp}})\Bigg]^2}} \nonumber\\
&&= \frac{E_{\bar{\nu}_{\rm e}}M^3_{\rm N}}{128\pi^3} \,\Bigg(\frac{E_{\rm
th}}{M_{\rm N}} \Bigg)^{\!\!7/2}\Bigg(\frac{2 m_{\rm e}}{E_{\rm
th}}\Bigg)^{\!\!3/2}\,(y-1)^2\,\Omega_{\rm p p e^-}(y),
\end{eqnarray}
where $T_{\rm e^-}$ is the kinetic energy of the electron, $E_{\rm th}$ is
the neutrino energy threshold of the reaction $\nu_{\rm e}$ + D $\to$ ${\rm
e}^-$ + p + p, and is given by $E_{\rm th}= \varepsilon_{\rm D} + m_{\rm e}
- (M_{\rm n} - M_{\rm p}) = (2.225 + 0.511 - 1.293) \, {\rm MeV} =
1.443\,{\rm MeV}$. The function $\Omega_{\rm p p e^-}(y)$, where
$y=E_{\nu_{\rm
e}}/E_{\rm th}$, is defined as
\begin{eqnarray}\label{label6.17}
&&\Omega_{\rm p p e^-}(y) = \int\limits^{1}_{0} dx \sqrt{x (1 - x)} \Bigg(1
+ \frac{E_{\rm th}}{m_{\rm e}}(y-1)(1-x)\Bigg) \sqrt{1 + \frac{E_{\rm
th}}{2 m_{\rm e}}(y-1)(1-x)}\nonumber\\
&&C^2(\sqrt{M_{\rm N}E_{\rm th}\,(y - 1)\,x}) \,F_{\rm D}(M_{\rm N}E_{\rm
th}\,(y - 1)\,x)\,F(Z,m_{\rm e} + E_{\rm th}(y - 1)\,(1-x))\nonumber\\
&&\Bigg\{(a^{\rm e}_{\rm pp})^2\,M_{\rm N}E_{\rm th}\,(y - 1)\,x
C^4(\sqrt{M_{\rm N}E_{\rm th}\,(y - 1)\,x}) + \nonumber\\
&&\Bigg[1 - \frac{1}{2}a^{\rm e}_{\rm pp} r^{\rm e}_{\rm pp}M_{\rm N}E_{\rm
th}\,(y - 1)\,x + \frac{a^{\rm e}_{\rm pp}}{r_{\rm C}}\,h(2 r_{\rm
C}\sqrt{M_{\rm N}E_{\rm th}\,(y - 1)\,x})\Bigg]^2\Bigg\}^{-1},
\end{eqnarray}
where we have changed the variable $T_{\rm pp} = (E_{\nu_{\rm e}} -
E_{\rm th})\,x$.

The cross section for $\nu_{\rm e}$ + D $\to$ ${\rm e}^-$ + p + p is defined
\begin{eqnarray}\label{label6.18}
\hspace{-0.5in}\sigma^{\nu_{\rm e} D}_{\rm cc}(E_{\nu_{\rm e}}) &=& S_{\rm pp}(0)\,\frac{|{\cal F}^{\rm e}_{\rm ppe^-}|^2}{|{\cal F}^{\rm e}_{\rm pp}|^2}\,
\frac{1280\,r_{\rm C}}{\pi \Omega_{\rm D e^+\nu_{\rm e}}}\Bigg(\frac{M_{\rm
N}}{E_{\rm th}}\Bigg)^{3/2}\Bigg(\frac{E_{\rm th}}{2m_{\rm
e}}\Bigg)^{7/2}\,(y-1)^2\,\Omega_{\rm p p e^-}(y)=\nonumber\\
&=&6.74\times10^5\,S_{\rm pp}(0)\,(y-1)^2\,\Omega_{\rm p p e^-}(y),
\end{eqnarray}
where $S_{\rm pp}(0)$ is measured in ${\rm MeV}\,{\rm cm}^2$. For $S_{\rm pp}(0) = 5.52\times 10^{-49}\,{\rm MeV}\,{\rm cm}^2$ Eq.~(\ref{label4.11}) the cross section
$\sigma^{\nu_{\rm e} D}_{\rm cc}(E_{\nu_{\rm e}})$ reads
\begin{eqnarray}\label{label6.19}
\sigma^{\nu_{\rm e} D}_{\rm cc}(E_{\nu_{\rm e}}) =
3.72\,(y-1)^2\,\Omega_{\rm p p e^-}(y)\,10^{-43}\,{\rm cm}^2.
\end{eqnarray}
The cross section Eq.~(\ref{label6.19}) should be compared with the cross section calculated in the PMA. The most recent PMA data on the cross section for the process $\nu_{\rm e}$ + D $\to$ e$^-$ + p + p have been obtained in Refs.~[42,43] and tabulated for the neutrino energies ranging the values from threshold up to 160$\,{\rm MeV}$. Since our result is restricted by the neutrino energies from threshold up to 10$\,{\rm MeV}$, we compute the cross section for $E_{\nu_{\rm
e}} = 3.25\,{\rm MeV}$ and $E_{\nu_{\rm e}} = 10\,{\rm MeV}$ and get
\begin{eqnarray}\label{label6.20}
\sigma^{\nu_{\rm e} D}_{\rm cc}(E_{\nu_{\rm e}})\Big|_{E_{\nu_{\rm
e}}=3.25\,{\rm MeV}} &=&  3.72\,\times 10^{-43}\,(y-1)^2\,\Omega_{\rm p p
e^-}(y)\Big|_{E_{\nu_{\rm e}}=3.25\,{\rm MeV}}\,{\rm cm}^2 =\nonumber\\
&=& 4.03\times 10^{-43}\,{\rm cm}^2,\nonumber\\
\sigma^{\nu_{\rm e} D}_{\rm cc}(E_{\nu_{\rm e}})\Big|_{E_{\nu_{\rm
e}}=10\,{\rm MeV}} &=&  3.72\,\times 10^{-43}\,(y-1)^2\,\Omega_{\rm p p
e^-}(y)\Big|_{E_{\nu_{\rm e}}=10\,{\rm MeV}}\,{\rm cm}^2 =\nonumber\\
&=&1.91\times 10^{-41}\,{\rm cm}^2,
\end{eqnarray}
The PMA data read [42,43]:
\begin{eqnarray}\label{label6.21}
\sigma^{\nu_{\rm e} D}_{\rm cc}(E_{\nu_{\rm e}})\Big|_{E_{\nu_{\rm e}} =
3.25\,{\rm MeV}} &=& 6.46\times 10^{-44}\,{\rm cm}^2,\nonumber\\
\sigma^{\nu_{\rm e} D}_{\rm cc}(E_{\nu_{\rm e}})\Big|_{E_{\nu_{\rm e}} =
10\,{\rm MeV}} &=& 2.55\times 10^{-42}\,{\rm cm}^2,
\end{eqnarray}
When matching the values we find a distinction between the predictions by a factor of 7 in average. Such a discrepancy should be a challenge to solar neutrino experiments planned by SNO [14].

In Conclusion we discuss the experimental data on the process $\nu_{\rm e}$ + D $\to$ e$^-$ + p + p induced by neutrinos of the $\mu$--meson decays and give a fit of the cross section for the process $\nu_{\rm e}$ + D $\to$ e$^-$ + p + p calculated in the PMA [42] for neutrino energies ranging the values $10\,{\rm MeV} \le E_{\nu_{\rm e}} \le 160\,{\rm MeV}$.

\section{Astrophysical factor for pep process}
\setcounter{equation}{0}

The cross section for the neutrino disintegration of
the deuteron $\nu_{\rm e}$ + D $\to$ e$^-$ + p + p calculated in Sect.~6 can be applied to the
computation of the astrophysical factor $S_{\rm pep}(0)$ for the process p + e$^-$ + p $\to$ D + $\nu_{\rm e}$ or the pep--process. In the RFMD the squared amplitudes of the processes p +  e$^-$ + p $\to$ D + $\nu_{\rm e}$ and $\nu_{\rm e}$ + D $\to$ e$^-$ + p + p, averaged and summed over polarization of interacting particles,  are related by
\begin{eqnarray}\label{label7.1}
\overline{|{\cal M}({\rm p} + {\rm e}^- + {\rm p} \to {\rm D} + \nu_{\rm
e})|^2} = \frac{3}{8}\,\frac{|{\cal F}^{\rm e}_{\rm pp}|^2}{|{\cal F}^{\rm e}_{\rm ppe^-}|^2}\,\overline{|{\cal M}(\nu_{\rm e} + {\rm D} \to {\rm
e}^- + {\rm p} + {\rm p})|^2}.
\end{eqnarray}
At low energies the cross section $\sigma_{\rm pep}(T_{\rm pp})$ for the pep--process can be defined as follows [31]:
\begin{eqnarray}\label{label7.2}
&&\sigma_{\rm pep}(T_{\rm pp}) = \frac{1}{v}\frac{1}{4M^2_{\rm N}}\int
\frac{d^3k_{\rm e^-}}{(2\pi)^3 2 E_{\rm e^-}}\,g\, n(\vec{k}_{\rm e^-})\int
\overline{|{\cal M}({\rm p} + {\rm e}^- + {\rm p} \to {\rm D} + \nu_{\rm
e})|^2}\nonumber\\
&&(2\pi)^4 \delta^{(4)}(k_{\rm D} + k_{\nu_{\rm e}} - p_1 - p_2 - k_{\rm e^-}) \frac{d^3k_{\rm D}}{(2\pi)^3 2M_{\rm D}}\frac{d^3k_{\nu_{\rm e}}}{(2\pi)^3 2E_{\nu_{\rm e}}},
\end{eqnarray}
where $g = 2$ is the number of the electron spin states and $v$ is a relative velocity of the protons. The electron
distribution function $n(\vec{k}_{\rm e^-})$ can be taken in the form  [41]
\begin{eqnarray}\label{label7.3}
n(\vec{k}_{\rm e^-}) = e^{\displaystyle \bar{\nu} - T_{\rm e^-}/kT_c},
\end{eqnarray}
where $k = 8.617\times 10^{-11}\,{\rm MeV\,K^{-1}}$, $T_c = 15.1\times
10^6\,{\rm K}$ is a temperature of the sun core (see Sect.~5). The distribution function $n(\vec{k}_{\rm e^-})$ is normalized by the condition
\begin{eqnarray}\label{label7.4}
g\int \frac{d^3k_{\rm e^-}}{(2\pi)^3}\,n(\vec{k}_{\rm e^-}) = n_{\rm e^-},
\end{eqnarray}
where $n_{\rm e^-}$ is the electron number density. From the normalization condition Eq.~(\ref{label7.4}) we obtain
\begin{eqnarray}\label{label7.5}
e^{\displaystyle \bar{\nu}} = \frac{\displaystyle  4\,\pi^3\, n_{\rm
e^-}}{\displaystyle (2\pi\,m_{\rm e}\,kT_c)^{3/2}}.
\end{eqnarray}
The astrophysical factor $S_{\rm pep}(0)$ is then defined
\begin{eqnarray}\label{label7.6}
S_{\rm pep}(0) = S_{\rm pp}(0)\,\frac{30}{\pi}\,\frac{1}{\Omega_{\rm D e^+
\nu_{\rm e}}}\,\frac{1}{m^3_{\rm e}}\,\Bigg(\frac{E_{\rm th}}{m_{\rm
e}}\Bigg)^2\,e^{\displaystyle \bar{\nu}}\,\int d^3k_{\rm e^-} \,e^{\displaystyle - T_{\rm e^-}/kT_c}\,F(Z, E_{\rm e^-}).
\end{eqnarray}
The r.h.s. of Eq.~(\ref{label7.6}) can be reduced to the form
\begin{eqnarray}\label{label7.7}
S_{\rm pep}(0) = \frac{2^{3/2}\pi^{5/2}}{f_{\rm pp}(0)}\,\Bigg(\frac{\alpha
Z n_{\rm e^-}}{m^3_{\rm e}}\Bigg)\,\Bigg(\frac{E_{\rm th}}{m_{\rm
e}}\Bigg)^2\,\sqrt{\frac{m_{\rm e}}{k T_c}}\,I\Bigg(Z\sqrt{\frac{2 m_{\rm e}}{k T_c}}\Bigg)\,S_{\rm pp}(0).
\end{eqnarray}
We have set $f_{\rm pp}(0) = \Omega_{\rm D e^+ \nu_{\rm e}}/30 = 0.149$ [41] and the function $I(x)$ having been introduced by Bahcall and May [41] reads
\begin{eqnarray}\label{label7.8}
I(x) = \int\limits^{\infty}_0 {\displaystyle \frac{\displaystyle
du\,e^{\displaystyle -u}}{\displaystyle 1 - e^{\displaystyle
-\pi\alpha\,x/\sqrt{u}}}}.
\end{eqnarray}
The relation between the astrophysical factors $S_{\rm pep}(0)$ and $S_{\rm pp}(0)$ given by Eq.~(\ref{label7.7}) is in complete agreement with that obtained by Bahcall and May [41].

\section{Cross section for $\bar{\nu}_{\rm e}$ + D $\to$ e$^+$ + n + n}
\setcounter{equation}{0}

In Ref.~[4] we have calculated the cross section for the disintegration of
the deuteron by reactor anti--neutrinos $\bar{\nu}_{\rm e}$ + D $\to$ e$^+$ + n + n. The strong low--energy transition n + n $\to$ n + n has been described by the local four--nucleon interaction Eq.~(\ref{label1.1}).  Since the investigation of the processes p + p $\to$ D + e$^+$ + $\nu_{\rm e}$ and
$\nu_{\rm e}$ + D $\to$ e$^-$ + p + p has demanded to involve a smeared potential instead of the $\delta$--function potential, we would like to revise here the result obtained in Ref.~[4] by applying the smeared four--neutron interaction (see Eq.~(\ref{label1.7})):
\begin{eqnarray}\label{label8.1}
&&{\cal L}^{\rm nn \to nn}_{\rm eff}(x) = \frac{1}{2}\,G_{\rm \pi
nn}\,\int d^3\rho\,\frac{M^2_{\pi}}{4\pi}\,{\displaystyle
\frac{\displaystyle e^{\displaystyle - M_{\pi}\rho}}{\rho}}\nonumber\\
&&\times \{[\bar{n}(t,\vec{x} - \frac{1}{2}\vec{\rho}\,)\gamma_{\mu}
\gamma^5 n^c(t,\vec{x} + \frac{1}{2}\vec{\rho}\,)]
[\bar{n^c}(t,\vec{x} - \frac{1}{2}\vec{\rho}\,)\gamma^{\mu}\gamma^5
n(t,\vec{x} + \frac{1}{2}\vec{\rho}\,)] \nonumber\\
&&+ (\gamma_{\mu}\gamma^5 \otimes \gamma^{\mu}\gamma^5 \to \gamma^5 \otimes
\gamma^5)\}.
\end{eqnarray}
The amplitude of the process $\bar{\nu}_{\rm e}$ + D $\to$ e$^+$ + n + n  contains the contribution of the n + n $\to$ n + n transition in the form of the vertex
\begin{eqnarray}\label{label8.2}
\int d^3\rho\,\psi^*_{\rm nn}(\vec{\rho}\,)_{\rm out}\,{\displaystyle
\frac{M^2_{\pi}}{4\pi}\,\frac{e^{\displaystyle -
M_{\pi}\rho}}{\rho}}\,\psi_{\rm nn}(\vec{\rho}\,)_{\rm in}.
\end{eqnarray}
The wave function $\psi_{\rm nn}(\vec{\rho}\,)_{\rm in}$ describes the relative movement of the neutrons in the one--nucleon loop diagrams and can be taken in the form of the plane wave. In turn the wave function $\psi_{\rm nn}(\vec{\rho}\,)_{\rm out}$ describes the relative movement of the neutrons due to strong interactions in the final state. As has been noted above,  the resultant integrals over $\rho$, derived after the calculation of the momentum integrals defining the one--nucleon loop diagrams, are concentrated in the region $0\le \rho \le 1/M_{\rm N}$, and only the irregular part of the wave function $\psi_{\rm nn}(\vec{\rho}\,)_{\rm out}$ gives a substantial contribution. Therefore, the wave function $\psi_{\rm nn}(\vec{\rho}\,)_{\rm out}$ can be written as follows
\begin{eqnarray}\label{label8.3}
\psi_{\rm nn}(\vec{\rho}\,)_{\rm out} = {\displaystyle e^{\displaystyle
i\delta_{\rm nn}(k)}\sin\delta_{\rm nn}(k)\,\frac{\displaystyle v_{\rm
nn}(k\rho)}{\displaystyle k\rho}}\,,
\end{eqnarray}
where $v_{\rm nn}(k\rho)$ is an irregular part of the wave function of strong low--energy nn interactions in the final state for distances $\rho \ge r_{\rm NF}$.

The amplitude of the process $\bar{\nu}_{\rm e}$ + D $\to$  e$^+$ + n + n can be computed by analogy with the amplitude of the process $\nu_{\rm e}$ + D $\to$ e$^-$ + p + p (see Appendix D) and reads
\begin{eqnarray}\label{label8.4}
&&i{\cal M}(\bar{\nu}_{\rm e} + {\rm D} \to {\rm e}^+ + {\rm n} + {\rm n} )
= - g_{\rm A} M_{\rm N} \frac{G_{\rm
V}}{\sqrt{2}}\,\frac{3g_{\rm V}}{2\pi^2}\,
G_{\rm \pi nn}\,e_{\mu}(Q)\,[\bar{v}(k_{\bar{\nu}_{\rm e}})\gamma^{\mu}(1-\gamma^5)
v(k_{{\rm e}^+})]\,\nonumber\\
&&\hspace{1in} \times \,[\bar{u}(p_1) \gamma^5 u^c(p_2)]\,\times\,{\cal
F}_{\rm nne^+}\,v_{\rm nn}(0)\,\times\,{\displaystyle e^{\displaystyle
i\delta_{\rm nn}(k)}\frac{\displaystyle \sin\delta_{\rm
nn}(k)}{\displaystyle a_{\rm nn}k}}\,,
\end{eqnarray}
where $\bar{v}(k_{\bar{\nu}_{\rm e}})$, $v(k_{{\rm e}^+})$, $\bar{u}(p_1)$ and $u^c(p_2)$ are the Dirac bispinors of the anti--neutrino, the positron and the neutrons, $e_{\mu}(Q)$ is the 4--vector
of the polarization of the deuteron. We have taken into account that
$\bar{u}(p_2)\gamma^5 u^c(p_1) = -\bar{u}(p_1) \gamma^5 u^c(p_2)$. The
factor ${\cal
F}_{\rm nne^+}$ can be obtained by analogy with the factor ${\cal F}^{\rm e}_{\rm ppe^-}$ (see Appendix D) and reads Eq.~({\rm D}.34):
\begin{eqnarray}\label{label8.5}
{\cal F}_{\rm nne^+} &=& - a_{\rm nn}\,\frac{44}{27}\,\Bigg[\frac{M^2_{\pi}}{\sqrt{M^2_{\rm N} - M^2_{\pi}}}\,{\rm arctg}\frac{\sqrt{M^2_{\rm N} - M^2_{\pi}}}{M_{\pi}} - \frac{8}{11}\,\frac{M^3_{\pi}}{M^2_{\rm N} - M^2_{\pi}}\nonumber\\
&&+ \frac{8}{11}\,\frac{M^4_{\pi}}{(M^2_{\rm N} - M^2_{\pi})^{3/2}}\,{\rm arctg}\frac{\sqrt{M^2_{\rm N} - M^2_{\pi}}}{M_{\pi}}\Bigg] = 3.56.
\end{eqnarray}
The numerical value ${\cal F}_{\rm nne^+} = 3.56$ is computed for $a_{\rm nn} = - 16.4\,{\rm fm}$,  $M_{\rm N} = 940\,{\rm MeV}$ and $M_{\pi} = 135\,{\rm MeV}$. Then, $v_{\rm nn}(0)$ is the value of the irregular wave function extrapolated to the region $\rho \sim 1/M_{\rm N}$ a computation of which in terms of a nuclear
potential goes beyond the scope of the RFMD. In the RFMD we consider  $v_{\rm nn}(0)$ as a free parameter and can fix it through the following consideration.

We would like to accentuate that the concentration of the integrals over relative distances $\rho$ to the region $\rho \sim 1/M_{\rm N}$, which is small compared with the range $\rho \sim 1/M_{\pi}$ of the Yukawa potential defined by the one--pion exchange,  confirms to some extent the use of the $\delta$--function potential for the description of strong low--energy NN interactions in the ${^1}{\rm S}_0$--state [2,4]. Therefore, we can formulate a low--energy theorem. For this aim, first, we suggest to denote the amplitude Eq.~(\ref{label8.4}) as ${\cal M}(\bar{\nu}_{\rm e} + {\rm D} \to {\rm e}^+ + {\rm n} + {\rm n})_{\rm s.p.}$ (smeared potential) and the amplitude calculated for the $\delta$--potential as ${\cal M}(\bar{\nu}_{\rm e} + {\rm D} \to {\rm e}^+ + {\rm n} + {\rm n})_{\rm \delta .p.}$ ($\delta$--potential) [4] (see Appendix E). Expressing then these amplitudes in terms of each other we get
\begin{eqnarray}\label{label8.6}
{\cal M}(\bar{\nu}_{\rm e} + {\rm D} \to {\rm e}^+ + {\rm n} + {\rm n}
)_{\rm s.p.} &=& {\cal M}(\bar{\nu}_{\rm e} + {\rm D} \to {\rm e}^+ + {\rm
n} + {\rm n} )_{\rm \delta .p.}\nonumber\\
&&\times\,{\cal F}_{\rm nne^+}\,v_{\rm
nn}(0)\,\times\,{\displaystyle e^{\displaystyle i\delta_{\rm
nn}(k)}\frac{\displaystyle \sin\delta_{\rm nn}(k)}{\displaystyle a_{\rm
nn}k}}\,.
\end{eqnarray}
Near threshold in the low--energy limit $k\to 0$, when the neutrons become
localized to the region of order of $O(1/k)$ being much larger than the  range of nuclear forces, a wave function of a relative movement
of the neutrons can be described well by a plane wave. Thereby, at $k \to 0$ the amplitude ${\cal M}(\bar{\nu}_{\rm e} + {\rm D} \to {\rm e}^+ + {\rm
n} + {\rm n})_{\rm s.p.}$ should coincide with ${\cal M}(\bar{\nu}_{\rm e}
+ {\rm D} \to {\rm e}^+ + {\rm n} + {\rm n})_{\rm \delta .p.}$ and the
low--energy theorem reads
\begin{eqnarray}\label{label8.7}
\lim_{\displaystyle k \to 0}{\cal M}(\bar{\nu}_{\rm e} + {\rm D} \to {\rm e}^+ + {\rm n} + {\rm n} )_{\rm s.p.} = {\cal M}(\bar{\nu}_{\rm e} + {\rm D} \to {\rm e}^+
+ {\rm n} + {\rm n} )_{\rm \delta .p.}.
\end{eqnarray}
Using the amplitude ${\cal M}(\bar{\nu}_{\rm e} + {\rm D} \to {\rm e}^+ +
{\rm n} + {\rm n})_{\rm s.p.}$ given by Eq.~(\ref{label8.6}) we derive the
relation
\begin{eqnarray}\label{label8.8}
{\cal F}_{\rm nne^+}\,v_{\rm nn}(0) =1.
\end{eqnarray}
At $a_{\rm nn} = - 16.4\,{\rm fm}$ [5] we obtain $v_{\rm nn}(0) = 0.28$\footnote{ This numerical value together with the relation Eq.~(\ref{label9.10}) caused by isotopical invariance of nuclear forces is supported by the constraint on the value of the astrophysical factor $S^{\rm NF}_{\rm pp}(0)$ for the solar proton burning defined by nuclear forces only (see discussion below  Eq.~(\ref{label12.2}))  and the fit of the cross section for the process $\nu_{\rm e}$ + D $\to$ e$^-$ + p + p calculated in the PMA  at high neutrino energies Eq.~(\ref{label12.7}).}.

Let us dwell a bit more on the justification of the possibility to regard $v_{\rm nn}(0)$ as a free parameter of the RFMD. The wave function $\psi_{\rm nn}(\vec{\rho}\,)_{\rm out}$ given by Eq.~(\ref{label8.3}) describes a relative movement of the neutrons at distances $\rho \ge r_{\rm NF}$. The information about strong low--energy interactions of the neutrons at distances $\rho \le r_{\rm NF}$ is represented in terms of the phase shift $\delta_{\rm nn}(k)$ depending on the S--wave scattering length $a_{\rm nn}$ and the effective range $r_{\rm nn}$ of the low--energy elastic nn scattering. In the PMA for the self--consistent definition of the parameters of the phase shift $\delta_{\rm nn}(k)$ the wave function Eq.~(\ref{label8.3}) supplemented by a regular part should be continued via boundary conditions at $\rho = r_{\rm NF}$ to the region $0 \le \rho \le r_{\rm NF}$. This gives the scattering length $a_{\rm nn}$ and the effective range $r_{\rm nn}$ expressed in terms of the parameters of the nn potential and the range of nuclear forces $r_{\rm NF}$ [40]. By tuning these parameters one can fit the experimental values of the scattering length $a_{\rm nn}$ and the effective range $r_{\rm nn}$ [40]. The continuation of $\psi_{\rm nn}(\vec{\rho}\,)_{\rm out}$ through a boundary at $\rho = r_{\rm NF}$ to the region $\rho \le r_{\rm NF}$ defines a shape of the wave function of the neutrons up to $\rho \to 0$. Thus, the explicit shape of the wave function of the neutrons in the whole region of relative distances $\rho$ depends on the parameters of the nn potential defined in the region  $0 \le \rho \le r_{\rm NF}$ and the value of the range of nuclear forces $r_{\rm NF}$.

Since in the RFMD nuclear forces in the region of relative distances $0 \le \rho \le r_{\rm NF}$ are described through the one--nucleon loop exchanges, one cannot, in principle,  fix a shape of the wave function $\psi_{\rm nn}(\vec{\rho}\,)_{\rm out}$ and, correspondingly, $v_{\rm nn}(k\rho)$ in terms of the parameters of the nn potential and the range of nuclear forces $r_{\rm NF}$. This leaves the wave function $v_{\rm nn}(k\rho)$ completely undetermined in the RFMD at distances $0 \le \rho \le r_{\rm NF}$ and  allows to consider $v_{\rm nn}(0)$  as a free parameter of the approach. Without loss of generality we can assume too that $v_{\rm nn}(0)$ does not depend on the relative momentum of the neutrons $k$ or this dependence is very smooth in comparison with the dependence induced by the phase shift $\sin\delta_{\rm nn}(k)/a_{\rm nn}k$. This means that in the RFMD for the description of the NN system coupled in the ${^1}{\rm S}_0$--state to the deuteron at low energies one does not need to get more detailed information about short--distance behaviour  of the NN system than that included in the phase shift $\delta_{\rm NN}(k)$ expressed in terms of the S--wave scattering length $a_{\rm NN}$ and the effective range $r_{\rm NN}$ of the low--energy elastic NN scattering.

Substituting Eq.~(\ref{label8.8}) in Eq.~(\ref{label8.4}) we bring  up the amplitude of the $\bar{\nu}_{\rm e}$ + D $\to$ e$^+$ + n + n  reaction to the form
\begin{eqnarray}\label{label8.9}
&&i{\cal M}(\bar{\nu}_{\rm e} + {\rm D} \to {\rm e}^+ + {\rm n} + {\rm
n})_{\rm s.p.} = g_{\rm A} M_{\rm N} \frac{G_{\rm
V}}{\sqrt{2}}\,\frac{3g_{\rm V}}{2\pi^2}\,
G_{\rm \pi nn}\,e_{\mu}(Q)\,[\bar{v}(k_{\bar{\nu}_{\rm
e}})\gamma^{\mu}(1-\gamma^5)
v(k_{{\rm e}^+})]\,\nonumber\\
&&\hspace{1in} \times \,[\bar{u}(p_1) \gamma^5 u^c(p_2)]\,{\displaystyle
e^{\displaystyle i\delta_{\rm nn}(k)}\frac{\displaystyle \sin\delta_{\rm
nn}(k)}{\displaystyle a_{\rm nn}k}}\,.
\end{eqnarray}
Further we omit the label s.p.. The amplitude Eq.~(\ref{label8.9}) squared,
averaged over polarizations of the deuteron and summed over polarizations
of the final particles reads
\begin{eqnarray}\label{label8.10}
&&\overline{|{\cal M}(\bar{\nu}_{\rm e} + {\rm D} \to {\rm e}^+ + {\rm n} + {\rm n})|^2} = \nonumber\\
&&=g^2_{\rm A}M^6_{\rm N}\frac{144 G^2_{\rm V}Q_{\rm
D}}{\pi^2}\frac{G^2_{\rm \pi nn}F_{\rm D}(k^2)}{\displaystyle \Bigg(1 -
\frac{1}{2}a_{\rm nn} r_{\rm nn} k^2 \Bigg)^2 + a^2_{\rm nn}k^2 }\Bigg(
E_{{\rm e}^+}
E_{\bar{\nu}_{\rm e}} - \frac{1}{3}\vec{k}_{{\rm e}^+}\cdot
\vec{k}_{\bar{\nu}_{\rm e}}\Bigg)\,.
\end{eqnarray}
We have used here the relation
\begin{eqnarray}\label{label8.11}
a_{\rm nn}k\,{\rm ctg}\delta_{\rm nn}(k) = - 1 + \frac{1}{2}\,a_{\rm
nn}r_{\rm nn}k^2,
\end{eqnarray}
where $r_{\rm nn} = (2.86 \pm 0.02)\,{\rm fm}$ is the effective range of the nn scattering in the ${^1}{\rm S}_0$--state [5]. The form factor $F_{\rm D}(k^2)$ defined by Eq.~(\ref{label6.11}) is introduced to describe a spatial smearing of the deuteron, where $k= \sqrt{M_{\rm N}T_{\rm nn}}$ is a relative momentum and $T_{\rm nn}$ is a kinetic energy of the relative movement of the nn system.

In the rest frame of the deuteron the cross section for the process $\bar{\nu}_{\rm e}$ + D $\to$ e$^+$ + n + n is defined by
\begin{eqnarray}\label{label8.12}
&&\sigma^{\bar{\nu}_{\rm e} D}_{\rm cc}(E_{\bar{\nu}_{\rm e}}) =
\frac{1}{4M_{\rm D}E_{\bar{\nu}_{\rm e}}}\int\,\overline{|{\cal
M}(\bar{\nu}_{\rm e} +
{\rm D} \to {\rm e}^+ + {\rm n} + {\rm n} )|^2}\nonumber\\
&&\frac{1}{2}\,(2\pi)^4\,\delta^{(4)}(Q + k_{{\bar{\nu}_{\rm e}}} - p_1 -
p_2 - k_{{\rm e}^+})\,
\frac{d^3p_1}{(2\pi)^3 2E_1}\frac{d^3 p_2}{(2\pi)^3 2E_2}\frac{d^3k_{{\rm
e}^+}}{(2\pi)^3
2E_{{\rm e}^+}},
\end{eqnarray}
where  $E_{\bar{\nu}_{\rm e}}$, $E_1$, $E_2$  and $E_{{\rm
e}^+}$ are the energies of the anti--neutrino, the neutrons and the 
positron. The integration over the phase volume of the (${\rm
n~n~e^+}$)--state we perform in the non--relativistic limit 
\begin{eqnarray}\label{label8.13}
&&\frac{1}{2}\,\int\frac{d^3p_1}{(2\pi)^3 2E_1}\frac{d^3p_2}{(2\pi)^3 2E_2}
\frac{d^3k_{{\rm e}^+}}{(2\pi)^3 2E_{{\rm e}^+}}(2\pi)^4\,\delta^{(4)}(Q +
k_{{\bar{\nu}_{\rm e}}} - p_1 - p_2 - k_{{\rm e}^+})\,\nonumber\\
&&\times\, \Bigg( E_{{\rm e}^+} E_{\bar{\nu}_{\rm e}} - \frac{1}{3}
\vec{k}_{{\rm e}^+}\cdot
\vec{k}_{\bar{\nu}_{\rm e}}\Bigg)\,\frac{F_{\rm D}(M_{\rm N}\,T_{\rm
nn})}{\displaystyle \Bigg(1 - \frac{1}{2}a_{\rm nn} r^{\rm e}_{\rm pp}
M_{\rm N}\,T_{\rm nn}\Bigg)^2 + a^2_{\rm nn}M_{\rm N}\,T_{\rm
nn}} =\nonumber\\
&&= \frac{E_{\bar{\nu}_{\rm e}}M^3_{\rm N}}{1024\pi^2}\,\Bigg(\frac{E_{\rm
th}}{M_{\rm N}}
\Bigg)^{\!\!7/2}\Bigg(\frac{2 m_{\rm e}}{E_{\rm
th}}\Bigg)^{\!\!3/2}\frac{8}{\pi E^2_{\rm th}}
\int\!\!\!\int dT_{\rm e^+} dT_{\rm nn}\,
\sqrt{T_{\rm e^+}T_{\rm nn}}\, \nonumber\\
&&\times\,\frac{F_{\rm D}(M_{\rm N}\,T_{\rm
nn})}{\displaystyle \Bigg(1 - \frac{1}{2}a_{\rm nn} r^{\rm e}_{\rm pp}
M_{\rm N}\,T_{\rm nn}\Bigg)^2 + a^2_{\rm nn}M_{\rm N}\,T_{\rm
nn}}\,\Bigg(1 + \frac{T_{\rm
e^+}}{m_{\rm e}}
\Bigg)\,{\displaystyle \sqrt{1 + \frac{T_{\rm e^+}}{2 m_{\rm e}} }}\nonumber\\
&&\hspace{1in} \times\,\delta\Big(E_{\bar{\nu}_{\rm e}}- E_{\rm th} -
T_{\rm e^+}
 - T_{\rm nn}\Big) = \nonumber\\
&&= \frac{E_{\bar{\nu}_{\rm e}}M^3_{\rm N}}{1024\pi^2}\,\Bigg(\frac{E_{\rm
th}}{M_{\rm N}}
\Bigg)^{\!\!7/2}\Bigg(\frac{2 m_{\rm e}}{E_{\rm
th}}\Bigg)^{\!\!3/2}\,(y-1)^2\,\Omega_{\rm n n e^+}(y),
\end{eqnarray}
where $T_{\rm e^+}$ and $m_{\rm e} = 0.511\,{\rm MeV}$ are the kinetic
energy and the mass of the positron, $E_{\rm th}$ is the anti--neutrino
energy threshold of the reaction $\bar{\nu}_{\rm e}$ + D $\to$ e$^+$ + n +
n , and is given by $E_{\rm th} = \varepsilon_{\rm D} + m_{\rm e} + (M_{\rm
n} - M_{\rm p}) = (2.225 + 0.511 + 1.293) \, {\rm MeV} = 4.029\,{\rm MeV}$.
The function $\Omega_{\rm n n e^+}(y)$, where $y = E_{\bar{\nu}_{\rm
e}}/E_{\rm th}$,
is defined as
\begin{eqnarray}\label{label8.14}
&&\Omega_{\rm n n e^+}(y) = \frac{8}{\pi}\,\int\limits^{1}_{0}
dx\,\frac{\sqrt{x\,(1 - x)}\,F_{\rm D}(M_{\rm N}E_{\rm th}\,(y -
1)\,x)}{\displaystyle \Bigg(1 - \frac{1}{2}a_{\rm nn} r_{\rm nn}
M_{\rm N}E_{\rm th}\,(y - 1)\,x\Bigg)^2 + a^2_{\rm nn}M_{\rm N}E_{\rm
th}\,(y - 1)\,x}\,\nonumber\\
&&\times\,\Bigg(1 + \frac{E_{\rm th}}{m_{\rm e}}(y-1)(1-x)\Bigg) \,\sqrt{1
+ \frac{E_{\rm th}}{2 m_{\rm e}}(y-1)(1-x)},
\end{eqnarray}
where we have changed the variable $T_{\rm nn} = (E_{\bar{\nu}_{\rm e}} -
E_{\rm th})\,x$. The function $\Omega_{\rm n n e^+}(y)$ is normalized to
unity at $y=1$, i.e., at threshold $E_{\bar{\nu}_{\rm e}}= E_{\rm th}$.
Thus, the cross section for the anti--neutrino disintegration of the deuteron
reads
\begin{eqnarray}\label{label8.15}
\sigma^{\rm \bar{\nu}_{\rm e} D}_{\rm cc}(E_{\bar{\nu}_{\rm e}}) =
\sigma_0\,(y - 1)^2\,\Omega_{\rm n n e^+}(y),
\end{eqnarray}
where $\sigma_0$ is defined by
\begin{eqnarray}\label{label8.16}
\sigma_0 = Q_{\rm D}\,G^2_{\rm \pi nn}\frac{9g^2_{\rm A}G^2_{\rm
V}M^8_{\rm N}}{512\pi^4} \Bigg(\frac{E_{\rm th}}{M_{\rm
N}}\Bigg)^{\!\!7/2}\Bigg(\frac{2m_{\rm e}}{E_{\rm th}}\Bigg)^{\!\!3/2} =
(3.88 \pm 0.74)\times 10^{-43}\,{\rm cm}^2.
\end{eqnarray}
Here $\pm 0.74$ describes the assumed theoretical uncertainty of our approach which is about 19\,$\%$ (see Appendix A). The value $\sigma_0 = (3.88 \pm 0.74)\times
10^{-43}\,{\rm cm}^2$ is  20$\%$ less than the value $\sigma_0 = (4.68 \pm 1.40) \times 10^{-43}\,{\rm cm}^2$ obtained in the PMA [44,45] (see Fig.~7 of Ref.~[45]). Such a decrease in comparison with our agreement obtained in Ref.~[4] is related to the change of the value of the effective coupling constant of the four--neutron interaction $G_{\rm \pi NN} = 3.27\times 10^{-3}\,{\rm MeV}^{-2} \to G_{\rm \pi nn} = 3.02\times 10^{-3}\,{\rm MeV}^{-2}$.

The experimental data on the anti--neutrino disintegration of the deuteron are given in terms of the cross section averaged over the anti--neutrino energy spectrum per anti--neutrino fission in the energy region of anti--neutrinos $E_{\rm th}
\le E_{\bar{\nu}_{\rm e}}
\le 10\,{\rm MeV}$. The
experimental data read 
$<\sigma^{\bar{\nu}_{\rm e}D}_{\rm
cc}(E_{\bar{\nu}_{\rm e}})>_{\exp} = 
(1.5\pm 0.4) \times 10^{-45}\,{\rm
cm}^2/{\bar{\nu}_{\rm e}}\,{\rm fission}$ [46], $<\sigma^{\bar{\nu}_{\rm e}D}_{\rm
cc}(E_{\bar{\nu}_{\rm e}})>_{\exp} = (0.9\pm 0.4) \times 10^{-45}\,{\rm
cm}^2/{\bar{\nu}_{\rm e}}\,{\rm fission}$  [47] and  $<\sigma^{\bar{\nu}_{\rm
e}D}_{\rm cc}(E_{\bar{\nu}_{\rm e}})>_{\exp} = (1.84 \pm 0.04) \times
10^{-45}\, {\rm cm}^2/
{\bar{\nu}_{\rm e}}\, {\rm fission}$ [48].

The cross section $<\sigma^{\rm \bar{\nu}_{\rm e} D}_{\rm
cc}(E_{\bar{\nu}_{\rm e}})>$, calculated in the RFMD and averaged over the anti--neutrino Avignone--Greenwood spectrum [49,50] in the energy region
$E_{\rm th}\le E_{\bar{\nu}_{\rm e}} \le 10\;{\rm
MeV}$, is given by
\begin{eqnarray}\label{label8.17}
<\sigma^{\rm \bar{\nu}_{\rm e} D}_{\rm cc}(E_{\bar{\nu}_{\rm e}})> &=&
\frac{a}{N_{\bar{\nu}_{\rm e}}}\int\limits^{2.482}_{1}dy\,e^{\displaystyle
-b\,y}\,
\sigma_0\,(y-1)^2\,\Omega_{\rm n n e^+}(y) = \nonumber\\
&=& (1.66 \pm 0.32)\times 10^{-45}\,{\rm cm}^2/\,\bar{\nu}_{\rm e}\,{\rm
fission},
\end{eqnarray}
where $a = 17.8\, E_{\rm th} = 71.72$, $b = 1.01\, E_{\rm th} = 4.07$, and $N_{\bar{\nu}_{\rm e}} = 6$ is the number of anti--neutrinos per fission [49,50].

The theoretical value $<\sigma^{\rm \bar{\nu}_{\rm e} D}_{\rm
cc}(E_{\bar{\nu}_{\rm e}})> = (1.66 \pm 0.32)\times 10^{-45}\,{\rm
cm}^2/\,\bar{\nu}_{\rm e}\,{\rm fission}$ agrees  with the experimental data given by the Reines$^{\prime}$s group: $<\sigma^{\rm \bar{\nu}_{\rm e}D}_{\rm cc}(E_{\bar{\nu}_{\rm e}})>_{\exp} = (1.5\pm 0.4) \times 10^{-45}\,{\rm cm}^2/ {\bar{\nu}_{\rm e}}\,{\rm
fission}$  [46], $<\sigma^{\rm \bar{\nu}_{\rm e}D}_{\rm
cc}(E_{\bar{\nu}_{\rm e}})>_{\exp} = (0.9\pm 0.4) \times 10^{-45}\,{\rm
cm}^2/ {\bar{\nu}_{\rm e}}\,{\rm
fission}$ [47] and Russian experimental groups [48]: $<\sigma^{\rm
\bar{\nu}_{\rm e}D}_{\rm cc}(E_{\bar{\nu}_{\rm e}})>_{\exp} = (1.84\pm
0.04) \times 10^{-45}\,{\rm cm}^2/ {\bar{\nu}_{\rm e}}\,{\rm fission}$.

A comparison of the cross section Eq.~(\ref{label8.15}) calculated in the RFMD with the PMA data we perform for the anti--neutrino energies $E_{\bar{\nu}_{\rm e}}=
6.5\,{\rm MeV}$ and  $E_{\bar{\nu}_{\rm e}}=
10\,{\rm MeV}$. In the RFMD we get
\begin{eqnarray}\label{label8.18}
\sigma^{\rm \bar{\nu}_{\rm e} D}_{\rm cc}(E_{\bar{\nu}_{\rm
e}})|_{E_{\bar{\nu}_{\rm e}}=
6.5\,{\rm MeV}}&=&1.07\,(1\pm 0.19) \times 10^{-43}\,{\rm cm}^2, \nonumber\\
\sigma^{\rm \bar{\nu}_{\rm e} D}_{\rm cc}(E_{\bar{\nu}_{\rm
e}})|_{E_{\bar{\nu}_{\rm e}}=
10\,{\rm MeV}}&=&0.79\,(1\pm 0.19) \times 10^{-42}\,{\rm cm}^2.
\end{eqnarray}
In turn the PMA data read [42,43]:
\begin{eqnarray}\label{label8.19}
\sigma^{\rm \bar{\nu}_{\rm e} D}_{\rm cc}(E_{\bar{\nu}_{\rm
e}})|_{E_{\bar{\nu}_{\rm e}}=
6.5\,{\rm MeV}} &=&1.71 \times 10^{-43}\,{\rm cm}^2, \nonumber\\
\sigma^{\rm \bar{\nu}_{\rm e} D}_{\rm cc}(E_{\bar{\nu}_{\rm
e}})|_{E_{\bar{\nu}_{\rm e}}=
10\,{\rm MeV}}&=&1.13 \times 10^{-42}\,{\rm cm}^2.
\end{eqnarray}
The numerical values disagree by a factor of 1.5 in average. Referring to the result of obtained in Ref.~[4] we explain such a disagreement by the change of the value of the effective coupling constant $G_{\rm \pi nn} = 3.02\times 10^{-3}\,{\rm MeV}^{-2}$ instead of $G_{\rm \pi NN} = 3.27\times 10^{-3}\,{\rm MeV}^{-2}$ [4], the change of the value of the S--wave scattering length $a_{\rm nn} = -16.4\,{\rm fm}$ instead of $a_{\rm nn} = -17.0\,{\rm fm}$ [4] and the use of the non--zero effective range of the low--energy elastic nn scattering. The change of the input parameters for the description of the process of the disintegration of the deuteron by anti--neutrinos $\bar{\nu}_{\rm e}$ + D $\to$ e$^+$ + n + n in the generalized RFMD is required by the phenomenology of the low--energy elastic nn scattering. Of course, such a change of the input parameters  has led to the disagreement with the PMA data, but the agreement with the experimental data has become much better [4].

\section{Cross section for $\nu_{\rm e}$ + D $\to$ $\nu_{\rm e}$ + n + p}
\setcounter{equation}{0}

For the calculation of the n + p $\to$ n + p transition entering to the amplitude of the process $\nu_{\rm e}$ + D $\to$ $\nu_{\rm e}$ + n + p we use the effective four--nucleon interaction Eq.~(\ref{label1.7})
\begin{eqnarray}\label{label9.1}
&&{\cal L}^{\rm np \to np}_{\rm eff}(x) = G_{\rm \pi
np}\,\int d^3\rho\,\frac{M^2_{\pi}}{4\pi}\,{\displaystyle
\frac{\displaystyle e^{\displaystyle - M_{\pi}\rho}}{\rho}}\nonumber\\
&&\times \{[\bar{n}(t,\vec{x} - \frac{1}{2}\vec{\rho}\,)\gamma_{\mu}
\gamma^5 p^c(t,\vec{x} + \frac{1}{2}\vec{\rho}\,)]
[\bar{p^c}(t,\vec{x} - \frac{1}{2}\vec{\rho}\,)\gamma^{\mu}\gamma^5
n(t,\vec{x} + \frac{1}{2}\vec{\rho}\,)] \nonumber\\
&&+ (\gamma_{\mu}\gamma^5 \otimes \gamma^{\mu}\gamma^5 \to \gamma^5 \otimes
\gamma^5)\}.
\end{eqnarray}
The process $\nu_{\rm e}$ + D $\to$ $\nu_{\rm e}$ + n + p should run via the intermediate Z--boson exchange. The effective Lagrangian describing the electroweak interactions of the Z--boson with nucleons and leptons reads
[51]
\begin{eqnarray}\label{label9.2}
&&{\cal L}^{\rm w}_{\rm eff}(x) = - \frac{g_{\rm W}}{4\cos\vartheta_{\rm
W}}\,\bar{p}(x)[\gamma^{\mu} (1 - 4\sin^2\vartheta_{\rm W}) - g_{\rm
A}\,\gamma^{\mu}\gamma^5]  p(x) Z_{\mu}(x) \nonumber\\
&&+ \frac{g_{\rm W}}{4\cos\vartheta_{\rm W}}[\bar{n}(x) \gamma^{\mu} (1 -
g_{\rm A}\,\gamma^5) n(x) - \bar{\psi}_{{\nu}_{\rm e}}(x) \gamma^{\mu} (1 -
\gamma^5) \psi_{{\nu}_{\rm e}}(x)] Z_{\mu}(x),
\end{eqnarray}
where $\psi_{{\nu}_{\rm e}}(x)$ is the operator of the neutrino field, the weak
angle $\vartheta_{\rm W}$ links the masses of the W-- and Z--bosons:
$M_{\rm Z}= M_{\rm W}/\cos\vartheta_{\rm W}$ [51].

For the calculation of the transition n + p $\to$ n + p we apply a
wave function
\begin{eqnarray}\label{label9.3}
\psi_{\rm np}(\vec{\rho}\,)_{\rm out} = {\displaystyle e^{\displaystyle
i\delta_{\rm np}(k)}\sin\delta_{\rm np}(k)\,\frac{\displaystyle v_{\rm
np}(0)}{\displaystyle k\rho}},
\end{eqnarray}
where $v_{\rm np}(0)$ is the irregular part of the wave function of a relative movement of the proton and the neutron in the ${^1}{\rm S}_0$--state. The phase shift $\delta_{\rm np}(k)$ is defined by the relation
\begin{eqnarray}\label{label9.4}
a_{\rm np}k\,{\rm ctg}\delta_{\rm np}(k) = - 1 + \frac{1}{2}\,a_{\rm
np}r_{\rm np}k^2,
\end{eqnarray}
where $a_{\rm np} = (-23.748\pm 0.010)\,{\rm fm}$ and
$r_{\rm np} = (2.75 \pm 0.05)\,{\rm fm}$ are the scattering length and the effective range of the np scattering in the ${^1}{\rm S}_0$--state [5], then $k= \sqrt{M_{\rm N}T_{\rm np}}$ is the relative momentum and $T_{\rm np}$ is the kinetic energy of the relative movement of the np system.

The amplitude of the process $\nu_{\rm e}$ + D $\to$ $\nu_{\rm e}$ + n + p is calculated in Appendix E and reads
\begin{eqnarray}\label{label9.5}
&&i{\cal M}(\nu_{\rm e} + {\rm D} \to \nu_{\rm e} + {\rm n} + {\rm p}) = g_{\rm A}\,  M_{\rm N}\,\frac{G_{\rm F}}{\sqrt{2}}\,\frac{3g_{\rm
V}}{4\pi^2}\,
G_{\rm \pi np}\,e_{\mu}(k_{\rm D})\,[\bar{u}(k^{\prime}_{\nu_{\rm
e}})\gamma^{\mu}(1-\gamma^5)
u(k_{\nu_{\rm e}})]\,\nonumber\\
&&\times [\bar{u}(p_1) \gamma^5 u^c(p_2)]\,\times\,{\cal F}_{\rm np\nu_{\rm e}}\,v_{\rm np}(0)\,\times\,{\displaystyle e^{\displaystyle i\delta_{\rm
np}(k)}\frac{\displaystyle \sin\delta_{\rm np}(k)}{\displaystyle a_{\rm
np}k}},
\end{eqnarray}
where $\bar{u}(k^{\prime}_{\nu_{\rm e}})$, $u(k_{\nu_{\rm e}})$,
$\bar{u}(p_1)$ and $u^c(p_2)$ are the Dirac bispinors of the initial and the final neutrinos, and the nucleons, $e_{\mu}(k_{\rm D})$ is the 4--vector of the polarization of the deuteron. The factor ${\cal F}_{\rm np\nu_{\rm e}}$ is related to the factor ${\cal F}_{\rm nne^+}$ as 
\begin{eqnarray}\label{label9.6}
{\cal F}_{\rm np\nu_{\rm e}} = \frac{a_{\rm np}}{a_{\rm nn}}\,{\cal F}_{\rm nne^+}.
\end{eqnarray}
Substituting Eq.~(\ref{label9.6}) in Eq.~({\ref{label9.5}) we obtain
\begin{eqnarray}\label{label9.7}
&&i{\cal M}(\nu_{\rm e} + {\rm D} \to \nu_{\rm e} + {\rm n} + {\rm p}) = g_{\rm A}\,  M_{\rm N}\,\frac{G_{\rm F}}{\sqrt{2}}\,\frac{3g_{\rm
V}}{4\pi^2}\,
G_{\rm \pi np}\,e_{\mu}(k_{\rm D})\,[\bar{u}(k^{\prime}_{\nu_{\rm
e}})\gamma^{\mu}(1-\gamma^5)
u(k_{\nu_{\rm e}})]\,\nonumber\\
&&\times [\bar{u}(p_1) \gamma^5 u^c(p_2)]\,\times\,\frac{a_{\rm np}}{a_{\rm nn}}\,\times\,{\cal F}_{\rm nne^+}\,v_{\rm np}(0)\times\,{\displaystyle e^{\displaystyle i\delta_{\rm
np}(k)}\frac{\displaystyle \sin\delta_{\rm np}(k)}{\displaystyle a_{\rm
np}k}},
\end{eqnarray}
In the RFMD the contributions of strong low--energy nn and np interactions to the amplitudes of the processes $\bar{\nu}_{\rm e}$ + D $\to$ e$^+$ + n + n and $\nu_{\rm e}$ + D $\to$ $\nu_{\rm e}$ + n + p are proportional to the amplitudes of low--energy elastic nn and np scattering determined as [6]:
\begin{eqnarray}\label{label9.8}
{\cal A}_{\rm nn}(k) &=& {\displaystyle e^{\displaystyle i\delta_{\rm
nn}(k)}\frac{\displaystyle \sin\delta_{\rm nn}(k)}{\displaystyle k}},\nonumber\\
{\cal A}_{\rm np}(k) &=& {\displaystyle e^{\displaystyle i\delta_{\rm
np}(k)}\frac{\displaystyle \sin\delta_{\rm np}(k)}{\displaystyle k}}.
\end{eqnarray}
In the low--energy limit $k\to 0$, i.e., near thresholds of the processes $\bar{\nu}_{\rm e}$ + D $\to$ e$^+$ + n + n and $\nu_{\rm e}$ + D $\to$ $\nu_{\rm e}$ + n + p, the amplitudes Eq.~(\ref{label9.8}) obey the relation
\begin{eqnarray}\label{label9.9}
\frac{{\cal A}_{\rm np}(0)}{{\cal A}_{\rm nn}(0)} = \frac{a_{\rm np}}{a_{\rm nn}}.
\end{eqnarray}
In order to hold such a threshold relation between strong parts of the amplitudes of the processes  $\bar{\nu}_{\rm e}$ + D $\to$ e$^+$ + n + n and $\nu_{\rm e}$ + D $\to$ $\nu_{\rm e}$ + n + p  it is sufficient only to follow the isotopical symmetry of nuclear forces [6,52] and set
\begin{eqnarray}\label{label9.10}
v_{\rm nn}(0) = v_{\rm pp}(0) = v_{\rm np}(0) = 0.28.
\end{eqnarray}
This assumption is not too strict. Indeed, as has been discussed in Ref.~[52] (Bethe) the discrepancy between phenomenological values of the S--wave scattering lengths of the low--energy elastic np, nn and pp scattering caused by nuclear forces can be arranged by varying the depth of the nuclear potential well within 3$\%$ of the magnitude and holding the value of the range fixed. The phenomenological data on the S--wave scattering lengths of the low--energy elastic NN scattering are represented in the wave function of the NN system in the form of the factor $e^{\textstyle i\delta_{\rm NN}(k)} \sin\delta_{\rm NN}(k)/k$ as
\begin{eqnarray*}
\psi_{\rm NN}(k\rho) = e^{\displaystyle i\delta_{\rm NN}(k)}\frac{\displaystyle \sin\delta_{\rm NN}(k)}{k}\times \frac{v_{\rm NN}(k\rho)}{\rho}.
\end{eqnarray*}
Thereby, the wave function $v_{\rm NN}(k\rho)$, defined as a solution of Schr\"odinger equation with the potential of nuclear forces, can be taken in the isotopically invariant form within an accuracy better than 3$\%$. The isotopical relation Eq.(\ref{label9.10}) and the numerical values of the wave functions $v_{\rm nn}(0) = v_{\rm pp}(0) = v_{\rm np}(0) = 0.28$ are justified by  the constraint on the value of the astrophysical factor $S^{\rm NF}_{\rm pp}(0)$ for the solar proton burning caused by nuclear forces only (see discussion below Eq.~(\ref{label12.2})) and the fit of the cross section for the process $\nu_{\rm e}$ + D $\to$ e$^-$ + p + p calculated in the PMA  at high neutrino energies Eq.~(\ref{label12.7}).

Due to the relation Eq.~(\ref{label9.10}) the amplitude of the process $\nu_{\rm e}$ + D $\to$ $\nu_{\rm e}$ + n + p  can be defined as follows
\begin{eqnarray}\label{label9.11}
&&i{\cal M}(\nu_{\rm e} + {\rm D} \to \nu_{\rm e} + {\rm n} + {\rm p}) = g_{\rm A}\,  M_{\rm N}\,\frac{G_{\rm F}}{\sqrt{2}}\,\frac{3g_{\rm
V}}{4\pi^2}\,
G_{\rm \pi np}\,e_{\mu}(k_{\rm D})\,[\bar{u}(k^{\prime}_{\nu_{\rm
e}})\gamma^{\mu}(1-\gamma^5)
u(k_{\nu_{\rm e}})]\,\nonumber\\
&&\times [\bar{u}(p_1) \gamma^5 u^c(p_2)]\,\times\,Ä\frac{a_{\rm np}}{a_{\rm nn}}\,\times\,{\cal F}_{\rm nne^+}\,v_{\rm nn}(0)\times\,{\displaystyle e^{\displaystyle i\delta_{\rm
np}(k)}\frac{\displaystyle \sin\delta_{\rm np}(k)}{\displaystyle a_{\rm
np}k}},
\end{eqnarray}
Using then the constraint  Eq.~(\ref{label8.8}) we get 
\begin{eqnarray}\label{label9.12}
&&i{\cal M}(\nu_{\rm e} + {\rm D} \to \nu_{\rm e} + {\rm n} + {\rm p}) =  g_{\rm A}\,  M_{\rm N}\,\frac{G_{\rm F}}{\sqrt{2}}\,\frac{3g_{\rm
V}}{4\pi^2}\,
G_{\rm \pi np}\,e_{\mu}(k_{\rm D})\,[\bar{u}(k^{\prime}_{\nu_{\rm
e}})\gamma^{\mu}(1-\gamma^5)
u(k_{\nu_{\rm e}})]\,\nonumber\\
&&\times [\bar{u}(p_1) \gamma^5 u^c(p_2)]\,\times\,\frac{a_{\rm np}}{a_{\rm nn}}\,\times\,{\displaystyle e^{\displaystyle i\delta_{\rm
np}(k)}\frac{\displaystyle \sin\delta_{\rm np}(k)}{\displaystyle a_{\rm
np}k}},
\end{eqnarray}
The amplitude Eq.~(\ref{label9.12}) squared, averaged over polarizations of the deuteron, summed over polarizations of the nucleons and extrapolated to the energies far from threshold, the energy of which $E_{\rm th}$ equals the binding energy of the deuteron $E_{\rm th} = \varepsilon_{\rm D} = 2.225\,{\rm MeV}$, reads
\begin{eqnarray}\label{label9.13}
&&\overline{|{\cal M}(\nu_{\rm e} + {\rm D} \to \nu_{\rm e} + {\rm n} +{\rm p})|^2} = \nonumber\\
&&= g^2_{\rm A}M^6_{\rm N}\frac{36 G^2_{\rm F}Q_{\rm
D}}{\pi^2}\,\frac{a^2_{\rm np}}{a^2_{\rm nn}}\,\frac{G^2_{\rm \pi np}F_{\rm
D}(k^2)}{\displaystyle \Bigg(1 - \frac{1}{2}a_{\rm np} r_{\rm np} k^2
\Bigg)^2 + a^2_{\rm np}k^2 }\Bigg( E^{\prime}_{\nu_{\rm e}} E_{\nu_{\rm e}}
- \frac{1}{3} \vec{k}^{\prime}_{\nu_{\rm e}} \cdot \vec{k}_{\nu_{\rm
e}}\Bigg).
\end{eqnarray}
In the rest frame of the deuteron the cross section for the process $\nu_{\rm e}$ + D $\to$ $\nu_{\rm e}$ + n + p is defined
\begin{eqnarray}\label{label9.14}
&&\sigma^{\nu_{\rm e} D}_{\rm nc}(E_{\bar{\nu}_{\rm e}}) =
\frac{1}{4M_{\rm D}E_{\nu_{\rm e}}}\int\,\overline{|{\cal
M}(\nu_{\rm e} +
{\rm D} \to \nu_{\rm e} + {\rm n} + {\rm p})|^2}\nonumber\\
&&(2\pi)^4\,\delta^{(4)}(k_{\rm D} + k_{\nu_{\rm e}} - p_1 -
p_2 - k^{\prime}_{\nu_{\rm e}})\,
\frac{d^3p_1}{(2\pi)^3 2E_1}\frac{d^3 p_2}{(2\pi)^3
2E_2}\frac{d^3k^{\prime}_{\nu_{\rm e}}}{(2\pi)^3
2E^{\prime}_{\nu_{\rm e}}}.
\end{eqnarray}
The abbreviation (nc) denotes the neutral current. The integration over the phase volume of the (${\rm n p \nu_{\rm e}}$)--state we perform in the non--relativistic limit and in the rest frame of the deuteron,
\begin{eqnarray}\label{label9.15}
&&\int\frac{d^3p_1}{(2\pi)^3 2E_1}\frac{d^3p_2}{(2\pi)^3 2E_2}
\frac{d^3k^{\prime}_{\nu_{\rm e}}}{(2\pi)^3 2E^{\prime}_{\nu_{\rm
e}}}(2\pi)^4\,\delta^{(4)}(k_{\rm D} + k_{\nu_{\rm e}} - p_1 - p_2 -
k^{\prime}_{\nu_{\rm e}})\,\nonumber\\
&&\Bigg( E_{\nu_{\rm e}} E^{\prime}_{\nu_{\rm e}} - \frac{1}{3}
\vec{k}_{\nu_{\rm e}}\cdot
\vec{k}^{\prime}_{\nu_{\rm e}}\Bigg)\,\frac{F_{\rm D}(M_{\rm N}T_{\rm
np})}{\displaystyle \Bigg(1 - \frac{1}{2}a_{\rm np} r_{\rm np}M_{\rm
N}T_{\rm np} \Bigg)^2 + a^2_{\rm np}M_{\rm N}T_{\rm np}} = \nonumber\\
&&=\frac{E_{\nu_{\rm e}}M^3_{\rm N}}{210\pi^3}\,\Bigg(\frac{E_{\rm
th}}{M_{\rm N}}\Bigg)^{\!\!7/2}\,(y - 1)^{7/2}\,\Omega_{\rm np\nu_{\rm e}}(y).
\end{eqnarray}
The function $\Omega_{\rm np\nu_{\rm e}}(y)$, where $y=E_{\bar{\nu}_{\rm e}}/E_{\rm th}$, is defined as
\begin{eqnarray}\label{label9.16}
\Omega_{\rm np\nu_{\rm e}}(y) = \frac{105}{16}\,\int\limits^{1}_{0} dx
\frac{\sqrt{x}\,(1 - x)^2}{\displaystyle \Bigg(1 - \frac{1}{2}\frac{a_{\rm
np} r_{\rm np}}{r^2_{\rm D}}(y-1) x \Bigg)^2 + \frac{a^2_{\rm np}}{r^2_{\rm
D}}(y-1) x}\frac{1}{1 + (y - 1) x},
\end{eqnarray}
where we have changed the variable $T_{\rm np} = (E_{\bar{\nu}_{\rm e}} - E_{\rm th})\,x$ and used the relation $M_{\rm N}E_{\rm th} = 1/r^2_{\rm D}$ at $E_{\rm th}=\varepsilon_{\rm D}$. The function $\Omega_{\rm np\nu_{\rm e}}(y)$ is normalized to unity at $y=1$, i.e., at threshold $E_{\bar{\nu}_{\rm e}} = E_{\rm th}$.

The cross section for the neutrino disintegration of the deuteron caused by the neutral weak current $\nu_{\rm e}$ + D $\to$ $\nu_{\rm e}$ + n + p reads
\begin{eqnarray}\label{label9.17}
\sigma^{\rm \nu_{\rm e} D}_{\rm nc}(E_{\nu_{\rm e}}) = \sigma_0\,(y -
1)^{7/2}\,\Omega_{\rm np\nu_{\rm e}}(y),
\end{eqnarray}
where $\sigma_0$ is defined by
\begin{eqnarray}\label{label9.18}
\sigma_0 = Q_{\rm D}\,g^2_{\rm A}\,\frac{3G^2_{\rm
V} G^2_{\rm \pi NN} M^8_{\rm N}}{140\pi^5}\,\frac{a^2_{\rm np}}{a^2_{\rm nn}}
\Bigg(\frac{E_{\rm th}}{M_{\rm N}}\Bigg)^{\!\!7/2}= (3.81 \pm 0.72)\times
10^{-43}\,{\rm cm}^2.
\end{eqnarray}
Here $\pm 0.72$ describes the assumed theoretical uncertainty of our approach which is about 19$\%$  (see Appendix A).

Since in our approach the cross section for the disintegration of the
deuteron by neutrinos $\nu_{\rm e}$ + D $\to$ $\nu_{\rm e}$ + n + p
coincides with the cross section for the disintegration of the deuteron by anti--neutrinos $\bar{\nu}_{\rm e}$ + D $\to$ $\bar{\nu}_{\rm e}$ + n + p, i.e., $\sigma^{\rm \nu_{\rm e}D}_{\rm nc}(E_{\nu_{\rm e}}) = \sigma^{\rm \bar{\nu}_{\rm e}D}_{\rm nc}(E_{\bar{\nu}_{\rm e}})$, we can compare our result with the experimental data on the disintegration of the deuteron by reactor anti--neutrinos [46--48]. For this aim we should average the cross section Eq.~(\ref{label9.17}) over the anti--neutrino Avignone--Greenwood spectrum [49,50] in the energy region $E_{\rm th}\le E_{\bar{\nu}_{\rm e}} \le 10\,{\rm MeV}$. The cross section $<\sigma^{\rm \bar{\nu}_{\rm e} D}_{\rm nc}(E_{\bar{\nu}_{\rm e}})>$ is given by 
\begin{eqnarray}\label{label9.19}
\hspace{-0.5in}<\sigma^{\rm \bar{\nu}_{\rm e} D}_{\rm nc}(E_{\bar{\nu}_{\rm e}})> &=&
\frac{a_1}{N_{\bar{\nu}_{\rm e}}}\int\limits^{1.528}_{1}dy\,e^{\displaystyle
-b_1\,y}\,
\sigma_0\,(y-1)^{7/2}\,\Omega_{\rm np\bar{\nu}_{\rm e}}(y) \nonumber\\
&+& \frac{a_2 }{N_{\bar{\nu}_{\rm e}}}\int\limits^{4.494}_{1.528}dy\,e^{\displaystyle
-b_2\,y}\,
\sigma_0\,(y-1)^{7/2}\,\Omega_{\rm np\bar{\nu}_{\rm e}}(y) = \nonumber\\
&=& (0.26 + 5.06)\times 10^{-45}\,{\rm cm}^2/\,\bar{\nu}_{\rm e}\,{\rm
fission} =\nonumber\\
&=&(5.32 \pm 1.01)\times 10^{-45}\,{\rm cm}^2/\,\bar{\nu}_{\rm e}\,{\rm fission},
\end{eqnarray}
where $a_1 = 3.63\, E_{\rm th} = 8.08$, $b_1 = 0.543\, E_{\rm th} = 1.208$, $a_2 = 17.8\, E_{\rm th} = 39.61$, $b_2 = 1.01\, E_{\rm th} = 2.247$, and $N_{\bar{\nu}_{\rm e}} = 6$ is the number of anti--neutrinos per fission [49,50].

As we have obtained $<\sigma^{\rm \bar{\nu}_{\rm e} D}_{\rm
cc}(E_{\bar{\nu}_{\rm e}})>  = (1.66 \pm 0.32)\times 10^{-45}\,{\rm
cm}^2/\,\bar{\nu}_{\rm e}\,{\rm fission}$ Eq.~(\ref{label8.17}), we predict a ratio
\begin{eqnarray}\label{label9.20}
r = \frac{\displaystyle <\sigma^{\rm \bar{\nu}_{\rm e} D}_{\rm
cc}(E_{\bar{\nu}_{\rm e}})>}{\displaystyle <\sigma^{\rm \bar{\nu}_{\rm e}
D}_{\rm nc}(E_{\bar{\nu}_{\rm e}})>} = 0.31\pm 0.06.
\end{eqnarray}
Our theoretical predictions   $<\sigma^{\rm \bar{\nu}_{\rm e} D}_{\rm
cc}(E_{\bar{\nu}_{\rm e}})> = (1.66 \pm 0.32)\times 10^{-45}\,{\rm cm}^2/\,\bar{\nu}_{\rm e}\,{\rm fission}$ and $<\sigma^{\rm \bar{\nu}_{\rm e} D}_{\rm
nc}(E_{\bar{\nu}_{\rm e}})> = (5.32 \pm 1.01)\times 10^{-45}\,{\rm cm}^2/\,\bar{\nu}_{\rm e}\,{\rm fission}$, and $r = 0.31\pm 0.06$ agree well with the experimental data by  the Reines$^{\prime}$s group: $<\sigma^{\rm \bar{\nu}_{\rm e} D}_{\rm
cc}(E_{\bar{\nu}_{\rm e}})>_{\exp}  = (1.5 \pm 0.4)\times 10^{-45}\,{\rm
cm}^2/\,\bar{\nu}_{\rm e}\,{\rm fission}$, $<\sigma^{\rm \bar{\nu}_{\rm e}
D}_{\rm nc}(E_{\bar{\nu}_{\rm e}})>_{\exp} = (3.8 \pm 0.9)\times
10^{-45}\,{\rm cm}^2/\,\bar{\nu}_{\rm e}\,{\rm fission}$ and $r_{\exp}=
0.39\pm 0.14$ [46], $<\sigma^{\rm \bar{\nu}_{\rm e} D}_{\rm
cc}(E_{\bar{\nu}_{\rm e}})>_{\exp}  = (0.9 \pm 0.4)\times 10^{-45}\,{\rm
cm}^2/\,\bar{\nu}_{\rm e}\,{\rm fission}$, $<\sigma^{\rm \bar{\nu}_{\rm e}
D}_{\rm nc}(E_{\bar{\nu}_{\rm e}})>_{\exp} = (5.3 \pm 0.8)\times
10^{-45}\,{\rm cm}^2/\,\bar{\nu}_{\rm e}\,{\rm fission}$ and $r_{\exp}=
0.17\pm 0.09$ [47] and  by Russian experimental groups [48]: $<\sigma^{\rm
\bar{\nu}_{\rm e} D}_{\rm cc}(E_{\bar{\nu}_{\rm e}})>_{\exp}  = (1.84 \pm
0.04)\times 10^{-45}\,{\rm cm}^2/\,\bar{\nu}_{\rm e}\,{\rm fission}$,
$<\sigma^{\rm \bar{\nu}_{\rm e} D}_{\rm nc}(E_{\bar{\nu}_{\rm e}})>_{\exp}
= (5.0 \pm 1.7)\times 10^{-45}\,{\rm cm}^2/\,\bar{\nu}_{\rm e}\,{\rm
fission}$, and $r_{\exp}= 0.37\pm 0.13$.

Now let us compare the cross section $\sigma^{\rm \nu_{\rm e}
D}_{\rm nc}(E_{\bar{\nu}_{\rm e}})$ calculated in the RFMD with the PMA data [42,43]. For this aim we suggest to compare the numerical values of the cross section for the anti--neutrino energies $E_{\bar{\nu}_{\rm e}}=
3.25\,{\rm MeV}$, $E_{\bar{\nu}_{\rm e}}=
4.25\,{\rm MeV}$ and $E_{\bar{\nu}_{\rm e}}=
10\,{\rm MeV}$. In the RFMD we get
\begin{eqnarray}\label{label9.21}
\sigma^{\rm \bar{\nu}_{\rm e} D}_{\rm nc}(E_{\bar{\nu}_{\rm
e}})|_{E_{\bar{\nu}_{\rm e}}=
3.25\,{\rm MeV}} &=& 5.12\,(1\pm 0.30) \times 10^{-45}\,{\rm cm}^2, \nonumber\\
\sigma^{\rm \bar{\nu}_{\rm e} D}_{\rm nc}(E_{\bar{\nu}_{\rm
e}})|_{E_{\bar{\nu}_{\rm e}}=
4.25\,{\rm MeV}}&=& 3.17\,(1\pm 0.30) \times 10^{-44}\,{\rm cm}^2, \nonumber\\
\sigma^{\rm \bar{\nu}_{\rm e} D}_{\rm cc}(E_{\bar{\nu}_{\rm
e}})|_{E_{\bar{\nu}_{\rm e}}=
10\,{\rm MeV}}&=& 0.91\,(1\pm 0.30) \times 10^{-42}\,{\rm cm}^2.
\end{eqnarray}
The PMA data read [42,43]:
\begin{eqnarray}\label{label9.22}
\sigma^{\rm \bar{\nu}_{\rm e} D}_{\rm nc}(E_{\bar{\nu}_{\rm
e}})|_{E_{\bar{\nu}_{\rm e}}=
3.25\,{\rm MeV}}&=&5.92\,\times 10^{-45}\,{\rm cm}^2, \nonumber\\
\sigma^{\rm \bar{\nu}_{\rm e} D}_{\rm cc}(E_{\bar{\nu}_{\rm
e}})|_{E_{\bar{\nu}_{\rm e}}=
4.25\,{\rm MeV}}&=&3.81 \times 10^{-45}\,{\rm cm}^2,\nonumber\\
\sigma^{\rm \bar{\nu}_{\rm e} D}_{\rm cc}(E_{\bar{\nu}_{\rm
e}})|_{E_{\bar{\nu}_{\rm e}}=
10\,{\rm MeV}}&=&1.00 \times 10^{-42}\,{\rm cm}^2.
\end{eqnarray}
It is seen that the RFMD and the PMA data agree themselves within an accuracy better than 17$\%$.

\section{Neutron--proton radiative capture}
\setcounter{equation}{0}

At very low energies the neutron--proton radiative capture n + p $\to$ D + $\gamma$ runs through the magnetic dipole (M1) transition [6]. In the
RFMD the cross section for the neutron--proton radiative capture has been calculated in Ref.~[2,4]. For the local four--nucleon interaction describing the low--energy transition n + p $\to$ n + p the amplitude of the process n + p $\to$ D + $\gamma$ has been obtained in the form [2,4] (the details of the calculation one can find in Appendix F):
\begin{eqnarray}\label{label10.1}
{\cal M}({\rm n}+ {\rm p} \to {\rm D} + \gamma) &=& (\mu_{\rm p}-\mu_{\rm
n})\,\frac{e}{2M_{\rm N}}\,\frac{g_{\rm V}}{4\pi^2}\,G_{\rm \pi
np}\,\varepsilon^{\alpha\beta\mu\nu} k_{\alpha}
e^{\ast}_{\beta}(k)\,e^{\ast}_{\mu}(k_{\rm D})\nonumber\\
&&[\bar{u^c}(p_2)\,(2 k_{\rm D\,\nu} - M_{\rm N}\,\gamma_{\nu})\gamma^5
u(p_1)],
\end{eqnarray}
where $\mu_{\rm p} = 2.793$ and $\mu_{\rm n} = - 1.913$ are the magnetic moments of the proton and the neutron, $e$ is the electric charge of the proton. 

The cross section of the neutron--proton radiative capture obtained in
Refs.~[2,4] reads
\begin{eqnarray}\label{label10.2}
\sigma({\rm n p} \to {\rm D\gamma}) = \frac{1}{v}\,(\mu_{\rm p}-\mu_{\rm
n})^2\,\frac{25}{64}\,\frac{\alpha}{\pi^2}\,\,Q_{\rm D}\,G^2_{\rm \pi
np}\,M_{\rm N}\varepsilon^3_{\rm D} = 276\,{\rm m b}.
\end{eqnarray}
The numerical value has been computed for $\epsilon_{\rm D} = 2.225\,{\rm MeV}$ and $v = 7.34\,\times 10^{-6}$ (the absolute value $v = 2.2\,\times 10^5\,{\rm cm}\,{\rm s}^{-1}$), the laboratory velocity of the neutron. The theoretical value $\sigma({\rm n p} \to {\rm D\gamma}) = 276\,{\rm m b}$ agrees within an accuracy better than 10$\%$ with the theoretical value [6]
\begin{eqnarray}\label{label10.3}
\sigma({\rm n p} \to {\rm D\gamma})_{\rm PMA}\,=\,(302.5\pm 4)\,{\rm mb}
\end{eqnarray}
calculated in the PMA for pure M1 transition ${^1}{\rm S}_0 \to {^3}{\rm S}_1$. It is compatible with the experimental value [53]
\begin{eqnarray}\label{label10.4}
\sigma({\rm n p} \to {\rm D\gamma})_{\exp}\,=\,(334.2\pm 0.5)\,{\rm mb}\,.
\end{eqnarray}
within an accuracy of 9.5$\%$.

Below we revise the process n + p $\to$ D + $\gamma$ in the generalized RFMD. We recalculate the cross section for the  process n + p $\to$ D + $\gamma$ by using the effective four--nucleon interaction defined by Eq.~(\ref{label1.7}).  As has been shown in Appendix F the amplitude of the neutron--proton radiative capture calculated for the effective Yukawa potential Eq.~(\ref{label1.7}) should coincide with the former result obtained for the $\delta^{(3)}(\vec{\rho}\,)$--potential Eq.~(\ref{label1.1}). In fact, in the low--energy limit $K \to 0$, where $K$ is a 3--momentum of a relative movement of the neutron and the proton, the np system is localized in the region of order $O(1/K)$ which is much larger than the range of nuclear forces. Thereby, the wave function of the relative movement of the neutron and the proton can be described by a plane wave, and the amplitude of the neutron--proton radiative capture should not depend on the shape and the range of the nuclear potential [6]. The same consideration has been used for the derivation of the low--energy theorem Eq.~(\ref{label8.7}).

In comparison with the experimental value Eq.~(\ref{label10.4}) the cross section for the neutron--proton radiative capture calculated in the RFMD is  about 18$\%$ of the experimental value less. In order to improve an agreement between the theoretical cross section and the experimental data one should include additional contributions coming from the exchanges by  heavy mesons like the $\rho(770)$ and the $\omega(780)$ mesons, the $\Delta(1230)$ resonance and pion--exchanges defining chiral corrections.  The contributions of the $\rho(770)$, the $\omega(780)$ and the $\Delta(1230)$ resonance exchanges have been taken into account in the PMA by Riska and Brown [54]. In the EFT approach with Chiral perturbation theory the calculation of the cross section for the neutron--proton radiative capture in agreement with the experimental data has been carried out by Park {\it et al.} [55].

In the RFMD the contributions of heavy meson exchanges are taken into account in the phenomenological coupling constant $G_{\rm \pi np}$ given by Eq.~(\ref{label1.8}). Therefore, we can add only the contributions of the $\Delta(1230)$ resonance exchange and chiral corrections induced by pion--exchanges. Unfortunately, the off--mass shell interaction of the $\Delta(1230)$ resonance is parameterized by the parameter $Z$ which ranges values from the region $-0.8 \le Z \le 0.3$ [56]. Therefore, at the present level of the definiteness of the off--mass shell coupling of the $\Delta(1230)$ resonance we cannot make any reliable predictions concerning its contribution to the amplitude of the neutron--proton radiative capture. The contributions of chiral corrections caused by pion--exchanges we are planning to take into account within Chiral perturbation theory incorporated into the RFMD in further development of the RFMD.

The obtained result Eq.~(\ref{label10.2}) can be directly extended on the description of the photomagnetic disintegration of the deuteron [4].

\section{Low--energy elastic NN scattering}
\setcounter{equation}{0}

In this section we would like to show that in the RFMD one can describe low--energy elastic NN scattering in complete agreement with low--energy nuclear phenomenology. For simplicity we suggest to consider the low--energy elastic np scattering in the ${^1}{\rm S}_0$--state. The amplitude of the transition n + p $\to$ n + p can be given as follows
\begin{eqnarray}\label{label11.1}
{\cal M}({\rm n p} \to {\rm n p})(k) = - {\cal A}_{\rm np}(k)\,\frac{4\pi}{M_{\rm N}}\,[\bar{u}(p^{\prime}_1)\gamma^5 u^c(p^{\prime}_2)]\,[\bar{u^c}(p_2)\gamma^5 u(p_1)],
\end{eqnarray}
as for low energies the interaction $\gamma^{\mu}\gamma^5\times \gamma_{\mu}\gamma^5$ reduces to $\gamma^5\times \gamma^5$, then $p_i$ and $p^{\prime}_i$ (i=1,2) are 4--momenta of the proton and the neutron in the initial and final states and $k$ is a relative 3--momentum of the np system. The phenomenological amplitude of the low--energy elastic np scattering ${\cal A}_{\rm np}(k)_{\rm ph}$ reads
\begin{eqnarray}\label{label11.2}
{\cal A}_{\rm np}(k)_{\rm ph} = e^{\displaystyle i \delta_{\rm np}(k)}\frac{\sin\delta_{\rm np}(k)}{k}= \frac{1}{\displaystyle k\,{\rm ctg}\delta_{\rm np}(k) - i k},
\end{eqnarray}
and satisfies the unitarity condition
\begin{eqnarray}\label{label11.3}
{\cal J}\!m{\cal A}_{\rm np}(k)_{\rm ph} = k |{\cal A}_{\rm np}(k)_{\rm ph}|^2.
\end{eqnarray}
Then, $k\,{\rm ctg}\delta_{\rm np}(k)$ obeys the relation Eq.~(\ref{label9.4}). At $k\to 0$ we get ${\cal A}_{\rm np}(0)_{\rm ph} = - 4\pi\,a_{\rm np}/M_{\rm N}$ which gives the cross section $\sigma({\rm np}\to {\rm np}) = 4\pi a^2_{\rm np}$ (see Appendix G).

In the RFMD due to the low--energy reduction
\begin{eqnarray}\label{label11.4}
[\bar{u}(p^{\prime}_1)\gamma_{\alpha}\gamma^5 u^c(p^{\prime}_2)]\,[\bar{u^c}(p_2)\gamma^{\alpha}\gamma^5 u(p_1)]\to - [\bar{u}(p^{\prime}_1)\gamma^5 u^c(p^{\prime}_2)]\,[\bar{u^c}(p_2)\gamma^5 u(p_1)]
\end{eqnarray}
the np scattering comes through the one--nucleon loop exchange. This makes the description of the low--energy elastic np scattering in the RFMD completely different to that in the PMA or in the EFT approach. 

Using the effective interaction Eq.~(\ref{label1.7}) we can write down the effective Lagrangian for the low--energy elastic np scattering:
\begin{eqnarray}\label{label11.5}
\hspace{-0.3in}&&\int d^4x\,{\cal L}^{\rm np \to np}_{\rm scattering}(x) = i\,G^2_{\rm \pi np}\int d^4x d^4z \int d^3\rho \,U(\rho )\int d^3 r\,U(r)\nonumber\\
\hspace{-0.3in}&&\times\,[\bar{n}(x_0,\vec{x} + \frac{1}{2}\vec{\rho}\,)\gamma_{\alpha}
\gamma^5 p^c(x_0,\vec{x}- \frac{1}{2}\vec{\rho}\,)] [\bar{p^c}(z_0,\vec{z}
+ \frac{1}{2}\vec{r}\,)\gamma_{\beta}\gamma^5 n(z_0,\vec{z} -
\frac{1}{2}\vec{r}\,)]\nonumber\\
\hspace{-0.3in}&&\times\,{\rm tr}\{\gamma^{\alpha}\gamma^5 S_{\rm F}(x_0 - z_0,\vec{x} - \vec{z} - \frac{1}{2}\,(\vec{\rho} + \vec{r}\,)) \gamma^{\beta}\gamma^5 S^c_{\rm F}(z_0 - x_0, \vec{z}  - \vec{x} - \frac{1}{2}\,(\vec{r} + \vec{\rho}\,))\}\nonumber\\
\hspace{-0.3in}&& + i\,G^2_{\rm \pi np}\int d^4x d^4z \int d^3\rho \,U(\rho )\int d^3 r\,U(r)\nonumber\\
\hspace{-0.3in}&&\times\,[\bar{n}(x_0,\vec{x} + \frac{1}{2}\vec{\rho}\,) 
\gamma^5 p^c(x_0,\vec{x} - \frac{1}{2}\vec{\rho}\,)] [\bar{p^c}(z_0,\vec{z}
+ \frac{1}{2}\vec{r}\,) \gamma^5 n(z_0,\vec{z} -
\frac{1}{2}\vec{r}\,)]\nonumber\\
\hspace{-0.3in}&&\times\,{\rm tr}\{\gamma^5 S_{\rm F}(x_0 - z_0,\vec{x} - \vec{z} - \frac{1}{2}\,(\vec{\rho}  + \vec{r})) \gamma^5 S^c_{\rm F}(z_0 - x_0, \vec{z} - \vec{x}  - \frac{1}{2}\,(\vec{r}  + \vec{\rho}\,))\}\nonumber\\
\hspace{-0.3in}&& + i\,G^2_{\rm \pi np}\int d^4x d^4z \int d^3\rho \,U(\rho )\int d^3 r\,U(r)\nonumber\\
\hspace{-0.3in}&&\times\,[\bar{n}(x_0,\vec{x} + \frac{1}{2}\vec{\rho}\,)\gamma_{\alpha}
\gamma^5 p^c(x_0,\vec{x}- \frac{1}{2}\vec{\rho}\,)] [\bar{p^c}(z_0,\vec{z}
+ \frac{1}{2}\vec{r}\,)\gamma^5 n(z_0,\vec{z} -
\frac{1}{2}\vec{r}\,)]\nonumber\\
\hspace{-0.3in}&&\times\,{\rm tr}\{\gamma^{\alpha}\gamma^5 S_{\rm F}(x_0 - z_0,\vec{x} - \vec{z} - \frac{1}{2}\,(\vec{\rho} + \vec{r}\,)) \gamma^5 S^c_{\rm F}(z_0 - x_0, \vec{z}  - \vec{x} - \frac{1}{2}\,(\vec{r} + \vec{\rho}\,))\}\nonumber\\
\hspace{-0.3in}&&+ i\,G^2_{\rm \pi np}\int d^4x d^4z \int d^3\rho \,U(\rho )\int d^3 r\,U(r)\nonumber\\
\hspace{-0.3in}&&\times\,[\bar{n}(x_0,\vec{x} + \frac{1}{2}\vec{\rho}\,) \gamma^5 p^c(x_0,\vec{x}- \frac{1}{2}\vec{\rho}\,)] [\bar{p^c}(z_0,\vec{z}
+ \frac{1}{2}\vec{r}\,)\gamma_{\beta}\gamma^5 n(z_0,\vec{z} -
\frac{1}{2}\vec{r}\,)]\nonumber\\
\hspace{-0.3in}&&\times\,{\rm tr}\{\gamma^5 S_{\rm F}(x_0 - z_0,\vec{x} - \vec{z} - \frac{1}{2}\,(\vec{\rho} + \vec{r}\,)) \gamma^{\beta}\gamma^5 S^c_{\rm F}(z_0 - x_0, \vec{z}  - \vec{x} - \frac{1}{2}\,(\vec{r} + \vec{\rho}\,))\}.
\end{eqnarray}
Passing to the momentum representation of the nucleon (anti--nucleon) Green function we obtain
\begin{eqnarray}\label{label11.6}
\hspace{-0.3in}&&\int d^4x\,{\cal L}^{\rm np \to np}_{\rm scattering}(x) = \nonumber\\
\hspace{-0.3in}&&=- \frac{G^2_{\rm \pi np}}{16\pi^2}\int d^4x\int \frac{d^4z d^4q}{(2\pi)^4}\,e^{\textstyle - i q\cdot (x - z)} \int d^3\rho\,U(\rho)\,e^{\displaystyle -i\vec{q}\cdot \vec{\rho}/2}\int d^3r\,U(r)\,e^{\displaystyle - i\vec{q}\cdot \vec{r}/2}\nonumber\\
\hspace{-0.3in}&&\times\,[\bar{n}(x_0,\vec{x} + \frac{1}{2}\vec{\rho}\,)\gamma_{\alpha}
\gamma^5 p^c(x_0,\vec{x} - \frac{1}{2}\vec{\rho}\,)] [\bar{p^c}(z_0,\vec{x}
+ \frac{1}{2}\vec{r}\,)\gamma_{\beta}\gamma^5 n(z_0,\vec{z} -
\frac{1}{2}\vec{r}\,)]\nonumber\\
\hspace{-0.3in}&&\times\,\int \frac{d^4Q}{\pi^2i}\,{\displaystyle e^{\displaystyle -i\vec{Q}\cdot(\vec{\rho} + \vec{r}\,)}}\,{\rm tr}\Bigg\{\gamma^{\alpha}\gamma^5 \frac{1}{M_{\rm N} - \hat{Q}  - \hat{q}}\gamma^{\beta}\gamma^5\frac{1}{M_{\rm N} - \hat{Q}}\Bigg\}\nonumber\\
\hspace{-0.3in}&& - \frac{G^2_{\rm \pi np}}{16\pi^2}\int d^4x \int \frac{d^4z d^4q}{(2\pi)^4}\,e^{\textstyle - i q \cdot (x - z)} \int d^3\rho\,U(\rho)\,e^{\displaystyle -i\vec{q}\cdot \vec{\rho}/2}\int d^3r\,U(r)\,e^{\displaystyle - i\vec{q}\cdot \vec{r}/2}\nonumber\\
\hspace{-0.3in}&&\times\,[\bar{n}(x_0,\vec{x} + \frac{1}{2}\vec{\rho}\,) \gamma^5 p^c(x_0,\vec{x} - \frac{1}{2}\vec{\rho}\,)] [\bar{p^c}(z_0,\vec{z}
+ \frac{1}{2}\vec{r}\,)\gamma^5 n(z_0,\vec{z} -
\frac{1}{2}\vec{r}\,)]\nonumber\\
\hspace{-0.3in}&&\times\,\int \frac{d^4Q}{\pi^2i}\,{\displaystyle e^{\displaystyle -i\vec{Q}\cdot(\vec{\rho} + \vec{r}\,)}}\,{\rm tr}\Bigg\{\gamma^5 \frac{1}{M_{\rm N} - \hat{Q} - \hat{q}}\gamma^5\frac{1}{M_{\rm N} - \hat{Q}}\Bigg\}\nonumber\\
\hspace{-0.3in}&& - \frac{G^2_{\rm \pi np}}{16\pi^2}\int d^4x\int \frac{d^4z d^4q}{(2\pi)^4}\,e^{\textstyle - i q\cdot (x - z)} \int d^3\rho\,U(\rho)\,e^{\displaystyle -i\vec{q}\cdot \vec{\rho}/2}\int d^3r\,U(r)\,e^{\displaystyle - i\vec{q}\cdot \vec{r}/2}\nonumber\\
\hspace{-0.3in}&&\times\,[\bar{n}(x_0,\vec{x} + \frac{1}{2}\vec{\rho}\,)\gamma_{\alpha}
\gamma^5 p^c(x_0,\vec{x} - \frac{1}{2}\vec{\rho}\,)] [\bar{p^c}(z_0,\vec{x}
+ \frac{1}{2}\vec{r}\,)\gamma^5 n(z_0,\vec{z} -
\frac{1}{2}\vec{r}\,)]\nonumber\\
\hspace{-0.3in}&&\times\,\int \frac{d^4Q}{\pi^2i}\,{\displaystyle e^{\displaystyle -i\vec{Q}\cdot(\vec{\rho} + \vec{r}\,)}}\,{\rm tr}\Bigg\{\gamma^{\alpha}\gamma^5 \frac{1}{M_{\rm N} - \hat{Q}  - \hat{q}}\gamma^5\frac{1}{M_{\rm N} - \hat{Q}}\Bigg\}\nonumber\\
\hspace{-0.3in}&&- \frac{G^2_{\rm \pi np}}{16\pi^2}\int d^4x\int \frac{d^4z d^4q}{(2\pi)^4}\,e^{\textstyle - i q\cdot (x - z)} \int d^3\rho\,U(\rho)\,e^{\displaystyle -i\vec{q}\cdot \vec{\rho}/2}\int d^3r\,U(r)\,e^{\displaystyle - i\vec{q}\cdot \vec{r}/2}\nonumber\\
\hspace{-0.3in}&&\times\,[\bar{n}(x_0,\vec{x} + \frac{1}{2}\vec{\rho}\,)
\gamma^5 p^c(x_0,\vec{x} - \frac{1}{2}\vec{\rho}\,)] [\bar{p^c}(z_0,\vec{x}
+ \frac{1}{2}\vec{r}\,)\gamma_{\beta}\gamma^5 n(z_0,\vec{z} -
\frac{1}{2}\vec{r}\,)]\nonumber\\
\hspace{-0.3in}&&\times\,\int \frac{d^4Q}{\pi^2i}\,{\displaystyle e^{\displaystyle -i\vec{Q}\cdot(\vec{\rho} + \vec{r}\,)}}\,{\rm tr}\Bigg\{\gamma^5 \frac{1}{M_{\rm N} - \hat{Q}  - \hat{q}}\gamma^{\beta}\gamma^5\frac{1}{M_{\rm N} - \hat{Q}}\Bigg\}.
\end{eqnarray}
It is convenient to rewrite the r.h.s. of Eq.~(\ref{label11.6}) in terms of the structure functions ${\cal J}^{\alpha\beta} (q; \vec{\rho} + \vec{r}\,)$, ${\cal J}(q; \vec{\rho} + \vec{r}\,)$, ${\cal J}^{\alpha } (q; \vec{\rho} + \vec{r}\,)$ and $\bar{{\cal J}}^{\beta} (q; \vec{\rho} + \vec{r}\,)$: 
\begin{eqnarray}\label{label11.7}
\hspace{-0.3in}&&\int d^4x\,{\cal L}^{\rm np \to np}_{\rm scattering}(x) = \nonumber\\
\hspace{-0.3in}&&=- \frac{G^2_{\rm \pi np}}{16\pi^2}\int d^4x \int \frac{d^4z d^4q}{(2\pi)^4}\,e^{\textstyle - i q\cdot (x - z)} \int d^3\rho\,U(\rho)\,e^{\displaystyle -i\vec{q}\cdot \vec{\rho}/2}\int d^3r\,U(r)\,e^{\displaystyle - i\vec{q}\cdot \vec{r}/2}\nonumber\\
\hspace{-0.3in}&&\times\,[\bar{n}(x_0,\vec{x} + \frac{1}{2}\vec{\rho}\,)\gamma_{\alpha}
\gamma^5 p^c(x_0,\vec{x} - \frac{1}{2}\vec{\rho}\,)] [\bar{p^c}(z_0,\vec{z}
+ \frac{1}{2}\vec{r}\,)\gamma_{\beta}\gamma^5 n(z_0,\vec{z} -
\frac{1}{2}\vec{r}\,)]\,{\cal J}^{\alpha\beta} (q;\vec{\rho} + \vec{r}\,)\nonumber\\
\hspace{-0.3in}&& - \frac{G^2_{\rm \pi np}}{16\pi^2}\int d^4x \int \frac{d^4z d^4q}{(2\pi)^4}\,e^{\textstyle - i q \cdot (x - z)} \int d^3\rho\,U(\rho)\,e^{\displaystyle -i\vec{q}\cdot \vec{\rho}/2}\int d^3r\,U(r)\,e^{\displaystyle - i\vec{q} \cdot \vec{r}/2}\nonumber\\
\hspace{-0.3in}&&\times\,[\bar{n}(x_0,\vec{x} + \frac{1}{2}\vec{\rho}\,) \gamma^5 p^c(x_0,\vec{x} - \frac{1}{2}\vec{\rho}\,)] [\bar{p^c}(z_0,\vec{z}
+ \frac{1}{2}\vec{r}\,)\gamma^5 n(z_0,\vec{z} -
\frac{1}{2}\vec{r}\,)]\,{\cal J}(q;\vec{\rho} + \vec{r}\,)\nonumber\\
\hspace{-0.3in}&&- \frac{G^2_{\rm \pi np}}{16\pi^2}\int d^4x \int \frac{d^4z d^4q}{(2\pi)^4}\,e^{\textstyle - i q\cdot (x - z)} \int d^3\rho\,U(\rho)\,e^{\displaystyle -i\vec{q}\cdot \vec{\rho}/2}\int d^3r\,U(r)\,e^{\displaystyle - i\vec{q}\cdot \vec{r}/2}\nonumber\\
\hspace{-0.3in}&&\times\,[\bar{n}(x_0,\vec{x} + \frac{1}{2}\vec{\rho}\,)\gamma_{\alpha}
\gamma^5 p^c(x_0,\vec{x} - \frac{1}{2}\vec{\rho}\,)] [\bar{p^c}(z_0,\vec{z}
+ \frac{1}{2}\vec{r}\,)\gamma^5 n(z_0,\vec{z} -
\frac{1}{2}\vec{r}\,)]\,{\cal J}^{\alpha} (q;\vec{\rho} + \vec{r}\,)\nonumber\\
\hspace{-0.3in}&&- \frac{G^2_{\rm \pi np}}{16\pi^2}\int d^4x \int \frac{d^4z d^4q}{(2\pi)^4}\,e^{\textstyle - i q\cdot (x - z)} \int d^3\rho\,U(\rho)\,e^{\displaystyle -i\vec{q}\cdot \vec{\rho}/2}\int d^3r\,U(r)\,e^{\displaystyle - i\vec{q}\cdot \vec{r}/2}\nonumber\\
\hspace{-0.3in}&&\times\,[\bar{n}(x_0,\vec{x} + \frac{1}{2}\vec{\rho}\,)
\gamma^5 p^c(x_0,\vec{x} - \frac{1}{2}\vec{\rho}\,)] [\bar{p^c}(z_0,\vec{z}
+ \frac{1}{2}\vec{r}\,)\gamma_{\beta}\gamma^5 n(z_0,\vec{z} -
\frac{1}{2}\vec{r}\,)]\,\bar{{\cal J}}^{\beta} (q;\vec{\rho} + \vec{r}\,),
\end{eqnarray}
where the structure functions ${\cal J}^{\alpha\beta} (q; \vec{\rho} + \vec{r}\,)$, ${\cal J}(q; \vec{\rho} + \vec{r}\,)$, ${\cal J}^{\alpha } (q; \vec{\rho} + \vec{r}\,)$ and $\bar{{\cal J}}^{\beta} (q; \vec{\rho} + \vec{r}\,)$  are defined by the momentum integrals
\begin{eqnarray}\label{label11.8}
{\cal J}^{\alpha\beta} (q; \vec{\rho} + \vec{r}\,) &=& \int \frac{d^4Q}{\pi^2i}\,{\displaystyle e^{\displaystyle -i\vec{Q}\cdot(\vec{\rho} + \vec{r}\,)}}\,{\rm tr}\Bigg\{\frac{1}{M_{\rm N} - \hat{Q}  - \hat{q}}\gamma^{\alpha}\gamma^5 \frac{1}{M_{\rm N} - \hat{Q}}\gamma^{\beta}\gamma^5\Bigg\},\nonumber\\
{\cal J}(q; \vec{\rho} + \vec{r}\,) &=& \int \frac{d^4q}{\pi^2i}\,{\displaystyle e^{\displaystyle -i\vec{Q}\cdot(\vec{\rho} + \vec{r}\,)}}\,{\rm tr}\Bigg\{\frac{1}{M_{\rm N} - \hat{Q}  - \hat{q}}\gamma^5 \frac{1}{M_{\rm N} - \hat{Q}}\gamma^5\Bigg\},\nonumber\\
{\cal J}^{\alpha} (q; \vec{\rho} + \vec{r}\,) &=& \int \frac{d^4Q}{\pi^2i}\,{\displaystyle e^{\displaystyle -i\vec{Q}\cdot(\vec{\rho} + \vec{r}\,)}}\,{\rm tr}\Bigg\{\frac{1}{M_{\rm N} - \hat{Q}  - \hat{q}}\gamma^{\alpha}\gamma^5 \frac{1}{M_{\rm N} - \hat{Q}}\gamma^5\Bigg\},\nonumber\\
\bar{{\cal J}}^{\beta} (q; \vec{\rho} + \vec{r}\,) &=& \int \frac{d^4Q}{\pi^2i}\,{\displaystyle e^{\displaystyle -i\vec{Q}\cdot(\vec{\rho} + \vec{r}\,)}}\,{\rm tr}\Bigg\{\frac{1}{M_{\rm N} - \hat{Q}  - \hat{q}} \gamma^5 \frac{1}{M_{\rm N} - \hat{Q}}\gamma^{\beta}\gamma^5\Bigg\}.
\end{eqnarray}
Due to the low--energy reduction
\begin{eqnarray}\label{label11.9}
\hspace{-0.3in}&&[\bar{n}(x_0,\vec{x} + \frac{1}{2}\vec{\rho}\,)\gamma_{\alpha}
\gamma^5 p^c(x_0,\vec{x} - \frac{1}{2}\vec{\rho}\,)] \to  g_{\alpha 0} [\bar{n}(x_0,\vec{x} + \frac{1}{2}\vec{\rho}\,) \gamma^5 p^c(x_0,\vec{x} - \frac{1}{2}\vec{\rho}\,)],\nonumber\\
\hspace{-0.3in}&&[\bar{p^c}(z_0,\vec{z}
+ \frac{1}{2}\vec{r}\,)\gamma_{\beta}\gamma^5 n(z_0,\vec{z} -
\frac{1}{2}\vec{r}\,)] \to  - g_{\beta 0}[\bar{p^c}(z_0,\vec{z}
+ \frac{1}{2}\vec{r}\,)\gamma^5 n(z_0,\vec{z} -
\frac{1}{2}\vec{r}\,)]
\end{eqnarray}
the r.h.s. of Eq.~(\ref{label11.7}) can be brought up to the form
\begin{eqnarray}\label{label11.10}
\hspace{-0.3in}&&\int d^4x\,{\cal L}^{\rm np \to np}_{\rm scattering}(x) = \nonumber\\
\hspace{-0.3in}&&= - \frac{G^2_{\rm \pi np}}{16\pi^2}\int d^4x \int \frac{d^4z d^4q}{(2\pi)^4}\,e^{\textstyle - i q \cdot (x - z)} \int d^3\rho\,U(\rho)\,e^{\displaystyle -i\vec{q}\cdot \vec{\rho}/2}\int d^3r\,U(r)\,e^{\displaystyle - i\vec{q} \cdot \vec{r}/2}\nonumber\\
\hspace{-0.3in}&&\times\,[\bar{n}(x_0,\vec{x} + \frac{1}{2}\vec{\rho}\,) \gamma^5 p^c(x_0,\vec{x} - \frac{1}{2}\vec{\rho}\,)] [\bar{p^c}(z_0,\vec{z}
+ \frac{1}{2}\vec{r}\,)\gamma^5 n(z_0,\vec{z} -
\frac{1}{2}\vec{r}\,)]\,\nonumber\\
\hspace{-0.3in}&&\times\,[ - {\cal J}^{00} (q; \vec{\rho} + \vec{r}\,) + {\cal J}(q;\vec{\rho} + \vec{r}\,) + {\cal J}^{0} (q; \vec{\rho} + \vec{r}\,) - \bar{{\cal J}}^{0} (q; \vec{\rho} + \vec{r}\,)]. 
\end{eqnarray}
The amplitude of the low--energy elastic np scattering we define as
\begin{eqnarray}\label{label11.11}
\hspace{-0.3in}&&\int d^4x\,<p(p^{\prime}_2) n(p^{\prime}_1)|{\cal L}^{\rm np \to np}_{\rm scattering}(x)|n(p_1) p(p_2)> = \nonumber\\
\hspace{-0.3in}&&= (2\pi)^4\delta^{(4)}(p^{\prime}_2 + p^{\prime}_1 - p_2 - p_1)\,\frac{{\cal M}({\rm n + p \to n + p})}{\displaystyle \sqrt{2E^{\prime}_1 V\,2 E^{\prime}_2 V \,2 E_1 V\,2 E_2 V}},
\end{eqnarray}
where $|n(p_1) p(p_2)>$ and $<p(p^{\prime}_2) n(p^{\prime}_1)|$ are the wave functions of the initial and the final states, $E_i(E^{\prime}_i)\,(i=1,2)$ are the energies of the initial (final) neutron and proton, and $V$ is the normalization volume.

The matrix element of the four--nucleon operator between the initial $|n(p_1) p(p_2)>$ and the final $<p(p^{\prime}_2) n(p^{\prime}_1)|$ states amounts to
\begin{eqnarray}\label{label11.12}
&&<p(p^{\prime}_2) n(p^{\prime}_1)|[\bar{n}(x_0,\vec{x} + \frac{1}{2}\vec{\rho}\,) \gamma^5 p^c(x_0,\vec{x} - \frac{1}{2}\vec{\rho}\,)] \nonumber\\
&&[\bar{p^c}(z_0,\vec{z}
+ \frac{1}{2}\vec{r}\,)\gamma^5 n(z_0,\vec{z} -
\frac{1}{2}\vec{r}\,)]|n(p_1) p(p_2)> = \nonumber\\
&&<p(p^{\prime}_2) n(p^{\prime}_1)|[\bar{n}(x_0,\vec{x} + \frac{1}{2}\vec{\rho}\,) \gamma^5 p^c(x_0,\vec{x} - \frac{1}{2}\vec{\rho}\,)]|0>\nonumber\\
&&<0|[\bar{p^c}(z_0,\vec{z}
+ \frac{1}{2}\vec{r}\,)\gamma^5 n(z_0,\vec{z} -
\frac{1}{2}\vec{r}\,)]|n(p_1) p(p_2)> = \nonumber\\
&&= [\bar{u}(p^{\prime}_1)\gamma^5 u^c(p^{\prime}_2)]\,[\bar{u^c}(p_2)\gamma^5 u(p_1)]\,\times\,\psi^*_{\rm np}(\rho)_{\rm out}\,\times\,\psi_{\rm np}(r)_{\rm in}\nonumber\\
&&\times\,\frac{\displaystyle e^{\displaystyle i(p^{\prime}_2 + p^{\prime}_1)\cdot x}e^{\displaystyle - i(p_2 + p_1)\cdot z}}{\displaystyle \sqrt{2E^{\prime}_1 V\,2 E^{\prime}_2 V \,2 E_1 V\,2 E_2 V}},
\end{eqnarray}
where $\psi_{\rm np}(r)_{\rm in}$ and $\psi^*_{\rm np}(\rho)_{\rm out}$ are the wave functions of the initial (in) and the final (out) ${^1}{\rm S}_0$--state of the np system which we set equal to
\begin{eqnarray}\label{label11.13}
\hspace{-0.3in}\psi_{\rm np}(r)_{\rm in} &=& e^{\displaystyle i\delta_{\rm np}(k)}\sin \delta_{\rm np}(k) \frac{v_{\rm np}(0)}{kr},\nonumber\\
\hspace{-0.3in}\psi^*_{\rm np}(\rho)_{\rm out} &=& \frac{\sin k\rho}{k\rho}.
\end{eqnarray}
For the justification of this choice of the wave functions of the initial and final states we refer to the quantum mechanical description of the np scattering. Indeed, suppose that the neutron and the proton couple through the potential $U_{\rm eff}(\rho)$.  Schr\"odinger equation with the potential $U_{\rm eff}(\rho)$ reads
\begin{eqnarray}\label{label11.14}
(\bigtriangleup + k^2)\,\psi_{\rm np}(\rho) = M_{\rm N}\,U_{\rm eff}(\rho)\,\psi_{\rm np}(\rho),
\end{eqnarray}
where $\psi_{\rm np}(\rho)$ is the solution of Eq.~(\ref{label11.14}). The amplitude of the np scattering is then defined as
\begin{eqnarray}\label{label11.15}
f_{\rm np}(k) = - \frac{M_{\rm N}}{4\pi}\int d^3\rho\,e^{\displaystyle - i\vec{k}^{\,\prime}\cdot \vec{\rho}}\,U_{\rm eff}(\rho)\,\psi_{\rm np}(\rho),
\end{eqnarray}
where $e^{\textstyle - i\vec{k}^{\,\prime}\cdot \vec{\rho}}$ is the wave function of the final state of the np scattering and $\vec{k}^{\,\prime}$ is a relative momentum of the np system in the final state. If finally the np system is in the ${^1}{\rm S}_0$--state, we should expand the exponential into spherical harmonics and hold only the S--wave contribution [8]. This changes the amplitude $f_{\rm np}(k)$ given by Eq.~(\ref{label11.15}) as follows
\begin{eqnarray}\label{label11.16}
f_{\rm np}(k) = - \frac{M_{\rm N}}{4\pi}\int d^3\rho\,\frac{\sin k\rho}{k\rho}\,U_{\rm eff}(\rho)\,\psi_{\rm np}(\rho),
\end{eqnarray}
where we have set $|\vec{k}^{\prime}\,|= k$. The  wave function $\psi_{\rm np}(\rho)$ we have taken in the form Eq.~(\ref{label9.3}) having been applied to the description of the process $\nu_{\rm e}$ + D $\to$ $\nu_{\rm e}$ + n + p.

Substituting Eq.~(\ref{label11.12}) in the l.h.s. of Eq.~(\ref{label11.11}) and integrating over $x$, $z$ and $q$ we arrive at the amplitude of the low--energy elastic np scattering 
\begin{eqnarray}\label{label11.17}
\hspace{-0.3in}&&{\cal M}({\rm n + p} \to {\rm n + p}) = - \,e^{\displaystyle i\delta_{\rm np}(k)}\frac{\sin \delta_{\rm np}(k)}{k}\,\frac{4\pi}{M_{\rm N}}\,[\bar{u}(p^{\prime}_1)\gamma^5 u^c(p^{\prime}_2)] \,[\bar{u^c}(p_2)\gamma^5 u(p_1)]\,\nonumber\\
\hspace{-0.3in}&&\times\,v_{\rm np}(0)\,\frac{M_{\rm N}G^2_{\rm \pi np}}{64\pi^3}\int d^3\rho\,U(\rho)\,\frac{\sin k\rho}{k\rho}\int \frac{d^3r}{r}\,U(r)\,[ - {\cal J}^{00} (P; \vec{\rho} + \vec{r}\,) + {\cal J}(P;\vec{\rho} + \vec{r}\,)\nonumber\\
\hspace{-0.3in}&&+ {\cal J}^{0} (P; \vec{\rho} + \vec{r}\,) - \bar{{\cal J}}^{0} (P; \vec{\rho} + \vec{r}\,)],
\end{eqnarray}
where $P = p_1 + p_2 = p^{\prime}_1 + p^{\prime}_2$ and in the center of mass frame $P^{\mu} = (2\sqrt{k^2 + M^2_{\rm N}}, \vec{0}\,)$.

The r.h.s. of Eq.~(\ref{label11.17}) can be represented in the more convenient form
\begin{eqnarray}\label{label11.18}
{\cal M}({\rm n + p} \to {\rm n + p}) = - {\cal A}_{\rm np}(k)_{\rm RFMD}\,\frac{4\pi}{M_{\rm N}}\,[\bar{u}(p^{\prime}_1)\gamma^5 u^c(p^{\prime}_2)] \,[\bar{u^c}(p_2)\gamma^5 u(p_1)].
\end{eqnarray}
where the amplitude ${\cal A}_{\rm np}(k)_{\rm RFMD}$ is defined by
\begin{eqnarray}\label{label11.19}
{\cal A}_{\rm np}(k)_{\rm RFMD} = \kappa_{\rm np}(k)\,{\cal A}_{\rm np}(k)_{\rm ph}.
\end{eqnarray}
Thus, the parameter $\kappa_{\rm np}(k)$, determined by the expression
\begin{eqnarray}\label{label11.20}
\hspace{-0.5in}&&\kappa_{\rm np}(k) = v_{\rm np}(0)\,\frac{M_{\rm N}G^2_{\rm \pi np}}{64\pi^3}\int d^3\rho\,U(\rho)\,\frac{\sin k\rho}{k\rho}\int \frac{d^3r}{r}\,U(r)\nonumber\\
\hspace{-0.3in}&&\times\,[ - {\cal J}^{00} (P; \vec{\rho} + \vec{r}\,) + {\cal J}(P;\vec{\rho} + \vec{r}\,) + {\cal J}^{0} (P; \vec{\rho} + \vec{r}\,) - \bar{{\cal J}}^{0} (P; \vec{\rho} + \vec{r}\,)],
\end{eqnarray}
distinguishes the amplitude of the low--energy elastic np scattering ${\cal A}_{\rm np}(k)_{\rm RFMD}$ from the phenomenological amplitude ${\cal A}_{\rm np}(k)_{\rm ph}$. Therefore, we have to focus on the analysis of $\kappa_{\rm np}(k)$.

It is well--known [25,4] that the structure functions ${\cal J}^{\alpha\beta}(P;\vec{0})$ and  ${\cal J}(P;\vec{0})$ are ambiguously defined under shifts of virtual momenta $Q\to Q + a P$, where $a$ is an arbitrary parameter, if the cut--off regularization is applied. Following the prescription suggested in Ref.~[25] we obtain [4]:
\begin{eqnarray}\label{label11.21}
\delta{\cal J}^{\alpha\beta}(P;\vec{0}) &=& {\cal J}^{\alpha\beta}(P;\vec{0}; a P) - {\cal J}^{\alpha\beta}(P;\vec{0}) = 2\,a(a+1)\,(2\,P^{\alpha}P^{\beta} - P^2\,g^{\alpha\beta}),\nonumber\\
\delta{\cal J}(P;\vec{0}) &=& {\cal J}(P;\vec{0}; a P) - {\cal J}(P;\vec{0})= -\,2\,a(a+1)\,P^2.
\end{eqnarray}
These are the exact non--perturbative relations [25,4].

In our case only the structure function  ${\cal J}^{00} (P; \vec{\rho} + \vec{r}\,)$ is ambiguously defined under the shift of the time component of the virtual momentum, i.e.,  $Q_0\to Q_0 + a P_0$. Due to the shift $Q_0\to Q_0 + a P_0$ the structure function ${\cal J}^{00} (P; \vec{\rho} + \vec{r}\,)$ acquires the contribution:
\begin{eqnarray}\label{label11.22}
\hspace{-0.5in}&&\frac{1}{4}\,\delta {\cal J}^{00} (P; \vec{\rho} + \vec{r}\,) = \int \frac{d^4Q}{\pi^2i}\,{\displaystyle e^{\displaystyle -i\vec{Q}\cdot(\vec{\rho} + \vec{r}\,)}}\nonumber\\
\hspace{-0.5in}
&&\times\,\frac{- 2M^2_{\rm N} + E^2_{\vec{Q}} + Q^2_0 + (2a+1)\,Q_0P_0 + a(a+1)\,P^2_0}{[E^2_{\vec{Q}} - Q^2_0 - 2(a+1)\,Q_0P_0 - (a+1)^2P^2_0 - i0][E^2_{\vec{Q}} - Q^2_0 - 2a\,Q_0P_0 - a^2P^2_0 - i0]}\nonumber\\
\hspace{-0.5in}
&&-\int \frac{d^4Q}{\pi^2i}\,{\displaystyle e^{\displaystyle -i\vec{Q}\cdot(\vec{\rho} + \vec{r}\,)}} \,
\frac{- 2M^2_{\rm N} + E^2_{\vec{Q}} + Q^2_0 + Q_0P_0}{[E^2_{\vec{Q}} - Q^2_0 - 2\,Q_0 P_0 - P^2_0 - i0]
[E^2_{\vec{Q}} - Q^2_0 - i0]}=\nonumber\\
\hspace{-0.5in}&& = \int \frac{d^4Q}{\pi^2i}\,{\displaystyle e^{\displaystyle -i\vec{Q}\cdot(\vec{\rho} + \vec{r}\,)}}\,\frac{- 2M^2_{\rm N} + E^2_{\vec{Q}} + Q^2_0 + (2a+1)\,Q_0P_0 + a(a+1)\,P^2_0}{[E^2_{\vec{Q}} - Q^2_0 - i0]^2}\nonumber\\
\hspace{-0.5in}
&&\times\,\Bigg\{1 + 2(2a+1)\frac{P_0Q_0}{[E^2_{\vec{Q}} - Q^2_0 - i0]} +  (2a^2 + 2a + 1)\frac{P^2_0}{[E^2_{\vec{Q}} - Q^2_0 - i0]}\nonumber\\
\hspace{-0.5in}&& + 4(2a+1)^2 \frac{P^2_0 Q^2_0}{[E^2_{\vec{Q}} - Q^2_0 - i0]^2}+\ldots\Bigg\}\nonumber\\
\hspace{-0.5in}&&- \int \frac{d^4Q}{\pi^2i}\,{\displaystyle e^{\displaystyle -i\vec{Q}\cdot(\vec{\rho} + \vec{r}\,)}}\,\frac{- 2M^2_{\rm N} + E^2_{\vec{Q}} + Q^2_0 + Q_0P_0}{[E^2_{\vec{Q}} - Q^2_0 - i0]^2}\nonumber\\
\hspace{-0.5in}&&
\times\,\Bigg\{1 + 2\,\frac{P_0Q_0}{[E^2_{\vec{Q}} - Q^2_0 - i0]} +  \frac{P^2_0}{[E^2_{\vec{Q}} - Q^2_0 - i0]} +4\,\frac{P^2_0 Q^2_0}{[E^2_{\vec{Q}} - Q^2_0 - i0]^2}+\ldots\Bigg\}=\nonumber\\
\hspace{-0.5in}&&= a\,(a + 1)\,P^2_0\int \frac{d^4Q}{\pi^2i}\,{\displaystyle e^{\displaystyle -i\vec{Q}\cdot(\vec{\rho} + \vec{r}\,)}}\,\frac{1}{\displaystyle [E^2_{\vec{Q}} - Q^2_0 - i0]^2}\nonumber\\
&&+ 8\,a(a+1)\,P^2_0\int \frac{d^4Q}{\pi^2i}\,{\displaystyle e^{\displaystyle -i\vec{Q}\cdot(\vec{\rho} + \vec{r}\,)}}\,\frac{Q^2_0}{\displaystyle [E^2_{\vec{Q}} - Q^2_0 - i0]^3}\nonumber\\
&&+ 2\,a(a+1)\,P^2_0 \int \frac{d^4Q}{\pi^2i}\,{\displaystyle e^{\displaystyle -i\vec{Q}\cdot(\vec{\rho} + \vec{r}\,)}}\,\frac{E^2_{\vec{Q}} + Q^2_0}{\displaystyle [E^2_{\vec{Q}} - Q^2_0 - i0]^3}\nonumber\\
&&+ 16\,a(a+1)\,P^2_0  \int \frac{d^4Q}{\pi^2i}\,{\displaystyle e^{\displaystyle -i\vec{Q}\cdot(\vec{\rho} + \vec{r}\,)}}\,\frac{Q^2_0(E^2_{\vec{Q}} + Q^2_0)}{\displaystyle [E^2_{\vec{Q}} - Q^2_0 - i0]^4}\nonumber\\
&&- 4\,a(a+1)\,M^2_{\rm N}\,P^2_0 \int \frac{d^4Q}{\pi^2i}\,{\displaystyle e^{\displaystyle -i\vec{Q}\cdot(\vec{\rho} + \vec{r}\,)}}\,\frac{1}{\displaystyle [E^2_{\vec{Q}} - Q^2_0 - i0]^3}\nonumber\\
&&- 32\,a(a+1)\,M^2_{\rm N}\,P^2_0\int \frac{d^4Q}{\pi^2i}\,{\displaystyle e^{\displaystyle -i\vec{Q}\cdot(\vec{\rho} + \vec{r}\,)}}\,\frac{Q^2_0}{\displaystyle [E^2_{\vec{Q}} - Q^2_0 - i0]^4}
\end{eqnarray}
Making a Wick rotation $Q_0 = iQ_4$ and integrating over $Q_4$ we get
\begin{eqnarray}\label{label11.23}
\delta {\cal J}^{00} (P; \vec{\rho} + \vec{r}\,) &=&  2\,a\,(a + 1)\,P^2_0 \int \frac{d^3Q}{\pi}\,{\displaystyle e^{\displaystyle -i\vec{Q}\cdot(\vec{\rho} + \vec{r}\,)}}\,\frac{M^2_{\rm N}}{\displaystyle E^5_{\vec{Q}}}=\nonumber\\
&=&  2\,a\,(a + 1)\,P^2_0 \int \frac{d^3Q}{\pi}\,{\displaystyle e^{\displaystyle -i\vec{Q}\cdot(\vec{\rho} + \vec{r}\,)}}\,\frac{M^2_{\rm N}}{\displaystyle (M^2_{\rm N} + \vec{Q}^{\,2})^{5/2}}.
\end{eqnarray}
Dropping divergent contributions appearing for the computation of the structure functions ${\cal J}^{00} (P; \vec{\rho} + \vec{r}\,)$ and ${\cal J}(P; \vec{\rho} + \vec{r}\,)$ we obtain
\begin{eqnarray}\label{label11.24}
\hspace{-0.5in}&&\kappa_{\rm np}(k) = v_{\rm np}(0)\,\frac{M_{\rm N}G^2_{\rm \pi np}}{64\pi^3}\int d^3\rho\,U(\rho)\,\frac{\sin k\rho}{k\rho}\int \frac{d^3r}{r}\,U(r)\,\Bigg[ (-2\,a^2 - 2\,a + 1)\,\nonumber\\
\hspace{-0.5in}
&&\times\,P^2_0 \int\frac{d^3Q}{\pi}\,{\displaystyle e^{\displaystyle -i\vec{Q}\cdot(\vec{\rho} + \vec{r}\,)}}\frac{M^2_{\rm N}}{\displaystyle E^5_{\vec{Q}}} + (2\,M_{\rm N} + P_0)^2 \int\frac{d^3Q}{\pi}\,{\displaystyle e^{\displaystyle -i\vec{Q}\cdot(\vec{\rho} + \vec{r}\,)}}\frac{1}{\displaystyle E^3_{\vec{Q}}}\Bigg],
\end{eqnarray}
where we have used that 
\begin{eqnarray}\label{label11.25}
{\cal J}^{0} (P; \vec{\rho} + \vec{r}\,) - \bar{{\cal J}}^{0} (P; \vec{\rho} + \vec{r}\,) = 4\,M_{\rm N}\,P_0 \int\frac{d^3Q}{\pi}\,{\displaystyle e^{\displaystyle -i\vec{Q}\cdot(\vec{\rho} + \vec{r}\,)}}\frac{1}{\displaystyle E^3_{\vec{Q}}}.
\end{eqnarray}
Since $k \ll M_{\rm N}$ and $|\vec{Q}\,| \sim M_{\rm N}$, the main regions of the integration over $\rho$ and $r$ are restricted by inequalities: $r, \rho \le 1/M_{\rm N}$. Hence, we can neglect the $k$--dependence of the parameter $\kappa_{\rm np}(k)$ and define it as follows:
\begin{eqnarray}\label{label11.26}
\hspace{-0.5in}&&\kappa_{\rm np}(k) \simeq \kappa_{\rm np} = v_{\rm np}(0)\,\frac{M^3_{\rm N}G^2_{\rm \pi np}}{16\pi^3}\int d^3\rho\,U(\rho)\int \frac{d^3r}{r}\,U(r)\,\Bigg[ (-2\,a^2 - 2\,a + 1)\,\nonumber\\
\hspace{-0.5in}
&&\times\,
\int\frac{d^3Q}{\pi}\,
{\displaystyle e^{\displaystyle 
-i\vec{Q}\cdot(\vec{\rho} + \vec{r}\,)}}\frac{M^2_{\rm N}\,}{\displaystyle E^5_{\vec{Q}}} + 4\int\frac{d^3Q}{\pi}\,{\displaystyle 
e^{\displaystyle -i\vec{Q}\cdot(\vec{\rho} + \vec{r}\,)}}
\frac{1}{\displaystyle E^3_{\vec{Q}}}\Bigg].
\end{eqnarray}
Integrating over $\vec{\rho}$ and $\vec{r}$ we get
\begin{eqnarray}\label{label11.27}
\hspace{-0.5in}\kappa_{\rm np} &=& v_{\rm np}(0)\,\frac{M^4_{\pi} G^2_{\rm \pi np}}{2\pi^3}\Bigg[\Bigg( - a^2 - a + \frac{1}{2}\Bigg)\,\int\limits^{\infty}_0\frac{dQ Q}{Q^2 + M^2_{\pi}}{\rm arctg}\frac{Q}{M_{\pi}}
\frac{M^5_{\rm N}}{\displaystyle (Q^2 + M^2_{\rm N})^{5/2}}
\nonumber\\
&&\hspace{2.1in}+ 2\int\limits^{\infty}_0\frac{dQ Q}{Q^2 + M^2_{\pi}}{\rm arctg}\frac{Q}{M_{\pi}}
\frac{M^3_{\rm N}}
{\displaystyle (Q^2 + M^2_{\rm N})^{3/2}}\Bigg].
\end{eqnarray}
Computing numerically the integrals over $Q$ which amount to
\begin{eqnarray}\label{label11.28}
\hspace{-0.5in}\int\limits^{\infty}_0\frac{dQ Q}{Q^2 + M^2_{\pi}}{\rm arctg}\frac{Q}{M_{\pi}}\frac{M^5_{\rm N}}{\displaystyle (Q^2 + M^2_{\rm N})^{5/2}}&=& 1.3925,\nonumber\\
\hspace{-0.5in}\int\limits^{\infty}_0\frac{dQ Q}{Q^2 + M^2_{\pi}}{\rm arctg}\frac{Q}{M_{\pi}}\frac{M^3_{\rm N}}{\displaystyle (Q^2 + M^2_{\rm N})^{3/2}}&=&1.8056
\end{eqnarray}
and using the numerical values of the parameters we get
\begin{eqnarray}\label{label11.29}
\kappa_{\rm np} = 22.331\,( - a^2 - a + 3.093).
\end{eqnarray}
The amplitude ${\cal A}_{\rm np}(k)_{\rm RFMD}$ as well as the phenomenological amplitude ${\cal A}_{\rm np}(k)_{\rm ph}$ should satisfy the unitarity condition
\begin{eqnarray}\label{label11.30}
{\cal J}\!m{\cal A}_{\rm np}(k)_{\rm RFMD} = k |{\cal A}_{\rm np}(k)_{\rm RFMD}|^2.
\end{eqnarray}
Substituting Eq.~(\ref{label11.2}) in Eq.~(\ref{label11.30}) and using the relation Eq.~(\ref{label11.3}) we can fix the value of the parameter $\kappa_{\rm np}$. We get the equation $\kappa_{\rm np}(\kappa_{\rm np} - 1) = 0$, the non--trivial solution of which gives $\kappa_{\rm np} = 1$. From Eq.~(\ref{label11.29}) we obtain that such a solution, $\kappa_{\rm np} = 1$, always exists, for example, for $a = 1.316$ or $a = -2.316$.

Thus, due to the unitarity condition we have fixed the ambiguities of the calculation of the momentum integrals coming form the one--nucleon loop diagrams and defining the amplitude of the low--energy elastic np scattering. As a result we obtain the amplitude of the low--energy elastic np scattering in the phenomenological form
\begin{eqnarray}\label{label11.31}
\hspace{-0.5in}{\cal A}_{\rm np}(k)_{\rm RFMD}= e^{\displaystyle i \delta_{\rm np}(k)}\frac{\sin\delta_{\rm np}(k)}{k}= \frac{1}{\displaystyle k\,{\rm ctg}\delta_{\rm np}(k) - i k}=\frac{1}{\displaystyle - \frac{1}{a_{\rm np}} + \frac{1}{2}\,r_{\rm np}\,k^2 - i\,k}.
\end{eqnarray}
By analogy with the amplitude of the low--energy elastic np scattering we can describe in the RFMD the amplitudes of the low--energy elastic pp scattering and nn scattering in full agreement with low--energy nuclear phenomenology. This completes the description of low--energy elastic NN scattering in the RFMD.

\section{Conclusion}
\setcounter{equation}{0}

We have shown that in comparison with the PMA and EFT approach the generalized RFMD applied to the description of the solar neutrino processes gives  completely new predictions for the processes containing two protons in the initial state, the solar proton burning p + p $\to$ D + e$^+$ + $\nu_{\rm e}$,  and in the final state, the neutrino disintegration of the deuteron $\nu_{\rm e}$ + D $\to$ e$^-$ + p + p.

For the astrophysical factor $S_{\rm pp}(0)$ of the solar proton burning p + p $\to$ D + e$^+$ + $\nu_{\rm e}$ we have obtained the value  $S_{\rm pp}(0) = 5.52\,\times 10^{-25}\,{\rm MeV\,\rm b}$ which is enhanced by a factor of $1.42\,$ with respect to the classical value $S^*_{\rm pp}(0) = 3.89\,\times 10^{-25}\,{\rm MeV\,\rm b}$ obtained by Kamionkowski and Bahcall in the PMA [7]. The Coulomb repulsion between two protons is taken into account in terms of the S--wave scattering length $a^{\rm e}_{\rm pp}$ of the low--energy elastic pp scattering in the ${^1}{\rm S}_0$--state and the Gamow penetration factor $C(\eta) = \textstyle \sqrt{2\pi\eta} \,\exp(- \pi\eta)$, where $\eta =\alpha/v$ and $\alpha =1/137$ and $v$ are the fine structure constant and a relative velocity of two protons.

We argue that the value of the astrophysical factor $S_{\rm pp}(0)$ calculated in the generalized RFMD is due to the explicit account for the Coulomb repulsion between the protons in terms of the phenomenological parameters of the low--energy elastic pp scattering and dynamics of low--energy nuclear forces induced by quantum fluctuations of nucleon fields through the one--nucleon loop exchanges.

Suppose, we have switched off the Coulomb repulsion. In this case the factor ${\cal F}^{\rm e}_{\rm pp}$ reduces itself to the factor ${\cal F}^{\rm NF}_{\rm pp}$, caused by the contribution of nuclear forces only, which reads
\begin{eqnarray}\label{label12.1}
\hspace{-0.5in}{\cal F}^{\rm NF}_{\rm pp} &=&-\sqrt{2}\,a_{\rm pp}\,v_{\rm pp}(0)\, \frac{28}{27}\,\Bigg[\frac{M^2_{\pi}}{\sqrt{M^2_{\rm N} - M^2_{\pi}}}\,{\rm arctg}\frac{\sqrt{M^2_{\rm N} - M^2_{\pi}}}{M_{\pi}} + \frac{8}{7}\,\frac{M^3_{\pi}}{M^2_{\rm N} - M^2_{\pi}}\nonumber\\
&&- \frac{8}{7}\,\frac{M^4_{\pi}}{(M^2_{\rm N} - M^2_{\pi})^{3/2}}\,{\rm arctg}\frac{\sqrt{M^2_{\rm N} - M^2_{\pi}}}{M_{\pi}}\Bigg] = 1.09,
\end{eqnarray}
where $a_{\rm pp} = - 17.1\,{\rm fm}$ [5] and $v_{\rm pp}(0) = 0.28$ given by Eq.~(\ref{label9.10}). This yields the astrophysical factor equal
\begin{eqnarray}\label{label12.2}
\hspace{-0.5in}S^{\rm NF}_{\rm pp}(0) &=&\alpha\,\frac{9g^2_{\rm A}G^2_{\rm V}Q_{\rm
D}M^4_{\rm N}}{2560\pi^4}\,G^2_{\rm \pi pp}\,|{\cal
F}^{\rm NF}_{\rm pp}|^2\,W^5\,f\Bigg(\frac{m_{\rm
e}}{W}\Bigg) =  2.07\,\times 10^{-25}\,{\rm MeV\,\rm b},
\end{eqnarray}
The value $S^{\rm NF}_{\rm pp}(0) = 2.07\,\times 10^{-25}\,{\rm MeV\,\rm b}$ makes up 53$\%$ of the classical value $S^*_{\rm pp}(0) = 3.89\times 10^{-25}\,{\rm MeV\,\rm b}$ by Kamionkowski and Bahcall [7]. However, as has been stated by Kamionkowski and Bahcall [7] the contribution of the strong interactions between two protons should be smaller than 40$\%$ of the classical value, i.e., $S^{* \rm NF}_{\rm pp}(0) <  1.56\times 10^{-25}\,{\rm MeV\,\rm b}$. Thus, the RFMD predicts the enhancement not only for the total value of the astrophysical factor of the solar proton burning but for the part of it caused by the contribution of nuclear forces only [4]. This underscores an important role of short--distance quantum fluctuations of nucleon fields contributing through the one--nucleon loop exchanges describing dynamics of low--energy nuclear forces in the RFMD. We should emphasize that the astrophysical factor value Eq.~(\ref{label12.2}) is decreased twice relative to that obtained in Ref.~[4]. The former is due to a partial cancellation  between the contributions of the interactions $[\bar{p}(x)\gamma_{\alpha}\gamma^5 p^c(x)][\bar{p^c}(x)\gamma^{\alpha}\gamma^5 p(x)]$ and $[\bar{p}(x) \gamma^5 p^c(x)][\bar{p^c}(x) \gamma^5 p(x)]$. In fact, in the generalized RFMD the interaction $[\bar{p}(x) \gamma^5 p^c(x)][\bar{p^c}(x) \gamma^5 p(x)]$ gives a convergent contribution to the amplitude of the solar proton burning which cancels partly the contribution of the interaction $[\bar{p}(x)\gamma_{\alpha}\gamma^5 p^c(x)][\bar{p^c}(x)\gamma^{\alpha}\gamma^5 p(x)]$ (see Appendix C). 

The estimate of the astrophysical factor $S^{\rm NF}_{\rm pp}(0)$, caused by the contribution of nuclear forces only, can be applied to the justification of our statement concerning the properties of the wave functions $v_{\rm nn}(0)$, $v_{\rm pp}(0)$ and $v_{\rm np}(0)$ and their values $v_{\rm nn}(0) = v_{\rm pp}(0) = v_{\rm np}(0) = 0.28$. Since in the RFMD the value of the astrophysical factor $S^{\rm NF}_{\rm pp}(0)$ cannot exceed the value $S_{\rm pp}(0) = 4.02\times 10^{-25}\,{\rm MeV\,\rm b}$ calculated for the local four--nucleon interaction Eq.~(\ref{label1.1}), the factor ${\cal F}^{\rm NF}_{\rm pp}$ should obey the constraint ${\cal F}^{\rm NF}_{\rm pp} \le \sqrt{2}$. This entails the constraint on the wave function $v_{\rm pp}(0)$, i.e., $v_{\rm pp}(0) \le 0.37$, which agrees well with the value $v_{\rm pp}(0) = 0.28$ imposed by the requirement of isotopical invariance of nuclear forces Eq.~(\ref{label9.10}) and the estimate  $v_{\rm nn}(0) = 0.28$ given by  Eq.~(\ref{label8.8}) due to the low--energy theorem Eq.~(\ref{label8.7}).

By virtue of the enhancement factor $1.42$ the solar neutrino fluxes become substantially reduced. This relaxes the Solar Neutrino Problem [38].  However, such an enhancement of the $S_{\rm pp}(0)$ factor is contradicting to the data on helioseismology [39], which presently allows deviations from the classical value $S^*_{\rm pp}(0) = 3.89\times 10^{-25}\,{\rm MeV\,\rm b}$ [7] less than 20$\%$ of the magnitude.

Our predictions for the cross section for the neutrino disintegration of the deuteron with two protons in the final state $\nu_{\rm e}$ + D $\to$ e$^-$ + p + p for the neutrino energies ranging the values from the region $E_{\rm th} \le E_{\nu_{\rm e}} \le 10\,{\rm MeV}$ is 7 times larger in average than the cross section  calculated in the PMA [42,43]. Such a discrepancy is very impressive and should be verified experimentally for the solar neutrino experiments at SNU [14]. In the end of Conclusion we show that such an enhancement of the cross section cannot be observed for the experimental investigation of the reaction $\nu_{\rm e}$ + D $\to$ e$^-$ + p + p induced by neutrinos from the $\mu$--meson decays. 

The astrophysical factor $S_{\rm pep}(0)$ for the pep--process, i.e., p + e$^-$ + p $\to$ p + p, which is the inverse process with respect to the disintegration of the deuteron by neutrinos  $\nu_{\rm e}$ + D $\to$ e$^-$ + p + p, has been calculated relative to the astrophysical factor $S_{\rm pp}(0)$ for the solar proton burning in complete agreement with the result obtained by Bahcall and May [41]. 

In the case of the disintegration of the deuteron by anti--neutrinos caused by the charged weak current $\bar{\nu}_{\rm e}$ + D $\to$ e$^+$ + n + n the cross section calculated in the generalized RFMD has been found 1.5 times less than the cross section calculated in the PMA [42,43]. Referring to our former result of the calculation of the cross section for the disintegration of the deuteron by anti--neutrinos  Ref.~[4], where we have got a good agreement with the PMA data, the obtained disagreement can be explained by the change of the value of the effective coupling constant $G_{\rm \pi nn} = 3.02\times 10^{-3}\,{\rm MeV}^{-2}$ instead of $G_{\rm \pi NN} = 3.27\times 10^{-3}\,{\rm MeV}^{-2}$ [4], the change of the value of the S--wave scattering length $a_{\rm nn} = - 16.4\,{\rm fm}$ instead of $a_{\rm nn} = -17.0\,{\rm fm}$ [4] and the use of the non--zero effective range for the nn scattering. The change of the input parameters for the description of the process of the disintegration of the deuteron by anti--neutrinos $\bar{\nu}_{\rm e}$ + D $\to$ e$^+$ + n + n in the generalized RFMD has been required by the statement to use more phenomenology of the low--energy elastic nn scattering. Of course, such a change of the input parameters  has led to the disagreement with the PMA data, but the agreement with the experimental data has become much better.

The cross section for the disintegration of the deuteron by anti--neutrinos (neutrinos) induced by the neutral weak current $\bar{\nu}_{\rm e}(\nu_{\rm e})$ + D $\to$ $\bar{\nu}_{\rm e}(\nu_{\rm e})$ + n + p the cross sections calculated in the RFMD agree with  the cross sections calculated in the PMA [42,43] with an accuracy better than 17$\%$. 

The experimental data on the processes $\bar{\nu}_{\rm e}$ + D $\to$ e$^+$ + n + n and $\bar{\nu}_{\rm e}$ + D $\to$ $\bar{\nu}_{\rm e}$ + n + p are given in the form of the cross sections averaged over the reactor anti--neutrino energy spectrum for the anti--neutrino energies ranging $E_{\rm th} \le E_{\bar{\nu}_{\rm e}} \le 10\,{\rm MeV}$.

The theoretical values of the cross sections for the processes $\bar{\nu}_{\rm e}$ + D
$\to$ e$^+$ + n + n and $\bar{\nu}_{\rm e}$ + D $\to$ $\bar{\nu}_{\rm e}$ +
n + p averaged over the reactor anti--neutrino energy spectrum amount to  $<\sigma^{\rm \bar{\nu}_{\rm e} D}_{\rm cc}(E_{\bar{\nu}_{\rm e}})>
= (1.66 \pm 0.32)\times 10^{-45}\,{\rm cm}^2/\,\bar{\nu}_{\rm e}\,{\rm
fission}$, $<\sigma^{\rm \bar{\nu}_{\rm e} D}_{\rm nc}(E_{\bar{\nu}_{\rm
e}})> = (5.32 \pm 1.01)\times 10^{-45}\,{\rm cm}^2/\,\bar{\nu}_{\rm
e}\,{\rm fission}$, respectively.  For the ratio of these cross sections we obtain $r = 0.31\pm 0.06$. These theoretical values agree well with the experimental data by  the Reines$^{\prime}$s group: $<\sigma^{\rm \bar{\nu}_{\rm e} D}_{\rm cc}(E_{\bar{\nu}_{\rm e}})>_{\exp}  = (1.5 \pm 0.4)\times 10^{-45}\,{\rm cm}^2/\,\bar{\nu}_{\rm e}\,{\rm fission}$, $<\sigma^{\rm \bar{\nu}_{\rm e} D}_{\rm nc}(E_{\bar{\nu}_{\rm e}})>_{\exp} = (3.8 \pm 0.9)\times
10^{-45}\,{\rm cm}^2/\,\bar{\nu}_{\rm e}\,{\rm fission}$ and $r_{\exp}= 0.39\pm 0.14$ [46],  $<\sigma^{\rm \bar{\nu}_{\rm e} D}_{\rm cc}(E_{\bar{\nu}_{\rm e}})>_{\exp}  = (0.9 \pm 0.4)\times
10^{-45}\,{\rm cm}^2/\,\bar{\nu}_{\rm e}\,{\rm fission}$, $<\sigma^{\rm
\bar{\nu}_{\rm e} D}_{\rm nc}(E_{\bar{\nu}_{\rm e}})>_{\exp} = (5.3 \pm
0.8)\times 10^{-45}\,{\rm cm}^2/\,\bar{\nu}_{\rm e}\,{\rm fission}$ and
$r_{\exp}= 0.17\pm 0.09$ [47] and Russian experimental groups [48]: $<\sigma^{\rm
\bar{\nu}_{\rm e} D}_{\rm cc}(E_{\bar{\nu}_{\rm e}})>_{\exp}  = (1.84 \pm
0.04)\times 10^{-45}\,{\rm cm}^2/\,\bar{\nu}_{\rm e}\,{\rm fission}$,
$<\sigma^{\rm \bar{\nu}_{\rm e} D}_{\rm nc}(E_{\bar{\nu}_{\rm e}})>_{\exp}
= (5.0 \pm 1.7)\times 10^{-45}\,{\rm cm}^2/\,\bar{\nu}_{\rm e}\,{\rm
fission}$, and $r_{\exp}= 0.37\pm 0.13$.

The discrepancy between the experimental values of the cross section $<\sigma^{\rm \bar{\nu}_{\rm e} D}_{\rm cc}(E_{\bar{\nu}_{\rm e}})>$ for the process $\bar{\nu}_{\rm e}$ + D $\to$ e$^+$ + n + n  with the theoretical predictions obtained in the PMA [44] has been valued by Reines {\it et al.} [47] as an experimental hint for the existence of neutrino oscillations [57]. However, the experimental value $<\sigma^{\rm \bar{\nu}_{\rm e} D}_{\rm cc}(E_{\bar{\nu}_{\rm e}})>_{\exp}  = (0.9 \pm 0.4)\times 10^{-45}\,{\rm cm}^2/\,\bar{\nu}_{\rm e}\,{\rm fission}$ has not
been confirmed in the experiments of Russian experimental groups [48]
represented in 1990. Therefore, the agreement of our theoretical
predictions for the cross sections $<\sigma^{\rm \bar{\nu}_{\rm e} D}_{\rm cc}(E_{\bar{\nu}_{\rm e}})>  = (1.66 \pm 0.32)\times 10^{-45}\,{\rm cm}^2/\,\bar{\nu}_{\rm e}\,{\rm fission}$, $<\sigma^{\rm \bar{\nu}_{\rm e} D}_{\rm nc}(E_{\bar{\nu}_{\rm e}})>  = (5.32 \pm 1.01)\times 10^{-45}\,{\rm cm}^2/\, \bar{\nu}_{\rm e}\,{\rm fission}$ with experimental data given by both the Reines$^{\prime}$s group [46,47] and Russian experimental groups [48] rules out a contribution of neutrino oscillations  to the processes $\bar{\nu}_{\rm e}$ + D $\to$ e$^+$ + n + n and $\bar{\nu}_{\rm e}$ + D $\to$ $\bar{\nu}_{\rm e}$ + n + p. 

The application of the RFMD to the computation of the cross section for the neutron--proton radiative capture n + p $\to$ D + $\gamma$ for thermal neutrons has given the value $\sigma({\rm np}\to {\rm D}\gamma)_{\rm RFMD} = 276\,{\rm m b}$ which agrees within an accuracy better than 10$\%$ with the theoretical value [6]: $\sigma({\rm n p} \to {\rm D\gamma})_{\rm PMA}\,=\,(302.5\pm 4)\,{\rm mb}$ calculated in the PMA for pure M1 transition ${^1}{\rm S}_0 \to {^3}{\rm S}_1$. In comparison with the experimental value [53]: $\sigma({\rm n p} \to {\rm D\gamma})_{\rm PMA}\,=\,(334.2\pm 0.5)\,{\rm mb}$ the cross section calculated in the RFMD is about 18$\%$ of the experimental value less. In turn the value of the cross section calculated in the PMA is only about 10$\%$ less than the experimental value. The contributions increasing the theoretical value of the cross section have been taken into account in the form the $\rho(770)$, the $\omega(780)$ and the $\Delta(1230)$ resonance exchanges   by Riska and Brown [54] in the PMA and by Park {\it et al.} [55] in the EFT with Chiral perturbation theory accounting for chiral corrections in the form of pion--exchanges. In the RFMD the contributions of  the $\rho(770)$ and the $\omega(780)$ meson exchanges are taken into account by the phenomenological coupling constant $G_{\rm \pi np}$ given by Eq.~(\ref{label1.8}). Therefore, we can add only the contribution of the $\Delta(1230)$ resonance and chiral corrections. Unfortunately, the off--mass interaction of the $\Delta(1230)$ resonance is parameterized by the parameter $Z$ ranging the  values from the region $-0.8 \le Z \le 0.3$ [56]. Therefore, at the present level of the definiteness of the off--mass shell coupling of the $\Delta(1230)$ resonance we cannot make any reliable predictions for the contribution of the $\Delta(1230)$ resonance to the cross section for the neutron--proton radiative capture for thermal neutrons. In turn, the contribution of chiral corrections within Chiral perturbation theory incorporated into the RFMD we are planning to analyse in our further development of the RFMD.

Finally we have shown that in the generalized RFMD one can describe low--energy elastic NN scattering in complete agreement with low--energy nuclear phenomenology. In the RFMD due to the low--energy cancellation between effective four--nucleon interactions $[\bar{N}(x)\gamma_{\alpha}\gamma^5 N^c(y)][\bar{N^c}(y)\gamma^{\alpha}\gamma^5 N(x)]$ and $[\bar{N}(x) \gamma^5 N^c(y)][\bar{N^c}(y) \gamma^5 N(x)]$ low--energy elastic NN scattering runs through the one--nucleon loop exchange. The computation of the amplitude of the low--energy elastic NN scattering encounters the problem of ambiguities induced by shifts of virtual momenta of the integrals describing one--nucleon loop diagrams. The arbitrariness introduced by these ambiguities can be expressed in the amplitude in terms of an arbitrary parameter $\kappa_{\rm NN}$ which differs the amplitude of the low--energy elastic NN scattering calculated in the RFMD  from the phenomenological one.  By virtue of the unitarity condition the parameter $\kappa_{\rm NN}$ is fixed to be equal to unity, $\kappa_{\rm NN} = 1$. This brings up the amplitude of the low--energy elastic NN scattering calculated in the RFMD to the  phenomenological form.

Our method of the calculation of the amplitude of the low--energy elastic NN scattering is similar to some extent to that accepted in the EFT approach (see Beane et al. [11]). However, in the RFMD due to the choice of the wave function of the nucleons coupled in the initial state we do not need to fit separately the scattering length and the effective range of the NN scattering (see Beane et al. [11]), but fix them simultaneously through the unitarity condition. 

The derivation of the amplitude ${\cal A}(k)_{\rm RFMD}$ in agreement with the phenomenological form refutes the statement by Bahcall and Kamionkowski [15] concerning inability of the RFMD to describe low--energy elastic NN scattering with a non--zero effective range. To the same extent as we have described the low--energy elastic np scattering we can describe within the RFMD the low--energy elastic pp scattering with the Coulomb repulsion.

Concluding the discussion of the RFMD we would like to emphasize that the RFMD is an effective field theory model. It does not contain any small parameter allowing to justify the one--nucleon loop dominance. We incline to consider the one--nucleon loop fit of input parameters of the RFMD in terms of the parameters of the physical deuteron to some extent as a variational procedure in quantum field theory, where one--nucleon loop diagrams play the role of trial functions realizing a minimal way of the transfer of flavour degrees of freedom from an initial state to a final one. On this way one does not need to include multi--nucleon loop corrections.

For the completeness of our investigation we would like to recur to the process $\nu_{\rm e}$ + D $\to$ e$^-$ + p + p. We discuss  the available experimental data  and give the fit of the cross section calculated in Refs.[42,43].

Experimentally the cross section for the process $\nu_{\rm e}$ + D $\to$ e$^-$ + p + p have been investigated by Willis {\it et al.} [58] following the decay of stopped muons in the LAMPF beamstop [42]. The experimental value of the cross section weighted with the neutrino energy distribution function reads [58]:
\begin{eqnarray}\label{label12.3}
<\sigma^{\nu_{\rm e} D}_{\rm cc}(E_{\nu_{\rm e}})>_{\exp} = (5.20\pm 1.80)\times 10^{-41}\,{\rm cm}^2.
\end{eqnarray}
For the comparison of our result with the experimental value Eq.~(\ref{label12.3}) we should average the cross section Eq.~(\ref{label6.19}) over the neutrino energy spectrum with the distribution function [42]
\begin{eqnarray}\label{label12.4}
\phi(E_{\nu_{\rm e}}) = \frac{192}{m^4_{\mu}}\,E^2_{\nu_{\rm e}}\,\Bigg(\frac{m_{\mu}}{2} - E_{\nu_{\rm e}}\Bigg) 
\end{eqnarray}
normalized to unity
\begin{eqnarray}\label{label12.5}
\int\limits^{m_{\mu}/2}_0 dE_{\nu_{\rm e}}\,\phi(E_{\nu_{\rm e}}) = 1,
\end{eqnarray}
where $m_{\mu} = 105.658\,{\rm MeV}$ is the mass of the $\mu$--meson. The distribution function has an end point at $m_{\mu}/2 \simeq 53\,{\rm MeV}$ and is maximal at $E_{\nu_{\rm e}}= m_{\mu}/3 \simeq 35\,{\rm MeV}$ [42].

The cross section for the process $\nu_{\rm e}$ + D $\to$ e$^-$ + p + p given by Eq.~(\ref{label6.19}) and weighted with the distribution function Eq.~(\ref{label12.4}) over the region $E_{\rm th} \le E_{\nu_{\rm e}} \le 10\,{\rm MeV}$ amounts to
\begin{eqnarray}\label{label12.6}
<\sigma^{\nu_{\rm e} D}_{\rm cc}(E_{\nu_{\rm e}})> = \int\limits^{\displaystyle E_{\nu_{\rm e}} =10\,{\rm MeV}}_{\displaystyle E_{\rm th}}\phi(E_{\nu_{\rm e}})\,\sigma^{\nu_{\rm e} D}_{\rm cc}(E_{\nu_{\rm e}})\, dE_{\nu_{\rm e}} = 0.02\,\times\,10^{-41}\,{\rm cm}^2.
\end{eqnarray}
The result makes up about 0.25$\%$  of the experimental value. This means that the region of the neutrino energies $E_{\rm th} \le E_{\nu_{\rm e}} \le 10\,{\rm MeV}$ is not important for the analysis of the cross section for the process $\nu_{\rm e}$ + D $\to$ e$^-$ + p + p induced by decay neutrinos of stopped muons in the LAMPF beamstop [42]. This also means that one cannot catch the obtained discrepancy between the RFMD and the PMA data for the experiments in the LAMPF beamstop [58]. The most important contribution to the cross section weighted with the distribution function Eq.~(\ref{label12.4}) comes form the neutrinos with energies around the maximum of the distribution function, i.e., of order of $E_{\nu_{\rm e}} \sim 35\,{\rm MeV}$. Since for such energies the Coulomb interactions between charged particles in the final state is not important, in order to understand the experimental and the PMA data we suggest to fit of the cross section calculated in the PMA [42]. 

Indeed, switching off the Coulomb interaction we leave with the parameters defined by strong and weak interactions only. Therefore, using the cross section given by Eq.~(\ref{label6.19}) and making changes $a^{\rm e}_{\rm pp} \to a_{\rm pp}= - 17.1\,{\rm fm}$, $r^{\rm e}_{\rm pp} \to r_{\rm pp}= (2.84\pm 0.03)\,{\rm fm}$ [5], $C(k) \to 1$, $F(Z, E_{\rm e^-}) \to 1$ and $h(2kr_C) \to 0$ we should get the cross section for the process $\nu_{\rm e}$ + D $\to$ e$^-$ + p + p valid for high neutrino energies 
\begin{eqnarray}\label{label12.7}
\sigma^{\nu_{\rm e} D}_{\rm cc}(E_{\nu_{\rm e}}) =
2.28\,(y-1)^2\,\Omega_{\rm p p e^-}(y)\,10^{-43}\,{\rm cm}^2.
\end{eqnarray}
with $\Omega_{\rm p p e^-}(y)$ defined by
\begin{eqnarray}\label{label12.8}
\Omega_{\rm p p e^-}(y) &=& \int\limits^{1}_{0} dx \sqrt{x (1 - x)} \Bigg(1
+ \frac{E_{\rm th}}{m_{\rm e}}(y-1)(1-x)\Bigg) \sqrt{1 + \frac{E_{\rm
th}}{2 m_{\rm e}}(y-1)(1-x)}\nonumber\\
&&\times\,\frac{\displaystyle F_{\rm D}(M_{\rm N}E_{\rm
th}\,(y - 1)\,x)}{\displaystyle \Bigg(1 - \frac{1}{2}\,a_{\rm pp} r_{\rm pp}M_{\rm N}E_{\rm
th}\,(y - 1)\,x \Bigg)^2 + a^2_{\rm pp} M_{\rm N}E_{\rm th}\,(y - 1)\,x}.
\end{eqnarray}
The coefficient $2.28$ has been obtained due to the following change
\begin{eqnarray}\label{label12.9}
3.72 \to 3.72 \times \frac{a_{\rm pp}}{a^{\rm e}_{\rm pp}}\times v_{\rm pp}(0) = 2.28
\end{eqnarray}
which should be carried out by switching off the Coulomb repulsion. 
The numerical values of the cross section Eq.~(\ref{label12.7}) calculated for energies $E_{\nu_{\rm e}} = 10\,{\rm MeV}$, $E_{\nu_{\rm e}}= 55\,{\rm MeV}$ and $E_{\nu_{\rm e}} = 160\,{\rm MeV}$
\begin{eqnarray}\label{label12.10}
\sigma^{\nu_{\rm e} D}_{\rm cc}(E_{\nu_{\rm e}})\Big|_{E_{\nu_{\rm
e}}=10\,{\rm MeV}} &=&  2.28\times 10^{-43}\,(y-1)^2\,\Omega_{\rm p p
e^-}(y)\Big|_{E_{\nu_{\rm e}}=10\,{\rm MeV}}\,{\rm cm}^2 = \nonumber\\
&=&3.00\times 10^{-42}\,{\rm cm}^2,\nonumber\\
\sigma^{\nu_{\rm e} D}_{\rm cc}(E_{\nu_{\rm e}})\Big|_{E_{\nu_{\rm
e}}=55\,{\rm MeV}} &=&  2.28\times 10^{-43}\,(y-1)^2\,\Omega_{\rm p p
e^-}(y)\Big|_{E_{\nu_{\rm e}}=55\,{\rm MeV}}\,{\rm cm}^2 =\nonumber\\
&=& 1.69\times 10^{-40}\,{\rm cm}^2,\nonumber\\
\sigma^{\nu_{\rm e} D}_{\rm cc}(E_{\nu_{\rm e}})\Big|_{E_{\nu_{\rm
e}}=160\,{\rm MeV}} &=&  2.28\times 10^{-43}\,(y-1)^2\,\Omega_{\rm p p
e^-}(y)\Big|_{E_{\nu_{\rm e}}=160\,{\rm MeV}}\,{\rm cm}^2 =\nonumber\\
&=& 1.64\times 10^{-40}\,{\rm cm}^2 
\end{eqnarray}
agree well with the numerical values of the cross section calculated in the PMA [42]:
\begin{eqnarray}\label{label12.11}
\sigma^{\nu_{\rm e} D}_{\rm cc}(E_{\nu_{\rm e}})\Big|_{E_{\nu_{\rm e}} =
10\,{\rm MeV}} &=& 2.55\times 10^{-42}\,{\rm cm}^2,\nonumber\\
\sigma^{\nu_{\rm e} D}_{\rm cc}(E_{\nu_{\rm e}})\Big|_{E_{\nu_{\rm e}} =
55\,{\rm MeV}} &=& 1.66\times 10^{-40}\,{\rm cm}^2,\nonumber\\
\sigma^{\nu_{\rm e} D}_{\rm cc}(E_{\nu_{\rm e}})\Big|_{E_{\nu_{\rm e}} =160\,{\rm MeV}} &=& 1.65\times 10^{-39}\,{\rm cm}^2.
\end{eqnarray}
Hence, with a reasonable accuracy the cross section Eq.~(\ref{label12.7}) fits the PMA data given in [42] for the neutrino energy ranging the region $10\,{\rm MeV} \le E_{\nu_{\rm e}} \le 160\,{\rm MeV}$. This agreement with the PMA data confirms too our statement concerning the properties of the wave functions $v_{\rm NN}(0)$ introduced in the generalized RFMD  as free parameters and fixed through a low--energy theorem for the amplitude of the reaction $\bar{\nu}_{\rm e}$ + D $\to$ e$^+$ + n + n and requirement of isotopical invariance of nuclear forces (see Eq.~(\ref{label8.8}), Eq.~(\ref{label9.10}) and Eq.~(\ref{label12.1})).

The cross section Eq.~(\ref{label12.7}) weighted with the distribution function Eq.~(\ref{label12.4}) 
\begin{eqnarray}\label{label12.12}
<\sigma^{\nu_{\rm e} D}_{\rm cc}(E_{\nu_{\rm e}})> = \int\limits^{\displaystyle m_{\mu/2}}_{E_{\nu_{\rm e}}=10\,{\rm MeV} }\,\phi(E_{\nu_{\rm e}})\,\sigma^{\nu_{\rm e} D}_{\rm cc}(E_{\nu_{\rm e}})\, dE_{\nu_{\rm e}}= 5.70\,\times\,10^{-41}\,{\rm cm}^2 
\end{eqnarray}
agrees well with the experimental value Eq.~(\ref{label12.3}).

Thus, in order to catch experimentally the discrepancy between the predictions of the RFMD and the PMA for the cross section for the process $\nu_{\rm e}$ + D $\to$ e$^-$ + p + p  in the region of the neutrino energies $E_{\rm th} \le E_{\nu_{\rm e}} \le 10\,{\rm MeV}$ it is necessary to use low--energy neutrino beams. The former is of relevance of experiments at SNO [14].

\noindent{\bf Perspectives and further applications of the RFMD.} We have shown that the RFMD can be successfully applied to the description of solar neutrino processes related to the process of the solar proton burning p + p $\to$ D + e$^+$ + $\nu_{\rm e}$ or the proton--proton (pp) fusion. In the main--sequence stars the pp fusion is the starting reaction of the proton--proton (p--p) chain of the nucleosynthesis. After the synthesis of the deuteron caused by the pp fusion the next step of the nucleosynthesis is the burning of the deuteron via the reactions [59]: p + D $\to$ ${^3}{\rm He}$ + $\gamma$, D + D $\to$ ${^4}{\rm He}$ + $\gamma$, D + D $\to$ ${^3}{\rm H}$ + p and D + D $\to$ ${^3}{\rm He}$ + n and so on. Between the listed reactions the reaction of the proton--deuteron radiative capture p + D $\to$ ${^3}{\rm He}$ + $\gamma$ is the predominant one. It is due to the lowest Coulomb barrier of all the reactions in the p--p chain [59]. The produced ${^3}{\rm He}$ with the likelihood of 86$\%$ [59] leads to the reaction ${^3}{\rm He}$ + ${^3}{\rm He}$ $\to$ 2\,p + ${^4}{\rm He}$ completing the chain I of the p--p chain. 

We see the nearest perspectives of the RFMD in the extension of the RFMD by the inclusion of three--nucleon bound states like ${^3}{\rm He}$ and the triton ${^3}{\rm H}$ with the structures (nnp) and (npp), respectively, and possessing the similar properties [60]. The extension of the RFMD by the inclusion of the low--energy interactions of the ${^3}{\rm He}$ and the triton ${^3}{\rm H}$ should convey more quantum field theory phenomena related to the nucleon loop exchanges to the physics of low--energy interactions of light nuclei.

The inclusion of the three--nucleon bound states ${^3}{\rm He}$ and ${^3}{\rm H}$ should give the possibility to continue the investigation of the reactions of the p--p chain and to apply the extended version of the RFMD to the description of the reactions p + D $\to$ ${^3}{\rm He}$ + $\gamma$ and the reaction of the deuteron burning like D + D $\to$ ${^3}{\rm H}$ + p and D + D $\to$ ${^3}{\rm He}$ + n. The RFMD extended by the inclusion of the three--nucleon bound states ${^3}{\rm He}$ and ${^3}{\rm H}$ should be able to describe the neutron--deuteron radiative capture n + D $\to$ ${^3}{\rm H}$ + $\gamma$ and the $\beta$--decay of the triton ${^3}{\rm H}$ $\to$ ${^3}{\rm He}$ + e$^-$ + $\bar{\nu}_{\rm e}$ [61].

For further applications of the RFMD extended by the inclusion of the three--nucleon bound states we are planning the calculation  of the cross sections for (i) elastic scattering of nucleons by the deuteron n + D $\to$ n + D and p + D $\to$ p + D, (ii) reactions of the low--energy disintegration of the deuteron by nucleons n + D $\to$ n + n + p and p + D $\to$ n + p + p and (iii) elastic scattering of nucleons by ${^3}{\rm H}$ and ${^3}{\rm He}$: p + ${^3}{\rm He}$ $\to$ p + ${^3}{\rm He}$ and  n + ${^3}{\rm H}$ $\to$ n + ${^3}{\rm H}$. Since on these processes there are enough experimental data and they are very good investigated theoretically in the PMA, the predictions obtained in the RFMD should be under strict control.

We are also planning to apply the RFMD to the computation of the electric and magnetic polarizabilities of the deuteron which can be obtained from the amplitude of the Compton scattering by the deuteron, the revision of our former computation of the S--wave scattering length of the elastic $\pi$D scattering [3] and the arrangement of the discrepancy between the experimental and theoretical values of the cross section for the neutron--proton radiative capture for thermal neutrons. The revision of the computation of the S--wave scattering length of the elastic $\pi$D scattering is required by the appearance of new experimental data [62] decreasing the former experimental value of the S--wave scattering length of the low--energy elastic ${\rm \pi D}$ scattering by two times. These applications of the RFMD are closely related to the inclusion of the $\Delta(1230)$ resonance and the incorporation of  Chiral perturbation theory into the RFMD.

\section*{Acknowledgment}

We thank the Fonds zur F\"orderung wissenschaftlichen Forschung in \"Osterreich (project P10361--PHY) for partial financial support of this work. Discussions with Prof.~I.~N.~Toptygin are appreciated.

\newpage

\section*{Appendix A. Binding energy of the deuteron and theoretical uncertainty of the RFMD}

The computation of the binding energy of the deuteron $\varepsilon_{\rm D} = 2.225\,{\rm MeV}$ in the RFMD has been carried out in Refs.~[1,2]. Below we adduce the improved computation of $\varepsilon_{\rm D}$ in order to specify the cut--off parameter and the theoretical uncertainty of the approach. Indeed, in Ref.~[1] we have estimated the theoretical uncertainty of the RFMD by computing two--nucleon loop contributions to the binding energy of the deuteron relative to the binding energy of the deuteron calculated in one--nucleon loop approximation.

The effective  Lagrangian of the unphysical deuteron field $D^{(0)}_{\mu}(x)$ with the mass $M_0 = 2 M_{\rm N}$ and the zero binding energy strongly coupled to the proton and the neutron reads [1,2]:
$$
{\cal L}_{\rm bare}(x) = -\frac{1}{2}\,D^{(0)\dagger}_{\mu\nu}(x)\, D^{(0)\mu\nu}(x) + M^2_0 D^{(0)\dagger}_{\mu}(x)\,D^{(0)\mu}(x)
$$
$$
- i g_{\rm V}[\bar{p}(x) \gamma^{\mu} n^c(x) - \bar{n}(x)\gamma^{\mu} p^c(x)]\,D^{(0)}_{\mu}(x) 
$$
$$= - i g_{\rm V}[\bar{p^c}(x) \gamma^{\mu} n(x) - \bar{n^c}(x)\gamma^{\mu} p(x)]\,D^{(0)\dagger}_{\mu}(x),\eqno({\rm A}.1)
$$
where $D^{(0)}_{\mu\nu}(x) = \partial_{\mu} D^{(0)}_{\nu}(x) - \partial_{\nu} D^{(0)}_{\mu}(x)$ is the field strength of the deuteron field. 

In order to obtain the effective Lagrangian of the physical deuteron field $D_{\mu}(x)$ we should calculate one--nucleon loop contributions [1,2]. The one--nucleon loop corrections can be represented by the Lagrangian [1,2]:
$$
\int d^4x\,\delta {\cal L}^{(0)}(x)_{\rm one-loop} = 
$$
$$
-\int d^4x\int\frac{d^4x_1 d^4k_1}{(2\pi)^4}\,e^{\displaystyle -ik_1\cdot (x - x_1)}D^{(0)\dagger}_{\mu}(x)\,D^{(0)}_{\nu}(x_1)\,\frac{g^2_{\rm V}}{4\pi^2}\,\Pi^{\mu\nu}(k_1), \eqno({\rm A}.2)
$$
where the structure function $\Pi^{\mu\nu}(k_1)$ is defined as
$$
\Pi^{\mu\nu}(k_1) = \int\frac{d^4k}{\pi^2i}\,{\rm  tr}\Bigg\{\frac{1}{M_{\rm N} - \hat{k} - \hat{k}_1}\gamma^{\mu}\frac{1}{M_{\rm N} - \hat{k}}\gamma^{\nu}\Bigg\}.\eqno({\rm A}.3)
$$
Keeping only leading terms in the $k_1$--momentum expansion we get 
$$
\Pi^{\mu\nu}(k_1) = \frac{4}{3}\,(k^2_1g^{\mu\nu} - k^{\mu}_1 k^{\nu}_1)\,J_2(M_{\rm N}) + 2\,g^{\mu\nu}\,[J_1(M_{\rm N}) + M^2_{\rm N} J_2(M_{\rm N})],\eqno({\rm A}.4)
$$
where $J_1(M_{\rm N})$ and $J_2(M_{\rm N})$ are quadratically and logarithmically divergent integrals
$$
J_1(M_{\rm N}) =\int\frac{d^4k}{\pi^2i}\frac{1}{M^2_{\rm N} - k^2}= 4\int\limits^{\Lambda_{\rm D}}_0 \frac{d|\vec{k}\,| \vec{k}^{\,2}}{(M^2_{\rm N} + \vec{k}^{\,2})^{1/2}},
$$
$$
J_2(M_{\rm N}) =\int\frac{d^4k}{\pi^2i}\frac{1}{(M^2_{\rm N} - k^2)^2}= 2\int\limits^{\Lambda_{\rm D}}_0 \frac{d|\vec{k}\,|\vec{k}^{\,2}}{(M^2_{\rm N} + \vec{k}^{\,2})^{2/2}}.\eqno({\rm A}.5)
$$
The cut--off  $\Lambda_{\rm D}$ restricts 3--momenta of fluctuations of virtual nucleons forming the physical deuteron. Since in the RFMD the cut--off $\Lambda_{\rm D}$ is much less than the mass of the nucleon [1,2], i.e., $M_{\rm N}Ê\gg \Lambda_{\rm D}$, we would use below the relation:
$$
J_1(M_{\rm N}) = 2\,M^2_{\rm N} J_2(M_{\rm N}) = \frac{4}{3}\,\frac{\Lambda^3_{\rm D}}{M_{\rm N}}.\eqno({\rm A}.6)
$$
The Lagrangian $\delta {\cal L}^{(0)}(x)_{\rm one-loop}$ reads
$$
\delta {\cal L}^{(0)}(x)_{\rm one-loop} = -\frac{1}{2}\,\frac{g^2_{\rm V}}{3\pi^2}\,J_2(M_{\rm N})\,D^{(0)\dagger}_{\mu\nu}(x)\,D^{(0)\mu\nu}(x) 
$$
$$- \frac{g^2_{\rm V}}{2\pi^2}\,[J_1(M_{\rm N}) + M^2_{\rm N} J_2(M_{\rm N})]\, D^{(0)\dagger}_{\mu}(x)\,D^{(0)\mu}(x),\eqno({\rm A}.7)
$$
where we have used the relation
$$
\int d^4x\int\frac{d^4x_1 d^4k_1}{(2\pi)^4}\,e^{\displaystyle -ik_1\cdot (x - x_1)}D^{(0)\dagger}_{\mu}(x)\,D^{(0)}_{\nu}(x_1)\,(k^2_1g^{\mu\nu} - k^{\mu}_1 k^{\nu}_1) =
$$
$$
= \int d^4x\,\frac{1}{2}\,D^{(0)\dagger}_{\mu\nu}(x)\, 
D^{(0)\mu\nu}(x).\eqno({\rm A}.8)
$$
The effective Lagrangian of the free physical deuteron field $D_{\mu}(x)$ reads [1,2]
$$
{\cal L}^{\rm kin}_{\rm eff}(x) = -\frac{1}{2}\,D^{\dagger}_{\mu\nu}(x)\, D^{\mu\nu}(x) + M^2_{\rm D} D^{\dagger}_{\mu}(x)\,D^{\mu}(x),\eqno({\rm A}.9)
$$
where we have renormalized the deuteron field
$$
D_{\mu}(x) = \Bigg(1 + \frac{g^2_{\rm V}}{3\pi^2}\,J_2(M_{\rm N})\Bigg)^{1/2}D^{(0)}_{\mu}(x),\eqno({\rm A}.10)
$$
and defined the mass of the physical deuteron as $M_{\rm D} = M_0 - \varepsilon_{\rm D}$. The binding energy of the deuteron $\varepsilon_{\rm D}$ reads then
$$
\varepsilon_{\rm D} = \frac{17}{48}\,\frac{g^2_{\rm V}}{\pi^2}\,\frac{J_1(M_{\rm N})}{M_{\rm N}},\eqno({\rm A}.11)
$$
where we have used the relation Eq.~({\rm A}.6). Using then the relation $g^2_{\rm V}/\pi^2 = 2\,Q_{\rm D} M^2_{\rm N}$ [1,2] we obtain the binding energy as a function of a cut--off $\Lambda_{\rm D}$:
$$
\varepsilon_{\rm D} = \frac{17}{18}\,Q_{\rm D}\,\Lambda^3_{\rm D}. \eqno({\rm A}.12)
$$
For the experimental values of the binding energy $\varepsilon_{\rm D} = 2.225\,{\rm MeV}$ and the electric quadrupole moment $Q_{\rm D} = 0.286\,{\rm fm}^2$ we estimate the value of the cut--off $\Lambda_{\rm D}$, which amounts to $\Lambda_{\rm D} = 68.452\,{\rm MeV}$. Due to the uncertainty relation $\Delta r\,\Lambda_{\rm D} \ge 1/2$ the spatial  region of virtual nucleon fluctuations forming the physical deuteron is defined by $\Delta r \ge 1.44\,{\rm fm}$. This estimate agrees with a range of nuclear forces (NF) caused by the one--pion exchange with the mass $M_{\pi} =135\,{\rm MeV}$: $r_{\rm N F} \sim 1/M_{\pi} = 1.46\,{\rm fm}$.

In order to estimate the theoretical uncertainty of the model we have suggested in Ref.~[1] to calculate two--nucleon loop contributions to the binding energy. Since we state the one--nucleon loop origin of the deuteron, the comparison of the two--nucleon loop contribution to the binding energy with the binding energy of the deuteron should be valued as the theoretical uncertainty of the model [1,2]. Following Ref.~[1] the effective Lagrangian describing two--nucleon loop contribution to the binding energy of the deuteron reads
$$
\int d^4x\,\delta {\cal L}^{(0)}(x)_{\rm two-loop} = 
$$
$$
=\int d^4x\int\frac{d^4x_1 d^4k_1}{(2\pi)^4}\,e^{\displaystyle -ik_1\cdot (x - x_1)}D^{\dagger}_{\mu}(x)\,D_{\nu}(x_1)\,\frac{g^2_{\rm \pi NN}}{4 M^2_{\pi}}\frac{3 g^2_{\rm V}}{64\pi^4}\,\bar{\Pi}^{\mu\nu}(k_1), \eqno({\rm A}.13)
$$
where the structure function $\bar{{\cal J}}^{\mu\nu}(k_1)$ is defined by
$$
\bar{\Pi}^{\mu\nu}(k_1) = \int\frac{d^4k}{\pi^2i}\,{\rm  tr}\Bigg\{\frac{1}{M_{\rm N} - \hat{k} - \hat{k}_1}\gamma^{\mu}\frac{1}{M_{\rm N} - \hat{k}}\gamma^{\alpha}\Bigg\}\int\frac{d^4q}{\pi^2i}\,{\rm  tr}\Bigg\{\frac{1}{M_{\rm N} - \hat{q} - \hat{k}_1}\gamma_{\alpha}\frac{1}{M_{\rm N} - \hat{q}}\gamma^{\nu}\Bigg\}=
$$
$$
=\Pi^{\mu\alpha}(k_1)\,{\Pi_{\alpha}}^{\nu}(k_1).\eqno({\rm A}.14)
$$
The calculation of the structure function ${{\cal J}}^{\mu\nu}(k_1)$ is obvious and the resultant expression for the effective Lagrangian reads
$$
\delta {\cal L}^{(0)}(x)_{\rm two-loop} = \frac{1}{2}\,\frac{g^2_{\rm \pi NN}}{4 M^2_{\pi}}\,\frac{g^2_{\rm V}}{4\pi^4}\,J_2(M_{\rm N})\,[J_1(M_{\rm N}) + M^2_{\rm N} J_2(M_{\rm N})]\,D^{\dagger}_{\mu\nu}(x)\,D^{\mu\nu}(x) 
$$
$$+ \frac{g^2_{\rm \pi NN}}{4 M^2_{\pi}}\,\frac{3 g^2_{\rm V}}{16\pi^4}\,[J_1(M_{\rm N}) + M^2_{\rm N} J_2(M_{\rm N})]^2\, D^{\dagger}_{\mu}(x)\,D^{\mu}(x),\eqno({\rm A}.15)
$$
The two--nucleon loop contribution the the binding energy of the deuteron is then given by [1]
$$
\delta \varepsilon_{\rm D}^{\rm two-loop} = - \frac{75}{64}\,\frac{g^2_{\rm \pi NN}}{4 M^2_{\pi}}\,\frac{g^2_{\rm V}}{4\pi^4}\,\frac{J^2_1(M_{\rm N})}{M_{\rm N}} = - \frac{25}{24}\,\frac{g^2_{\rm \pi NN}}{4 M^2_{\pi}}\,\frac{Q_{\rm D}}{\pi^2}\,\frac{\Lambda^6_{\rm D}}{M_{\rm N}} = - 0.209\,{\rm MeV}.\eqno({\rm A}.16)
$$
This correction makes up 9.5$\%$ of the binding energy of the deuteron $\varepsilon_{\rm D}= 2.225\,{\rm MeV}$. Following the statement of Ref.~[1] the theoretical uncertainty of the RFMD should make up 9.5$\%$ for amplitudes and, correspondingly, 19$\%$ for cross sections. Thus, for an estimate of a theoretical uncertainty of  cross sections calculated in the RFMD one can use the value about $\Delta = \pm 19\%$.

We, of course, should emphasize that as we fit all input parameters of the model in one--nucleon loop approximation, the computation of the two--nucleon loop contribution to the binding energy of the deuteron does not have so much physical meaning. The value of this correction can serve to some extent as a hint to an expected uncertainty of the approach. Of course, it cannot assure completely a true calculation of a theoretical uncertainty of the approach. The predicted theoretical uncertainties of cross sections calculated in RFMD, $\Delta = \pm 19\%$, can turn out to be much smaller in reality.

\section*{Appendix B.  Effective four--nucleon potential}

In this Appendix we give the derivation of the effective potential Eq.~(\ref{label1.1}) for the proton--proton interaction.
We start with the standard ${\rm \pi^0 pp}$ interaction [63]
$$
{\cal L}_{\rm pp\pi^0}(x) = g_{\rm \pi NN}\,\bar{p}(x)\,i\,\gamma^5 p(x)\,\pi^0(x).\eqno({\rm B}.1)
$$
The effective Lagrangian describing the transition p + p $\to$ p + p  through the one--pion exchange is given by
$$
\int d^4x\,{\cal L}^{\rm pp \to pp}_{\rm eff}(x)_{\rm one-pion} = 
$$
$$
=\frac{i}{2}\,g^2_{\rm \pi NN}\int\!\!\!\int d^4x_1 d^4x_2 [\bar{p}(x_1)\,i\,\gamma^5 p(x_1)]\,<0|{\rm T}(\pi^0(x_1) \pi^0(x_2))|0>[\bar{p}(x_2)\,i\,\gamma^5 p(x_2)]
$$
$$
=\frac{1}{2}\,g^2_{\rm \pi NN}\int\!\!\!\int d^4x_1 d^4x_2 [\bar{p}(x_1)\,\gamma^5 p(x_1)]\,\Delta\,(x_1 - x_2)\,[\bar{p}(x_2)\,\gamma^5 p(x_2)],\eqno({\rm B}.2)
$$
where $\Delta\,(x_1 - x_2)$ is the Green function of the $\pi^0$--field
$$
\Delta\,(x_1 - x_2) = \int \frac{d^4q}{(2\pi)^4}\frac{\displaystyle e^{\displaystyle - iq\cdot (x_1 - x_2)}}{\displaystyle q^2 - M^2_{\pi} + i\,0}=
$$
$$
\int \frac{d^3q}{(2\pi)^3}\int\limits^{\infty}_{-\infty}
 \frac{dq_0}{2\pi}
 \frac{\displaystyle e^{\displaystyle 
 -iq_0 (t_1 - t_2) + i\vec{q}\cdot (\vec{x}_1 - \vec{x}_2)}}{\displaystyle q^2_0 - \vec{q}^{\,2} - M^2_{\pi} + i\,0}.\eqno({\rm B}.3)
$$
Since the interacting protons are non--relativistic, we can set $q_0 = 0$ [11, 63] in the denominator and reduce the r.h.s. of Eq.~({\rm B}.3) to the form
$$
\Delta\,(x_1 - x_2) = - \delta\,(t_1 - t_2)\int \frac{d^3q}{(2\pi)^3}\,\frac{\displaystyle e^{\displaystyle  i\vec{q}\cdot (\vec{x}_1 - \vec{x}_2)}}{\displaystyle  \vec{q}^{\,2} + M^2_{\pi}}.\eqno({\rm B}.4)
$$
In the coordinate representation the momentum integral gives a standard Yukawa potential [6,63]:
$$
Y(|\vec{x}_1 - \vec{x}_2|) = \int \frac{d^3q}{(2\pi)^3}\,\frac{\displaystyle e^{\displaystyle  i\vec{q}\cdot (\vec{x}_1 - \vec{x}_2)}}{\displaystyle  \vec{q}^{\,2} + M^2_{\pi}} = \frac{1}{4\pi}\,{\displaystyle \frac{\displaystyle e^{\displaystyle - M_{\pi}|\vec{x}_1 - \vec{x}_2|}}{|\vec{x}_1 - \vec{x}_2|}}.\eqno({\rm B}.5)
$$
The effective Lagrangian ${\cal L}^{\rm pp \to pp}_{\rm eff}(x)_{\rm one-pion}$ reads
$$
\int d^4x\,{\cal L}^{\rm pp \to pp}_{\rm eff}(x)_{\rm one-pion} = -\frac{1}{2}\,\int\limits^{\infty}_{-\infty} dt\int\!\!\!\int d^3x_1 d^3x_2 
$$
$$
[\bar{p}(t,\vec{x}_1)\,\gamma^5 p(t,\vec{x}_1)]\,\frac{g^2_{\rm \pi NN}}{4\pi}\,{\displaystyle \frac{\displaystyle e^{\displaystyle - M_{\pi}|\vec{x}_1 - \vec{x}_2|}}{|\vec{x}_1 - \vec{x}_2|}}[\bar{p}(t,\vec{x}_2)\,\gamma^5 p(t,\vec{x}_2)].\eqno({\rm B}.6)
$$
In the case of squares of transferred momenta small compared with $M^2_{\pi}$, i.e., $\vec{q}^{\,2} \ll M^2_{\pi}$, the momentum integral gives a $\delta$--function
$$
\lim_{M_{\pi}\to \infty} Y(|\vec{x}_1 - \vec{x}_2|) = \frac{1}{M^2_{\pi}}\,\delta\,(\vec{x}_1 - \vec{x}_2).\eqno({\rm B}.7)
$$
In the case $\vec{q}^{\,2} \ll M^2_{\pi}$ the effective Lagrangian ${\cal L}^{\rm pp \to pp}_{\rm eff}(x)_{\rm one-pion}$ takes the form
$$
\int d^4x\,{\cal L}^{\rm pp \to pp}_{\rm eff}(x)_{\rm one-pion} = -\frac{1}{2}\,\int\limits^{\infty}_{-\infty} dt \int\!\!\!\int d^3x_1 d^3x_2 
$$
$$
[\bar{p}(t,\vec{x}_1)\,\gamma^5 p(t,\vec{x}_1)]\,\frac{g^2_{\rm \pi NN}}{M^2_{\pi}}\,\delta\,(\vec{x}_1 - \vec{x}_2)\,[\bar{p}(t,\vec{x}_2)\,\gamma^5 p(t,\vec{x}_2)].\eqno({\rm B}.8)
$$
When denoting $U(|\vec{x}_1 - \vec{x}_2|) = M^2_{\pi}Y(|\vec{x}_1 - \vec{x}_2|)$ we can rewrite the effective Lagrangian ${\cal L}^{\rm pp \to pp}_{\rm eff}(x)_{\rm one-pion}$ as follows
$$
\int d^4x\,{\cal L}^{\rm pp \to pp}_{\rm eff}(x)_{\rm one-pion} = -\frac{1}{2}\,\frac{g^2_{\rm \pi NN}}{M^2_{\pi}}\,\int\limits^{\infty}_{-\infty} dt \int\!\!\!\int d^3 x_1 d^3 x_2 
$$
$$
[\bar{p}(t,\vec{x}_1)\,\gamma^5 p(t,\vec{x}_1)]\,U(|\vec{x}_1 - \vec{x}_2|)\,[\bar{p}(t,\vec{x}_2)\,\gamma^5 p(t,\vec{x}_2)].\eqno({\rm B}.9)
$$
Now it is convenient to pass to the center of mass frame: $\vec{x} = \frac{1}{2}\,(\vec{x}_1 + \vec{x}_2)$ and $\vec{\rho} = \vec{x}_1 - \vec{x}_2$ or $\vec{x}_1 = \vec{x} + \frac{1}{2}\,\vec{\rho}$ and $\vec{x}_2 = \vec{x} - \frac{1}{2}\,\vec{\rho}$. This gives
$$
\int d^4x\,{\cal L}^{\rm pp \to pp}_{\rm eff}(x)_{\rm one-pion} = -\frac{1}{2}\,\frac{g^2_{\rm \pi NN}}{M^2_{\pi}}\,\int d^4x 
$$
$$
\int d^3\rho \,[\bar{p}(t,\vec{x} + \frac{1}{2}\,\vec{\rho})\,\gamma^5 p(t,\vec{x} + \frac{1}{2}\,\vec{\rho})]\,U(\rho)\,[\bar{p}(t,\vec{x} - \frac{1}{2}\,\vec{\rho})\,\gamma^5 p(t,\vec{x} - \frac{1}{2}\,\vec{\rho})] \eqno({\rm B}.10)
$$
and 
$$
{\cal L}^{\rm pp \to pp}_{\rm eff}(x)_{\rm one-pion} = -\frac{1}{2}\,\frac{g^2_{\rm \pi NN}}{M^2_{\pi}}\,\int d^3\rho \,U(\rho)\,
$$
$$
[\bar{p}(t,\vec{x} + \frac{1}{2}\,\vec{\rho}\,)\,\gamma^5 p(t,\vec{x} + \frac{1}{2}\,\vec{\rho}\,)]\,[\bar{p}(t,\vec{x} - \frac{1}{2}\,\vec{\rho}\,)\,\gamma^5 p(t,\vec{x} - \frac{1}{2}\,\vec{\rho}\,)].\eqno({\rm B}.11)
$$
The effective interaction in Eq.~({\rm B}.11) is defined for the $t$--channel of the pp scattering. In order to find the interaction in the $s$--channel of the pp scattering we have to perform Fierz transformation:
$$
\gamma^5 \otimes \gamma^5 = \frac{1}{4}\,C \otimes C + \frac{1}{4}\,\gamma^5 C\otimes C\,\gamma^5  + \frac{1}{4}\,\gamma^{\mu}\,C \otimes C\,\gamma_{\mu} + \frac{1}{4}\,\gamma^{\mu}\,\gamma^5\,C \otimes C\,\gamma_{\mu}\,\gamma^5
$$
$$
+ \frac{1}{8}\,\sigma^{\mu\nu}\,C \otimes C\,\sigma_{\mu\nu},\eqno({\rm B}.12)
$$
where $C$ is the matrix of a charge conjugation and $\sigma^{\mu\nu}= \frac{1}{2}\,(\gamma^{\mu}\gamma^{\nu} - \gamma^{\nu}\gamma^{\mu})$. This brings the effective Lagrangian Eq.~({\rm B}.12) to the form

$$
{\cal L}^{\rm pp \to pp}_{\rm eff}(x)_{\rm one-pion} = \frac{1}{2}\,\frac{g^2_{\rm \pi NN}}{4 M^2_{\pi}}\,\int d^3\rho \,U(\rho)\,
$$
$$
\{[\bar{p}(t,\vec{x} + \frac{1}{2}\,\vec{\rho}\,)\,p^c (t,\vec{x} - \frac{1}{2}\,\vec{\rho}\,)]\,[\bar{p^c}(t,\vec{x} + \frac{1}{2}\,\vec{\rho}\,)\,p(t,\vec{x} - \frac{1}{2}\,\vec{\rho}\,)]
$$
$$
+[\bar{p}(t,\vec{x} + \frac{1}{2}\,\vec{\rho}\,)\,
\gamma^5 p^c (t,\vec{x} - \frac{1}{2}\,\vec{\rho}\,)]\,
[\bar{p^c}(t,\vec{x} + \frac{1}{2}\,\vec{\rho}\,)\,
\gamma^5 p(t,\vec{x} - \frac{1}{2}\,\vec{\rho}\,)]
$$
$$
+[\bar{p}(t,\vec{x} + \frac{1}{2}\,\vec{\rho}\,)\,\gamma^{\mu} p^c (t,\vec{x} - \frac{1}{2}\,\vec{\rho}\,)]\,[\bar{p^c}(t,\vec{x} + \frac{1}{2}\,\vec{\rho}\,)\,\gamma_{\mu} p(t,\vec{x} - \frac{1}{2}\,\vec{\rho}\,)]
$$
$$
+[\bar{p}(t,\vec{x} + \frac{1}{2}\,\vec{\rho}\,)\,\gamma^{\mu}\gamma^5 p^c (t,\vec{x} - \frac{1}{2}\,\vec{\rho}\,)]\,[\bar{p^c}(t,\vec{x} + \frac{1}{2}\,\vec{\rho}\,)\,\gamma_{\mu}\gamma^5 p(t,\vec{x} - \frac{1}{2}\,\vec{\rho}\,)]
$$
$$
+ \frac{1}{2}\,[\bar{p}(t,\vec{x} + \frac{1}{2}\,\vec{\rho}\,)\,\sigma^{\mu\nu} p^c (t,\vec{x} - \frac{1}{2}\,\vec{\rho}\,)]\,[\bar{p^c}(t,\vec{x} + \frac{1}{2}\,\vec{\rho}\,)\,\sigma_{\mu\nu} p(t,\vec{x} - \frac{1}{2}\,\vec{\rho}\,)]\},\eqno({\rm B}.13)
$$
where $p^c (t,\vec{x} \mp \frac{1}{2}\,\vec{\rho}\,) = C\,\bar{p}^T(t,\vec{x} \mp \frac{1}{2}\,\vec{\rho}\,)$ and $\bar{p^c}(t,\vec{x} \mp \frac{1}{2}\,\vec{\rho}\,) = p^T(t,\vec{x} \mp \frac{1}{2}\,\vec{\rho}\,)\,C$. Leaving only terms $\gamma^{\mu} \gamma^5 \otimes \gamma_{\mu}\gamma^5$ and $\gamma^5 \otimes \gamma^5$ describing the interactions of the pp system in the ${^1}{\rm S}_0$--state, we arrive at the effective Lagrangian
$$
{\cal L}^{\rm pp \to pp}_{\rm eff}(x)_{\rm one-pion} = \frac{1}{2}\,\frac{g^2_{\rm \pi NN}}{4 M^2_{\pi}}\,\int d^3\rho \,U(\rho)\,
$$
$$
\{[\bar{p}(t,\vec{x} + \frac{1}{2}\,\vec{\rho}\,)\,\gamma^{\mu}\gamma^5 p^c (t,\vec{x} - \frac{1}{2}\,\vec{\rho}\,)]\,[\bar{p^c}(t,\vec{x} + \frac{1}{2}\,\vec{\rho}\,)\,\gamma_{\mu}\gamma^5 p(t,\vec{x} - \frac{1}{2}\,\vec{\rho}\,)]
$$
$$
+ [\bar{p}(t,\vec{x} + \frac{1}{2}\,\vec{\rho}\,)\,
\gamma^5 p^c (t,\vec{x} - \frac{1}{2}\,\vec{\rho}\,)]\,[\bar{p^c}(t,\vec{x} + \frac{1}{2}\,\vec{\rho}\,)\,\gamma^5 p(t,\vec{x} - \frac{1}{2}\,\vec{\rho}\,)]\}.\eqno({\rm B}.14)
$$
Thus, we have taken into account the contribution of the one--pion 
exchange. Now we should add the contact term, proportional to the S--wave scattering length $a_{\rm pp}$ of the low--energy elastic pp scattering [2,4]:
$$
{\cal L}^{\rm pp 
\to pp}_{\rm eff}
(x)_{\rm cont} = -\,\frac{1}{2}\,\frac{2\pi\,a_{\rm pp}}{M_{\rm N}}\,\int d^3\rho \,\delta^{(3)}(\vec{\rho}\,)\,
$$
$$
\{[\bar{p}(t,\vec{x} + \frac{1}{2}\,\vec{\rho}\,)\,\gamma^{\mu}\gamma^5 p^c (t,\vec{x} - \frac{1}{2}\,\vec{\rho}\,)]\,[\bar{p^c}(t,\vec{x} + \frac{1}{2}\,\vec{\rho}\,)\,\gamma_{\mu}\gamma^5 p(t,\vec{x} - \frac{1}{2}\,\vec{\rho}\,)]
$$
$$
+ [\bar{p}(t,\vec{x} + \frac{1}{2}\,\vec{\rho}\,)\,
\gamma^5 p^c (t,\vec{x} - \frac{1}{2}\,\vec{\rho}\,)]\,
[\bar{p^c}(t,\vec{x} + \frac{1}{2}\,\vec{\rho}\,)
\,\gamma^5 p(t,\vec{x} - \frac{1}{2}\,\vec{\rho}\,)]\}.
\eqno({\rm B}.15)
$$
Summing up the contributions Eq.~({\rm B}.14) and 
Eq.~({\rm B}.15) we obtain the effective 
Lagrangian describing in the RFMD 
the p + p $\to$ p + p transition:
$$
{\cal L}^{\rm pp \to pp}_{\rm eff}(x) = \frac{1}{2}\,G_{\rm \pi pp}\,\int d^3\rho \,\delta^{(3)}(\vec{\rho}\,)\,
$$
$$
\{[\bar{p}(t,\vec{x} + \frac{1}{2}\,\vec{\rho}\,)\,\gamma^{\mu}\gamma^5 p^c (t,\vec{x} - \frac{1}{2}\,\vec{\rho}\,)]\,[\bar{p^c}(t,\vec{x} + \frac{1}{2}\,\vec{\rho}\,)\,\gamma_{\mu}\gamma^5 p(t,\vec{x} - \frac{1}{2}\,\vec{\rho}\,)]
$$
$$
+ [\bar{p}(t,\vec{x} + \frac{1}{2}\,\vec{\rho}\,)\,
\gamma^5 p^c (t,\vec{x} - \frac{1}{2}\,\vec{\rho}\,)]\,[\bar{p^c}(t,\vec{x} + \frac{1}{2}\,\vec{\rho}\,)\,\gamma^5 p(t,\vec{x} - \frac{1}{2}\,\vec{\rho}\,)]\}.\eqno({\rm B}.16)
$$
Using nuclear phenomenology data  on low--energy elastic np, nn and pp scattering  [6] we postulate the interaction
$$
{\cal L}^{\rm NN \to NN}_{\rm eff}(x)= \int d^3\rho\,\delta^{(3)}(\vec{\rho}\,)
$$
$$
\times \{G_{\rm \pi np}\, [\bar{n}(t,\vec{x} + \frac{1}{2}\vec{\rho}\,)\gamma_{\mu} \gamma^5 p^c(t,\vec{x} - \frac{1}{2}\vec{\rho}\,)] [\bar{p^c}(t,\vec{x}
+ \frac{1}{2}\vec{\rho}\,)\gamma^{\mu}\gamma^5 n(t,\vec{x} -
\frac{1}{2}\vec{\rho}\,)]
$$
$$
+\frac{1}{2}\,G_{\rm \pi nn}\, [\bar{n}(t,\vec{x} + \frac{1}{2}\vec{\rho}\,) \gamma_{\mu} \gamma^5 n^c(t,\vec{x}
- \frac{1}{2}\vec{\rho}\,)] [\bar{n^c}(t,\vec{x} + \frac{1}{2}\vec{\rho}\,)
\gamma^{\mu}\gamma^5
n(t,\vec{x} - \frac{1}{2}\vec{\rho}\,)]
$$
$$
+ \frac{1}{2}\,G_{\rm \pi pp}\, [\bar{p}(t,\vec{x} + \frac{1}{2}\vec{\rho}\,) \gamma_{\mu} \gamma^5 p^c(t,\vec{x} - \frac{1}{2}\vec{\rho}\,)]
[\bar{p^c}(t,\vec{x} + \frac{1}{2}\vec{\rho}\,)\gamma^{\mu}\gamma^5 p(t,\vec{x}
- \frac{1}{2}\vec{\rho}\,)] 
$$
$$
+ (\gamma_{\mu}\gamma^5 \otimes \gamma^{\mu}\gamma^5 \to \gamma^5 \otimes \gamma^5)\}.\eqno({\rm B}.17)
$$
This completes the derivation of the effective potential Eq.~(\ref{label1.1}) for the squares of transferred  momenta of the interacting nucleons much less than $M^2_{\pi}$.

\section*{Appendix C. Computation of the matrix element of the solar proton burning}

In order to acquaint readers with the machinery of the RFMD we give below the detailed derivation of the amplitude Eq.~(\ref{label4.1}).

The process p + p $\to$ D + e$^+$ + $\nu_{\rm e}$ runs through the intermediate W--boson exchange, i.e., p + p $\to$ D + W$^+$ $\to$ D + e$^+$ + $\nu_{\rm e}$. The RFMD defines the transition in terms of the following effective interactions
$$
{\cal L}_{\rm npD}(x) = -ig_{\rm V}[\bar{p^c}(x)\gamma^{\mu}n(x) - \bar{n^c}(x)\gamma^{\mu}p(x)]\,D^{\dagger}_{\mu}(x),
$$
$$
\hspace{-0.5in}{\cal L}^{\rm pp \to pp}_{\rm eff}(x) = \frac{1}{2}\,G_{\rm \pi pp}\,\int d^3\rho \,U(\rho)\,
$$
$$
\{[\bar{p}(t,\vec{x} + \frac{1}{2}\,\vec{\rho}\,)\,\gamma^{\mu}\gamma^5 p^c (t,\vec{x} - \frac{1}{2}\,\vec{\rho}\,)]\,[\bar{p^c}(t,\vec{x} + \frac{1}{2}\,\vec{\rho}\,)\,\gamma_{\mu}\gamma^5 p(t,\vec{x} - \frac{1}{2}\,\vec{\rho}\,)]
$$
$$
 + [\bar{p}(t,\vec{x} + \frac{1}{2}\,\vec{\rho}\,)\,
 \gamma^5 p^c (t,\vec{x} - \frac{1}{2}\,\vec{\rho}\,)]\,[\bar{p^c}(t,\vec{x} + \frac{1}{2}\,\vec{\rho}\,)\,\gamma^5 p(t,\vec{x} - \frac{1}{2}\,\vec{\rho}\,)]\},
$$
$$
\hspace{-0.5in}{\cal L}_{\rm npW}(x) = - 
\frac{g_{\rm W}}{2\sqrt{2}}\,\cos\vartheta_C\,
[\bar{n}(x)\gamma^{\nu}
(1 - g_{\rm A}\gamma^5) p(x)]\,W^-_{\nu}(x).
\eqno({\rm C}.1)
$$
Then, the transition W$^+$ $\to$ e$^+$ + $\nu_{\rm e}$ is defined by the Lagrangian
$$
{\cal L}_{\rm \nu_{\rm e} e^+ W}(x) = - \frac{g_{\rm W}}{2\sqrt{2}}\,[\bar{\psi}_{\nu_{\rm e}}(x)\gamma^{\nu}(1 - \gamma^5) \psi_{\rm e}(x)]\,W^+_{\nu}(x). \eqno({\rm C}.2)
$$
The electroweak coupling constant $g_{\rm W}$ is connected with the Fermi weak constant $G_{\rm F}$ and the mass of the W--boson $M_{\rm W}$ through the relation
$$
\frac{g^2_{\rm W}}{8M^2_{\rm W}} = \frac{G_{\rm F}}{\sqrt{2}}.\eqno({\rm C}.3)
$$
In order not to deal with the intermediate coupling constant $g_{\rm W}$ it is convenient to apply to the computation of the matrix element of the transition p + p $\to$ D + W$^+$ the interaction
$$
{\cal L}_{\rm npW}(x) = [\bar{n}(x)\gamma^{\nu}(1 - g_{\rm A}\gamma^5) p(x)]\,W^-_{\nu}(x),\eqno({\rm C}.4)
$$
and for the description of the subsequent weak transition  W$^+$ $\to$ e$^+$ + $\nu_{\rm e}$ to replace the operator of the W--boson field by the operator of the leptonic weak current
$$
W^-_{\nu}(x) \to -\frac{G_{\rm V}}{\sqrt{2}}\,[\bar{\psi}_{\nu_{\rm e}}(x)\gamma_{\nu}(1 - \gamma^5) \psi_{\rm e}(x)].\eqno({\rm C}.5)
$$
The S matrix describing the transitions like  p + p $\to$ D + W$^+$ is defined
$$
{\rm S} = {\rm T}e^{\displaystyle i\int d^4x\,[{\cal L}_{\rm npD}(x)  + {\cal L}_{\rm npW}(x) + {\cal L}^{\rm pp \to pp}_{\rm eff}(x) + \ldots]},\eqno({\rm C}.6)
$$
where T is the time--ordering operator and the ellipses denote the contribution of interactions irrelevant to the computation of the transition p + p $\to$ D + W$^+$.

For the computation of the transition p + p $\to$ D + W$^+$ we have to consider the third order term of the S matrix which reads
$$
{\rm S}^{(3)} = \frac{i^3}{3!}\int d^4x_1 d^4x_2 d^4x_3\,{\rm T}([{\cal L}_{\rm npD}(x_1)  + {\cal L}_{\rm npW}(x_1) + {\cal L}^{\rm pp\to pp}_{\rm eff}(x_1) + \ldots]
$$
$$
\times\,[{\cal L}_{\rm npD}(x_2)  + {\cal L}_{\rm npW}(x_2) + {\cal L}^{\rm pp\to pp}_{\rm eff}(x_2) + \ldots]
$$
$$
\times\,[{\cal L}_{\rm npD}(x_3)  + {\cal L}_{\rm npW}(x_3) + {\cal L}^{\rm pp\to pp}_{\rm eff}(x_3) + \ldots]) =
$$
$$
= -i \int d^4x_1 d^4x_2 d^4x_3\,{\rm T}({\cal L}^{\rm pp\to pp}_{\rm eff}(x_1){\cal L}_{\rm npD}(x_2){\cal L}_{\rm npW}(x_3)) + \ldots \eqno({\rm C}.7)
$$
The ellipses denote the terms which do not contribute to the matrix element of the transition p + p $\to$ D + W$^+$ and the interaction ${\cal L}_{\rm npW}(x)$ is given by Eq.~({\rm C}.4).  The S matrix element ${\rm S}^{(3)}_{\rm pp\to DW^+}$ contributing to the transition p + p $\to$ D + W$^+$ we determine as follows
$$
{\rm S}^{(3)}_{\rm pp\to DW^+} = -i \int d^4x_1 d^4x_2 d^4x_3\,{\rm T}({\cal L}^{\rm pp\to pp}_{\rm eff}(x_1){\cal L}_{\rm npD}(x_2){\cal L}_{\rm npW}(x_3)).\eqno({\rm C}.8)
$$
For the derivation of the effective Lagrangian ${\cal L}_{\rm pp\to DW^+}(x)$ containing only the fields of the initial and the final particles we should make all necessary contractions of the operators of the proton and the neutron fields. These contractions we denote by the brackets as
$$
<{\rm S}^{(3)}_{\rm pp\to DW^+}> = -i \int d^4x_1 d^4x_2 d^4x_3\,<{\rm T}({\cal L}^{\rm pp\to pp}_{\rm eff}(x_1){\cal L}_{\rm npD}(x_2){\cal L}_{\rm npW}(x_3))>.\eqno({\rm C}.9)
$$
Now the effective Lagrangian ${\cal L}_{\rm pp\to DW^+}(x)$ related to the S matrix element $<{\rm S}^{(3)}_{\rm pp\to DW^+}>$ can be defined as 
$$
<{\rm S}^{(3)}_{\rm pp\to DW^+}> = i\int d^4x\,{\cal L}_{\rm pp\to DW^+}(x) = 
$$
$$
= -i \int d^4x_1 d^4x_2 d^4x_3\,<{\rm T}({\cal L}^{\rm pp\to pp}_{\rm eff}(x_1){\cal L}_{\rm npD}(x_2){\cal L}_{\rm npW}(x_3))>. \eqno({\rm C}.10)
$$
In terms of the operators of the interacting fields the effective Lagrangian ${\cal L}_{\rm pp\to DW^+}(x)$ reads
$$
\int d^4x\,{\cal L}_{\rm pp\to DW^+}(x) =  - \int d^4x_1 d^4x_2 d^4x_3\,<{\rm T}({\cal L}^{\rm pp\to pp}_{\rm eff}(x_1){\cal L}_{\rm npD}(x_2){\cal L}_{\rm npW}(x_3))>
$$
$$
= -\,\frac{1}{2}\,G_{\rm \pi pp}\,\times\,(-ig_{\rm V})\,\times \,(-g_{\rm A})\int d^4x_1 d^4x_2 d^4x_3\,\int d^3\rho\,U(\rho)\,
$$
$$
\times\,{\rm T}([\bar{p^c}(t_1,\vec{x}_1 + \frac{1}{2}\,\vec{\rho}\,)\,\gamma_{\alpha}\gamma^5 p(t_1,\vec{x}_1 - \frac{1}{2}\,\vec{\rho}\,)]\,D^{\dagger}_{\mu}(x_2)\,W^-_{\nu}(x_3))
$$
$$
\times <0|{\rm T}([\bar{p}(t_1,\vec{x}_1 + \frac{1}{2}\,\vec{\rho}\,)\,\gamma^{\alpha}\gamma^5 p^c (t_1,\vec{x}_1 - \frac{1}{2}\,\vec{\rho}\,)][\bar{p^c}(x_2)\gamma^{\mu}n(x_2) - \bar{n^c}(x_2)\gamma^{\mu}p(x_2)]\,
$$
$$
\times\,[\bar{n}(x_3)\gamma^{\nu}\gamma^5 p(x_3)])|0> -\,\frac{1}{2}\,G_{\rm \pi NN}\,\times\,(-ig_{\rm V})\,\times \,(-g_{\rm A})\int d^4x_1 d^4x_2 d^4x_3\,\int d^3\rho \,U(\rho)\,
$$
$$
\times\,{\rm T}([\bar{p^c}(t_1,\vec{x}_1 + \frac{1}{2}\,\vec{\rho}\,) \gamma^5 p(t_1,\vec{x}_1 - \frac{1}{2}\,\vec{\rho}\,)]\,D^{\dagger}_{\mu}(x_2)\,W^-_{\nu}(x_3))
$$
$$
\times <0|{\rm T}([\bar{p}(t_1,\vec{x}_1 + \frac{1}{2}\,\vec{\rho}\,) \gamma^5 p^c (t_1,\vec{x}_1 - \frac{1}{2}\,\vec{\rho}\,)][\bar{p^c}(x_2)\gamma^{\mu}n(x_2) - \bar{n^c}(x_2)\gamma^{\mu}p(x_2)]\,
$$
$$
\times\,[\bar{n}(x_3)\gamma^{\nu}\gamma^5 p(x_3)])|0>.\eqno({\rm C}.11)
$$
Since p + p $\to$ D + W$^+$ is the Gamow--Teller transition, we have taken into account the W--boson coupled with the axial nucleon current.

Due to the relation $\bar{n^c}(x_2)\gamma^{\mu}p(x_2) = - \bar{p^c}(x_2)\gamma^{\mu}n(x_2)$ the r.h.s. of Eq.~({\rm C}.11) can be simplified as follows
$$
\int d^4x\,{\cal L}_{\rm pp\to DW^+}(x) =  - \int d^4x_1 d^4x_2 d^4x_3\,<{\rm T}({\cal L}^{\rm pp\to pp}_{\rm eff}(x_1){\cal L}_{\rm npD}(x_2){\cal L}_{\rm npW}(x_3))>
$$
$$
=  G_{\rm \pi pp}\,\times\,(-ig_{\rm V})\,\times \,g_{\rm A}\int d^4x_1 d^4x_2 d^4x_3\,\int d^3\rho \,U(\rho)\,
$$
$$
\times\,{\rm T}([\bar{p^c}(t_1,\vec{x}_1 + \frac{1}{2}\,\vec{\rho}\,)\,\gamma_{\alpha}\gamma^5 p(t_1,\vec{x}_1 - \frac{1}{2}\,\vec{\rho}\,)]\,D^{\dagger}_{\mu}(x_2)\,W^-_{\nu}(x_3))
$$
$$
\times<0|{\rm T}([\bar{p}(t_1,\vec{x}_1 + \frac{1}{2}\,\vec{\rho}\,)\,\gamma^{\alpha}\gamma^5 p^c (t_1,\vec{x}_1 - \frac{1}{2}\,\vec{\rho}\,)][\bar{p^c}(x_2)\gamma^{\mu}n(x_2)][\bar{n}(x_3)\gamma^{\nu}\gamma^5 p(x_3)])|0> 
$$
$$
 + G_{\rm \pi pp}\,\times\,(-ig_{\rm V})\,\times \,g_{\rm A}\int d^4x_1 d^4x_2 d^4x_3\,\int d^3\rho \,U(\rho)\,
$$
$$
\times\,{\rm T}([\bar{p^c}(t_1,\vec{x}_1 + \frac{1}{2}\,\vec{\rho}\,) \gamma^5 p(t_1,\vec{x}_1 - \frac{1}{2}\,\vec{\rho}\,)]\,D^{\dagger}_{\mu}(x_2)\,W^-_{\nu}(x_3))
$$
$$
\times<0|{\rm T}([\bar{p}(t_1,\vec{x}_1 + \frac{1}{2}\,\vec{\rho}\,) \gamma^5 p^c (t_1,\vec{x}_1 - \frac{1}{2}\,\vec{\rho}\,)][\bar{p^c}(x_2)\gamma^{\mu}n(x_2)][\bar{n}(x_3)\gamma^{\nu}\gamma^5 p(x_3)])|0>.\eqno({\rm C}.12)
$$
Making the necessary contractions we arrive at the expression
$$
\int d^4x\,{\cal L}_{\rm pp\to DW^+}(x) =  - \int d^4x_1 d^4x_2 d^4x_3\,<{\rm T}({\cal L}^{\rm pp\to pp}_{\rm eff}(x_1){\cal L}_{\rm npD}(x_2){\cal L}_{\rm npW}(x_3))> 
$$
$$
= 2\,\times\,G_{\rm \pi pp}\,\times\,(-ig_{\rm V})\,\times \,g_{\rm A}\int d^4x_1 d^4x_2 d^4x_3\,\int d^3\rho \,U(\rho)\,
$$
$$
\times\,{\rm T}([\bar{p^c}(t_1,\vec{x}_1 + \frac{1}{2}\,\vec{\rho}\,)\,\gamma_{\alpha}\gamma^5 p(t_1,\vec{x}_1 - \frac{1}{2}\,\vec{\rho}\,)]\,D^{\dagger}_{\mu}(x_2)\,W^-_{\nu}(x_3))
$$
$$
\times\,(-1)\,{\rm tr}\{\gamma^{\alpha}\gamma^5 (-i) S^c_F(t_1 - t_2,\vec{x}_1 - \vec{x}_2 - \frac{1}{2}\,\vec{\rho}\,) \gamma^{\mu} (-i) S_F(x_2 - x_3) \gamma^{\nu}\gamma^5 
$$
$$
\times\,(-i) S_F(t_3 - t_1, \vec{x}_3  - \vec{x}_1 - \frac{1}{2}\,\vec{\rho}\,)\}
$$
$$
+ 2\,\times\,G_{\rm \pi pp}\,\times\,(-ig_{\rm V})\,\times \,g_{\rm A}\int d^4x_1 d^4x_2 d^4x_3\,\int d^3\rho \,U(\rho)\,
$$
$$
\times\,{\rm T}([\bar{p^c}(t_1,\vec{x}_1 + \frac{1}{2}\,\vec{\rho}\,) \gamma^5 p(t_1,\vec{x}_1 - \frac{1}{2}\,\vec{\rho}\,)]\,D^{\dagger}_{\mu}(x_2)\,W^-_{\nu}(x_3))
$$
$$
\times\,(-1)\,{\rm tr}\{\gamma^5 (-i) S^c_F(t_1 - t_2,\vec{x}_1 - \vec{x}_2 - \frac{1}{2}\,\vec{\rho}\,) \gamma^{\mu} (-i) S_F(x_2 - x_3) \gamma^{\nu}\gamma^5 
$$
$$
\times\,(-i) S_F(t_3 - t_1, \vec{x}_3  - \vec{x}_1 - \frac{1}{2}\,\vec{\rho}\,)\},\eqno({\rm C}.13)
$$
where the combinatorial factor 2 takes into account the fact that the protons are identical particles in the nucleon loop. This is resulted by the contribution of two diagrams depicted in Fig.~1.  

Let us confirm the appearance of the factor 2 by a direct calculation:
$$
\int d^4x\,{\cal L}_{\rm pp\to DW^+}(x) =  - \int d^4x_1 d^4x_2 d^4x_3\,<{\rm T}({\cal L}^{\rm pp\to pp}_{\rm eff}(x_1){\cal L}_{\rm npD}(x_2){\cal L}_{\rm npW}(x_3))> 
$$
$$
= G_{\rm \pi pp}\,\times\,(-ig_{\rm V})\,\times \,g_{\rm A}\int d^4x_1 d^4x_2 d^4x_3\,\int d^3\rho \,U(\rho)\,
$$
$$
\times\,{\rm T}([\bar{p^c}(t_1,\vec{x}_1 + \frac{1}{2}\,\vec{\rho}\,)\,\gamma_{\alpha}\gamma^5 p(t_1,\vec{x}_1 - \frac{1}{2}\,\vec{\rho}\,)]\,D^{\dagger}_{\mu}(x_2)\,W^-_{\nu}(x_3))
$$
$$
\times\,<0|{\rm T}([\bar{p}_{\alpha_1}(t_1,\vec{x}_1 + \frac{1}{2}\,\vec{\rho}\,)\,(\gamma^{\alpha}\gamma^5 C)_{\alpha_1\beta_1} \bar{p}_{\beta_1}(t_1,\vec{x}_1 - \frac{1}{2}\,\vec{\rho}\,)][p_{\alpha_2}(x_2)(C\gamma^{\mu})_{\alpha_2\beta_2}n_{\beta_2}(x_2)]
$$
$$
\times\,[\bar{n}_{\alpha_3}(x_3)(\gamma^{\nu}\gamma^5)_{\alpha_3\beta_3} p_{\beta_3}(x_3)])|0>
+ (\gamma_{\alpha}\gamma^5 \otimes \gamma^{\alpha}\gamma^5 \to \gamma^5 \otimes \gamma^5) =
$$
$$
= G_{\rm \pi pp}\,\times\,(-ig_{\rm V})\,\times \,g_{\rm A}\int d^4x_1 d^4x_2 d^4x_3\,\int d^3\rho \,U(\rho)\,
$$
$$
\times\,{\rm T}([\bar{p^c}(t_1,\vec{x}_1 + \frac{1}{2}\,\vec{\rho}\,)\,\gamma_{\alpha}\gamma^5 p(t_1,\vec{x}_1 - \frac{1}{2}\,\vec{\rho}\,)]\,D^{\dagger}_{\mu}(x_2)\,W^-_{\nu}(x_3))
$$
$$
\times\,\Big\{(\gamma^{\alpha}\gamma^5 C)_{\alpha_1\beta_1} (-i) S_F(t_2 - t_1,\vec{x}_2 - \vec{x}_1 + \frac{1}{2}\vec{\rho}\,)_{\alpha_2\beta_1}(C\gamma^{\mu})_{\alpha_2\beta_2} (-i) S_F(x_2 - x_3)_{\beta_2\alpha_3}
$$
$$
\times\,(\gamma^{\nu}\gamma^5)_{\alpha_3\beta_3} (-i) S_F(t_3 - t_1,\vec{x}_3 - \vec{x}_1 - \frac{1}{2}\vec{\rho}\,)_{\beta_3\alpha_1} -
$$
$$
- (-i) S_F(t_2 - t_1,\vec{x}_2 - \vec{x}_1 - \frac{1}{2}\vec{\rho}\,)_{\alpha_2\alpha_1} (\gamma^{\alpha}\gamma^5 C)_{\alpha_1\beta_1} (C\gamma^{\mu})_{\alpha_2\beta_2} (-i) S_F(x_2 - x_3)_{\beta_2\alpha_3} 
$$
$$
\times\,(\gamma^{\nu}\gamma^5)_{\alpha_3\beta_3} (-i) S_F(t_3 - t_1,\vec{x}_3 - \vec{x}_1 + \frac{1}{2}\vec{\rho}\,)_{\beta_3\beta_1}\Big\}
+ (\gamma_{\alpha}\gamma^5 \otimes \gamma^{\alpha}\gamma^5 \to \gamma^5 \otimes \gamma^5) =
$$
$$
= G_{\rm \pi pp}\,\times\,(-ig_{\rm V})\,\times \,g_{\rm A}\int d^4x_1 d^4x_2 d^4x_3\,\int d^3\rho \,U(\rho)\,
$$
$$
\times\,{\rm T}([\bar{p^c}(t_1,\vec{x}_1 + \frac{1}{2}\,\vec{\rho}\,)\,\gamma_{\alpha}\gamma^5 p(t_1,\vec{x}_1 - \frac{1}{2}\,\vec{\rho}\,)]\,D^{\dagger}_{\mu}(x_2)\,W^-_{\nu}(x_3))
$$
$$
\times\,\Big\{{\rm tr}\{\gamma^{\alpha}\gamma^5 C (-i) S^T_F(t_2 - t_1,\vec{x}_2 - \vec{x}_1 + \frac{1}{2}\vec{\rho}\,) C \gamma^{\mu} (-i) S_F(x_2 - x_3) \gamma^{\nu}\gamma^5 
$$
$$
\times\,(-i) S_F(t_3 - t_1,\vec{x}_3 - \vec{x}_1 - \frac{1}{2}\vec{\rho}\,)\}
$$
$$
- (-i) [S_F(t_2 - t_1,\vec{x}_2 - \vec{x}_1 - \frac{1}{2}\vec{\rho}\,) \gamma^{\alpha}\gamma^5 C]_{\alpha_2\beta_1} [C \gamma^{\mu} (-i) S_F(x_2 - x_3) \gamma^{\nu}\gamma^5 
$$
$$
\times\,(-i) S_F(t_3 - t_1,\vec{x}_3 - \vec{x}_1 + \frac{1}{2}\vec{\rho}\,)]_{\alpha_2\beta_1}\Big\}
+ (\gamma_{\alpha}\gamma^5 \otimes \gamma^{\alpha}\gamma^5 \to \gamma^5 \otimes \gamma^5) =
$$
$$
= G_{\rm \pi pp}\,\times\,(-ig_{\rm V})\,\times \,g_{\rm A}\int d^4x_1 d^4x_2 d^4x_3\,\int d^3\rho \,U(\rho)\,
$$
$$
\times\,{\rm T}([\bar{p^c}(t_1,\vec{x}_1 + \frac{1}{2}\,\vec{\rho}\,)\,\gamma_{\alpha}\gamma^5 p(t_1,\vec{x}_1 - \frac{1}{2}\,\vec{\rho}\,)]\,D^{\dagger}_{\mu}(x_2)\,W^-_{\nu}(x_3))
$$
$$
\times\,\Big\{(-1) {\rm tr}\{\gamma^{\alpha}\gamma^5 [C^T S^T_F(t_2 - t_1,\vec{x}_2 - \vec{x}_1 + \frac{1}{2}\vec{\rho}\,) C] \gamma^{\mu}(-i) S_F(x_2 - x_3) \gamma^{\nu}\gamma^5 
$$
$$
\times\,(-i) S_F(t_3 - t_1,\vec{x}_3 - \vec{x}_1 - \frac{1}{2}\vec{\rho}\,)\}
$$
$$
+ (-1) {\rm tr}\{(-i) [S^T_F(t_2 - t_1,\vec{x}_2 - \vec{x}_1 + \frac{1}{2}\vec{\rho}\,) \gamma^{\alpha}\gamma^5 C]^T \gamma^{\mu}(-i) S_F(x_2 - x_3) \gamma^{\nu}\gamma^5 
$$
$$
\times\,(-i) S_F(t_3 - t_1,\vec{x}_3 - \vec{x}_1 + \frac{1}{2}\vec{\rho}\,)\}\Big\}
+ (\gamma_{\alpha}\gamma^5 \otimes \gamma^{\alpha}\gamma^5 \to \gamma^5 \otimes \gamma^5) =
$$
$$
= G_{\rm \pi pp}\,\times\,(-ig_{\rm V})\,\times \,g_{\rm A}\int d^4x_1 d^4x_2 d^4x_3\,\int d^3\rho \,U(\rho)\,
$$
$$
\times\,{\rm T}([\bar{p^c}(t_1,\vec{x}_1 + \frac{1}{2}\,\vec{\rho}\,)\,\gamma_{\alpha}\gamma^5 p(t_1,\vec{x}_1 - \frac{1}{2}\,\vec{\rho}\,)]\,D^{\dagger}_{\mu}(x_2)\,W^-_{\nu}(x_3))
$$
$$
\times\,\Big\{(-1) {\rm tr}\{\gamma^{\alpha}
\gamma^5 (-i) S^c_F(t_1 - t_2,\vec{x}_1 - \vec{x}_2 - \frac{1}{2}\vec{\rho}\,) \gamma^{\mu} (-i) S_F(x_2 - x_3) \gamma^{\nu}\gamma^5 
$$
$$
\times\,(-i) S_F(t_3 - t_1,\vec{x}_3 - \vec{x}_1 - \frac{1}{2}\vec{\rho}\,)\}
$$
$$
+ (-1) {\rm tr}\{ 
C^T (\gamma^{\alpha}\gamma^5)^T 
C (-i) S^c_F
(t_1 - t_2,\vec{x}_1 - \vec{x}_2 + \frac{1}{2}\vec{\rho}\,) \gamma^{\mu} (-i) S_F(x_2 - x_3) 
\gamma^{\nu}\gamma^5 
$$
$$
\times\,(-i) S_F(t_3 - t_1,\vec{x}_3 + \vec{x}_1 + \frac{1}{2}\vec{\rho}\,)\}\Big\}
+ (\gamma_{\alpha}\gamma^5 \otimes \gamma^{\alpha}\gamma^5 \to \gamma^5 \otimes \gamma^5).\eqno({\rm C}.14)
$$
Here we have used the relation $C = -C^T$. Then, by applying the relation $C^T (\gamma^{\alpha}\gamma^5)^T C = 
\gamma^{\alpha}\gamma^5$ we obtain the following expression
$$
\int d^4x\,{\cal L}_{\rm pp\to DW^+}(x) =  - 
\int d^4x_1 d^4x_2 d^4x_3\,<{\rm T}(
{\cal L}^{\rm pp\to pp}_{\rm eff}(x_1)
{\cal L}_{\rm npD}(x_2){\cal L}_{\rm npW}(x_3))> 
$$
$$
= G_{\rm \pi pp}\,\times\,(-ig_{\rm V})\,\times 
\,g_{\rm A}\int d^4x_1 d^4x_2 d^4x_3\,\int d^3\rho \,U(\rho)\,
$$
$$
\times\,\Big\{{\rm T}([\bar{p^c}(t_1,\vec{x}_1 + \frac{1}{2}\,\vec{\rho}\,)\,\gamma_{\alpha}
\gamma^5 p(t_1,\vec{x}_1 - \frac{1}{2}\,\vec{\rho}\,)]\,
D^{\dagger}_{\mu}(x_2)\,W^-_{\nu}(x_3))
$$
$$
\times\,(-1) {\rm tr}\{\gamma^{\alpha}\gamma^5 (-i) 
S^c_F(t_1 - t_2,\vec{x}_1 - \vec{x}_2 - 
\frac{1}{2}\vec{\rho}\,) 
\gamma^{\mu} (-i) S_F(x_2 - x_3) \gamma^{\nu}\gamma^5
$$
$$
\times\,(-i) S_F(t_3 - t_1,\vec{x}_3 - \vec{x}_1 - \frac{1}{2}\vec{\rho}\,)\}
$$
$$
+ {\rm T}([\bar{p^c}(t_1,\vec{x}_1 + \frac{1}{2}\,\vec{\rho}\,)\,\gamma_{\alpha}\gamma^5 
p(t_1,\vec{x}_1 - \frac{1}{2}\,\vec{\rho}\,)]\,
D^{\dagger}_{\mu}(x_2)
\,W^-_{\nu}(x_3))
$$
$$
\times\,(-1) {\rm tr}\{ \gamma^{\alpha}\gamma^5 (-i) 
S^c_F(t_1 - t_2,\vec{x}_1 - \vec{x}_2 
+ \frac{1}{2}\vec{\rho}\,) C \gamma^{\mu} 
(-i) S_F(x_2 - x_3) \gamma^{\nu}\gamma^5
$$
$$
\times\,(-i) S_F(t_3 - t_1,\vec{x}_3 + \vec{x}_1 + \frac{1}{2}\vec{\rho}\,)\}\Big\}
+ (\gamma_{\alpha}\gamma^5 \otimes \gamma^{\alpha}
\gamma^5 \to \gamma^5 \otimes \gamma^5).
\eqno({\rm C}.15)
$$
Using the property of the operators
$$
[\bar{p^c}(t_1,\vec{x}_1 + \frac{1}{2}\,\vec{\rho}\,)\,
\Gamma p(t_1,\vec{x}_1 - \frac{1}{2}\,\vec{\rho}\,)] = [\bar{p^c}(t_1,\vec{x}_1 - \frac{1}{2}\,\vec{\rho}\,)\,
\Gamma p(t_1,\vec{x}_1 + \frac{1}{2}\,\vec{\rho}\,)] 
\eqno({\rm C}.16)
$$
for $\Gamma=\gamma^{\alpha}\gamma^5$ and $\gamma^5$, we get
$$
\int d^4x\,{\cal L}_{\rm pp\to DW^+}(x) =  - 
\int d^4x_1 d^4x_2 d^4x_3\,<{\rm T}(
{\cal L}^{\rm pp\to pp}_{\rm eff}(x_1)
{\cal L}_{\rm npD}(x_2){\cal L}_{\rm npW}(x_3))> 
$$
$$
= G_{\rm \pi pp}\,\times\,(-ig_{\rm V})\,\times 
\,g_{\rm A}\int d^4x_1 d^4x_2 d^4x_3\,\int d^3\rho 
\,U(\rho)\,
$$
$$
\times\,\Big\{{\rm T}([\bar{p^c}(t_1,\vec{x}_1 + \frac{1}{2}\,\vec{\rho}\,)\,\gamma_{\alpha}\gamma^5 
p(t_1,\vec{x}_1 - \frac{1}{2}\,\vec{\rho}\,)]\,
D^{\dagger}_{\mu}(x_2)
\,W^-_{\nu}(x_3))
$$
$$
\times\,(-1) {\rm tr}\{\gamma^{\alpha}\gamma^5 
S^c_F(t_1 - t_2,\vec{x}_1 - \vec{x}_2 - \frac{1}{2}\vec{\rho}\,) \gamma^{\mu} (-i) S_F(x_2 - x_3) \gamma^{\nu}\gamma^5 
$$
$$
\times\,(-i) S_F(t_3 - t_1,\vec{x}_3 - \vec{x}_1 - \frac{1}{2}\vec{\rho}\,)\}
$$
$$
+ {\rm T}([\bar{p^c}(t_1,\vec{x}_1 - \frac{1}{2}\,\vec{\rho}\,)\,\gamma_{\alpha}
\gamma^5 p(t_1,\vec{x}_1 + \frac{1}{2}\,\vec{\rho}\,)]\,
D^{\dagger}_{\mu}(x_2)\,W^-_{\nu}(x_3))
$$
$$
\times\,(-1) {\rm tr}\{(-i) \gamma^{\alpha}
\gamma^5 S^c_F(t_1 - t_2,\vec{x}_1 - \vec{x}_2 + \frac{1}{2}\vec{\rho}\,) C \gamma^{\mu} 
(-i) S_F(x_2 - x_3) \gamma^{\nu}\gamma^5 
$$
$$
\times\,(-i) S_F(t_3 - t_1,\vec{x}_3 + \vec{x}_1 + \frac{1}{2}\vec{\rho}\,)\}\Big\}
+ (\gamma_{\alpha}\gamma^5 \otimes 
\gamma^{\alpha}\gamma^5 \to \gamma^5 \otimes \gamma^5).
\eqno({\rm C}.17)
$$
Making a change of variables 
$\vec{\rho}\,Ê\to - \vec{\rho}\,$ in the last term, we arrive at the expression Eq.~({\rm C}.13).

Then, $S^c_F(x)$ and $S_F(x)$ are the Green functions of the free anti--nucleon and nucleon field, respectively:
$$
S^c_F(x) = CS^T_F(-x) C^T = S_F(x) = 
\int\frac{d^4k}{(2\pi)^4}
\frac{\displaystyle 
e^{\displaystyle - i k\cdot x}}{M_{\rm N} - \hat{k}}.
\eqno({\rm C}.18)
$$
Passing to the momentum representation of the Green functions we get
$$
\int d^4x\,{\cal L}_{\rm pp\to DW^+}(x) = 
$$
$$
= - i\,g_{\rm A}G_{\rm \pi pp}\frac{g_{\rm V}}{8\pi^2}
\int d^4x_1 \int \frac{d^4x_2 d^4k_2}{(2\pi)^4}\frac{d^4x_3 d^4k_3}{(2\pi)^4}\,e^{\displaystyle -i k_2\cdot (x_2-x_1)}e^{\displaystyle -i k_3\cdot (x_3-x_1)}
$$
$$
\times\,\int d^3\rho \,U(\rho)\,
{\rm T}([\bar{p^c}(t_1,\vec{x}_1 + \frac{1}{2}\,\vec{\rho}\,)\,
\gamma_{\alpha}\gamma^5 p(t_1,\vec{x}_1 - \frac{1}{2}\,\vec{\rho}\,)]\,
D^{\dagger}_{\mu}(x_2)\,W^-_{\nu}(x_3))
$$
$$
\times\,\int\frac{d^4k_1}{\pi^2i}\,
e^{\displaystyle i\vec{q}\cdot \vec{\rho}}\,
{\rm tr}\Bigg\{\gamma^{\alpha}\gamma^5
\frac{1}{M_{\rm N} - \hat{k}_1 + \hat{k}_2}
\gamma^{\mu}\frac{1}{M_{\rm N} - \hat{k}_1}\gamma^{\nu}
\gamma^5 \frac{1}{M_{\rm N} -
\hat{k}_1 - \hat{k}_3}\Bigg\}
$$
$$
- i\,g_{\rm A}G_{\rm \pi pp}\frac{g_{\rm V}}{8\pi^2}
\int d^4x_1 \int 
\frac{d^4x_2 d^4k_2}{(2\pi)^4}\frac{d^4x_3 d^4k_3}{(2\pi)^4}\,e^{\displaystyle -i k_2\cdot (x_2-x_1)}e^{\displaystyle -i k_3\cdot (x_3-x_1)}
$$
$$
\times\,\int d^3\rho \,U(\rho)\,
{\rm T}([\bar{p^c}(t_1,\vec{x}_1 + \frac{1}{2}\,\vec{\rho}\,)\,\gamma^5 
p(t_1,\vec{x}_1 - \frac{1}{2}\,\vec{\rho}\,)]\, D^{\dagger}_{\mu}(x_2)\,W^-_{\nu}(x_3))
$$
$$
\times\,\int\frac{d^4k_1}{\pi^2i}\,
e^{\displaystyle i\vec{q}\cdot \vec{\rho}}\,{\rm tr}\Bigg\{\gamma^5\frac{1}{M_{\rm N} - \hat{k}_1 + \hat{k}_2}\gamma^{\mu}\frac{1}{M_{\rm N} - \hat{k}_1}\gamma^{\nu}\gamma^5 
\frac{1}{M_{\rm N} - \hat{k}_1 - 
\hat{k}_3}\Bigg\},\eqno({\rm C}.19)
$$
where $\vec{q} = \vec{k}_1 + (\vec{k}_3 - \vec{k}_2)/2$.

In order to obtain the effective Lagrangian describing the process p + p $\to$ D + e$^+$ + $\nu_{\rm e}$ we have to replace the operator of the W--boson field by the operator of the leptonic weak current Eq.~({\rm C}.5):
$$
\int d^4x\,{\cal L}_{\rm pp\to D e^+ \nu_{\rm e}}(x) = 
$$
$$
= i\,g_{\rm A}G_{\rm \pi pp}\frac{G_{\rm V}}{\sqrt{2}} \frac{g_{\rm V}}{8\pi^2}\int d^4x_1 \int \frac{d^4x_2 d^4k_2}{(2\pi)^4}\frac{d^4x_3 d^4k_3}{(2\pi)^4}\,e^{\displaystyle -i k_2\cdot (x_2-x_1)}e^{\displaystyle -i k_3\cdot (x_3-x_1)}
$$
$$
\times\,\int d^3\rho \,U(\rho)\,{\rm T}([\bar{p^c}(t_1,\vec{x}_1 + \frac{1}{2}\,\vec{\rho}\,)\,\gamma_{\alpha}\gamma^5 p(t_1,\vec{x}_1 - \frac{1}{2}\,\vec{\rho}\,)]\,D^{\dagger}_{\mu}(x_2)\,[\bar{\psi}_{\nu_{\rm e}}(x_3)\gamma_{\nu}(1 - \gamma^5) \psi_{\rm e}(x_3)])
$$
$$
\times\,\int\frac{d^4k_1}{\pi^2i}\,e^{\displaystyle i\vec{q}\cdot \vec{\rho}}\,{\rm tr}\Bigg\{\gamma^{\alpha}\gamma^5\frac{1}{M_{\rm N} - \hat{k}_1 + \hat{k}_2}\gamma^{\mu}\frac{1}{M_{\rm N} - \hat{k}_1}\gamma^{\nu}\gamma^5 \frac{1}{M_{\rm N} - \hat{k}_1 - \hat{k}_3}\Bigg\}
$$
$$
+ i\,g_{\rm A}G_{\rm \pi pp}\frac{G_{\rm V}}{\sqrt{2}}\frac{g_{\rm V}}{8\pi^2}\int d^4x_1 \int \frac{d^4x_2 d^4k_2}{(2\pi)^4}\frac{d^4x_3 d^4k_3}{(2\pi)^4}\,e^{\displaystyle -i k_2\cdot (x_2-x_1)}e^{\displaystyle -i k_3\cdot (x_3-x_1)}
$$
$$
\times\,\int d^3\rho \,U(\rho)\,{\rm T}([\bar{p^c}(t_1,\vec{x}_1 + \frac{1}{2}\,\vec{\rho}\,)\,\gamma^5 p(t_1,\vec{x}_1 - \frac{1}{2}\,\vec{\rho}\,)]\, D^{\dagger}_{\mu}(x_2)\,[\bar{\psi}_{\nu_{\rm e}}(x_3)\gamma_{\nu}(1 - \gamma^5) \psi_{\rm e}(x_3)])
$$
$$
\times\,\int\frac{d^4k_1}{\pi^2i}\,e^{\displaystyle i\vec{q}\cdot \vec{\rho}}\,{\rm tr}\Bigg\{\gamma^5\frac{1}{M_{\rm N} - \hat{k}_1 + \hat{k}_2}\gamma^{\mu}\frac{1}{M_{\rm N} - \hat{k}_1}\gamma^{\nu}\gamma^5 \frac{1}{M_{\rm N} - \hat{k}_1 - \hat{k}_3}\Bigg\}.\eqno({\rm C}.20)
$$
Now we are able to determine the matrix element of the  process p + p $\to$ D + e$^+$ + $\nu_{\rm e}$ as
$$
\int d^4x\,<{\rm D}(k_{\rm D}){\rm e}^+(k_{\rm e^+})\nu_{\rm e}(k_{\nu_{\rm e}})|{\cal L}_{\rm pp\to D e^+ \nu_{\rm e}}(x)|p(p_1)p(p_2)> = 
$$
$$
=(2\pi)^4\delta^{(4)}(k_{\rm D} + k_{\ell} - p_1 - p_2)\,\frac{{\cal M}({\rm p} + {\rm p} \to {\rm D} + {\rm e}^+ + \nu_{\rm e})}{\displaystyle 
\sqrt{2E_1V\,2E_2V\,2E_{\rm D}V\,
2E_{\rm e^+}V\,2E_{\nu_{\rm e}}V}}, 
\eqno({\rm C}.21)
$$
where $k_{\ell} = k_{\rm e^+} + k_{\nu_{\rm e}}$ is the 4--momentum of the leptonic pair, $E_i\,(i =1,2,{\rm D},{\rm e},\nu_{\rm e})$ are the energies of the protons, the deuteron, positron and neutrino, $V$ is the normalization volume.

Taking the r.h.s. of Eq.~({\rm C}.20) between the wave functions of the initial $|p(p_1)p(p_2)>$ and the final $<{\rm D}(k_{\rm D}){\rm e}^+(k_{\rm e^+})\nu_{\rm e}(k_{\nu_{\rm e}})|$ states we get
$$
(2\pi)^4\delta^{(4)}(k_{\rm D} + k_{\ell} - p_1 - p_2)\,\frac{{\cal M}({\rm p} + {\rm p} \to {\rm D} + {\rm e}^+ + \nu_{e})}
{\displaystyle \sqrt{2E_1V\,2E_2V
\,2E_{\rm D}V\,2E_{\rm e^+}V\,2E_{\nu_{\rm e}}V}}=
$$
$$
=  i\,g_{\rm A}G_{\rm \pi pp}\frac{G_{\rm V}}{\sqrt{2}} \frac{g_{\rm V}}{8\pi^2}\int d^4x_1 \int \frac{d^4x_2 d^4k_2}{(2\pi)^4}\frac{d^4x_3 d^4k_3}{(2\pi)^4}\,e^{\displaystyle -i k_2\cdot (x_2-x_1)}e^{\displaystyle -i k_3\cdot (x_3-x_1)}
$$
$$
\times\,\int d^3\rho \,U(\rho)\,<{\rm D}(k_{\rm D}){\rm e}^+(k_{\rm e^+})\nu_{\rm e}(k_{\nu_{\rm e}})|{\rm T}([\bar{p^c}(t_1,\vec{x}_1 + \frac{1}{2}\,\vec{\rho}\,)\,\gamma_{\alpha}\gamma^5 p(t_1,\vec{x}_1 - \frac{1}{2}\,\vec{\rho}\,)]\,D^{\dagger}_{\mu}(x_2)\,
$$
$$
\times\,[\bar{\psi}_{\nu_{\rm e}}(x_3)\gamma_{\nu}(1 - \gamma^5) \psi_{\rm e}(x_3)])|p(p_1) p(p_2)>
$$
$$
\times\,\int\frac{d^4k_1}{\pi^2i}\,e^{\displaystyle i\vec{q}\cdot \vec{\rho}}\,{\rm tr}\Bigg\{\gamma^{\alpha}\gamma^5\frac{1}{M_{\rm N} - \hat{k}_1 + \hat{k}_2}\gamma^{\mu}\frac{1}{M_{\rm N} - \hat{k}_1}\gamma^{\nu}\gamma^5 \frac{1}{M_{\rm N} - \hat{k}_1 - \hat{k}_3}\Bigg\}
$$
$$
+ i\,g_{\rm A}G_{\rm \pi pp}\frac{G_{\rm V}}{\sqrt{2}}\frac{g_{\rm V}}{8\pi^2}\int d^4x_1 \int \frac{d^4x_2 d^4k_2}{(2\pi)^4}\frac{d^4x_3 d^4k_3}{(2\pi)^4}\,e^{\displaystyle -i k_2\cdot (x_2-x_1)}e^{\displaystyle -i k_3\cdot (x_3-x_1)}
$$
$$
\times\,\int d^3\rho \,U(\rho)\,<{\rm D}(k_{\rm D}){\rm e}^+(k_{\rm e^+})\nu_{\rm e}(k_{\nu_{\rm e}})|{\rm T}([\bar{p^c}(t_1,\vec{x}_1 + \frac{1}{2}\,\vec{\rho}\,)\,\gamma^5 p(t_1,\vec{x}_1 - \frac{1}{2}\,\vec{\rho}\,)]\, D^{\dagger}_{\mu}(x_2)\,
$$
$$
\times\,[\bar{\psi}_{\nu_{\rm e}}(x_3)\gamma_{\nu}(1 - \gamma^5) \psi_{\rm e}(x_3)])|p(p_1) p(p_2)>
$$
$$
\times\,\int\frac{d^4k_1}{\pi^2i}\,e^{\displaystyle i\vec{q}\cdot \vec{\rho}}\,{\rm tr}\Bigg\{\gamma^5\frac{1}{M_{\rm N} - \hat{k}_1 + \hat{k}_2}\gamma^{\mu}\frac{1}{M_{\rm N} - \hat{k}_1}\gamma^{\nu}\gamma^5 \frac{1}{M_{\rm N} - \hat{k}_1 - \hat{k}_3}\Bigg\}.\eqno({\rm C}.22)
$$
Between the initial $|p(p_1)p(p_2)>$ and the final $<{\rm D}(k_{\rm D}){\rm e}^+(k_{\rm e^+})\nu_{\rm e}(k_{\nu_{\rm e}})|$ states the matrix elements are defined
$$
<{\rm D}(k_{\rm D}){\rm e}^+(k_{\rm e^+})\nu_{\rm e}(k_{\nu_{\rm e}})|{\rm T}([\bar{p^c}(t_1,\vec{x}_1 + \frac{1}{2}\,\vec{\rho}\,)\,\gamma_{\alpha}\gamma^5 p(t_1,\vec{x}_1 - \frac{1}{2}\,\vec{\rho}\,)]\,D^{\dagger}_{\mu}(x_2)\,
$$
$$
\times\,[\bar{\psi}_{\nu_{\rm e}}(x_3)\gamma_{\nu}(1 - \gamma^5) \psi_{\rm e}(x_3)])|p(p_1) p(p_2)>= \sqrt{2}\,[\bar{u^c}(p_2)\gamma_{\alpha}\gamma^5 u(p_1)][\bar{u}(k_{\nu_{\rm e}})\gamma_{\nu}
(1-\gamma^5)v(k_{\rm e^+})]\,
$$
$$
\times\,e^*_{\mu}(k_{\rm D})\,\psi_{\rm pp}(\vec{\rho})_{\rm in}\,\frac{\displaystyle e^{\displaystyle -i(p_1+p_2)\cdot x_1}\,e^{\displaystyle ik_{\rm D}\cdot x_2}\,e^{\displaystyle ik_{{\ell}}\cdot x_3}}
{\displaystyle \sqrt{2E_1V\,2E_2V\,2E_{\rm D}V\,
2E_{\rm e^+}V\,2E_{\nu_{\rm e}}V}},
$$
$$
<{\rm D}(k_{\rm D}){\rm e}^+(k_{\rm e^+})\nu_{\rm e}(k_{\nu_{\rm e}})|{\rm T}([\bar{p^c}(t_1,\vec{x}_1 + \frac{1}{2}\,\vec{\rho}\,)\,\gamma^5 p(t_1,\vec{x}_1 - \frac{1}{2}\,\vec{\rho}\,)]\,D^{\dagger}_{\mu}(x_2)\,
$$
$$
\times\,[\bar{\psi}_{\nu_{\rm e}}(x_3)\gamma_{\nu}(1 - \gamma^5) \psi_{\rm e}(x_3)])|p(p_1) p(p_2)>= \sqrt{2}\,[\bar{u^c}(p_2)\gamma^5 u(p_1)][\bar{u}(k_{\nu_{\rm e}})\gamma_{\nu}
(1-\gamma^5)v(k_{\rm e^+})]\,
$$
$$
\times\,e^*_{\mu}(k_{\rm D})\,\psi_{\rm pp}(\vec{\rho})_{\rm in}\,\frac{\displaystyle e^{\displaystyle -i(p_1+p_2)\cdot x_1}\,e^{\displaystyle ik_{\rm D}\cdot x_2}\,e^{\displaystyle ik_{{\ell}}\cdot x_3}}{\displaystyle 
\sqrt{2E_1V\,2E_2V\,2E_{\rm D}V\,
2E_{\rm e^+}V\,2E_{\nu_{\rm e}}V}}.
\eqno({\rm C}.23)
$$
where $\psi_{\rm pp}(\vec{\rho})_{\rm in}$ is the wave function of the relative movement of the protons normalized per unit density [8]. At low--energies the wave function $\psi_{\rm pp}(\vec{\rho})_{\rm in}$ is given by Eq.~(\ref{label3.2}). 

Substituting Eq.~(\ref{label3.2}) in Eq.~({\rm C}.23) we obtain the matrix elements in the form
$$
<{\rm D}(k_{\rm D}){\rm e}^+(k_{\rm e^+})\nu_{\rm e}(k_{\nu_{\rm e}})|{\rm T}([\bar{p^c}(t_1,\vec{x}_1 + \frac{1}{2}\,\vec{\rho}\,)\,\gamma_{\alpha}\gamma^5 p(t_1,\vec{x}_1 - \frac{1}{2}\,\vec{\rho}\,)]\,D^{\dagger}_{\mu}(x_2)\,
$$
$$
\times\,[\bar{\psi}_{\nu_{\rm e}}(x_3)\gamma_{\nu}(1 - \gamma^5) \psi_{\rm e}(x_3)])|p(p_1) p(p_2)>= -\sqrt{2}\,a^{\rm e}_{\rm pp}\,C(\eta)\,\frac{\Phi(\rho)}{\rho}\,[\bar{u^c}(p_2)\gamma_{\alpha}\gamma^5 u(p_1)]
$$
$$
\times\,[\bar{u}(k_{\nu_{\rm e}})
 \gamma_{\nu}
(1-\gamma^5) v(k_{\rm e^+})]
\,e^*_{\mu}(k_{\rm D})\,
\frac{\displaystyle 
e^{\displaystyle -i(p_1+p_2)\cdot x_1}\,
e^{\displaystyle ik_{\rm D}\cdot x_2}\,
e^{\displaystyle ik_{{\ell}}\cdot x_3}}{\displaystyle \sqrt{2E_1V\,2E_2V\,
2E_{\rm D}V\,
 2E_{\rm e^+}V\,
 2E_{\nu_{\rm e}}V}},
$$
$$
<{\rm D}(k_{\rm D}){\rm e}^+(k_{\rm e^+})
\nu_{\rm e}(k_{\nu_{\rm e}})|{\rm T}
([\bar{p^c}(t_1,\vec{x}_1 + \frac{1}{2}\,\vec{\rho}\,)\,
\gamma^5 p(t_1,\vec{x}_1 - \frac{1}{2}\,\vec{\rho}\,)]\,D^{\dagger}_{\mu}(x_2)\,
$$
$$
\times\,[\bar{\psi}_{\nu_{\rm e}}(x_3)
\gamma_{\nu}(1 - \gamma^5) \psi_{\rm e}(x_3)])|p(p_1) p(p_2)>=-\sqrt{2}\,a^{\rm e}_{\rm pp}\,C(\eta)\,\frac{\Phi(\rho)}{\rho}\,
[\bar{u^c}(p_2)\gamma^5 u(p_1)]
$$
$$
\times\,[\bar{u}(k_{\nu_{\rm e}})\gamma_{\nu}
(1-\gamma^5)
v(k_{\rm e^+})]
\,e^*_{\mu}(k_{\rm D})\,
\frac{\displaystyle e^{\displaystyle -i(p_1+p_2)\cdot x_1}\,e^{\displaystyle ik_{\rm D}\cdot x_2}\,
e^{\displaystyle ik_{{\ell}}\cdot x_3}}{\displaystyle \sqrt{2E_1V\,2E_2V\,
2E_{\rm D}V\,
2E_{\rm e^+}V\,
2E_{\nu_{\rm e}}V}}. 
\eqno({\rm C}.24)
$$
The interacting protons are in the ${^1}{\rm S}_0$--state. This means that the spinorial wave function of the protons should be antisymmetric under the permutation. In our approach the spinorial wave function of the protons is described by $[\bar{u^c}(p_2)\gamma_{\alpha}\gamma^5 u(p_1)]$ and $[\bar{u^c}(p_2)\gamma^5 u(p_1)]$, antisymmetric under permutations of the protons: $[\bar{u^c}(p_2) \gamma_{\alpha} \gamma^5 u(p_1)] = - [\bar{u^c}(p_1) \gamma_{\alpha} \gamma^5 u(p_2)]$ and $[\bar{u^c}(p_2) \gamma^5  u(p_1)] = - [\bar{u^c}(p_1) \gamma^5 u(p_2)]$.

Now let us discuss in details the computation of the matrix elements:
$$
<0|\bar{p^c}(t_1,\vec{x}_1 + \frac{1}{2}\,\vec{\rho}\,)\,\Gamma\, p(t_1,\vec{x}_1 - \frac{1}{2}\,\vec{\rho}\,)|p(p_1) p(p_2)>, \eqno({\rm C}.25)
$$
where we have denoted $\Gamma = \gamma_{\alpha}\gamma^5$ or $\gamma^5$.

In the quantum field theory approach the wave function $|p(p_1) p(p_2)>$ should be described in terms of the operators of the creation of the protons $a^{\dagger}(\vec{p}_1,\sigma_1)$ and $a^{\dagger}(\vec{p}_2,\sigma_2)$, where $\vec{p}_i$ and $\sigma_i\,(i=1,2)$ are the 3--momenta and the polarizations of the protons. Therefore, $|p(p_1) p(p_2)>$ reads
$$
|p(p_1) p(p_2)> =\frac{1}{\sqrt{2}} a^{\dagger}(\vec{p}_1,\sigma_1)\, a^{\dagger}(\vec{p}_2,\sigma_2)|0>.\eqno({\rm C}.26)
$$
The wave function Eq.~({\rm C}.26) is taken in the standard form [31]. It is antisymmetric under permutations of the protons due to the anti--commutation relation 
\[a^{\dagger}(\vec{p}_1,\sigma_1) \,a^{\dagger}(\vec{p}_2,\sigma_2) = - a^{\dagger}(\vec{p}_2,\sigma_2) \,a^{\dagger}(\vec{p}_1,\sigma_1)\]
and normalized to unity. The factor $1/\sqrt{2}$ takes into account that the protons are correlated in the initial state.

The operators of the proton fields $\bar{p^c}(t_1,\vec{x}_1 + \frac{1}{2}\,\vec{\rho}\,)$ and $p(t_1,\vec{x}_1 - \frac{1}{2}\,\vec{\rho}\,)$ we represent, first, in terms of the plane--wave expansions
$$
\bar{p^c}(t_1,\vec{x}_1 + \frac{1}{2}\,\vec{\rho}\,) = \sum_{\vec{q}_1,\alpha_1}\frac{1}{\displaystyle \sqrt{2E_{\vec{q}_1}V}}\Bigg[ a(\vec{q}_1,\alpha_1)\,\bar{u^c}(q_1)\,e^{\displaystyle -iE_{\vec{q}_1}t_1 + i\vec{q}_1\cdot (\vec{x}_1 + \vec{\rho}/2)}
$$
$$
\hspace{0.5in}+  b^{\dagger}(\vec{q}_1,\alpha_1)\,\bar{v^c}(q_1)\,e^{\displaystyle iE_{\vec{q}_1}t_1 - i\vec{q}_1\cdot (\vec{x}_1 + \vec{\rho}/2)}\Bigg],
$$
$$
p(t_1,\vec{x}_1 - \frac{1}{2}\,\vec{\rho}\,) = \sum_{\vec{q}_2,\alpha_2}\frac{1}{\displaystyle \sqrt{2E_{\vec{q}_2}V}}\Bigg[ a(\vec{q}_2,\alpha_2)\,u(q_2)\,e^{\displaystyle -iE_{\vec{q}_2}t_1 + i\vec{q}_2\cdot (\vec{x}_1 - \vec{\rho}/2)}
$$
$$
 +  b^{\dagger}(\vec{q}_2,\alpha_2)\,v(q_2)\,e^{\displaystyle iE_{\vec{q}_2}t_1 - i\vec{q}_2\cdot (\vec{x}_1 - \vec{\rho}/2)}\Bigg],\eqno({\rm C}.27)
$$
where $a(\vec{q}_i,\alpha_i)\,(i=1,2)$ and $b^{\dagger}(\vec{q}_i,\alpha_i)\,(i=1,2)$ are the operators of the annihilation and the creation of protons and ani-protons, respectively. The computation of the matrix element Eq.~({\rm C}.25) runs the following way. Keeping only the terms containing the operators of the annihilation of the protons we get
$$
<0|\bar{p^c}(t_1,\vec{x}_1 + \frac{1}{2}\,\vec{\rho}\,)\,\Gamma p(t_1,\vec{x}_1 - \frac{1}{2}\,\vec{\rho}\,)|p(p_1) p(p_2)> =
$$
$$
=\sum_{\vec{q}_1,\alpha_1}\sum_{\vec{q}_2,\alpha_2}\frac{1}{\displaystyle \sqrt{2E_{\vec{q}_1}V}}\frac{1}{\displaystyle \sqrt{2E_{\vec{q}_2}V}}\,e^{\displaystyle -i(q_1 + q_2)\cdot x_1 + i(\vec{q}_1 - \vec{q}_2)\cdot \vec{\rho}/2}
$$
$$
\times\,[\bar{u^c}(q_1)\,\Gamma\,u(q_2)]\,
\frac{1}{\sqrt{2}}<0|a(\vec{q}_1,\alpha_1) 
a(\vec{q}_2,\alpha_2)\,
a^{\dagger}(\vec{p}_1,\sigma_1)
\,a^{\dagger}(\vec{p}_2,\sigma_2)|0>.\eqno({\rm C}.28)
$$
The vacuum expectation value 
$<0|a(\vec{q}_1,\alpha_1) \,a(\vec{q}_1,\alpha_1) 
\,a^{\dagger}(\vec{p}_1,\sigma_1)\,
a^{\dagger}(\vec{p}_2,\sigma_2)|0>$ reads:
$$
<0|a(\vec{q}_1,\alpha_1) 
\,a(\vec{q}_1,\alpha_1) \,
a^{\dagger}(\vec{p}_1,\sigma_1)
\,a^{\dagger}(\vec{p}_2,\sigma_2)|0> =
$$
$$
=-\delta_{\vec{q}_1\vec{p}_1}
\,\delta_{\alpha_1\sigma_1}\,
\delta_{\vec{q}_2\vec{p}_2}
\,\delta_{\alpha_2\sigma_2} + \delta_{\vec{q}_2\vec{p}_1}
\,\delta_{\alpha_2\sigma_1}
\,\delta_{\vec{q}_1\vec{p}_2}
\,\delta_{\alpha_1\sigma_2},\eqno({\rm C}.29)
$$
where we have used the anti--commutation relations
$$
a(\vec{q},\alpha) \,a^{\dagger}(\vec{p},\sigma) + a^{\dagger}(\vec{p},\sigma)a(\vec{q},\alpha) = \delta_{\vec{q}\vec{p}}\,\delta_{\alpha\sigma}\eqno({\rm C}.30)
$$
and the properties of the operators of the creation and the annihilation: $<0|a^{\dagger}(\vec{p},\sigma) = 0$ and $a(\vec{q},\alpha)|0> = 0$.

Substituting Eq.~({\rm C}.29)in Eq.~({\rm C}.28) and summing up the momenta and the spinorial indices we arrive at the expression
$$
<0|\bar{p^c}(t_1,\vec{x}_1 + \frac{1}{2}\,\vec{\rho}\,)\,\Gamma p(t_1,\vec{x}_1 - \frac{1}{2}\,\vec{\rho}\,)|p(p_1) p(p_2)> =-\frac{\displaystyle e^{\displaystyle -i(p_1 + p_2)\cdot x_1}}{\displaystyle \sqrt{2E_1V\,2E_2V}}\,
$$
$$
\times\,\frac{1}{\sqrt{2}}\,\Bigg([\bar{u^c}(p_1)\,\Gamma\,u(p_2)]
\,e^{\displaystyle i(\vec{p}_1 - \vec{p}_2)\cdot \vec{\rho}/2} -  [\bar{u^c}(p_2)\,\Gamma\,u(p_1)]
\,e^{\displaystyle - i(\vec{p}_1 - \vec{p}_2)\cdot \vec{\rho}/2}\Bigg) = 
$$
$$
= \frac{\displaystyle e^{\displaystyle -i(p_1 + p_2)\cdot x_1}}{\displaystyle \sqrt{2E_1V\,2E_2V}}\,\sqrt{2}\,[\bar{u^c}(p_2)\,\Gamma\,u(p_1)]
\,\frac{1}{2}\,\Bigg(e^{\displaystyle i(\vec{p}_1 - \vec{p}_2)\cdot \vec{\rho}/2} +  e^{\displaystyle - i(\vec{p}_1 - \vec{p}_2)\cdot \vec{\rho}/2}\Bigg),\eqno({\rm C}.31)
$$
where the relation $[\bar{u^c}(p_1)\,\Gamma\,u(p_2)] = - [\bar{u^c}(p_2)\,\Gamma\,u(p_1)]$ has been used. The sum of the exponentials
$$
\frac{1}{2}\,\Bigg(e^{\displaystyle i(\vec{p}_1 - \vec{p}_2)\cdot \vec{\rho}/2} +  e^{\displaystyle - i(\vec{p}_1 - \vec{p}_2)\cdot \vec{\rho}/2}\Bigg) \eqno({\rm C}.32)
$$
describes the spatial part of the wave function of the relative movement of the free protons. This wave function is  symmetric  under permutations of the protons and normalized per unit density [8]. Since the protons should be in the ${^1}{\rm S}_0$--state, expanding into spherical harmonics and keeping only the S--wave contribution we obtain [8]:
$$
\frac{1}{2}\,\Bigg(e^{\displaystyle i \vec{k}\cdot \vec{\rho}} +   e^{\displaystyle -i \vec{k}\cdot \vec{\rho}}\Bigg) = \frac{\sin k\rho}{k\rho}+ \ldots,\eqno({\rm C}.33)
$$
 where  $\vec{k}=(\vec{p}_1 - \vec{p}_2)/2$ is the relative momentum of the protons. In order to take into account the Coulomb repulsion between protons we should merely replace
$$
\frac{\sin k\rho}{k\rho} \to \psi_{\rm pp}(\rho),\eqno({\rm C}.34)
$$
where $\psi_{\rm pp}(\rho) = \psi_{\rm pp}(\vec{\rho}\,)_{\rm in}$ is the Coulomb wave function of the protons in the ${^1}{\rm S}_0$--state. In the low--energy limit  $\psi_{\rm pp}(\rho)$ is given by Eq.~(\ref{label3.2}). This completes the explanation of the derivation of the matrix elements in Eq.~({\rm C}.23) and Eq.~({\rm C}.24).

Substituting the matrix elements  Eq.~({\rm C}.24) in the r.h.s. of  Eq.~({\rm C}.22) we obtain the matrix element of the p + p $\to$ D + e$^+$ + $\nu_{\rm e}$ process  in the following form
$$
(2\pi)^4\delta^{(4)}(k_{\rm D} + k_{\ell} - p_1 - p_2)\,i{\cal M}({\rm p} + {\rm p} \to {\rm D} + {\rm e}^+ + \nu_{e}) = $$
$$
= \sqrt{2}\,C(\eta)\,a^{\rm e}_{\rm pp}g_{\rm A}G_{\rm \pi pp}\frac{G_{\rm V}}{\sqrt{2}} \frac{g_{\rm V}}{8\pi^2}[\bar{u^c}(p_2)\gamma_{\alpha}\gamma^5 u(p_1)][\bar{u}(k_{\nu_{\rm e}})\gamma_{\nu}
(1-\gamma^5)
v(k_{\rm e^+})]
\,e^*_{\mu}(k_{\rm D})
$$
$$
\times 
\int d^4x_1 \int \frac{d^4x_2 d^4k_2}{(2\pi)^4}
\frac{d^4x_3 d^4k_3}{(2\pi)^4}\,
e^{\displaystyle i (k_2 + k_3 - p_1 - p_2)\cdot x_1}
\,e^{\displaystyle i(k_{\rm D} - k_2)\cdot x_2}
\,e^{\displaystyle i(k_{\ell} - k_3)\cdot x_3}
$$
$$
\times\int d^3\rho   U(\rho)\frac{\Phi(\rho)}{\rho}\int\frac{d^4k_1}{\pi^2i} e^{\displaystyle i\vec{q}\cdot \vec{\rho}} {\rm tr}\Bigg\{\gamma^{\alpha}\gamma^5\frac{1}{M_{\rm N} - \hat{k}_1 + \hat{k}_2}\gamma^{\mu}\frac{1}{M_{\rm N} - \hat{k}_1}\gamma^{\nu}\gamma^5 \frac{1}{M_{\rm N} - \hat{k}_1 - \hat{k}_3}\Bigg\}
$$
$$
+  \sqrt{2}\,C(\eta)\,a^{\rm e}_{\rm pp}g_{\rm A}G_{\rm \pi pp}\frac{G_{\rm V}}{\sqrt{2}} \frac{g_{\rm V}}{8\pi^2}[\bar{u^c}(p_2)\gamma^5 u(p_1)][\bar{u}(k_{\nu_{\rm e}})\gamma_{\nu}
(1-\gamma^5)
v(k_{\rm e^+})]\,
e^*_{\mu}(k_{\rm D})
$$
$$
\times\int d^4x_1 \int \frac{d^4x_2 d^4k_2}{(2\pi)^4}
\frac{d^4x_3 d^4k_3}{(2\pi)^4}\,
e^{\displaystyle i (k_2 + k_3 - p_1 - p_2)\cdot x_1}\,
e^{\displaystyle i(k_{\rm D} - k_2)\cdot x_2}\,
e^{\displaystyle i(k_{\ell} - k_3)\cdot x_3}
$$
$$
\times\int d^3\rho U(\rho)\frac{\Phi(\rho)}{\rho} \int\frac{d^4k_1}{\pi^2i}e^{\displaystyle i\vec{q}\cdot \vec{\rho}}{\rm tr}\Bigg\{\gamma^5
\frac{1}{M_{\rm N} - \hat{k}_1 + \hat{k}_2}
\gamma^{\mu}
\frac{1}{M_{\rm N} - \hat{k}_1}\gamma^{\nu}
\gamma^5
\frac{1}{M_{\rm N} - \hat{k}_1 
- \hat{k}_3}\Bigg\}.
\eqno({\rm C}.35)
$$
Integrating over $x_1$, $x_2$, $x_3$, 
$k_2$ and $k_3$ we obtain in the r.h.s. 
of Eq.~({\rm C}.35) the $\delta$--function 
describing the 4--momentum conservation. Then, the matrix element of the  p + p $\to$ D + e$^+$ + $\nu_{\rm e}$ 
process becomes equal
$$
i{\cal M}({\rm p} + {\rm p} \to {\rm D} + {\rm e}^+ + \nu_{e}) =  \sqrt{2}\,C(\eta)\,a^{\rm e}_{\rm pp}
g_{\rm A}G_{\rm \pi pp}
\frac{G_{\rm V}}{\sqrt{2}} 
\frac{g_{\rm V}}{8\pi^2}
$$
$$
\times\,[\bar{u^c}(p_2)\gamma_{\alpha}
\gamma^5 u(p_1)][\bar{u}(k_{\nu_{\rm e}})
\gamma_{\nu}(1-\gamma^5)
v(k_{\rm e^+})]\,e^*_{\mu}(k_{\rm D})
\int d^3\rho\,  U(\rho)\,
\frac{\Phi(\rho)}{\rho}
\int\frac{d^4k_1}{\pi^2i} 
e^{\displaystyle i\vec{q}\cdot \vec{\rho}} 
$$
$$
\times\,{\rm tr}\Bigg\{
\gamma^{\alpha}\gamma^5
\frac{1}{M_{\rm N} - \hat{k}_1 + \hat{k}_2}
\gamma^{\mu}\frac{1}{M_{\rm N} - \hat{k}_1}
\gamma^{\nu}\gamma^5 
\frac{1}{M_{\rm N} - \hat{k}_1 
- \hat{k}_3}\Bigg\}
+ \sqrt{2}\,C(\eta)\,
a^{\rm e}_{\rm pp}
g_{\rm A}G_{\rm \pi pp}
$$
$$
\times\,\frac{G_{\rm V}}{\sqrt{2}}
\frac{g_{\rm V}}{8\pi^2}
[\bar{u^c}(p_2)\gamma^5 u(p_1)]
[\bar{u}(k_{\nu_{\rm e}})
\gamma_{\nu}(1-\gamma^5)v(k_{\rm e^+})]
\,e^*_{\mu}(k_{\rm D})\int d^3\rho\, 
U(\rho)\,\frac{\Phi(\rho)}{\rho}
$$
$$
\times \int\frac{d^4k_1}{\pi^2i}\,
e^{\displaystyle i\vec{q}\cdot \vec{\rho}}
{\rm tr}\Bigg\{\gamma^5
\frac{1}{M_{\rm N} - \hat{k}_1 + \hat{k}_{\rm D}}\gamma^{\mu}\frac{1}{M_{\rm N} - \hat{k}_1}
\gamma^{\nu}\gamma^5
\frac{1}{M_{\rm N} - \hat{k}_1
- \hat{k}_{\ell}}\Bigg\},\eqno({\rm C}.36)
$$
where $\vec{q}= \vec{k} + (\vec{k}_{\ell} - \vec{k}_{\rm D})/2$.

It is convenient to represent the matrix element Eq.~({\rm C}.35) in terms of the structure functions ${\cal J}^{\alpha\mu\nu}_{\rm pp}(k_{\rm D}, k_{\ell})$ and ${\cal J}^{\mu\nu}_{\rm pp}(k_{\rm D}, k_{\ell})$:
$$
i{\cal M}({\rm p} + {\rm p} \to {\rm D} + {\rm e}^+ + \nu_{e}) =
$$
$$
= - \,C(\eta)\,g_{\rm A}G_{\rm \pi pp}\frac{G_{\rm V}}{\sqrt{2}} \frac{g_{\rm V}}{8\pi^2}[\bar{u^c}(p_2)\gamma_{\alpha}
\gamma^5 u(p_1)][\bar{u}(k_{\nu_{\rm e}})\gamma_{\nu}
(1-\gamma^5)
v(k_{\rm e^+})]\,e^*_{\mu}(k_{\rm D})\,
{\cal J}^{\alpha\mu\nu}_{\rm pp}(k_{\rm D}, k_{\ell})
$$
$$
- \,C(\eta)\,g_{\rm A}G_{\rm \pi pp}\frac{G_{\rm V}}{\sqrt{2}} \frac{g_{\rm V}}{8\pi^2}[\bar{u^c}(p_2)\gamma^5 u(p_1)][\bar{u}(k_{\nu_{\rm e}})\gamma_{\nu}(1-\gamma^5)
v(k_{\rm e^+})]
\,e^*_{\mu}(k_{\rm D}){\cal J}^{\mu\nu}_{\rm pp}
(k_{\rm D}, k_{\ell}),\eqno({\rm C}.37)
$$
where the structure functions 
${\cal J}^{\alpha\mu\nu}_{\rm pp}
(k_{\rm D}, k_{\ell})$ and 
${\cal J}^{\mu\nu}_{\rm pp}
(k_{\rm D}, k_{\ell})$ are defined as
$$
{\cal J}^{\alpha\mu\nu}_{\rm pp}(k_{\rm D}, k_{\ell}) = -\,\sqrt{2}\,a^{\rm e}_{\rm pp}\int d^3\rho\,  U(\rho)\frac{\Phi(\rho)}{\rho}\int\frac{d^4k_1}{\pi^2i} e^{\displaystyle i\vec{q}\cdot \vec{\rho}}
$$
$$
\times\,{\rm tr} \Bigg\{\gamma^{\alpha}\gamma^5\frac{1}{M_{\rm N} - \hat{k}_1 + \hat{k}_{\rm D}}\gamma^{\mu}\frac{1}{M_{\rm N} - \hat{k}_1}\gamma^{\nu}\gamma^5 \frac{1}{M_{\rm N} - \hat{k}_1 - \hat{k}_{\ell}}\Bigg\},
$$
$$
{\cal J}^{\mu\nu}_{\rm pp}(k_{\rm D}, k_{\ell}) = -\,\sqrt{2}\,a^{\rm e}_{\rm pp}\int d^3\rho\,  U(\rho)\frac{\Phi(\rho)}{\rho}\int\frac{d^4k_1}{\pi^2i} e^{\displaystyle i\vec{q}\cdot \vec{\rho}}
$$
$$
\times\,
{\rm tr}\Bigg\{\gamma^5
\frac{1}{M_{\rm N} - \hat{k}_1 
+ \hat{k}_{\rm D}}
\gamma^{\mu}\frac{1}{M_{\rm N}
- \hat{k}_1}
\gamma^{\nu}\gamma^5 \frac{1}{M_{\rm N}
- \hat{k}_1 - \hat{k}_{\ell}}\Bigg\}.
\eqno({\rm C}.38)
$$
Thus, the problem of the computation of the matrix element 
of the p + p $\to$ D + e$^+$ 
+ $\nu_{\rm e}$ process reduces to 
the problem of the computation of the structure functions Eq.~({\rm C}.38). Integrating over directions of the relative radius--vector $\vec{\rho}$ we get
$$
{\cal J}^{\alpha\mu\nu}_{\rm pp}
(k_{\rm D}, k_{\ell}) = -\,\sqrt{2}\,a^{\rm e}_{\rm pp}\,
4\pi\int\limits^{\infty}_0 d\rho\,  
\rho\, U(\rho)\,\Phi(\rho)\int
\frac{d^4k_1}{\pi^2i}
\frac{\sin|\vec{q}\,|\rho}{|\vec{q}\,|\rho} 
$$
$$
\times\,{\rm tr} \Bigg\{\gamma^{\alpha}
\gamma^5
\frac{1}{M_{\rm N} - \hat{k}_1 + \hat{k}_{\rm D}} \gamma^{\mu}
\frac{1}{M_{\rm N} - \hat{k}_1}
\gamma^{\nu}\gamma^5 
\frac{1}{M_{\rm N} - \hat{k}_1 
- \hat{k}_{\ell}}\Bigg\},
$$
$$
{\cal J}^{\mu\nu}_{\rm pp}
(k_{\rm D}, k_{\ell}) = 
-\,\sqrt{2}\,a^{\rm e}_{\rm pp}\,
4\pi\int\limits^{\infty}_0 d\rho\,  \rho\, U(\rho)\,\Phi(\rho)\int\frac{d^4k_1}{\pi^2i}
\frac{\sin|\vec{q}\,|\rho}{|\vec{q}\,|\rho}
$$
$$
\times\,{\rm tr}\Bigg\{\gamma^5
\frac{1}{M_{\rm N} - \hat{k}_1 + \hat{k}_{\rm D}}\gamma^{\mu}\frac{1}{M_{\rm N} - \hat{k}_1}
\gamma^{\nu}\gamma^5 \frac{1}{M_{\rm N} - \hat{k}_1 - \hat{k}_{\ell}}\Bigg\}.
\eqno({\rm C}.39)
$$
We calculate the astrophysical factor at zero relative kinetic energy of the protons. This allows to simplify the calculation of the structure functions and set $\vec{k}_{\rm D} = k^{\mu}_{\ell} = 0$ [5]. In this limit the structure functions read
$$
{\cal J}^{\alpha\mu\nu}_{\rm pp}(k_{\rm D}, k_{\ell}) = -\,\sqrt{2}\,a^{\rm e}_{\rm pp}\,4\pi\int\limits^{\infty}_0 d\rho\,  \rho\, U(\rho)\,\Phi(\rho)\int\frac{d^4k_1}{\pi^2i}\frac{\sin|\vec{k}_1\,|\rho}{|\vec{k_1}\,|\rho} 
$$
$$
\times\,{\rm tr} \Bigg\{\gamma^{\alpha}\gamma^5\frac{1}{M_{\rm N} - \hat{k}_1 + \hat{k}_{\rm D}}\gamma^{\mu}\frac{1}{M_{\rm N} - \hat{k}_1}\gamma^{\nu}\gamma^5 \frac{1}{M_{\rm N} - \hat{k}_1}\Bigg\},
$$
$$
{\cal J}^{\mu\nu}_{\rm pp}(k_{\rm D}, k_{\ell}) = -\,\sqrt{2}\,a^{\rm e}_{\rm pp}\,4\pi\int\limits^{\infty}_0 d\rho\,  \rho\, U(\rho)\,\Phi(\rho)\int\frac{d^4k_1}{\pi^2i}\frac{\sin|\vec{k}_1\,|\rho}{|\vec{k}_1\,|\rho}
$$
$$
\times\,{\rm tr}\Bigg\{\gamma^5\frac{1}{M_{\rm N} - \hat{k}_1 + \hat{k}_{\rm D}}\gamma^{\mu}\frac{1}{M_{\rm N} - \hat{k}_1 }\gamma^{\nu}\gamma^5 \frac{1}{M_{\rm N} - \hat{k}_1}\Bigg\}.\eqno({\rm C}.40)
$$
The computation of the momentum integrals defining the structure functions cannot be carried out by a Lorentz covariant manner [2,4]. The obvious Lorentz covariance has been lost due to the description of the pp interaction in terms of the potential. Therefore, for the computation of the momentum integrals it is convenient to follow only the components which give the main contribution in the low--energy limit. 
For the calculation of 
${\cal J}^{\alpha\mu\nu}
(k_{\rm D}, k_{\ell})$ we should notice that in the 
low--energy limit only the time--component of the current $[\bar{u^c}(p_2)
\gamma_{\alpha}\gamma^5 u(p_1)]$ survives and obeys the relation
$$
[\bar{u^c}(p_2)\gamma_{\alpha}\gamma^5 u(p_1)] =  [u^T(p_2)\,C\,\gamma_{\alpha}\gamma^5 u(p_1)] \to  g_{\alpha 0}[u^T(p_2)\,C\,\gamma_0\gamma^5 u(p_1)] = 
$$
$$
=g_{\alpha 0}[u^T(p_2)\,C\,\gamma_0\,C^T C \gamma^5 u(p_1)] =
- g_{\alpha 0}[u^T(p_2)\,\gamma^T_0\,C\,\gamma^5 u(p_1)] =
$$
$$
=- g_{\alpha 0}[u^T(p_2)\,\gamma_0\,C\,\gamma^5 u(p_1)] = - g_{\alpha 0}[u^T(p_2)\,C\,\gamma^5 u(p_1)] =
$$
$$
=- g_{\alpha 0}[\bar{u^c}(p_2)\gamma^5 u(p_1)],
$$
$$
[\bar{u^c}(p_2)\gamma_{\alpha}\gamma^5 u(p_1)] \to  - g_{\alpha 0}[\bar{u^c}(p_2)\gamma^5 u(p_1)],\eqno({\rm C}.41)
$$
where we have used the relation $u^T(p_2)\,\gamma_0 = u^T(p_2)$, which is valid in the non--relativistic limit due to the dominance  of the large components of the Dirac bispinors.

Then, since the time--component of the polarization vector $e^*_{\mu}(k_{\rm D})$ of the deuteron is unphysical and does not contribute to the observed quantities like the cross section, we should follow only the spatial part of the polarization vector $e^*_{\mu}(k_{\rm D})$ for $\mu$ running over $\mu = 1,2,3$. 

Now it is rather clear that only the spatial part of the leptonic weak current, when the index $\nu$ runs over $\nu=1,2,3$, can give a non--trivial contribution. The former is caused by the property of the matrix element to be the scalar under spatial rotations of the Lorentz group. This leads to the contraction of indices $\mu$ and  $\nu$. 

Thus, the matrix element Eq.~({\rm C}.37) reduces to the form
$$
{\cal M}({\rm p} + {\rm p} \to {\rm D} + {\rm e}^+ + \nu_{e}) = - \,C(\eta)\,g_{\rm A}G_{\rm \pi pp}\frac{G_{\rm V}}{\sqrt{2}} \frac{g_{\rm V}}{8\pi^2}[\bar{u^c}(p_2)\gamma^5 u(p_1)]
$$
$$
\times\,[\bar{u}(k_{\nu_{\rm e}})\gamma_{\nu}(1-\gamma^5) 
v(k_{\rm e^+})]\,e^*_{\mu}(k_{\rm D})\,
[- {\cal J}^{0\mu\nu}_{\rm pp}(k_{\rm D}, k_{\ell}) +  
{\cal J}^{\mu\nu}_{\rm pp}(k_{\rm D}, k_{\ell})],
\eqno({\rm C}.42)
$$
where the structure functions 
${\cal J}^{0\mu\nu}_{\rm pp}
(k_{\rm D}, k_{\ell})$  and 
${\cal J}^{\mu\nu}_{\rm pp}
(k_{\rm D}, k_{\ell})$ are given in 
Eq.~({\rm C}.40).

For the calculation of the momentum integrals we would follow the philosophy of the derivation of Effective Chiral Lagrangians within effective quark models motivated by QCD [16--19], in particularly, Chiral perturbation theory at the quark level (CHPT)$_q$ [18] formulated on the basis of the ENJL model induced by the effective low--energy QCD with linearly rising confinement potential [64]. In (CHPT)$_q$ all low--energy vertices of meson interactions are determined by one--constituent quark loop diagrams with point--like quark--meson vertices and the Green functions of the free constituent quarks with constant masses $M_q = 330\,{\rm MeV}$ [18]. To the computation of the momentum integrals one applies a generalized hypothesis of Vector Dominance [20,28] postulating a smooth dependence of low--energy vertices of meson interactions on squared 4--momenta of interacting mesons. Due to this hypothesis one can hold all external particles off--mass shell at squared 4--momenta $p^2$ much less than $M^2_q$, i.e., $M^2_q\gg p^2$. Then, after the computation of the momentum integrals at leading order in long--wavelength expansion, i.e., in powers of external momenta, the resultant expression should be continued on--mass shell of interacting particles. Within the framework of this procedure one can restore completely all variety of phenomenological vertices of low--energy meson interactions predicted by Effective Chiral Lagrangians [16--19,20,21]. It is important to emphasize that this procedure works good not only for light mesons like $\pi$--meson, which mass is less than the mass of constituent quarks, but for vector mesons like $\rho(770)$, $\omega(780)$ and so on, which masses are twice larger than the constituent quark mass. Since the former resembles the RFMD, where the mass of the deuteron amounts to twice the mass of virtual nucleons, we expect that the long--wavelength approximation should work in the RFMD as well as in effective quark models with chiral $U(3) \times U(3)$ symmetry applied to the derivation of Effective Chiral Lagrangians.

Thus, for the computation of the momentum integrals we assume that the deuteron is off--mass shell and $M_{\rm N} \gg \sqrt{k^2_{\rm D}}$. Then, we expand the integrand of the structure functions Eq.~({\rm A}.39) in powers of $k_{\rm D}$ keeping only leading contributions. The result of the computation we continue on--mass shell of the deuteron $k^2_{\rm D} \to M^2_{\rm D}$ [2,4].
\vspace{0.2in}

\noindent{\bf The computation of 
${\cal J}^{0\mu\nu}_{\rm pp}
(k_{\rm D}, k_{\ell})$}. 
For the computation of 
${\cal J}^{0\mu\nu}_{\rm pp}
(k_{\rm D}, k_{\ell})$ we should integrate, first, over the virtual momentum. It is convenient 
to replace $k_1 \to k$:
$$
\int\frac{d^4k}{\pi^2i}\frac{\sin|\vec{k}\,|\rho}{|\vec{k}\,|\rho} \,{\rm tr} \Bigg\{\gamma^0\gamma^5
\frac{1}{M_{\rm N} - \hat{k} + \hat{k}_{\rm D}}\gamma^{\mu}\frac{1}{M_{\rm N} - \hat{k}}\gamma^{\nu}\gamma^5
\frac{1}{M_{\rm N} - \hat{k}}\Bigg\}=
$$
$$
=\int\frac{d^4k}{\pi^2i}\frac{\sin|\vec{k}\,|\rho}{|\vec{k}\,|\rho} \,\frac{{\rm tr}\{\gamma^0\gamma^5
(M_{\rm N} + \hat{k} - \hat{k}_{\rm D})
\gamma^{\mu}(M_{\rm N} + \hat{k})\gamma^{\nu}
\gamma^5(M_{\rm N} + 
\hat{k})\}}{[M^2_{\rm N} - (k - k_{\rm D})^2
- i0][M^2_{\rm N} - k^2 - i0]^2}=
$$
$$
=\int\frac{d^4k}{\pi^2i}\frac{\sin|\vec{k}\,|\rho}{|\vec{k}\,|\rho} \,\frac{{\rm tr}\{\gamma^0\gamma^5(M_{\rm N} + \hat{k} - \hat{k}_{\rm D})\gamma^{\mu}(M_{\rm N} + \hat{k})\gamma^{\nu}\gamma^5(M_{\rm N} + \hat{k})\}}{[M^2_{\rm N} - k^2 - i0]^3}
$$
$$
\times\,\Bigg\{1 - 
\frac{2 k\cdot k_{\rm D}}{[M^2_{\rm N}
- k^2 - i0]}\Bigg\}=
$$
$$
=\int\frac{d^4k}{\pi^2i}\frac{\sin|\vec{k}\,|\rho}{|\vec{k}\,|\rho} \,\frac{{\rm tr}\{\gamma^0\hat{k}_{\rm D}\gamma^{\mu}(M_{\rm N} + \hat{k})\gamma^{\nu}(M_{\rm N} - 
\hat{k})\}}{[M^2_{\rm N} - k^2 - i0]^3}
$$
$$
+\int\frac{d^4k}{\pi^2i}\frac{\sin|\vec{k}\,|\rho}{|\vec{k}\,|\rho} \,\frac{2 k\cdot k_{\rm D}\,{\rm tr}\{\gamma^0(M_{\rm N} + \hat{k})\gamma^{\mu}(M_{\rm N} + \hat{k})\gamma^{\nu}(M_{\rm N} - \hat{k})\}}{[M^2_{\rm N} - k^2 - i0]^4}=
$$
$$
= J^{0\mu\nu}_1 + J^{0\mu\nu}_2.\eqno({\rm C}.43)
$$
The computation of $J^{0\mu\nu}_1$ runs as follows
$$
 J^{0\mu\nu}_1=\int\frac{d^4k}{\pi^2i}\frac{\sin|\vec{k}\,|\rho}{|\vec{k}\,|\rho} \,\frac{{\rm tr}\{\gamma^0\hat{k}_{\rm D}\gamma^{\mu}(M_{\rm N} + \hat{k})\gamma^{\nu}(M_{\rm N} - \hat{k})\}}{[M^2_{\rm N} - k^2 - i0]^3}
$$
$$
= k^0_{\rm D}\int\frac{d^4k}{\pi^2i}\frac{\sin|\vec{k}\,|\rho}{|\vec{k}\,|\rho} \,\frac{{\rm tr}\{\gamma^{\mu}(M_{\rm N} + \hat{k})\gamma^{\nu}(M_{\rm N} - \hat{k})\}}{[M^2_{\rm N} - k^2 - i0]^3}=
$$
$$
=4 k^0_{\rm D}\int\frac{d^4k}{\pi^2i}\frac{\sin|\vec{k}\,|\rho}{|\vec{k}\,|\rho} \,\frac{M^2_{\rm N}\,g^{\mu\nu} - 2\, k^{\mu}k^{\nu} + k^2 g^{\mu\nu}}{[E^2_{\vec{k}} - k^2_0 - i0]^3},\eqno({\rm C}.44)
$$
where $k^0_{\rm D}$ is the time--component of the 4--vector $k^{\mu}_{\rm D} = (k^0_{\rm D}, \vec{0}\,)$ and $E_{\vec{k}}=\sqrt{\vec{k}^{\,2} + M^2_{\rm N}}$. Since the indices $\mu$ and $\nu$ are the spatial ones, the integration over directions of the vector $k^{\mu}$ gives
$$
 k^{\mu}k^{\nu} \to \frac{1}{3}\,\vec{k}^{\,2}\,\delta^{\mu\nu}= - \frac{1}{3}\, \vec{k}^{\,2}\,g^{\mu\nu}.\eqno({\rm C}.45)
$$
Substituting Eq.~({\rm C}.45) in Eq.~({\rm C}.44) we get
$$
J^{0\mu\nu}_1=4\,k^0_{\rm D}\,g^{\mu\nu}
\int\frac{d^4k}{\pi^2i}
\frac{\sin|\vec{k}\,|\rho}{|\vec{k}\,|\rho} 
\,\frac{\displaystyle M^2_{\rm N}
+ \frac{2}{3}\,\vec{k}^{\,2} + k^2_0 
- \vec{k}^{\,2} }{[E^2_{\vec{k}} 
- k^2_0 - i0]^3}
$$
$$
=4\,k^0_{\rm D}\,g^{\mu\nu}\int\frac{d^4k}{\pi^2i}
\frac{\sin|\vec{k}\,|\rho}{|\vec{k}\,|\rho} 
\,\frac{\displaystyle 
\frac{4}{3}\,M^2_{\rm N}- 
\frac{1}{3}\,E^2_{\vec{k}} 
+ k^2_0}{[E^2_{\vec{k}} 
- k^2_0 - i0]^3}=
$$
$$
=\frac{4}{3}\,k^0_{\rm D}\,g^{\mu\nu}
\int\frac{d^4k}{\pi^2i}
\frac{\sin|\vec{k}\,|\rho}{|\vec{k}\,|\rho}
\,\frac{\displaystyle 4\,M^2_{\rm N}-
E^2_{\vec{k}} + 3\, k^2_0}{[E^2_{\vec{k}}
- k^2_0 - i0]^3}.
\eqno({\rm C}.46)
$$
Making the Wick rotation we obtain
$$
 J^{0\mu\nu}_1=\frac{4}{3}\,k^0_{\rm D}\,g^{\mu\nu}\int\frac{d^3k}{\pi^2}
 \frac{\sin|\vec{k}\,|\rho}{|\vec{k}\,|\rho}
 \int\limits^{\infty}_{-\infty}
 dk_4 
 \frac{\displaystyle 4\,M^2_{\rm N}- 
 E^2_{\vec{k}} - 3\, k^2_4}{[E^2_{\vec{k}} 
 + k^2_4]^3}.
 \eqno({\rm C}.47)
$$
Integration over $k_4$ gives
$$
 J^{0\mu\nu}_1=\frac{4}{3}\,k^0_{\rm D}\,g^{\mu\nu}\int\frac{d^3k}{\pi^2}
 \frac{\sin|\vec{k}\,|\rho}{|\vec{k}\,|\rho}
 \left[\frac{\displaystyle 4\,M^2_{\rm N}- E^2_{\vec{k}}}{E^5_{\vec{k}}}
 \frac{\displaystyle \Gamma\Bigg(\frac{1}{2}\Bigg)
 \Gamma\Bigg(\frac{5}{2}\Bigg)}{\displaystyle 
 \Gamma(3)} - \frac{3}{E^3_{\vec{k}}}
 \frac{\displaystyle \Gamma\Bigg(\frac{3}{2}\Bigg)
 \Gamma\Bigg(\frac{3}{2}\Bigg)}{\displaystyle
 \Gamma(3)}
 \right]=
$$
$$
=\frac{4}{3}\,k^0_{\rm D}\,g^{\mu\nu}
\int\frac{d^3k}{\pi^2}
\frac{\sin|\vec{k}\,|\rho}{|\vec{k}\,|\rho}
\frac{3\pi}{4}
\frac{\displaystyle 2\,M^2_{\rm N}- E^2_{\vec{k}}}{E^5_{\vec{k}}}=\,k^0_{\rm D}\,g^{\mu\nu}\int\frac{d^3k}{\pi}
\frac{\sin|\vec{k}\,|\rho}{|\vec{k}\,|\rho}
\frac{\displaystyle 2\,M^2_{\rm N}
- E^2_{\vec{k}}}{E^5_{\vec{k}}}=
$$
$$
=4 k^0_{\rm D}\,g^{\mu\nu}\frac{1}{\rho}
\left[2M^2_{\rm N}\int\limits^{\infty}_0
\frac{\displaystyle dk\,k\,\cos k\rho}{\displaystyle 
(M^2_{\rm N} + k^2)^{5/2}}-
\int\limits^{\infty}_0\frac{\displaystyle dk\,k\,
\cos k\rho}{\displaystyle (M^2_{\rm N} + k^2)^{3/2}}
\right]=
$$
$$
=4 k^0_{\rm D}\,g^{\mu\nu}
\left[\frac{2}{3}\,M^2_{\rm N}
\int\limits^{\infty}_0
\frac{\displaystyle dk\,\cos k\rho}{\displaystyle 
(M^2_{\rm N} + k^2)^{3/2}}-\int\limits^{\infty}_0
\frac{\displaystyle dk\,\cos k\rho}{\displaystyle 
(M^2_{\rm N} + k^2)^{1/2}}\right].
\eqno({\rm C}.48)
$$
We have performed the integration by parts over 
$|\vec{k}\,|=k$. As a result we obtain
$$
J^{0\mu\nu}_1=\int\frac{d^4k}{\pi^2i}
\frac{\sin|\vec{k}\,|\rho}{|\vec{k}\,|\rho} 
\,\frac{{\rm tr}\{\gamma^0\hat{k}_{\rm D}
\gamma^{\mu}(M_{\rm N} + \hat{k})
\gamma^{\nu}(M_{\rm N} - \hat{k})\}}{[M^2_{\rm N}
- k^2 - i0]^3}
$$
$$
=4 k^0_{\rm D}\,g^{\mu\nu}\left[\frac{2}{3}
\,M^2_{\rm N}\int\limits^{\infty}_0
\frac{\displaystyle dk\,\cos k\rho}{\displaystyle 
(M^2_{\rm N} + k^2)^{3/2}}-\int\limits^{\infty}_0
\frac{\displaystyle dk\,\cos k\rho}{\displaystyle 
(M^2_{\rm N} + k^2)^{1/2}}\right].
\eqno({\rm C}.49)
$$
The computation of $J^{0\mu\nu}_2$ runs as follows:
$$
J^{0\mu\nu}_2=\int\frac{d^4k}{\pi^2i}
\frac{\sin|\vec{k}\,|\rho}{|\vec{k}\,|\rho} 
\,\frac{2 k\cdot k_{\rm D}\,
{\rm tr}\{\gamma^0(M_{\rm N} + \hat{k})
\gamma^{\mu}(M_{\rm N} + \hat{k})
\gamma^{\nu}(M_{\rm N} - 
\hat{k})\}}{[M^2_{\rm N} - k^2 - i0]^4}=
$$
$$
= 2 k^0_{\rm D}\int\frac{d^4k}{\pi^2i}
\frac{\sin|\vec{k}\,|\rho}{|\vec{k}\,|\rho} 
\,\frac{k_0\,{\rm tr}\{\gamma^0(M_{\rm N} + \hat{k})\gamma^{\mu}(M_{\rm N} + \hat{k})
\gamma^{\nu}(M_{\rm N} 
- \hat{k})\}}{[E^2_{\vec{k}}
- k^2_0 - i0]^4}.\eqno({\rm C}.50)
$$
The computation of the trace over Dirac matrices
$$
{\rm tr}\{\gamma^0(M_{\rm N} + \hat{k})\gamma^{\mu}(M_{\rm N} + \hat{k})\gamma^{\nu}(M_{\rm N} - \hat{k})\}=
$$
$$
={\rm tr}\{(M_{\rm N} - \hat{k})\gamma^0(M_{\rm N} + \hat{k})\gamma^{\mu}(M_{\rm N} + \hat{k})\gamma^{\nu}\}=
$$
$$
={\rm tr}\{(M_{\rm N}\gamma^0 - \hat{k}\gamma^0)(M^2_{\rm N}\gamma^{\mu}\gamma^{\nu} + 2\,M_{\rm N}k^{\mu}\gamma^{\nu} + \hat{k}\gamma^{\mu}\hat{k}\gamma^{\nu})\}=
$$
$$
 ={\rm tr}\{- M^2_{\rm N}\hat{k}\gamma^0\gamma^{\mu}\gamma^{\nu}- \hat{k}\gamma^0\hat{k}\gamma^{\mu}\hat{k}\gamma^{\nu}\}=
$$
$$
={\rm tr}\{- M^2_{\rm N}\hat{k}\gamma^0\gamma^{\mu}\gamma^{\nu}+ k^2\gamma^0\gamma^{\mu}\hat{k}\gamma^{\nu}- 2k_0\hat{k}\gamma^{\mu}\hat{k}\gamma^{\nu}\}=
$$
$$
=4\,(-\,k_0\, M^2_{\rm N}\,g^{\mu\nu} - k_0\, k^2\,g^{\mu\nu}- 2k_0\,2 k^{\mu}k^{\nu} + 2k_0\,k^2\,g^{\mu\nu})=
$$
$$
= -\,4\,k_0\,(M^2_{\rm N}\,g^{\mu\nu} -  k^2\,g^{\mu\nu} + 4\, k^{\mu}k^{\nu}).\eqno({\rm C}.51)
$$
Substituting Eq.~({\rm C}.51) in Eq.~({\rm C}.50) we obtain
$$
J^{0\mu\nu}_2 =- 8\, k^0_{\rm D}\int\frac{d^4k}{\pi^2i}\frac{\sin|\vec{k}\,|\rho}{|\vec{k}\,|\rho} \,\frac{k^2_0
(M^2_{\rm N}\,
g^{\mu\nu} -  k^2\,g^{\mu\nu} 
+ 4\, k^{\mu}k^{\nu})}{[E^2_{\vec{k}} 
- k^2_0 - i0]^4}=
$$
$$
=- 8\, k^0_{\rm D}\int\frac{d^4k}{\pi^2i}\frac{\sin|\vec{k}\,|\rho}{|\vec{k}\,|\rho} \,\frac{\displaystyle k^2_0(M^2_{\rm N}\,g^{\mu\nu} -  k^2\,g^{\mu\nu} - \frac{4}{3}\,\vec{k}^{\,2}\,g^{\mu\nu})}
{[E^2_{\vec{k}} - k^2_0 - i0]^4}=
$$
$$
=- 8\, k^0_{\rm D}\,g^{\mu\nu}\int\frac{d^4k}{\pi^2i}\frac{\sin|\vec{k}\,|\rho}{|\vec{k}\,|\rho} \,\frac{\displaystyle k^2_0(M^2_{\rm N}
- k^2_0 + \vec{k}^{\,2}
- \frac{4}{3}\,
\vec{k}^{\,2})}
{[E^2_{\vec{k}} - k^2_0 - i0]^4}=
$$
$$
=- 8\, k^0_{\rm D}\,g^{\mu\nu}\int\frac{d^4k}{\pi^2i}\frac{\sin|\vec{k}\,|\rho}{|\vec{k}\,|\rho} \,\frac{\displaystyle k^2_0(M^2_{\rm N}- k^2_0 - \frac{1}{3}\,\vec{k}^{\,2})}{[E^2_{\vec{k}} 
- k^2_0 - i0]^4}=
$$
$$
=- 8\, k^0_{\rm D}\,g^{\mu\nu}\int\frac{d^4k}{\pi^2i}\frac{\sin|\vec{k}\,|\rho}{|\vec{k}\,|\rho} \,\frac{\displaystyle k^2_0(\frac{4}{3}\,M^2_{\rm N}- \frac{1}{3}\,E^2_{\vec{k}}
- k^2_0)}{[E^2_{\vec{k}} - 
k^2_0 - i0]^4}=
$$
$$
=- \frac{8}{3}\, k^0_{\rm D}\,g^{\mu\nu}\int\frac{d^4k}{\pi^2i}\frac{\sin|\vec{k}\,|\rho}{|\vec{k}\,|\rho} \,\frac{\displaystyle k^2_0(4\,M^2_{\rm N}- 
E^2_{\vec{k}} - 3\, k^2_0)}{[E^2_{\vec{k}} 
- k^2_0 - i0]^4}=
$$
$$
=\frac{8}{3}\, k^0_{\rm D}\,g^{\mu\nu}\int\frac{d^3k}{\pi^2}
\frac{\sin|\vec{k}\,|\rho}{|\vec{k}\,|\rho}\int
\limits^{\infty}_{-\infty}dk_4
\frac{\displaystyle k^2_4
(4\,M^2_{\rm N}- E^2_{\vec{k}} 
+ 3\, k^2_4)}{[E^2_{\vec{k}} + k^2_4]^4}=
$$
$$
=\frac{8}{3}\, k^0_{\rm D}\,g^{\mu\nu}
\int\frac{d^3k}{\pi^2}
\frac{\sin|\vec{k}\,|\rho}{|\vec{k}\,|\rho}
\left[\frac{\displaystyle 4\,M^2_{\rm N}- E^2_{\vec{k}}}{E^5_{\vec{k}}}\frac{\displaystyle \Gamma\Bigg(\frac{3}{2}\Bigg)\Gamma
\Bigg(\frac{5}{2}\Bigg)}{\displaystyle \Gamma(4)} + \frac{3}{E^3_{\vec{k}}}
\frac{\displaystyle \Gamma\Bigg(\frac{5}{2}\Bigg)
\Gamma\Bigg(\frac{3}{2}\Bigg)}{\displaystyle 
\Gamma(4)}\right]=
$$
$$
=\frac{1}{3}\, k^0_{\rm D}\,g^{\mu\nu}\int\frac{d^3k}{\pi}\frac{\sin|\vec{k}\,|\rho}{|\vec{k}\,|\rho}\,\frac{\displaystyle 2\,M^2_{\rm N}+ E^2_{\vec{k}}}{E^5_{\vec{k}}}=
$$
$$
=\frac{4}{3}\, k^0_{\rm D}\,g^{\mu\nu}
\left[\frac{2}{3}\,M^2_{\rm N}
\int\limits^{\infty}_0
\frac{\displaystyle dk\,
\cos k\rho}{\displaystyle (M^2_{\rm N} + k^2)^{3/2}} + \int\limits^{\infty}_0\frac{\displaystyle dk\,\cos k\rho}{\displaystyle 
(M^2_{\rm N} + k^2)^{1/2}}\right].
\eqno({\rm C}.52)
$$
As a result $J^{0\mu\nu}_2$ reads
$$
J^{0\mu\nu}_2=\int\frac{d^4k}{\pi^2i}\frac{\sin|\vec{k}\,|\rho}{|\vec{k}\,|\rho} \,\frac{2 k\cdot k_{\rm D}\,{\rm tr}\{\gamma^0(M_{\rm N} + \hat{k})\gamma^{\mu}(M_{\rm N} + \hat{k})\gamma^{\nu}(M_{\rm N} - \hat{k})\}}{[M^2_{\rm N} - k^2 - i0]^4}=
$$
$$
= \frac{4}{3}\, k^0_{\rm D}\,g^{\mu\nu}\left[\frac{2}{3}\,M^2_{\rm N}\int\limits^{\infty}_0\frac{\displaystyle dk\,\cos k\rho}{\displaystyle (M^2_{\rm N} + k^2)^{3/2}} + \int\limits^{\infty}_0\frac{\displaystyle dk\,\cos k\rho}{\displaystyle (M^2_{\rm N} + k^2)^{1/2}}\right].\eqno({\rm C}.53)
$$
Summing up the contributions given by Eq.~({\rm C}.49) and Eq.~({\rm C}.53) we obtain the structure function ${\cal J}^{0\mu\nu}(k_{\rm D}, k_{\ell})$:
$$
{\cal J}^{0\mu\nu}_{\rm pp}
(k_{\rm D}, k_{\ell})=
$$
$$
=- \frac{8}{3}\, k^0_{\rm D}\,g^{\mu\nu}\,\sqrt{2}\,a^{\rm e}_{\rm pp}\,4\pi\int\limits^{\infty}_0d\rho\,\rho\,U(\rho)\,\Phi(\rho)\left[\frac{4}{3}\,M^2_{\rm N}\int\limits^{\infty}_0\frac{\displaystyle dk\,\cos k\rho}{\displaystyle (M^2_{\rm N} + k^2)^{3/2}} - \int\limits^{\infty}_0\frac{\displaystyle dk\,\cos k\rho}{\displaystyle (M^2_{\rm N} + k^2)^{1/2}}\right]=
$$
$$
=- \frac{8}{3}\, k^0_{\rm D}\,g^{\mu\nu}\,\sqrt{2}\,a^{\rm e}_{\rm pp}\,4\pi\int\limits^{\infty}_0d\rho\,\rho\,U(\rho)\,\Phi(\rho)\left[\frac{4}{3}\,M_{\rm N}\rho\,K_1(M_{\rm N}\rho) - K_0(M_{\rm N}\rho)\right],\eqno({\rm C}.54)
$$
where $K_1(M_{\rm N}\rho)$ and $K_0(M_{\rm N}\rho)$ are the McDonald functions.

Now the structure function should be continued on--mass shell of the deuteron. For this aim we should only set $k^0_{\rm D} = M_{\rm D} \simeq 2\,M_{\rm N}$:
$$
{\cal J}^{0\mu\nu}_{\rm pp}
(k_{\rm D}, k_{\ell}) =
- \frac{16}{3}\, M_{\rm N}\,g^{\mu\nu}\,\sqrt{2}\,a^{\rm e}_{\rm pp}\,4\pi\int\limits^{\infty}_0d\rho\,\rho\,U(\rho)\,\Phi(\rho)\,\left[\frac{4}{3}\,M_{\rm N}\rho\,K_1(M_{\rm N}\rho) - K_0(M_{\rm N}\rho)\right].\eqno({\rm C}.55)
$$
The structure function ${\cal J}^{0\mu\nu}_{\rm pp}
(k_{\rm D}, k_{\ell})$ given by Eq.~({\rm C}.55) should be applied to the computation of the matrix element of the solar proton burning.
\vspace{0.2in}

\noindent{\bf The computation of ${\cal J}^{\mu\nu}_{\rm pp}
(k_{\rm D}, k_{\ell})$}. 
The computation of 
${\cal J}^{\mu\nu}_{\rm pp}
(k_{\rm D}, k_{\ell})$ is analogous to 
${\cal J}^{0\mu\nu}_{\rm pp}
(k_{\rm D}, k_{\ell})$ and runs as follows:
$$
{\cal J}^{\mu\nu}_{\rm pp}
(k_{\rm D}, k_{\ell}) = -\,\sqrt{2}\,a^{\rm e}_{\rm pp}\,4\pi\int\limits^{\infty}_0 d\rho\,  \rho\, U(\rho)\,\Phi(\rho)\int\frac{d^4k}{\pi^2i}\frac{\sin|\vec{k}\,|\rho}{|\vec{k}\,|\rho}
$$
$$
\times\,{\rm tr}\Bigg\{\gamma^5\frac{1}{M_{\rm N} - \hat{k} + \hat{k}_{\rm D}}\gamma^{\mu}\frac{1}{M_{\rm N} - \hat{k}}\gamma^{\nu}\gamma^5 \frac{1}{M_{\rm N} - \hat{k}}\Bigg\}
$$
$$
= -\,\sqrt{2}\,a^{\rm e}_{\rm pp}\,4\pi\int\limits^{\infty}_0 d\rho\,  \rho\, U(\rho)\,\Phi(\rho)\int\frac{d^4k}{\pi^2i}\frac{\sin|\vec{k}\,|\rho}{|\vec{k}\,|\rho}{\rm tr}\Bigg\{\gamma^5\frac{1}{M_{\rm N} - \hat{k}}\gamma^{\mu}\frac{1}{M_{\rm N} - \hat{k}}\gamma^{\nu}\gamma^5 \frac{1}{M_{\rm N} - \hat{k}}\Bigg\}
$$
$$
= -\sqrt{2}a^{\rm e}_{\rm pp}4\pi\int\limits^{\infty}_0 d\rho\,  \rho\, U(\rho)\,\Phi(\rho)\int\frac{d^4k}{\pi^2i}\frac{\sin|\vec{k}\,|\rho}{|\vec{k}\,|\rho}\frac{{\rm tr}\{\gamma^5(M_{\rm N} +  \hat{k})\gamma^{\mu}(M_{\rm N} + \hat{k})\gamma^{\nu}\gamma^5 (M_{\rm N} + \hat{k})\}}{[M^2_{\rm N} - k^2 - i0]^3}
$$
$$
= -\,\sqrt{2}\,a^{\rm e}_{\rm pp}4\pi\int\limits^{\infty}_0 d\rho\,  \rho\, U(\rho)\,\Phi(\rho)\int\frac{d^4k}{\pi^2i}\frac{\sin|\vec{k}\,|\rho}{|\vec{k}\,|\rho}\frac{{\rm tr}\{(M_{\rm N} +  \hat{k})\gamma^{\mu}(M_{\rm N} + \hat{k})\gamma^{\nu}(M_{\rm N} - \hat{k})\}}{[M^2_{\rm N} - k^2 - i0]^3}
$$
$$
= -\,\sqrt{2}\,a^{\rm e}_{\rm pp}\,4\pi\int\limits^{\infty}_0 d\rho\,  \rho\, U(\rho)\,\Phi(\rho)\int\frac{d^4k}{\pi^2i}\frac{\sin|\vec{k}\,|\rho}{|\vec{k}\,|\rho}\,\frac{{\rm tr}\{\gamma^{\mu}(M_{\rm N} + \hat{k})\gamma^{\nu}\}}{[M^2_{\rm N} - k^2 - i0]^2}
$$
$$
= -\,4\,M_{\rm N}\,g^{\mu\nu}\,\sqrt{2}\,a^{\rm e}_{\rm pp}\,4\pi\int\limits^{\infty}_0 d\rho\,  \rho\, U(\rho)\,\Phi(\rho)\int \frac{d^4k}{\pi^2i}\frac{\sin|\vec{k}\,|\rho}{|\vec{k}\,|\rho}\,\frac{1}{[M^2_{\rm N} - k^2 - i0]^2}
$$
$$
= -\,4\,M_{\rm N}\,g^{\mu\nu}\,\sqrt{2}\,a^{\rm e}_{\rm pp}\,4\pi\int\limits^{\infty}_0 d\rho\,  \rho\, U(\rho)\,\Phi(\rho)\int \frac{d^3k}{\pi^2}\frac{\sin|\vec{k}\,|\rho}{|\vec{k}\,|\rho}\int\limits^{\infty}_{-\infty}\frac{d k_4}{[E^2_{\vec{k}} + k^2_4]^2}
$$
$$
= -\,8\,M_{\rm N}\,g^{\mu\nu}\,\sqrt{2}\,a^{\rm e}_{\rm pp}\,4\pi\int\limits^{\infty}_0 d\rho\,  \rho\, U(\rho)\,\Phi(\rho)\,K_0(M_{\rm N}\rho).\eqno({\rm C}.56)
$$
Thus, the structure function 
${\cal J}^{\mu\nu}_{\rm pp}
(k_{\rm D},k_{\ell})$ is given by
$$
{\cal J}^{\mu\nu}_{\rm pp}
(k_{\rm D}, k_{\ell}) = -\,8\,M_{\rm N}\,g^{\mu\nu}\,
\sqrt{2}\,a^{\rm e}_{\rm pp}\,4\pi
\int\limits^{\infty}_0 d\rho\,  \rho\, U(\rho)\,
\Phi(\rho)\,K_0(M_{\rm N}\rho).\eqno({\rm C}.57)
$$
The structure function 
${\cal J}^{\mu\nu}_{\rm pp}
(k_{\rm D},k_{\ell})$ does not depend on the 4--momentum of the deuteron. Therefore, it does not change itself due to the continuation on--mass shell of the deuteron.

We represent the structure function defining the amplitude of the solar proton burning 
Eq.~({\rm C}.42) as follows
$$
-{\cal J}^{0\mu\nu}_{\rm pp}(k_{\rm D}, k_{\ell}) 
+ {\cal J}^{\mu\nu}_{\rm pp}(k_{\rm D}, k_{\ell}) = 
6\,M_{\rm N}\,g^{\mu\nu}\,{\cal F}^{\rm e}_{\rm pp}.
\eqno({\rm C}.58)
$$
The factor ${\cal F}^{\rm e}_{\rm pp}$ is given by
$$
{\cal F}^{\rm e}_{\rm pp}  = \sqrt{2}\,a^{\rm e}_{\rm pp}\,\frac{32}{27}\,4\pi\int\limits^{\infty}_0 d\rho\,  \rho\, U(\rho)\,\Phi(\rho)\,\left[ M_{\rm N}\rho \,K_1(M_{\rm N}\rho) - \frac{15}{8}\,K_0(M_{\rm N}\rho)\right].\eqno({\rm C}.59)
$$
Due to the McDonald functions the integral over $\rho$ is concentrated in the region $0 < \rho \sim 1/M_{\rm N}$, one can set with a good accuracy $\Phi(\rho) = 1$. This signifies that the main contribution comes from the irregular part of the Coulomb wave function. As a result in the RFMD the contribution of the Coulomb repulsion to the amplitude of the solar proton burning reduces itself to the appearance of the S--wave scattering length $a^{\rm e}_{\rm pp} = (- 7.828\pm 0.008)\,{\rm fm}$ of the low--energy elastic pp scattering and the Gamow penetration factor $C(\eta)$. Setting $\Phi(\rho) = 1$ we get
$$
{\cal F}^{\rm e}_{\rm pp} = \sqrt{2}\,a^{\rm e}_{\rm pp}\,\frac{32}{27}\,4\pi\int\limits^{\infty}_0 d\rho\,  \rho\, U(\rho)\,\left[ M_{\rm N}\rho \,K_1(M_{\rm N}\rho) - \frac{15}{8}\,K_0(M_{\rm N}\rho)\right]=
$$
$$
= \sqrt{2}\,a^{\rm e}_{\rm pp}\,\frac{32}{27}\,M^2_{\pi}\int\limits^{\infty}_0 d\rho\,e^{\displaystyle - M_{\pi}\rho}\left[ M_{\rm N}\rho \,K_1(M_{\rm N}\rho) - \frac{15}{8}\,K_0(M_{\rm N}\rho)\right],\eqno({\rm C}.60)
$$
where we have substituted the Yukawa potential Eq.~(\ref{label1.4}).

For the calculation of the integral over $\rho$ we suggest to use auxiliary formulae
$$
\int\limits^{\infty}_0 dx\,e^{\displaystyle - \lambda\,x}\,K_0(x) = \frac{1}{\sqrt{1 - \lambda^2}}\,{\rm arctg}\frac{\sqrt{1 - \lambda^2}}{\lambda},
$$
$$
\int\limits^{\infty}_0 dx\,x\,e^{\displaystyle - \lambda\,x}\,K_0(x) = \frac{1}{1 - \lambda^2} - \frac{\lambda}{(1 - \lambda^2)^{3/2}}\,{\rm arctg}\frac{\sqrt{1 - \lambda^2}}{\lambda},
$$
$$
\int\limits^{\infty}_0 dx\,x\,e^{\displaystyle - \lambda\,x}\,K_1(x) = \int\limits^{\infty}_0 dx\,e^{\displaystyle - \lambda\,x}\,K_0(x) - \lambda \int\limits^{\infty}_0 dx\,x\,e^{\displaystyle - \lambda\,x}\,K_0(x).\eqno({\rm C}.61)
$$
By applying these formulae we compute the factor ${\cal F}^{\rm e}_{\rm pp}$:
$$
{\cal F}^{\rm e}_{\rm pp}  = -\sqrt{2}\,a^{\rm e}_{\rm pp}\,\frac{28}{27}\,\Bigg[\frac{M^2_{\pi}}{\sqrt{M^2_{\rm N} - M^2_{\pi}}}\,{\rm arctg}\frac{\sqrt{M^2_{\rm N} - M^2_{\pi}}}{M_{\pi}} + \frac{8}{7}\,\frac{M^3_{\pi}}{M^2_{\rm N} - M^2_{\pi}}
$$
$$
- \frac{8}{7}\,\frac{M^4_{\pi}}{(M^2_{\rm N} - M^2_{\pi})^{3/2}}\,{\rm arctg}\frac{\sqrt{M^2_{\rm N} - M^2_{\pi}}}{M_{\pi}}\Bigg] = 1.78.\eqno({\rm C}.62)
$$
This, completes the explanation of the derivation of the matrix element of the solar proton burning.

\section*{Appendix D. Computation of the matrix element of the process $\nu_{\rm e}+ {\rm D} \to {\rm e}^- + {\rm p} + {\rm p}$}

The process $\nu_{\rm e}$ + D $\to$ e$^-$ + p + p runs through the intermediate W--boson exchange as follows $\nu_{\rm e}$ + D $\to$ e$^-$ + W$^+$ + D $\to$ e$^-$ + p + p. In the RFMD the matrix element of the transition $\nu_{\rm e}$ + D $\to$ e$^-$ + W$^+$ + D $\to$ e$^-$ + p + p is defined by the effective interactions
$$
{\cal L}^{\dagger}_{\rm npD}(x) = -ig_{\rm V}[\bar{p}(x)\gamma^{\mu}n^c(x) - \bar{n}(x)\gamma^{\mu} p^c(x)]\,D_{\mu}(x),\eqno({\rm D}.1)
$$
${\cal L}^{\rm pp \to pp}_{\rm eff}(x)$ given by Eq.({\rm C}.1) and 
$$
{\cal L}^{\dagger}_{\rm npW}(x) = [\bar{p}(x)\gamma^{\nu}(1 - g_{\rm A}\gamma^5) n(x)]\,W^+_{\nu}(x).\eqno({\rm D}.2)
$$
For the description of the transition $\nu_{\rm e}$ $\to$ e$^-$ + W$^+$ we replace the operator of the W--boson field by the operator of the leptonic weak current
$$
W^+_{\nu}(x) \to -\frac{G_{\rm V}}{\sqrt{2}}\,[\bar{\psi}_{\rm e}(x)\gamma_{\nu}(1 - \gamma^5) \psi_{\nu_{\rm e}}(x)].\eqno({\rm D}.3)
$$
The S matrix element ${\rm S}^{(3)}_{\rm W^+D \to pp}$ responsible for the transition W$^+$ + D $\to$ p + p can be obtained by analogy with the S matrix element ${\rm S}^{(3)}_{\rm pp\to DW^+}$ describing the transition p + p $\to$ D + W$^+$ (see Eq.~({\rm C}.8)) and reads
$$
{\rm S}^{(3)}_{\rm W^+D\to pp} = -i \int d^4x_1 d^4x_2 d^4x_3\,{\rm T}({\cal L}^{\rm pp\to pp}_{\rm eff}(x_1){\cal L}^{\dagger}_{\rm npD}(x_2){\cal L}^{\dagger}_{\rm npW}(x_3)).\eqno({\rm D}.4)
$$
For the derivation of the effective Lagrangian 
${\cal L}_{\rm W^+D\to pp}(x)$ 
containing only the fields of the initial and the final particles we should make all necessary contractions of the operators of the proton and the neutron fields. Denoting these contractions by brackets we get
$$
<{\rm S}^{(3)}_{\rm W^+D\to pp}> = -i \int d^4x_1 d^4x_2 d^4x_3\,<{\rm T}({\cal L}^{\rm pp\to pp}_{\rm eff}(x_1){\cal L}^{\dagger}_{\rm npD}(x_2){\cal L}^{\dagger}_{\rm npW}(x_3))>.\eqno({\rm D}.5)
$$
The effective Lagrangian 
${\cal L}_{\rm W^+D\to pp}(x)$
related to the S matrix element
$<{\rm S}^{(3)}_{\rm W^+D\to pp}>$ is defined as 
$$
<{\rm S}^{(3)}_{\rm W^+D\to pp}> = i\int d^4x\,
{\cal L}_{\rm W^+D\to pp}(x) = 
$$
$$
= -i \int d^4x_1 d^4x_2 d^4x_3\,
<{\rm T}({\cal L}^{\rm pp\to pp}_{\rm eff}(x_1)
{\cal L}^{\dagger}_{\rm npD}(x_2)
{\cal L}^{\dagger}_{\rm npW}(x_3))>. 
\eqno({\rm D}.6)
$$
In terms of the operators of the interacting fields the effective Lagrangian 
${\cal L}_{\rm W^+D\to pp}(x)$ reads
$$
\int d^4x\,{\cal L}_{\rm W^+D\to pp}(x) =  - \int d^4x_1 d^4x_2 d^4x_3\,<{\rm T}({\cal L}^{\rm pp\to pp}_{\rm eff}(x_1)
{\cal L}^{\dagger}_{\rm npD}(x_2)
{\cal L}^{\dagger}_{\rm npW}(x_3))>
$$
$$
= \frac{1}{2}\,G_{\rm \pi pp}\,\times\,(-ig_{\rm V})\,\times 
\,g_{\rm A}\int d^4x_1 d^4x_2 d^4x_3\,\int d^3\rho\,U(\rho)\,
$$
$$
\times\,{\rm T}([\bar{p}(t_1,\vec{x}_1 + \frac{1}{2}\,\vec{\rho}\,)\,\gamma^{\alpha}
\gamma^5 p^c (t_1,\vec{x}_1 - \frac{1}{2}\,\vec{\rho}\,)]\,D_{\mu}(x_2)
\,W^+_{\nu}(x_3))
$$
$$
\times <0|{\rm T}([\bar{p^c}(t_1,\vec{x}_1 + \frac{1}{2}\,\vec{\rho}\,)\,\gamma_{\alpha}\gamma^5 p(t_1,\vec{x}_1 - \frac{1}{2}\,\vec{\rho}\,)][\bar{p}(x_2)\gamma^{\mu}n^c(x_2) - \bar{n}(x_2)
\gamma^{\mu}p^c(x_2)]\,
$$
$$
\times\,[\bar{p}(x_3)\gamma^{\nu}\gamma^5 n(x_3)])|0> +\frac{1}{2}\,G_{\rm \pi pp}\,\times\,(-ig_{\rm V})\,\times \,g_{\rm A} \int d^4x_1 d^4x_2 d^4x_3\,\int d^3\rho \,U(\rho)\,
$$
$$
\times\,{\rm T}([\bar{p}(t_1,\vec{x}_1 + \frac{1}{2}\,\vec{\rho}\,) \gamma^5 p^c (t_1,\vec{x}_1 - \frac{1}{2}\,\vec{\rho}\,)]\,D_{\mu}(x_2)\,W^+_{\nu}(x_3))
$$
$$
\times <0|{\rm T}([\bar{p^c}(t_1,\vec{x}_1 + \frac{1}{2}\,\vec{\rho}\,) \gamma^5 p(t_1,\vec{x}_1 - \frac{1}{2}\,\vec{\rho}\,)][\bar{p}(x_2)\gamma^{\mu}n^c(x_2) - \bar{n}(x_2)
\gamma^{\mu}p^c(x_2)]\,
$$
$$
\times\,[\bar{p}(x_3)\gamma^{\nu}\gamma^5 n(x_3)])|0>.\eqno({\rm D}.7)
$$
Since W$^+$ + D $\to$ p + p is the Gamow--Teller transition for the protons in the ${^1}{\rm S}_0$--state, we have taken into account the W--boson coupled with the axial nucleon current.

Due to the relation $\bar{p}(x_2)\gamma^{\mu}n^c(x_2) = -  \bar{n}(x_2)\gamma^{\mu}p^c(x_2)$ the r.h.s. of Eq.~({\rm D}.7) reduces to the form
$$
\int d^4x\,{\cal L}_{\rm W^+D\to pp}(x) =  - \int d^4x_1 d^4x_2 d^4x_3\,<{\rm T}({\cal L}^{\rm pp\to pp}_{\rm eff}(x_1){\cal L}^{\dagger}_{\rm npD}(x_2){\cal L}^{\dagger}_{\rm npW}(x_3))>
$$
$$
= G_{\rm \pi pp}\,\times\,ig_{\rm V}\,\times \,g_{\rm A}\int d^4x_1 d^4x_2 d^4x_3\,\int d^3\rho\,U(\rho)\,
$$
$$
\times\,{\rm T}([\bar{p}(t_1,\vec{x}_1 + \frac{1}{2}\,\vec{\rho}\,)\,\gamma_{\alpha}\gamma^5 p^c (t_1,\vec{x}_1 - \frac{1}{2}\,\vec{\rho}\,)]\,D_{\mu}(x_2)\,W^+_{\nu}(x_3))
$$
$$
\times <0|{\rm T}([\bar{p^c}(t_1,\vec{x}_1 + \frac{1}{2}\,\vec{\rho}\,)\,\gamma^{\alpha}\gamma^5 p(t_1,\vec{x}_1 - \frac{1}{2}\,\vec{\rho}\,)]
[\bar{n}(x_2)
\gamma^{\mu}p^c(x_2)]\,
$$
$$
\times\,[\bar{p}(x_3)\gamma^{\nu}\gamma^5 n(x_3)])|0> + G_{\rm \pi pp}\,\times\,ig_{\rm V}\,\times \,g_{\rm A} \int d^4x_1 d^4x_2 d^4x_3\,\int d^3\rho \,U(\rho)\,
$$
$$
\times\,{\rm T}([\bar{p}(t_1,\vec{x}_1 + \frac{1}{2}\,\vec{\rho}\,) \gamma^5 p^c (t_1,\vec{x}_1 - \frac{1}{2}\,\vec{\rho}\,)]\,D_{\mu}(x_2)\,W^+_{\nu}(x_3))
$$
$$
\times <0|{\rm T}([\bar{p^c}(t_1,\vec{x}_1 + \frac{1}{2}\,\vec{\rho}\,) \gamma^5 p(t_1,\vec{x}_1 - \frac{1}{2}\,\vec{\rho}\,)]
[\bar{n}(x_2)
\gamma^{\mu}p^c(x_2)]\,
$$
$$
\times\,[\bar{p}(x_3)\gamma^{\nu}\gamma^5 n(x_3)])|0>.
\eqno({\rm D}.8)
$$
After the necessary contractions we arrive at the expression
$$
\int d^4x\,{\cal L}_{\rm pp\to DW^+}(x) =  - \int d^4x_1 d^4x_2 d^4x_3\,<{\rm T}({\cal L}^{\rm pp\to pp}_{\rm eff}(x_1){\cal L}^{\dagger}_{\rm npD}(x_2){\cal L}^{\dagger}_{\rm npW}(x_3))> 
$$
$$
= 2\,\times\,G_{\rm \pi pp}\,\times\,ig_{\rm V}\,\times \,g_{\rm A}\int d^4x_1 d^4x_2 d^4x_3\,\int d^3\rho \,U(\rho)\,
$$
$$
\times\,{\rm T}([\bar{p}(t_1,\vec{x}_1 + \frac{1}{2}\,\vec{\rho}\,)\,\gamma_{\alpha}\gamma^5 p^c (t_1,\vec{x}_1 - \frac{1}{2}\,\vec{\rho}\,)]\,D_{\mu}(x_2)\,W^+_{\nu}(x_3))
$$
$$
\times\,(-1)\,{\rm tr}\{\gamma^{\alpha}\gamma^5 (-i) S_F(t_1 - t_3,\vec{x}_1 - \vec{x}_3 - \frac{1}{2}\,\vec{\rho}\,) \gamma^{\nu}\gamma^5 (-i) S_F(x_3 - x_2) \gamma^{\mu} 
$$
$$
\times\,(-i) S^c_F(t_2 - t_1, \vec{x}_2  - \vec{x}_1 - \frac{1}{2}\,\vec{\rho}\,)\}
$$
$$
+ 2\,\times\,G_{\rm \pi pp}\,\times\,ig_{\rm V}\,\times \,g_{\rm A}\int d^4x_1 d^4x_2 d^4x_3\,\int d^3\rho \,U(\rho)\,
$$
$$
\times\,{\rm T}([\bar{p}(t_1,\vec{x}_1 + \frac{1}{2}\,\vec{\rho}\,) \gamma^5 p^c (t_1,\vec{x}_1 - \frac{1}{2}\,\vec{\rho}\,)]\,D_{\mu}(x_2)\,W^+_{\nu}(x_3))
$$
$$
\times\,(-1)\,{\rm tr}\{\gamma^5 (-i) S_F(t_1 - t_3,\vec{x}_1 - \vec{x}_3 - \frac{1}{2}\,\vec{\rho}\,) \gamma^{\nu}\gamma^5 (-i) S_F(x_3 - x_2) \gamma^{\mu} 
$$
$$
\times\,(-i) S^c_F(t_2 - t_1, \vec{x}_2  - \vec{x}_1 - \frac{1}{2}\,\vec{\rho}\,)\}.\eqno({\rm D}.9)
$$
The combinatorial factor 2 takes into account that the protons are identical particles in the nucleon loop (see Eqs.~({\rm C}.13) -- ({\rm C}.17)).

In the momentum representation of the nucleon Green functions we define the effective Lagrangian ${\cal L}_{\rm W^+D \to pp}(x)$ as follows
$$
\int d^4x\,{\cal L}_{\rm W^+D \to pp}(x) = 
$$
$$
= i\,g_{\rm A}G_{\rm \pi pp}
\frac{g_{\rm V}}{8\pi^2}\int d^4x_1
\int \frac{d^4x_2 d^4k_2}{(2\pi)^4}
\frac{d^4x_3 d^4k_3}{(2\pi)^4}\,
e^{\displaystyle i k_2\cdot (x_2-x_1)}
e^{\displaystyle i k_3\cdot (x_3-x_1)}
$$
$$
\times\,\int d^3\rho \,U(\rho)\,
{\rm T}([\bar{p}(t_1,\vec{x}_1 + \frac{1}{2}\,\vec{\rho}\,)\,\gamma_{\alpha}
\gamma^5 p^c (t_1,\vec{x}_1 - \frac{1}{2}\,\vec{\rho}\,)]\,D_{\mu}(x_2)
\,W^+_{\nu}(x_3))
$$
$$
\times\,\int\frac{d^4k_1}{\pi^2i}\,
e^{\displaystyle - i\vec{q}\cdot \vec{\rho}}\,{\rm tr}\Bigg\{\gamma^{\alpha}\gamma^5
\frac{1}{M_{\rm N} - \hat{k}_1 - \hat{k}_3}
\gamma^{\nu}\gamma^5
\frac{1}{M_{\rm N} - \hat{k}_1}\gamma^{\mu}
\frac{1}{M_{\rm N} - \hat{k}_1 
+ \hat{k}_2}\Bigg\}
$$
$$
+ i\,g_{\rm A}G_{\rm \pi pp}
\frac{g_{\rm V}}{8\pi^2}\int d^4x_1 
\int \frac{d^4x_2 d^4k_2}{(2\pi)^4}
\frac{d^4x_3 d^4k_3}{(2\pi)^4}\,
e^{\displaystyle i k_2\cdot (x_2-x_1)}
e^{\displaystyle i k_3\cdot (x_3-x_1)}
$$
$$
\times\,\int d^3\rho \,U(\rho)\,
{\rm T}([\bar{p}(t_1,\vec{x}_1 + 
\frac{1}{2}\,\vec{\rho}\,)\,
\gamma^5 p^c (t_1,\vec{x}_1 - 
\frac{1}{2}\,\vec{\rho}\,)]\, 
D_{\mu}(x_2)\,W^+_{\nu}(x_3))
$$
$$
\times\,\int\frac{d^4k_1}{\pi^2i}\,
e^{\displaystyle - i\vec{q}\cdot \vec{\rho}}\,
{\rm tr}\Bigg\{\gamma^5
\frac{1}{M_{\rm N} - \hat{k}_1 - \hat{k}_3}
\gamma^{\nu}\gamma^5 \frac{1}{M_{\rm N} - \hat{k}_1}
\gamma^{\mu}
\frac{1}{M_{\rm N} - \hat{k}_1 + \hat{k}_2}\Bigg\},
\eqno({\rm D}.10)
$$
where $\vec{q} = \vec{k}_1 + 
(\vec{k}_3 - \vec{k}_2)/2$.

In order to obtain the effective Lagrangian describing the matrix element of the process $\nu_{\rm e}$ +  D 
$\to$ e$^-$ + p + p  we replace the operator of 
the W--boson field by the operator of the 
leptonic weak current Eq.~({\rm D}.3):
$$
\int d^4x\,{\cal L}_{\rm \nu_{\rm e}D\to e^-pp}(x) = 
$$
$$
= -\,i\,g_{\rm A}G_{\rm \pi pp}
\frac{G_{\rm V}}{\sqrt{2}}
\frac{g_{\rm V}}{8\pi^2}\int d^4x_1 \int 
\frac{d^4x_2 d^4k_2}{(2\pi)^4}
\frac{d^4x_3 d^4k_3}{(2\pi)^4}\,
e^{\displaystyle i k_2\cdot (x_2-x_1)}
e^{\displaystyle i k_3\cdot (x_3-x_1)}
$$
$$
\times\,\int d^3\rho \,U(\rho)\,
{\rm T}([\bar{p}(t_1,\vec{x}_1 + \frac{1}{2}\,\vec{\rho}\,)\,\gamma_{\alpha}
\gamma^5 p^c (t_1,\vec{x}_1 - \frac{1}{2}\,\vec{\rho}\,)]\,D_{\mu}(x_2)\,
[\bar{\psi}_{\rm e}(x_3)
\gamma_{\nu}(1 - \gamma^5) 
\psi_{\nu_{\rm e}}(x_3)])
$$
$$
\times\,\int\frac{d^4k_1}{\pi^2i}\,
e^{\displaystyle - i\vec{q}\cdot \vec{\rho}}
\,{\rm tr}\Bigg\{\gamma^{\alpha}\gamma^5
\frac{1}{M_{\rm N} - \hat{k}_1 - \hat{k}_3}
\gamma^{\nu}\gamma^5
\frac{1}{M_{\rm N} - \hat{k}_1}\gamma^{\mu}
\frac{1}{M_{\rm N} - \hat{k}_1 
+ \hat{k}_2}\Bigg\}
$$
$$
-\,i\,g_{\rm A}G_{\rm \pi pp}
\frac{G_{\rm V}}{\sqrt{2}}
\frac{g_{\rm V}}{8\pi^2}\int d^4x_1 
\int \frac{d^4x_2 d^4k_2}{(2\pi)^4}
\frac{d^4x_3 d^4k_3}{(2\pi)^4}\,
e^{\displaystyle i k_2\cdot (x_2-x_1)}
e^{\displaystyle i k_3\cdot (x_3-x_1)}
$$
$$
\times\,\int d^3\rho \,U(\rho)\,
{\rm T}([\bar{p}(t_1,\vec{x}_1 + 
\frac{1}{2}\,\vec{\rho}\,)\,\gamma^5 
p^c (t_1,\vec{x}_1 - \frac{1}{2}\,\vec{\rho}\,)]\, D_{\mu}(x_2)\,[\bar{\psi}_{\rm e}(x_3)
\gamma_{\nu}(1 - \gamma^5) 
\psi_{\nu_{\rm e}}(x_3)])
$$
$$
\times\,\int\frac{d^4k_1}{\pi^2i}\,
e^{\displaystyle - i\vec{q}\cdot \vec{\rho}}\,
{\rm tr}\Bigg\{\gamma^5
\frac{1}{M_{\rm N} - \hat{k}_1 - \hat{k}_3}
\gamma^{\nu}\gamma^5 \frac{1}{M_{\rm N} - \hat{k}_1}
\gamma^{\mu}
\frac{1}{M_{\rm N} - \hat{k}_1 + \hat{k}_2}\Bigg\},
\eqno({\rm D}.11)
$$
The matrix element of the process $\nu_{\rm e}$ + 
D $\to$ e$^-$ + p + p we define by a usual way as
$$
\int d^4x\,<p(p_2)p(p_1){\rm e}^-(k_{\rm e^-}) |
{\cal L}_{\rm \nu_{\rm e}{\rm D}\to e^-pp}(x)|
{\rm D}(k_{\rm D})\nu_{\rm e}(k_{\nu_{\rm e}})> = 
$$
$$
=(2\pi)^4\delta^{(4)}( p_2 + p_1 + k_{\rm e^-} - k_{\rm D} - k_{\nu_{\rm e}})\,\frac{{\cal M}(\nu_{\rm e} + {\rm D} 
\to {\rm e^-} + {\rm p} + {\rm p})}{\displaystyle 
\sqrt{2E_1V\,2E_2V\,2E_{\rm e^-}V\,
2E_{\rm D}V\,2E_{\nu_{\rm e}}V}}, 
\eqno({\rm D}.12)
$$
where  $E_i\,(i =1,2,{\rm D},{\rm e^-},
\nu_{\rm e})$ are the energies of the protons, 
the deuteron, electron and neutrino, $V$ is the normalization volume.

Now we should take the r.h.s. of Eq.~({\rm D}.11) between the wave functions of the initial $|{\rm D}(k_{\rm D})
\nu_{\rm e}(k_{\nu_{\rm e}})>$ and the final
$<p(p_2)p(p_1){\rm e^-}(k_{\rm e^-})|$ states.  
This gives
$$
(2\pi)^4\delta^{(4)}( p_2 + p_1 + k_{\rm e^-} - k_{\rm D} - k_{\nu_{\rm e}})\,\frac{{\cal M}(\nu_{\rm e} + {\rm D} \to {\rm e^-} + {\rm p} + {\rm p})}{\displaystyle \sqrt{2E_1V\,2E_2V\,2E_{\rm e^-}V\,2E_{\rm D}V\,2E_{\nu_{\rm e}}V}}=
$$
$$
= -\, i\,g_{\rm A}G_{\rm \pi pp}\frac{G_{\rm V}}{\sqrt{2}} \frac{g_{\rm V}}{8\pi^2}\int d^4x_1 \int \frac{d^4x_2 d^4k_2}{(2\pi)^4}\frac{d^4x_3 d^4k_3}{(2\pi)^4}\,e^{\displaystyle i k_2\cdot (x_2-x_1)}e^{\displaystyle i k_3\cdot (x_3-x_1)}
$$
$$
\times\,\int d^3\rho \,U(\rho)\,<p(p_2)p(p_1){\rm e^-}(k_{\rm e^-})|{\rm T}([\bar{p}(t_1,\vec{x}_1 + \frac{1}{2}\,\vec{\rho}\,)\,\gamma_{\alpha}\gamma^5 p^c (t_1,\vec{x}_1 - \frac{1}{2}\,\vec{\rho}\,)]\,D_{\mu}(x_2)\,
$$
$$
\times\,[\bar{\psi}_{\rm e}(x_3)\gamma_{\nu}(1 - \gamma^5) \psi_{\nu_{\rm e}}(x_3)])|{\rm D}(k_{\rm D})\nu_{\rm e}(k_{\nu_{\rm e}})>
$$
$$
\times\,\int\frac{d^4k_1}{\pi^2i}\,e^{\displaystyle - i\vec{q}\cdot \vec{\rho}}\,{\rm tr}\Bigg\{\gamma^{\alpha}\gamma^5\frac{1}{M_{\rm N} - \hat{k}_1 - \hat{k}_3}\gamma^{\nu}\gamma^5\frac{1}{M_{\rm N} - \hat{k}_1}\gamma^{\mu} \frac{1}{M_{\rm N} - \hat{k}_1 + \hat{k}_2}\Bigg\}
$$
$$
- i\,g_{\rm A}G_{\rm \pi pp}\frac{G_{\rm V}}{\sqrt{2}}\frac{g_{\rm V}}{8\pi^2}\int d^4x_1 \int \frac{d^4x_2 d^4k_2}{(2\pi)^4}\frac{d^4x_3 d^4k_3}{(2\pi)^4}\,e^{\displaystyle i k_2\cdot (x_2-x_1)}e^{\displaystyle i k_3\cdot (x_3-x_1)}
$$
$$
\times\,\int d^3\rho \,U(\rho)\,<p(p_2)p(p_1){\rm e^-}(k_{\rm e^-})|{\rm T}([\bar{p}(t_1,\vec{x}_1 + \frac{1}{2}\,\vec{\rho}\,)\,\gamma^5 p^c (t_1,\vec{x}_1 - \frac{1}{2}\,\vec{\rho}\,)]\, D_{\mu}(x_2)\,
$$
$$
\times\,[\bar{\psi}_{\rm e}(x_3)\gamma_{\nu}(1 - \gamma^5) \psi_{\nu_{\rm e}}(x_3)])|{\rm D}(k_{\rm D})\nu_{\rm e}(k_{\nu_{\rm e}})>
$$
$$
\times\,\int\frac{d^4k_1}{\pi^2i}\,e^{\displaystyle - i\vec{q}\cdot \vec{\rho}}\,{\rm tr}\Bigg\{\gamma^5\frac{1}{M_{\rm N} - \hat{k}_1 - \hat{k}_2}\gamma^{\nu}\gamma^5\frac{1}{M_{\rm N} - \hat{k}_1}\gamma^{\mu} \frac{1}{M_{\rm N} - \hat{k}_1 + \hat{k}_2}\Bigg\}.\eqno({\rm D}.13)
$$
Between the initial $|{\rm D}(k_{\rm D})\nu_{\rm e}(k_{\nu_{\rm e}})>$ and the final $<p(p_2)p(p_1){\rm e^-}(k_{\rm e^-})|$ states the matrix elements are defined (see Eq.~({\rm C}.23) and Eqs.~({\rm C}.25) -- ({\rm C}.34)):
$$
<p(p_2)p(p_1){\rm e^-}(k_{\rm e^-})|{\rm T}([\bar{p}(t_1,\vec{x}_1 + \frac{1}{2}\,\vec{\rho}\,)\,\gamma_{\alpha}\gamma^5 p^c (t_1,\vec{x}_1 - \frac{1}{2}\,\vec{\rho}\,)]\,D_{\mu}(x_2)\,
$$
$$
\times\,[\bar{\psi}_{\rm e}(x_3)\gamma_{\nu}(1 - \gamma^5) \psi_{\nu_{\rm e}}(x_3)])|{\rm D}(k_{\rm D})\nu_{\rm e}(k_{\nu_{\rm e}})> = [\bar{u}(p_1)\gamma_{\alpha}\gamma^5 u^c(p_2)][\bar{u}(k_{\nu_{\rm e}})\gamma_{\nu}
(1-\gamma^5)u(k_{\rm e^-})]\,
$$
$$
\times\,e_{\mu}(k_{\rm D})\,\times\,2\times\,\psi_{\rm pp}(\rho)\,\frac{\displaystyle 
e^{\displaystyle i(p_1+p_2)\cdot x_1}\,
e^{\displaystyle -ik_{\rm D}\cdot x_2}\,
e^{\displaystyle - 
ik_{{\ell}}\cdot x_3}}{\displaystyle 
\sqrt{2E_1V\,2E_2V\,2E_{\rm e^-}V\,
2E_{\rm D}V\,2E_{\nu_{\rm e}}V}},
$$
$$
<p(p_2)p(p_1){\rm e^-}(k_{\rm e^-})|{\rm T}(
[\bar{p^c}(t_1,\vec{x}_1 + 
\frac{1}{2}\,\vec{\rho}\,)\,
\gamma^5 p(t_1,\vec{x}_1 - 
\frac{1}{2}\,\vec{\rho}\,)]\,D_{\mu}(x_2)\,
$$
$$
\times\,[\bar{\psi}_{\rm e}(x_3) \gamma_{\nu}(1 - \gamma^5) \psi_{\nu_{\rm e}}(x_3)])|{\rm D}(k_{\rm D})
\nu_{\rm e}(k_{\nu_{\rm e}})> = 
[\bar{u}(p_1)\gamma^5 u^c(p_2)]
[\bar{u}(k_{\nu})\gamma_{\nu}
(1-\gamma^5)u(k_{\rm e^-})]\,
$$
$$
\times\,e_{\mu}(k_{\rm D})\,\times\,2\times\,\psi_{\rm pp}(\rho)\,\frac{\displaystyle 
e^{\displaystyle i(p_1+p_2)\cdot x_1}\,
e^{\displaystyle - ik_{\rm D}\cdot x_2}\,
e^{\displaystyle - ik_{{\ell}}\cdot x_3}}{\displaystyle \sqrt{2E_1V\,2E_2V\,2E_{\rm e^-}V\,
2E_{\rm D}V\,2E_{\nu_{\rm e}}V}}.
\eqno({\rm D}.14)
$$
where $k_{\ell} = k_{\rm e^-} - k_{\nu_{\rm e}}$ and 
$\psi_{\rm pp}(\rho)$ is the wave function of the relative movement of the protons normalized per unit density [8]. It is given by Eq.~(\ref{label6.3}). The wave function of the protons $<p(p_2)p(p_1)|$ is determined in terms of the operators of the annihilation in the standard form [31]
$$
<p(p_2)p(p_1)| = <0|a(\vec{p}_2,\sigma_2) a(\vec{p}_1,\sigma_1),\eqno({\rm D}.15)
$$
where as usually for the identical particles the factor $1/\sqrt{2}$ will be taken into account for the computation of the phase volume of the final state (ppe$^-$) in the squared form $(1/\sqrt{2})^2 =1/2$ [31]. The factor 2 is caused by the normalization of the wave function $\psi_{\rm pp}(\rho)$ per unit density (see Eqs.~({\rm C}.31) -- ({\rm C}.34) and Eqs.~({\rm G}.6) -- ({\rm G}.9)).

Inserting the matrix elements  Eq.~({\rm D}.14) to the r.h.s. of  Eq.~({\rm D}.13) we obtain the matrix element of the $\nu_{\rm e}$ + D $\to$ e$^-$ + p + p process in the following form
$$
(2\pi)^4\delta^{(4)}(p_2 + p_1 + k_{\rm e^-} - k_{\rm D} - k_{\nu_{\rm e}})\,i{\cal M}(\nu_{\rm e} + {\rm D}\to {\rm e}^- + {\rm p} + {\rm p}) = $$
$$
= g_{\rm A}G_{\rm \pi pp}\frac{G_{\rm V}}{\sqrt{2}}
\frac{g_{\rm V}}{4\pi^2}[\bar{u}(p_1)\gamma_{\alpha}\gamma^5 u^c(p_2)][\bar{u}(k_{\rm e^-})\gamma_{\nu}
(1-\gamma^5)u(k_{\nu_{\rm e}})]\,e_{\mu}(k_{\rm D})
\int d^4x_1 
$$
$$
\times \int \frac{d^4x_2 d^4k_2}{(2\pi)^4}\frac{d^4x_3 d^4k_3}{(2\pi)^4}\,e^{\displaystyle i (p_2 + p_1 - k_2 - k_3)\cdot x_1}\,e^{\displaystyle i(k_2 - k_{\rm D})\cdot x_2}\,e^{\displaystyle i(k_3 - k_{\ell})\cdot x_3}
$$
$$
\times\int d^3\rho   U(\rho)\psi_{\rm pp}(\rho) \int\frac{d^4k_1}{\pi^2i} e^{\displaystyle - i\vec{q}\cdot \vec{\rho}} {\rm tr}\Bigg\{\gamma^{\alpha}\gamma^5\frac{1}{M_{\rm N} - \hat{k}_1 - \hat{k}_3}\gamma^{\nu}\gamma^5\frac{1}{M_{\rm N} - \hat{k}_1}\gamma^{\mu}\frac{1}{M_{\rm N} - \hat{k}_1 + \hat{k}_2}\Bigg\}
$$
$$
+ g_{\rm A}G_{\rm \pi pp}\frac{G_{\rm V}}{\sqrt{2}} \frac{g_{\rm V}}{4\pi^2}[\bar{u}(p_1)\gamma^5 u^c(p_2)]
[\bar{u}(k_{\rm e^-})\gamma_{\nu}
(1-\gamma^5) u(k_{\nu_{\rm e}})]\,e_{\mu}(k_{\rm D}) \int d^4x_1 
$$
$$
\times \int \frac{d^4x_2 d^4k_2}{(2\pi)^4}\frac{d^4x_3 d^4k_3}{(2\pi)^4}\,e^{\displaystyle i (k_2 + k_3 - p_1 - p_2)\cdot x_1}\,e^{\displaystyle i(k_2 - k_{\rm D})\cdot x_2}\,
e^{\displaystyle i(k_3 - k_{\ell})\cdot x_3}
$$
$$
\times\int d^3\rho U(\rho) \,\psi_{\rm pp}(\rho) \int\frac{d^4k_1}{\pi^2i}e^{\displaystyle -i\vec{q}\cdot \vec{\rho}}{\rm tr}\Bigg\{\gamma^5\frac{1}{M_{\rm N} - \hat{k}_1 - \hat{k}_3}\gamma^{\nu}\gamma^5\frac{1}{M_{\rm N} - \hat{k}_1}\gamma^{\mu} \frac{1}{M_{\rm N} - \hat{k}_1 + \hat{k}_2}\Bigg\}.\eqno({\rm D}.16)
$$
Integrating over $x_1$, $x_2$, $x_3$, $k_2$ and $k_3$ we obtain in the r.h.s. of Eq.~({\rm D}.16) the $\delta$--function describing the 4--momentum conservation. Then, the matrix element of the  $\nu_{\rm e}$ + D $\to$ e$^-$ + p + p process becomes equal
$$
i{\cal M}({\rm p} + {\rm p} \to {\rm D} + {\rm e}^+ + \nu_{e}) =  g_{\rm A}G_{\rm \pi pp}\frac{G_{\rm V}}{\sqrt{2}} \frac{g_{\rm V}}{4\pi^2}
$$
$$
\times\,[\bar{u}(p_1)\gamma_{\alpha}\gamma^5 u^c(p_2)]
[\bar{u}(k_{\rm e^-})\gamma_{\nu}
(1-\gamma^5)u(k_{\nu_{\rm e}})]\,e_{\mu}(k_{\rm D})\int d^3\rho\,  U(\rho) \,\psi_{\rm pp}(\rho) 
$$
$$
\times\,\int\frac{d^4k}{\pi^2i} e^{\displaystyle -i\vec{q}\cdot \vec{\rho}}\,{\rm tr}\Bigg\{\gamma^{\alpha}\gamma^5\frac{1}{M_{\rm N} - \hat{k} -  \hat{k}_{\ell}}\gamma^{\nu}
\gamma^5\frac{1}{M_{\rm N} - \hat{k}}
\gamma^{\mu} \frac{1}{M_{\rm N} - \hat{k} + \hat{k}_{\rm D}}\Bigg\} + g_{\rm A}G_{\rm \pi pp}
$$
$$
\times\,\frac{G_{\rm V}}{\sqrt{2}} \frac{g_{\rm V}}{4\pi^2}[\bar{u}(p_1)\gamma^5 u^c(p_2)]
[\bar{u}(k_{\rm e^-})\gamma_{\nu}
(1-\gamma^5)u(k_{\nu_{\rm e}})]\,e_{\mu}(k_{\rm D})
\int d^3\rho\,  U(\rho) \,\psi_{\rm pp}(\rho)
$$
$$
\times \int\frac{d^4k}{\pi^2i}e^{\displaystyle -i\vec{q}\cdot \vec{\rho}}
{\rm tr}
\Bigg\{\gamma^5\frac{1}{M_{\rm N} - \hat{k} - \hat{k}_{\ell}}\gamma^{\nu}
\gamma^5 
\frac{1}{M_{\rm N} - \hat{k}}
\gamma^{\mu} \frac{1}{M_{\rm N} - \hat{k} + \hat{k}_{\rm D}}\Bigg\},\eqno({\rm D}.17)
$$
where $\vec{q} = \vec{k} + (\vec{k}_{\ell} - \vec{k}_{\rm D})/2$.

It is convenient to represent the matrix element Eq.~({\rm D}.17) in terms of the structure functions $\bar{{\cal J}}^{\alpha\nu\mu}_{\rm pp}(k_{\rm D}, k_{\ell})$ and $\bar{{\cal J}}^{\nu\mu}_{\rm pp}(k_{\rm D}, k_{\ell})$:
$$
i{\cal M}(\nu_{\rm e} + {\rm D} \to {\rm e^-} + {\rm p} + {\rm p}) =
$$
$$
= - g_{\rm A}G_{\rm \pi pp}\frac{G_{\rm V}}{\sqrt{2}} \frac{g_{\rm V}}{4\pi^2}[\bar{u}(p_1)\gamma_{\alpha}\gamma^5 u^c(p_2)][\bar{u}(k_{\rm e^-})
\gamma_{\nu}(1-\gamma^5)u(k_{\nu_{\rm e}})]
\,e_{\mu}(k_{\rm D})\,
\bar{{\cal J}}^{\alpha\nu\mu}_{\rm pp}
(k_{\rm D}, k_{\ell})
$$
$$
- g_{\rm A}G_{\rm \pi pp}\frac{G_{\rm V}}{\sqrt{2}} \frac{g_{\rm V}}{4\pi^2}[\bar{u}(p_1)\gamma^5 u^c(p_2)]
[\bar{u}(k_{\rm e^-})\gamma_{\nu}
(1-\gamma^5)u(k_{\nu_{\rm e}})]\,e_{\mu}(k_{\rm D})
\bar{{\cal J}}^{\nu\mu}_{\rm pp}
(k_{\rm D}, k_{\ell}),\eqno({\rm D}.18)
$$
where the structure functions $\bar{{\cal J}}^{\alpha\nu\mu}_{\rm pp}(k_{\rm D}, k_{\ell})$ and $\bar{{\cal J}}^{\nu\mu}_{\rm pp}(k_{\rm D}, k_{\ell})$ are defined as
$$
\bar{{\cal J}}^{\alpha\nu\mu}_{\rm pp}(k_{\rm D}, k_{\ell}) = -\int d^3\rho\,  U(\rho)\,\psi_{\rm pp}(\rho)\int\frac{d^4k}{\pi^2i} e^{\displaystyle  - i\vec{q}\cdot \vec{\rho}}
$$
$$
\times\,
{\rm tr} 
\Bigg\{\gamma^{\alpha}\gamma^5
\frac{1}{M_{\rm N} - \hat{k} - \hat{k}_{\ell}}
\gamma^{\nu}
\gamma^5 \frac{1}{M_{\rm N} - \hat{k}}
\gamma^{\mu}
\frac{1}{M_{\rm N} - \hat{k} + 
\hat{k}_{\rm D}}\Bigg\},
$$
$$
\bar{{\cal J}}^{\nu\mu}_{\rm pp}(k_{\rm D}, k_{\ell}) = - \int d^3\rho\,  U(\rho)\,\psi_{\rm pp}(\rho)\int\frac{d^4k}{\pi^2i} e^{\displaystyle - i\vec{q}\cdot \vec{\rho}}
$$
$$
\times\,{\rm tr}
\Bigg\{\gamma^5
\frac{1}{M_{\rm N} - \hat{k} - \hat{k}_{\ell}} 
\gamma^{\nu}\gamma^5 \frac{1}{M_{\rm N} - \hat{k}}
\gamma^{\mu} \frac{1}{M_{\rm N} - \hat{k} + \hat{k}_{\rm D}}\Bigg\}.\eqno({\rm D}.19)
$$
Thus, the problem of the computation of the matrix element of the process $\nu_{\rm e}$ + D $\to$ e$^-$ + p + p reduces to the problem of the computation of the structure functions Eq.~({\rm D}.19). Integrating over directions of the relative radius--vector $\vec{\rho}$ we get
$$
 \bar{{\cal J}}^{\alpha\nu\mu}_{\rm pp}
 (k_{\rm D}, k_{\ell}) = -
 \,4\pi \int\limits^{\infty}_0 
 d\rho\,\rho^2\,U(\rho)\,
  \psi_{\rm pp}(\rho)\int\frac{d^4k}{\pi^2i}\,   \frac{\sin|\vec{q}\,|\rho}{|\vec{q}\,|\rho}
$$
$$
\times\,
{\rm tr} 
 \Bigg\{\gamma^{\alpha}\gamma^5
 \frac{1}{M_{\rm N} - \hat{k} - \hat{k}_{\ell}}
 \gamma^{\nu}
 \gamma^5 \frac{1}{M_{\rm N} - \hat{k}}\gamma^{\mu}
 \frac{1}{M_{\rm N} - \hat{k} 
 + \hat{k}_{\rm D}}\Bigg\},
$$
$$
 \bar{{\cal J}}^{\nu\mu}_{\rm pp}
  (k_{\rm D}, k_{\ell}) = - \,4\pi \int\limits^{\infty}_0  d\rho\,\rho^2\, U(\rho)\,
   \psi_{\rm pp}(\rho)  \int\frac{d^4k}{\pi^2i}\, \frac{\sin|\vec{q}\,|\rho}{|\vec{q}\,|\rho}
$$
$$
\times\,
{\rm tr}
 \Bigg\{\gamma^5
 \frac{1}{M_{\rm N} - \hat{k} - \hat{k}_{\ell}} 
 \gamma^{\nu}\gamma^5 
 \frac{1}{M_{\rm N} - \hat{k}}
 \gamma^{\mu} 
 \frac{1}{M_{\rm N} - \hat{k} + \hat{k}_{\rm D}}
 \Bigg\}.\eqno({\rm D}.20)
$$
Since the energy  of 
the incident neutrino $E_{\nu_{\rm e}}$
ranges the region $E_{\rm th} \le  E_{\nu_{\rm e}} 
 \le  10\,{\rm MeV}$ and the deuteron in the rest 
frame 
$k^{\mu}_{\rm D} = 
(k^0_{\rm D}, \vec{0}\,)$, we would calculate 
the structure functions setting $k^{\mu}_{\ell} = 
\vec{k}_{\rm D} = 0$:
$$
\bar{{\cal J}}^{\alpha\nu\mu}_{\rm pp}(k_{\rm D}, k_{\ell})
= -\,4\pi \int\limits^{\infty}_0 d\rho\,\rho^2\,U(\rho)\,
\psi_{\rm pp}(\rho)\int\frac{d^4k}{\pi^2i}\, \frac{\sin|\vec{k}\,|\rho}{|\vec{k}\,|\rho}
$$
$$
\times\,{\rm tr} 
\Bigg\{\gamma^{\alpha}\gamma^5
\frac{1}{M_{\rm N} - \hat{k}}\gamma^{\nu}
\gamma^5 \frac{1}{M_{\rm N} - \hat{k}}
\gamma^{\mu} \frac{1}{M_{\rm N} - \hat{k} 
+ \hat{k}_{\rm D}}\Bigg\},
$$
$$
\bar{{\cal J}}^{\nu\mu}_{\rm pp}(k_{\rm D}, k_{\ell})
= - \,4\pi \int\limits^{\infty}_0 d\rho\,\rho^2\,
U(\rho)\,\psi_{\rm pp}(\rho)\int\frac{d^4k}{\pi^2i}\,
\frac{\sin|\vec{k}\,|\rho}{|\vec{k}\,|\rho}
$$
$$
\times\,{\rm tr}
\Bigg\{\gamma^5\frac{1}{M_{\rm N} - \hat{k}} 
\gamma^{\nu}\gamma^5
\frac{1}{M_{\rm N} - \hat{k}}
\gamma^{\mu} \frac{1}{M_{\rm N} - 
\hat{k} + \hat{k}_{\rm D}}\Bigg\}.
\eqno({\rm D}.21)
$$
The computation of the momentum integrals defining the structure functions cannot be carried out by a Lorentz covariant manner [2,4]. The obvious Lorentz covariance has been lost due to the description of the pp interaction in terms of the potential. Therefore, for the computation of the momentum integrals it is convenient to follow only the components which give the main contribution in the low--energy limit. For the calculation of 
${\cal J}^{\alpha\nu\mu}
(k_{\rm D}, k_{\ell})$ we should notice that in the low--energy limit only the time--component of the current 
$[\bar{u}(p_1)\gamma_{\alpha}
\gamma^5 u^c(p_2)]$ survives and obeys the relation
$$
[\bar{u}(p_1)\gamma_{\alpha}\gamma^5 u^c(p_2)] = g_{\alpha 0}[\bar{u}(p_1)\gamma^0\gamma^5 u^c(p_2)] = g_{\alpha 0}[\bar{u}(p_1)\gamma^5 u^c(p_2)],
$$
$$
[\bar{u}(p_1)\gamma_{\alpha}\gamma^5 u^c(p_2)] \to  g_{\alpha 0}[\bar{u}(p_1)\gamma^5 u^c(p_2)],\eqno({\rm D}.22)
$$
where we have used the relation $\bar{u}(p_1)\gamma^0 = \bar{u}(p_1)$, which is valid in the non--relativistic limit due to the dominance  of the large components of the Dirac bispinors. As has been discussed in Appendix~B the indices $\mu$ and $\nu$ should run over $\mu$ (or $\nu$) = 1,2,3. Thus, the matrix element Eq.~({\rm D}.18) reduces to the form
$$
i{\cal M}(\nu_{\rm e} + {\rm D} \to {\rm e^-} + {\rm p} + {\rm p}) =
$$
$$
= - g_{\rm A}G_{\rm \pi pp}\frac{G_{\rm V}}{\sqrt{2}} \frac{g_{\rm V}}{2\pi^2}[\bar{u}(p_1)\gamma^5 u^c(p_2)]
[\bar{u}(k_{\rm e^-})\gamma_{\nu}
(1-\gamma^5)u(k_{\nu_{\rm e}})]\,e_{\mu}(k_{\rm D})\,
$$
$$
\times\,[\bar{{\cal J}}^{0\nu\mu}_{\rm pp}(k_{\rm D}, k_{\ell}) + \bar{{\cal J}}^{\nu\mu}_{\rm pp}(k_{\rm D}, k_{\ell})],\eqno({\rm D}.23)
$$
where the structure functions 
$\bar{{\cal J}}^{0\nu\mu}_{\rm pp}
(k_{\rm D}, k_{\ell})$  and 
$\bar{{\cal J}}^{\mu\nu}_{\rm pp}
(k_{\rm D}, k_{\ell})$ 
are given 
in  Eq.~({\rm D}.21).

The computation of the momentum integrals we perform following the prescription of the RFMD, that is, we assume that the deuteron is off--mass shell and $M_{\rm N} \gg \sqrt{k^2_{\rm D}}= k^0_{\rm D}$. Then, we expand the integrand of the structure functions Eq.~({\rm C}.40) in powers of $k_{\rm D}$ keeping only leading contributions.
\vspace{0.2in}

\noindent{\bf The computation of 
${\cal J}^{0\nu\mu}_{\rm pp}
(k_{\rm D}, k_{\ell})$}. First, we should integrate over the virtual momentum $k$:
$$
\int\frac{d^4k}{\pi^2i}\frac{\sin|\vec{k}\,|\rho}{|\vec{k}\,|\rho} \,{\rm tr} \Bigg\{\gamma^0\gamma^5\frac{1}{M_{\rm N} - \hat{k}}\gamma^{\nu}\gamma^5\frac{1}{M_{\rm N} - \hat{k}}\gamma^{\mu}\frac{1}{M_{\rm N} - \hat{k} + \hat{k}_{\rm D}}\Bigg\}=
$$
$$
=\int\frac{d^4k}{\pi^2i}\frac{\sin|\vec{k}\,|\rho}{|\vec{k}\,|\rho} \,\frac{{\rm tr}\{\gamma^0\gamma^5(M_{\rm N} + \hat{k})\gamma^{\nu}\gamma^5 (M_{\rm N} + \hat{k})\gamma^{\mu}(M_{\rm N} + \hat{k} - \hat{k}_{\rm D})\}}{[M^2_{\rm N} - k^2 - i0]^2
[M^2_{\rm N} - (k - k_{\rm D})^2 - i0]}=
$$
$$
=\int\frac{d^4k}{\pi^2i}\frac{\sin|\vec{k}\,|\rho}{|\vec{k}\,|\rho} \,\frac{{\rm tr}\{\gamma^0\gamma^5(M_{\rm N} + \hat{k})\gamma^{\nu}\gamma^5 (M_{\rm N} + \hat{k})\gamma^{\mu} (M_{\rm N} + \hat{k} - \hat{k}_{\rm D})\}}{[M^2_{\rm N} - k^2
- i0]^3}
$$
$$
\times\,\Bigg\{1 - 
\frac{2 k\cdot k_{\rm D}}
{[M^2_{\rm N} - k^2 - i0]}\Bigg\}=
$$
$$
=\int\frac{d^4k}{\pi^2i}\frac{\sin|\vec{k}\,|\rho}{|\vec{k}\,|\rho} \,\frac{{\rm tr}\{\hat{k}_{\rm D}\gamma^0\gamma^{\nu}(M_{\rm N} + \hat{k})\gamma^{\mu}(M_{\rm N} - \hat{k})\}}{[M^2_{\rm N} - k^2 - i0]^3}
$$
$$
+\int\frac{d^4k}{\pi^2i}\frac{\sin|\vec{k}\,|\rho}{|\vec{k}\,|\rho} \,\frac{2 k\cdot k_{\rm D}\,{\rm tr}\{\gamma^0(M_{\rm N} - \hat{k})\gamma^{\nu}(M_{\rm N} + \hat{k})\gamma^{\mu}(M_{\rm N} + \hat{k})\}}{[M^2_{\rm N} - k^2
- i0]^4}=
$$
$$
= \bar{J}^{0\nu\mu}_1 + \bar{J}^{0\nu\mu}_2.
\eqno({\rm D}.24)
$$
Since $\hat{k}_{\rm D} = \gamma_0k^0_{\rm D}$, the structure function $\bar{J}^{0\nu\mu}_1$ coincides with the structure function $J^{0\nu\mu}_1$ given by Eq.~({\rm C}.49) and reads
$$
\bar{J}^{0\nu\mu}_1=4 k^0_{\rm D}\,g^{\mu\nu}\left[\frac{2}{3}\,M^2_{\rm N}\int\limits^{\infty}_0\frac{\displaystyle dk\,\cos k\rho}{\displaystyle (M^2_{\rm N} + k^2)^{3/2}}-\int\limits^{\infty}_0\frac{\displaystyle dk\,\cos k\rho}{\displaystyle (M^2_{\rm N} + k^2)^{1/2}}\right].
\eqno({\rm D}.25)
$$
The computation of $\bar{J}^{0\nu\mu}_2$ runs as follows:
$$
\bar{J}^{0\nu\mu}_2=\int\frac{d^4k}{\pi^2i}\frac{\sin|\vec{k}\,|\rho}{|\vec{k}\,|\rho} \,\frac{2 k\cdot k_{\rm D}\,{\rm tr}\{\gamma^0(M_{\rm N} - \hat{k})\gamma^{\nu}(M_{\rm N} + \hat{k})\gamma^{\mu}(M_{\rm N} + \hat{k})\}}{[M^2_{\rm N} - k^2 - i0]^4}=
$$
$$
= 2 k^0_{\rm D}\int\frac{d^4k}{\pi^2i}\frac{\sin|\vec{k}\,|\rho}{|\vec{k}\,|\rho} \,\frac{k_0\,{\rm tr}\{\gamma^0(M_{\rm N} - \hat{k})\gamma^{\nu}(M_{\rm N} + \hat{k})\gamma^{\mu}(M_{\rm N} + \hat{k})\}}
{[E^2_{\vec{k}} - k^2_0 - i0]^4}.
\eqno({\rm D}.26)
$$
Computing the trace over Dirac matrices
$$
{\rm tr}\{\gamma^0(M_{\rm N} - \hat{k})\gamma^{\nu}(M_{\rm N} + \hat{k})\gamma^{\mu}(M_{\rm N} + \hat{k})\}=
$$
$$
={\rm tr}\{(M_{\rm N}\gamma^0 - \gamma^0\hat{k})(M^2_{\rm N}\gamma^{\nu}\gamma^{\mu} + 2\,M_{\rm N}k^{\mu}\gamma^{\nu} + \gamma^{\nu}\hat{k}\gamma^{\mu}\hat{k})\}=
$$
$$
 ={\rm tr}\{- M^2_{\rm N}\gamma^0\hat{k}\gamma^{\nu}\gamma^{\mu}- \gamma^0\hat{k}\gamma^{\nu}\hat{k}\gamma^{\mu}\hat{k}\}=
$$
$$
={\rm tr}\{- M^2_{\rm N}\hat{k}\gamma^0\gamma^{\nu}\gamma^{\mu}+ k^2\gamma^0\gamma^{\nu}\gamma^{\nu}\hat{k}- 2k^{\nu}\gamma^0\hat{k}\gamma^{\mu}\hat{k}\}=
$$
$$
=4\,(-\,k_0\, M^2_{\rm N}\,g^{\mu\nu} + k_0\, k^2\,g^{\mu\nu}- 4 k_0\,k^{\mu}k^{\nu}) =
$$
$$
= -\,4\,k_0\,(M^2_{\rm N}\,g^{\mu\nu} -  k^2\,g^{\mu\nu} + 4\, k^{\mu}k^{\nu}).\eqno({\rm D}.27)
$$
we find that the result of the computation of the trace Eq.~({\rm D}.27) amounts to Eq.~({\rm C}.51). Thereby, the structure function $\bar{J}^{0\nu\mu}_2$ should coincide with the structure function $bar{J}^{0\nu\mu}_2$ too and reads:
$$
\bar{J}^{0\nu\mu}_2 = \frac{4}{3}\, k^0_{\rm D}\,g^{\mu\nu}\left[\frac{2}{3}\,M^2_{\rm N}\int\limits^{\infty}_0\frac{\displaystyle dk\,\cos k\rho}{\displaystyle (M^2_{\rm N} + k^2)^{3/2}} + \int\limits^{\infty}_0\frac{\displaystyle dk\,\cos k\rho}{\displaystyle (M^2_{\rm N} + k^2)^{1/2}}\right].\eqno({\rm D}.28)
$$
Summing up the contributions given by Eq.~({\rm D}.25) and Eq.~({\rm D}.28) we obtain the structure function $\bar{{\cal J}}^{0\nu\mu}_{\rm pp}(k_{\rm D}, k_{\ell})$:
$$
\bar{{\cal J}}^{0\nu\mu}_{\rm pp}(k_{\rm D}, k_{\ell})=
$$
$$
=- \frac{8}{3}\, k^0_{\rm D}\,g^{\mu\nu}\,
4\pi\int\limits^{\infty}_0
d\rho\,\rho^2\,U(\rho)\,\psi_{\rm pp}(\rho)
\left[\frac{4}{3}\,M^2_{\rm N}\int\limits^{\infty}_0\frac{\displaystyle dk\,\cos k\rho}{\displaystyle (M^2_{\rm N} + k^2)^{3/2}} - \int\limits^{\infty}_0\frac{\displaystyle dk\,\cos k\rho}{\displaystyle (M^2_{\rm N} + k^2)^{1/2}}\right]=
$$
$$
=- \frac{8}{3}\, k^0_{\rm D}\,g^{\mu\nu}\,
4\pi\int\limits^{\infty}_0
d\rho\,\rho^2\,U(\rho)\,\psi_{\rm pp}(\rho)
\left[\frac{4}{3}\,M_{\rm N}\rho\,K_1(M_{\rm N}\rho) 
- K_0(M_{\rm N}\rho)\right],\eqno({\rm D}.29)
$$
For the continuation of the structure function Eq.~({\rm D}.29) on--mass shell of the deuteron  we should only set $k^0_{\rm D} = M_{\rm D} \simeq 2\,M_{\rm N}$:
$$
{\cal J}^{0\nu\mu}_{\rm pp}
(k_{\rm D}, k_{\ell}) =
- \frac{16}{3}\, M_{\rm N}\,g^{\mu\nu}\,4\pi
\int\limits^{\infty}_0
d\rho\,\rho\,U(\rho)\,\psi_{\rm pp}(\rho)\,
\left[\frac{4}{3}\,M_{\rm N}\rho\,K_1(M_{\rm N}\rho) 
- K_0(M_{\rm N}\rho)\right].\eqno({\rm D}.30)
$$
Now let us proceed to the computation of the structure function $\bar{\cal J}^{\nu\mu}_{\rm pp}
(k_{\rm D}, k_{\ell})$.

\noindent{\bf The computation of
${\cal J}^{\mu\nu}_{\rm pp}
(k_{\rm D}, k_{\ell})$}. It is to show that the structure function $\bar{\cal J}^{\nu\mu}_{\rm pp}
(k_{\rm D}, k_{\ell})$ is proportional to the structure function ${\cal J}^{\mu\nu}_{\rm pp}
(k_{\rm D}, k_{\ell})$ given Eq.~({\rm C}.40) and, therefore, reads (see Eq.~({\rm C}.57)):
$$
\bar{{\cal J}}^{\nu\mu}_{\rm pp}(k_{\rm D}, k_{\ell}) 
= -\,8\,M_{\rm N}\,g^{\mu\nu}\,4\pi
\int\limits^{\infty}_0
d\rho\,  \rho\, U(\rho)\,
\psi_{\rm pp}(\rho)\,
K_0(M_{\rm N}\rho).\eqno({\rm D}.31)
$$
The structure function $\bar{{\cal J}}^{\nu\mu}_{\rm pp}(k_{\rm D},k_{\ell})$ does not depend on the 4--momentum of the deuteron and retains the form when it is continued on--mass shell of the deuteron.

For the computation of the structure function defining the amplitude of the process $\nu_{\rm e}$ + D $\to$ e$^-$ + p + p given by Eq.~({\rm D}.23) we use the wave function $\psi_{\rm pp}(\rho)$ in the form of Eq.~(\ref{label6.7}) and set
$$
\bar{{\cal J}}^{0\nu\mu}_{\rm pp}(k_{\rm D}, k_{\ell}) + \bar{{\cal J}}^{\nu\mu}_{\rm pp}(k_{\rm D}, k_{\ell}) = e^{\displaystyle
i\delta^{\rm e}_{\rm pp}(k)}\frac{\displaystyle \sin\delta^{\rm e}_{\rm
pp}(k)}{\displaystyle a^{\rm e}_{\rm pp} k C(k)}\,6\,M_{\rm N}\,g^{\mu\nu}\,{\cal F}^{\rm e}_{\rm ppe^-}.\eqno({\rm D}.32)
$$
The factor ${\cal F}^{\rm e}_{\rm ppe^-}$ amounts to
$$
{\cal F}^{\rm e}_{\rm ppe^-}  = -\,a^{\rm e}_{\rm pp}\,\frac{32}{27}\,4\pi\int\limits^{\infty}_0 
d\rho\, \rho\, U(\rho)\,\left[ M_{\rm N}\rho 
\,K_1(M_{\rm N}\rho) + \frac{3}{8}\,K_0(M_{\rm N}\rho)\right].\eqno({\rm D}.33)
$$
Inserting the Yukawa potential Eq.~(\ref{label1.4}) and using the formulae Eq.~({\rm C}.61) we calculate the factor ${\cal F}^{\rm e}_{\rm ppe^-}$ in the appropriate from
$$
{\cal F}^{\rm e}_{\rm ppe^-}  = -a^{\rm e}_{\rm pp}\,\frac{44}{27}\,\Bigg[
\frac{M^2_{\pi}}{\sqrt{M^2_{\rm N} - M^2_{\pi}}}\,
{\rm arctg}\frac{\sqrt{M^2_{\rm N} - M^2_{\pi}}}{M_{\pi}} - \frac{8}{11}\,\frac{M^3_{\pi}}{M^2_{\rm N} - M^2_{\pi}}
$$
$$
+ \frac{8}{11}\,\frac{M^4_{\pi}}{(M^2_{\rm N} - M^2_{\pi})^{3/2}}\,{\rm arctg}\frac{\sqrt{M^2_{\rm N} - M^2_{\pi}}}{M_{\pi}}\Bigg] = 1.70.\eqno({\rm D}.34)
$$
The numerical value has been obtained for $a^{\rm e}_{\rm pp} =- 7.828\,{\rm fm}$,  $M_{\rm N} = 940\,{\rm MeV}$ and $M_{\pi} = 135\,{\rm MeV}$.

The amplitude of the process $\nu_{\rm e}$ + D $\to$ e$^-$ + p + p is then given by
$$
i{\cal M}(\nu_{\rm e} + {\rm D} \to {\rm e^-} + {\rm p} + {\rm p}) =
- g_{\rm A}G_{\rm \pi pp}M_{\rm N}\frac{G_{\rm V}}{\sqrt{2}} \frac{3g_{\rm V}}{2\pi^2}\,{\cal F}^{\rm e}_{\rm ppe^-}\,
$$
$$
\times\,[\bar{u}(p_1)\gamma^5 u^c(p_2)][\bar{u}(k_{\rm e^-}) \gamma^{\mu} (1-\gamma^5) u(k_{\nu_{\rm e}})]\,e_{\mu}(k_{\rm D})\, e^{\displaystyle i\delta^{\rm e}_{\rm pp}(k)} \frac{\displaystyle \sin\delta^{\rm e}_{\rm
pp}(k)}{\displaystyle a^{\rm e}_{\rm pp} k C(k)}\eqno({\rm D}.35)
$$
This completes the computation of the matrix element of the process of 
$\nu_{\rm e}$ + D $\to$ e$^-$ + p + p. The computation of the matrix element of the process $\bar{\nu}_{\rm e}$ + D $\to$ e$^+$ + n + n is very analogous to that elaborated above.

\section*{Appendix E. Computation of the matrix element of the process $\nu_{\rm e}+ {\rm D} \to \nu_{\rm e} + {\rm n} + {\rm p}$}

The process $\nu_{\rm e}$ + D $\to$ $\nu_{\rm e}$ + n + p runs through the intermediate Z--boson exchange: $\nu_{\rm e}$ + D $\to$ $\nu_{\rm e}$ +Z + D $\to$ $\nu_{\rm e}$ + n + p. The matrix element of the transition $\nu_{\rm e}$ + D $\to$ $\nu_{\rm e}$ + Z + D $\to$ $\nu_{\rm e}$ + p + p is defined by the effective interactions
$$
{\cal L}^{\dagger}_{\rm npD}(x) = -ig_{\rm V}[\bar{p}(x)\gamma^{\mu}n^c(x) - \bar{n}(x)\gamma^{\mu} p^c(x)]\,D_{\mu}(x),
$$
$$
{\cal L}^{\rm np \to np}_{\rm eff}(x) =  G_{\rm \pi
np}\,\int d^3\rho\,U(\rho)
$$
$$
\times\,\{[\bar{n}(t,\vec{x} + \frac{1}{2}\vec{\rho}\,)\gamma_{\mu}
\gamma^5 p^c(t,\vec{x} - \frac{1}{2}\vec{\rho}\,)]
[\bar{p^c}(t,\vec{x} + \frac{1}{2}\vec{\rho}\,)\gamma^{\mu}\gamma^5
n(t,\vec{x} - \frac{1}{2}\vec{\rho}\,)]
$$
$$
+ (\gamma_{\mu}\gamma^5 \otimes \gamma^{\mu}\gamma^5 \to \gamma^5 \otimes
\gamma^5)\},
$$
$$
{\cal L}_{\rm NNZ}(x) = g_{\rm A} \,[\bar{p}(x)\gamma^{\nu}\gamma^5 p(x) - \bar{n}(x)\gamma^{\nu}\gamma^5 n(x)]\,Z_{\nu}(x).\eqno({\rm E}.1)
$$
For the description of the transition $\nu_{\rm e}$ $\to$  $\nu_{\rm e}$ + Z we replace the operator of the Z--boson field by the operator of the neutrino weak current
$$
Z_{\nu}(x) \to \frac{G_{\rm F}}{2\sqrt{2}}\,[\bar{\psi}_{\nu_{\rm e}}(x)\gamma_{\nu}(1 - \gamma^5) \psi_{\nu_{\rm e}}(x)].\eqno({\rm E}.2)
$$
The S matrix element ${\rm S}^{(3)}_{\rm ZD \to np}$ responsible for the transition Z + D $\to$ n + p can be obtained by analogy with the S matrix element ${\rm S}^{(3)}_{\rm pp\to DW^+}$ (see Eq.~({\rm C}.8)) describing the transition p + p $\to$ D + W$^+$ and the S matrix element  ${\rm S}^{(3)}_{\rm W^+D\to pp}$ (see Eq.~({\rm D}.4)) describing the transition W$^+$ + D $\to$ p + p and reads
$$
{\rm S}^{(3)}_{\rm ZD\to np} = -i \int d^4x_1 d^4x_2 d^4x_3\,{\rm T}({\cal L}^{\rm np\to np}_{\rm eff}(x_1){\cal L}^{\dagger}_{\rm npD}(x_2){\cal L}_{\rm NNZ}(x_3)).\eqno({\rm E}.3)
$$
For the derivation of the effective Lagrangian ${\cal L}_{\rm ZD\to np}(x)$ containing only the fields of the initial and the final particles we should make all necessary contractions of the operators of the proton and the neutron fields. These contractions we denote by brackets and obtain
$$
<{\rm S}^{(3)}_{\rm ZD\to np}> = -i \int d^4x_1 d^4x_2 d^4x_3\,<{\rm T}({\cal L}^{\rm np\to np}_{\rm eff}(x_1){\cal L}^{\dagger}_{\rm npD}(x_2){\cal L}_{\rm NNZ}(x_3))>.\eqno({\rm E}.4)
$$
The effective Lagrangian ${\cal L}_{\rm ZD\to np}(x)$ related to the S matrix element $<{\rm S}^{(3)}_{\rm ZD\to np}>$ is defined as 
$$
<{\rm S}^{(3)}_{\rm ZD\to np}> = i\int d^4x\,{\cal L}_{\rm ZD\to np}(x) = 
$$
$$
= -i \int d^4x_1 d^4x_2 d^4x_3\,<{\rm T}({\cal L}^{\rm np\to np}_{\rm eff}(x_1){\cal L}^{\dagger}_{\rm npD}(x_2){\cal L}_{\rm NNZ}(x_3))>. \eqno({\rm E}.5)
$$
In terms of the operators of the interacting fields the effective Lagrangian ${\cal L}_{\rm ZD\to np}(x)$ reads
$$
\int d^4x\,{\cal L}_{\rm ZD\to np}(x) =  - \int d^4x_1 d^4x_2 d^4x_3\,<{\rm T}({\cal L}^{\rm np\to np}_{\rm eff}(x_1){\cal L}^{\dagger}_{\rm npD}(x_2){\cal L}_{\rm NNZ}(x_3))>
$$
$$
= G_{\rm \pi np}\,\times\,(-ig_{\rm V})\,\times \,g_{\rm A}\int d^4x_1 d^4x_2 d^4x_3\,\int d^3\rho\,U(\rho)\,
$$
$$
\times\,{\rm T}([\bar{n}(t_1,\vec{x}_1 + \frac{1}{2}\,\vec{\rho}\,)\,\gamma^{\alpha}\gamma^5 p^c (t_1,\vec{x}_1 - \frac{1}{2}\,\vec{\rho}\,)]\,D_{\mu}(x_2)\,Z_{\nu}(x_3))
$$
$$
\times <0|{\rm T}([\bar{p^c}(t_1,\vec{x}_1 + \frac{1}{2}\,\vec{\rho}\,)\,\gamma_{\alpha}\gamma^5 n(t_1,\vec{x}_1 - \frac{1}{2}\,\vec{\rho}\,)][\bar{p}(x_2)\gamma^{\mu}n^c(x_2) - \bar{n}(x_2)
\gamma^{\mu}p^c(x_2)]\,
$$
$$
\times\,[\bar{p}(x_3)\gamma^{\nu}\gamma^5 p(x_3) - \bar{n}(x_3)\gamma^{\nu}\gamma^5 n(x_3)])|0> 
$$
$$
+ G_{\rm \pi np}\,\times\,(-ig_{\rm V})\,\times \,g_{\rm A} \int d^4x_1 d^4x_2 d^4x_3\,\int d^3\rho \,U(\rho)\,
$$
$$
\times\,{\rm T}([\bar{n}(t_1,\vec{x}_1 + \frac{1}{2}\,\vec{\rho}\,)\,\gamma^5 p^c (t_1,\vec{x}_1 - \frac{1}{2}\,\vec{\rho}\,)]\,D_{\mu}(x_2)\,Z_{\nu}(x_3))
$$
$$
\times <0|{\rm T}([\bar{p^c}(t_1,\vec{x}_1 + \frac{1}{2}\,\vec{\rho}\,)\,\gamma^5 n(t_1,\vec{x}_1 - \frac{1}{2}\,\vec{\rho}\,)][\bar{p}(x_2)\gamma^{\mu}n^c(x_2) - \bar{n}(x_2)
\gamma^{\mu}p^c(x_2)]\,
$$
$$
\times\,[\bar{p}(x_3)\gamma^{\nu}\gamma^5 p(x_3) - \bar{n}(x_3)\gamma^{\nu}\gamma^5 n(x_3)])|0>.\eqno({\rm E}.6) 
$$
Since we assume that for the ${^1}{\rm S}_0$--state of the np system the transition Z + D $\to$ n + p is mainly the Gamow--Teller one, we have taken into account the Z--boson coupled with the axial nucleon current.

The r.h.s. of Eq.~({\rm E}.6) can be reduced to the more convenient form if to apply the relations $\bar{p}(x_2) \gamma^{\mu}
n^c(x_2) = -  \bar{n}(x_2) \gamma^{\mu}
p^c(x_2)$ and  $\bar{p}(x) \gamma_{\alpha} 
\gamma^5 p(x) =  \bar{p^c}(x) 
\gamma_{\alpha} 
\gamma^5 p^c(x)$:  
$$
\int d^4x\,{\cal L}_{\rm ZD\to np}(x) =  
- \int d^4x_1 d^4x_2 d^4x_3\,<{\rm T}(
{\cal L}^{\rm np\to np}_{\rm eff}(x_1)
{\cal L}^{\dagger}_{\rm npD}(x_2)
{\cal L}_{\rm NNZ}(x_3))>
$$
$$
= 2\,G_{\rm \pi np}\,\times\,ig_{\rm V}\,
\times \,g_{\rm A}\int d^4x_1 d^4x_2 d^4x_3\,
\int d^3\rho\,U(\rho)\,
$$
$$
\times\,{\rm T}([\bar{n}(t_1,\vec{x}_1 + \frac{1}{2}\,\vec{\rho}\,)\,\gamma^{\alpha}
\gamma^5 p^c (t_1,\vec{x}_1 - \frac{1}{2}\,\vec{\rho}\,)]\,D_{\mu}(x_2)\,
Z_{\nu}(x_3))
$$
$$
\times <0|{\rm T}([\bar{p^c}(t_1,\vec{x}_1 + \frac{1}{2}\,\vec{\rho}\,)\,
\gamma_{\alpha}\gamma^5 n(t_1,\vec{x}_1
- \frac{1}{2}\,\vec{\rho}\,)][ \bar{n}(x_2)
\gamma^{\mu}p^c(x_2)]
[\bar{p^c}(x_3)\gamma^{\nu}
\gamma^5 p^c(x_3)])|0> 
$$
$$
+ 2\,G_{\rm \pi np}\,\times\,(-ig_{\rm V})\,
\times \,g_{\rm A} \int d^4x_1 d^4x_2 d^4x_3\,
\int d^3\rho \,U(\rho)\,
$$
$$
\times\,{\rm T}([\bar{n}(t_1,\vec{x}_1 + \frac{1}{2}\,\vec{\rho}\,)\,\gamma_{\alpha} 
\gamma^5 p^c (t_1,\vec{x}_1 - 
\frac{1}{2}\,\vec{\rho}\,)]\,
D_{\mu}(x_2)\,Z_{\nu}(x_3))
$$
$$
\times <0|{\rm T}([\bar{p^c}(t_1,\vec{x}_1 + \frac{1}{2}\,\vec{\rho}\,)\,
\gamma^{\alpha}\gamma^5 n(t_1,\vec{x}_1 - \frac{1}{2}\,\vec{\rho}\,)]\,[\bar{n}(x_3)
\gamma^{\nu}\gamma^5 n(x_3)]
[\bar{n}(x_2)
\gamma^{\mu}p^c(x_2)])|0>
$$
$$
+ 2\,G_{\rm \pi np}\,\times\,ig_{\rm V}\,\times
\,g_{\rm A} \int d^4x_1 d^4x_2 d^4x_3\,\int d^3
\rho \,U(\rho)\,
$$
$$
\times\,{\rm T}([\bar{n}(t_1,\vec{x}_1 + \frac{1}{2}\,\vec{\rho}\,)\,\gamma^5 
p^c (t_1,\vec{x}_1 - \frac{1}{2}\,\vec{\rho}\,)]
\,D_{\mu}(x_2)\,Z_{\nu}(x_3))
$$
$$
\times <0|{\rm T}([\bar{p^c}(t_1,\vec{x}_1 + \frac{1}{2}\,\vec{\rho}\,)\,
\gamma^5 n(t_1,\vec{x}_1 - \frac{1}{2}\,\vec{\rho}\,)]
[\bar{n}(x_2)
\gamma^{\mu}p^c(x_2)]
[\bar{p^c}(x_3)
\gamma^{\nu}\gamma^5 p^c(x_3)])|0> 
$$
$$
+ 2\,G_{\rm \pi np}\,\times\,(- ig_{\rm V})\,
\times \,g_{\rm A} \int d^4x_1 d^4x_2 d^4x_3\,
\int d^3\rho \,U(\rho)\,
$$
$$
\times\,{\rm T}([\bar{n}(t_1,\vec{x}_1 + \frac{1}{2}\,\vec{\rho}\,)\,\gamma^5
p^c (t_1,\vec{x}_1 - \frac{1}{2}\,\vec{\rho}\,)]
\,D_{\mu}(x_2)\,Z_{\nu}(x_3))
$$
$$
\times <0|{\rm T}([\bar{p^c}(t_1,\vec{x}_1 + \frac{1}{2}\,\vec{\rho}\,)\,
\gamma^5 n(t_1,\vec{x}_1 - 
\frac{1}{2}\,\vec{\rho}\,)]\,
[\bar{n}(x_3)\gamma^{\nu}\gamma^5 n(x_3)]
[\bar{n}(x_2)
\gamma^{\mu}
p^c(x_2)])|0>.
\eqno({\rm E}.7) 
$$
The vacuum expectation values of the nucleon current can be reduced to the equivalent form by applying the charge conjugation and the change $\vec{\rho} \to - \vec{\rho}$:
$$
\int d^4x\,{\cal L}_{\rm ZD\to np}(x) =  - 
\int d^4x_1 d^4x_2 d^4x_3\,
<{\rm T}({\cal L}^{\rm np\to np}_{\rm eff}(x_1)
{\cal L}^{\dagger}_{\rm npD}(x_2)
{\cal L}_{\rm NNZ}(x_3))>
$$
$$
= 2\,G_{\rm \pi np}\,\times\,ig_{\rm V}\,
\times \,g_{\rm A}\int d^4x_1 d^4x_2 d^4x_3\,\int 
d^3\rho\,U(\rho)\,
$$
$$
\times\,{\rm T}([\bar{n}(t_1,\vec{x}_1 + 
\frac{1}{2}\,\vec{\rho}\,)\,
\gamma^{\alpha}\gamma^5 p^c (t_1,\vec{x}_1 - \frac{1}{2}\,\vec{\rho}\,)]
\,D_{\mu}(x_2)\,Z_{\nu}(x_3))
$$
$$
\times <0|{\rm T}([\bar{p^c}(t_1,\vec{x}_1 + \frac{1}{2}\,\vec{\rho}\,)\,
\gamma_{\alpha}\gamma^5 
n(t_1,\vec{x}_1 - \frac{1}{2}\,\vec{\rho}\,)]
[ \bar{n}(x_2)
\gamma^{\mu}
p^c(x_2)]
[\bar{p^c}(x_3)\gamma^{\nu}
\gamma^5 p^c(x_3)])|0> 
$$
$$
+ 2\,G_{\rm \pi np}\,\times\,ig_{\rm V}\,\times 
\,g_{\rm A} \int d^4x_1 d^4x_2 d^4x_3\,\int 
d^3\rho \,U(\rho)\,
$$
$$
\times\,{\rm T}([\bar{p}(t_1,\vec{x}_1 + \frac{1}{2}\,\vec{\rho}\,)\,\gamma_{\alpha} 
\gamma^5 n^c (t_1,\vec{x}_1 - \frac{1}{2}\,\vec{\rho}\,)]\,
D_{\mu}(x_2)\,Z_{\nu}(x_3))
$$
$$
\times <0|{\rm T}([\bar{n^c}(t_1,\vec{x}_1 + \frac{1}{2}\,\vec{\rho}\,)\,
\gamma^{\alpha}\gamma^5 p(t_1,\vec{x}_1 - \frac{1}{2}\,\vec{\rho}\,)]\,
[\bar{p}(x_2)\gamma^{\mu}n^c(x_2)]
[\bar{n^c}(x_3)\gamma^{\nu}\gamma^5 n^c(x_3)]
)|0>
$$
$$
+ 2\,G_{\rm \pi np}\,\times\,ig_{\rm V}\,\times 
\,g_{\rm A} \int d^4x_1 d^4x_2 d^4x_3\,\int d^3\rho \,U(\rho)\,
$$
$$
\times\,{\rm T}([\bar{n}(t_1,\vec{x}_1 + \frac{1}{2}\,\vec{\rho}\,)\,\gamma^5 p^c (t_1,\vec{x}_1 - \frac{1}{2}\,\vec{\rho}\,)]\,D_{\mu}(x_2)\,Z_{\nu}(x_3))
$$
$$
\times <0|{\rm T}([\bar{p^c}(t_1,\vec{x}_1 + \frac{1}{2}\,\vec{\rho}\,)\,\gamma^5 n(t_1,\vec{x}_1 - \frac{1}{2}\,\vec{\rho}\,)]
[\bar{n}(x_2)
\gamma^{\mu}
p^c(x_2)]
[\bar{p^c}(x_3)
\gamma^{\nu}
\gamma^5 p^c(x_3)])|0> 
$$
$$
+ 2\,G_{\rm \pi np}\,\times\,ig_{\rm V}\,\times 
\,g_{\rm A} \int d^4x_1 d^4x_2 d^4x_3\,\int
d^3\rho \,U(\rho)\,
$$
$$
\times\,{\rm T}([\bar{p}(t_1,\vec{x}_1 + \frac{1}{2}\,\vec{\rho}\,)\,\gamma^5 
n^c (t_1,\vec{x}_1 - \frac{1}{2}\,\vec{\rho}\,)]\,
D_{\mu}(x_2)\,Z_{\nu}(x_3))
$$
$$
\times <0|{\rm T}([\bar{n^c}(t_1,\vec{x}_1 + \frac{1}{2}\,\vec{\rho}\,)\,\gamma^5 p(t_1,\vec{x}_1 - \frac{1}{2}\,\vec{\rho}\,)]\,
[\bar{p}(x_2)\gamma^{\mu}n^c(x_2)]
[\bar{n^c}(x_3)\gamma^{\nu}
\gamma^5 n^c(x_3)])|0>.
\eqno({\rm E}.8) 
$$
Since up to the interchange 
$p \longleftrightarrow n$ the vacuum expectation values of the nucleon currents are equivalent, we can rewrite Eq.~({\rm E}.8) as follows
$$
\int d^4x
\,{\cal L}_{\rm ZD\to np}(x) =  
- \int d^4x_1 d^4x_2 d^4x_3\,
<{\rm T}({\cal L}^{\rm np\to np}_{\rm eff}(x_1)
{\cal L}^{\dagger}_{\rm npD}(x_2)
{\cal L}_{\rm NNZ}(x_3))>
$$
$$
= 2\,G_{\rm \pi np}\,\times\,ig_{\rm V}\,\times \,
g_{\rm A}\int d^4x_1 d^4x_2 d^4x_3\,\int d^3\rho\,U(\rho)\,
$$
$$
\times\,{\rm T}([\bar{n}(t_1,\vec{x}_1 + \frac{1}{2}\,\vec{\rho}\,)\, \gamma^{\alpha}\gamma^5 p^c (t_1,\vec{x}_1 - \frac{1}{2}\,\vec{\rho}\,) + \bar{p}(t_1,\vec{x}_1 + \frac{1}{2}\,\vec{\rho}\,)\,\gamma_{\alpha} \gamma^5 n^c (t_1,\vec{x}_1 - \frac{1}{2}\,\vec{\rho}\,)] 
$$
$$
\times\,D_{\mu}(x_2)\,Z_{\nu}(x_3))
$$
$$
\times <0|{\rm T}([\bar{p^c}(t_1,\vec{x}_1 + \frac{1}{2}\,\vec{\rho}\,)\,\gamma_{\alpha}\gamma^5 n(t_1,\vec{x}_1 - \frac{1}{2}\,\vec{\rho}\,)][ \bar{n}(x_2)\gamma^{\mu} p^c(x_2)][\bar{p^c}(x_3)\gamma^{\nu}\gamma^5 p^c(x_3)])|0> 
$$
$$
+ 2\,G_{\rm \pi np}\,\times\,ig_{\rm V}\,\times \,g_{\rm A} \int d^4x_1 d^4x_2 d^4x_3\,\int d^3\rho \,U(\rho)\,
$$
$$
\times\,{\rm T}([\bar{n}(t_1,\vec{x}_1 + \frac{1}{2}\,\vec{\rho}\,) \,\gamma^5 p^c (t_1,\vec{x}_1 - \frac{1}{2}\,\vec{\rho}\,) + \bar{p}(t_1,\vec{x}_1 + \frac{1}{2}\,\vec{\rho}\,)\,\gamma^5 n^c (t_1,\vec{x}_1 - \frac{1}{2}\,\vec{\rho}\,)] 
$$
$$
\times\,D_{\mu}(x_2)\,Z_{\nu}(x_3))
$$
$$
\times <0|{\rm T}([\bar{p^c}(t_1,\vec{x}_1 + \frac{1}{2}\,\vec{\rho}\,) \,\gamma^5 n(t_1,\vec{x}_1 - \frac{1}{2}\,\vec{\rho}\,)] 
[\bar{n}(x_2)
\gamma^{\mu}p^c(x_2)] 
[\bar{p^c}(x_3)
\gamma^{\nu}
\gamma^5 p^c(x_3)])|0>.
\eqno({\rm E}.9) 
$$
Making all contractions we obtain
$$
\int d^4x\,
{\cal L}_{\rm ZD\to np}(x) =  
- \int d^4x_1 d^4x_2 d^4x_3\,
<{\rm T}({\cal L}^{\rm np\to np}_{\rm eff}(x_1)
{\cal L}^{\dagger}_{\rm npD}(x_2){\cal L}_{\rm NNZ}(x_3))>
$$
$$
= 2\,G_{\rm \pi np}\,\times\,ig_{\rm V}\,\times \,g_{\rm A}\int d^4x_1 d^4x_2 d^4x_3\,\int d^3\rho\,U(\rho)\,
$$
$$
\times\,{\rm T}([\bar{n}(t_1,\vec{x}_1 + \frac{1}{2}\,\vec{\rho}\,)\, \gamma^{\alpha}\gamma^5 p^c (t_1,\vec{x}_1 - \frac{1}{2}\,\vec{\rho}\,) + \bar{p}(t_1,\vec{x}_1 + \frac{1}{2}\,\vec{\rho}\,)\,\gamma_{\alpha} \gamma^5 n^c (t_1,\vec{x}_1 - \frac{1}{2}\,\vec{\rho}\,)] 
$$
$$
\times\,D_{\mu}(x_2)\,Z_{\nu}(x_3))
$$
$$
\times \frac{1}{i}\,{\rm tr}\{S^c_F(t_3 - t_1, \vec{x}_3 - \vec{x}_1 -\frac{1}{2}\,\vec{\rho}\,) \gamma_{\alpha}\gamma^5 S_F(t_1 - t_2,\vec{x}_1 - \vec{x}_2 - \frac{1}{2}\,\vec{\rho}\,) \gamma^{\mu} S^c_F(x_2 - x_3)\gamma^{\nu}\gamma^5 \}
$$
$$
+ 2\,G_{\rm \pi np}\,\times\,ig_{\rm V}\,\times \,g_{\rm A} \int d^4x_1 d^4x_2 d^4x_3\,\int d^3\rho \,U(\rho)\,
$$
$$
\times\,{\rm T}([\bar{n}(t_1,\vec{x}_1 + \frac{1}{2}\,\vec{\rho}\,) \,\gamma^5 p^c (t_1,\vec{x}_1 - \frac{1}{2}\,\vec{\rho}\,) + \bar{p}(t_1,\vec{x}_1 + \frac{1}{2}\,\vec{\rho}\,)\,\gamma^5 n^c (t_1,\vec{x}_1 - \frac{1}{2}\,\vec{\rho}\,)] 
$$
$$
\times\,D_{\mu}(x_2)\,Z_{\nu}(x_3))
$$
$$
\times \frac{1}{i}\,{\rm tr}\{S^c_F(t_3 - t_1, \vec{x}_3 - \vec{x}_1 -\frac{1}{2}\,\vec{\rho}\,) \gamma^5 S_F(t_1 - t_2,\vec{x}_1 - \vec{x}_2 - \frac{1}{2}\,\vec{\rho}\,) \gamma^{\mu} S^c_F(x_2 - x_3)\gamma^{\nu}\gamma^5 \}.\eqno({\rm E}.10) 
$$
In the momentum representation of the nucleon Green functions 
we define the effective Lagrangian ${\cal L}_{\rm ZD \to np}(x)$
as follows
$$
\int d^4x
\,{\cal L}_{\rm ZD \to np}(x) = 
$$
$$
= i\,g_{\rm A}G_{\rm \pi np}\frac{g_{\rm V}}{8\pi^2}\int d^4x_1 \int \frac{d^4x_2 d^4k_2}{(2\pi)^4}\frac{d^4x_3 d^4k_3}{(2\pi)^4}\,e^{\displaystyle i k_2\cdot (x_2-x_1)}e^{\displaystyle i k_3\cdot (x_3-x_1)}\int d^3\rho \,U(\rho)
$$
$$
\times\,{\rm T}([\bar{n}(t_1,\vec{x}_1 + \frac{1}{2}\,\vec{\rho}\,) \gamma_{\alpha}\gamma^5 p^c (t_1,\vec{x}_1 - \frac{1}{2}\,\vec{\rho}\,) + \bar{p}(t_1,\vec{x}_1 + \frac{1}{2}\,\vec{\rho}\,) \gamma_{\alpha}\gamma^5 n^c (t_1,\vec{x}_1 - \frac{1}{2}\,\vec{\rho}\,)]\,
$$
$$
\times\,D_{\mu}(x_2)\,Z_{\nu}(x_3))
$$
$$
\times\,\int\frac{d^4k_1}{\pi^2i}\,e^{\displaystyle - i\vec{q}\cdot \vec{\rho}}\,{\rm tr}\Bigg\{\gamma^{\alpha}\gamma^5\frac{1}{M_{\rm N} - \hat{k}_1 - \hat{k}_2}\gamma^{\mu}\frac{1}{M_{\rm N} - \hat{k}_1}\gamma^{\nu}\gamma^5\frac{1}{M_{\rm N} - \hat{k}_1 + \hat{k}_3}\Bigg\}
$$
$$
+ i\,g_{\rm A}G_{\rm \pi np}\frac{g_{\rm V}}{8\pi^2}\int d^4x_1 \int \frac{d^4x_2 d^4k_2}{(2\pi)^4}\frac{d^4x_3 d^4k_3}{(2\pi)^4}\,e^{\displaystyle i k_2\cdot (x_2-x_1)}e^{\displaystyle i k_3\cdot (x_3-x_1)}\int d^3\rho \,U(\rho)
$$
$$
\times \,{\rm T}
([\bar{n}(t_1,\vec{x}_1 + \frac{1}{2}\,\vec{\rho}\,) \gamma_{\alpha}\gamma^5
p^c (t_1,\vec{x}_1 - \frac{1}{2}\,\vec{\rho}\,) + \bar{p}(t_1,\vec{x}_1 + \frac{1}{2}\,\vec{\rho}\,) \gamma_{\alpha}\gamma^5 
n^c (t_1,\vec{x}_1 - \frac{1}{2}\,\vec{\rho}\,)] 
$$
$$
\times\,D_{\mu}(x_2)\,Z_{\nu}(x_3))
$$
$$
\times\,\int\frac{d^4k_1}{\pi^2i}\,
e^{\displaystyle - i\vec{q}\cdot \vec{\rho}}\,
{\rm tr}
\Bigg\{\gamma^5
\frac{1}{M_{\rm N} - \hat{k}_1 - \hat{k}_2}
\gamma^{\mu}
\frac{1}{M_{\rm N} - \hat{k}_1}
\gamma^{\nu}\gamma^5
\frac{1}{M_{\rm N} - \hat{k}_1 
+ \hat{k}_3}\Bigg\},\eqno({\rm E}.11)
$$
where $\vec{q} = \vec{k}_1 + (\vec{k}_3 - \vec{k}_2)/2$.

In order to obtain the effective Lagrangian describing the matrix element of the process $\nu_{\rm e}$ +  D $\to$  
$\nu_{\rm e}$ + n + p  we replace the operator of the Z--boson field by the operator of the neutrino weak current Eq.~({\rm E}.2):
$$
\int d^4x\,{\cal L}_{\rm \nu_{\rm e}{\rm D}\to 
\nu_{\rm e}np}(x) = 
$$
$$
= i\,g_{\rm A}G_{\rm \pi np}\frac{G_{\rm F}}{\sqrt{2}}
\frac{g_{\rm V}}{16\pi^2}\int d^4x_1 \int 
\frac{d^4x_2 d^4k_2}{(2\pi)^4}\frac{d^4x_3 d^4k_3}{(2\pi)^4}\,e^{\displaystyle i k_2\cdot (x_2-x_1)}e^{\displaystyle i k_3\cdot (x_3-x_1)}
\int d^3\rho \,U(\rho)
$$
$$
\times\,{\rm T}([\bar{n}(t_1,\vec{x}_1 + 
\frac{1}{2}\,\vec{\rho}\,) 
\gamma_{\alpha}
\gamma^5 p^c (t_1,\vec{x}_1 
- \frac{1}{2}\,\vec{\rho}\,) + 
\bar{p}(t_1,\vec{x}_1 + \frac{1}{2}\,\vec{\rho}\,) \gamma_{\alpha}\gamma^5 n^c (t_1,\vec{x}_1 - \frac{1}{2}\,\vec{\rho}\,)]\,
$$
$$
D_{\mu}(x_2)\,
[\bar{\psi}_{\nu_{\rm e}}(x_3)
\gamma_{\nu}(1 - \gamma^5) 
\psi_{\nu_{\rm e}}(x_3)])
$$
$$
\times\,\int\frac{d^4k_1}{\pi^2i}\,
e^{\displaystyle - i\vec{q}\cdot \vec{\rho}}\,{\rm tr}\Bigg\{\gamma^{\alpha}
\gamma^5\frac{1}{M_{\rm N} - \hat{k}_1 
- \hat{k}_2}\gamma^{\mu}
\frac{1}{M_{\rm N} - \hat{k}_1}
\gamma^{\nu}\gamma^5
\frac{1}{M_{\rm N} - \hat{k}_1 
+ \hat{k}_3}\Bigg\}
$$
$$
+ i\,g_{\rm A}G_{\rm \pi np}\frac{G_{\rm F}}{\sqrt{2}}
\frac{g_{\rm V}}{16\pi^2}\int d^4x_1 
\int \frac{d^4x_2 d^4k_2}{(2\pi)^4}
\frac{d^4x_3 d^4k_3}{(2\pi)^4}\,
e^{\displaystyle i k_2\cdot (x_2-x_1)}
e^{\displaystyle i k_3\cdot (x_3-x_1)}
\int d^3\rho \,U(\rho)
$$
$$
\times \,{\rm T}([\bar{n}(t_1,\vec{x}_1 + 
\frac{1}{2}\,\vec{\rho}\,) \gamma_{\alpha}
\gamma^5 p^c (t_1,\vec{x}_1 - 
\frac{1}{2}\,\vec{\rho}\,) + 
\bar{p}(t_1,\vec{x}_1 + \frac{1}{2}\,\vec{\rho}\,) \gamma_{\alpha}\gamma^5 
n^c (t_1,\vec{x}_1 - \frac{1}{2}\,\vec{\rho}\,)] 
$$
$$
D_{\mu}(x_2)\,[\bar{\psi}_{\nu_{\rm e}}(x_3)
\gamma_{\nu}(1 - \gamma^5) 
\psi_{\nu_{\rm e}}(x_3)])
$$
$$
\times\,\int\frac{d^4k_1}{\pi^2i}\,
e^{\displaystyle - i\vec{q}\cdot \vec{\rho}}\,
{\rm tr}\Bigg\{\gamma^5
\frac{1}{M_{\rm N} - \hat{k}_1 - \hat{k}_2}
\gamma^{\mu}\frac{1}{M_{\rm N} - \hat{k}_1}
\gamma^{\nu}\gamma^5 
\frac{1}{M_{\rm N} - \hat{k}_1 
+ \hat{k}_3}\Bigg\},\eqno({\rm E}.12)
$$
The matrix element of the process 
$\nu_{\rm e}$ + D $\to$ $\nu_{\rm e}$ + n + p
we define by a usual way as
$$
\int d^4x\,<p(p_2)n(p_1)\nu_{\rm e}
(k^{\prime}_{\nu_{\rm e}}) |
{\cal L}_{\rm \nu_{\rm e}{\rm D}\to \nu_{\rm e}np}(x)|
{\rm D}(k_{\rm D})\nu_{\rm e}(k_{\nu_{\rm e}})> = 
$$
$$
=(2\pi)^4\delta^{(4)}( p_2 + p_1 + k^{\prime}_{\nu_{\rm e}}
- k_{\rm D} - k_{\nu_{\rm e}})\,
\frac{{\cal M}(\nu_{\rm e} + {\rm D} 
\to \nu_{\rm e} + {\rm n} + {\rm p})}{\displaystyle \sqrt{2E_1V\,2E_2V\,
2E^{\prime}_{\nu_{\rm e}}V\,
2E_{\rm D}V\,2E_{\nu_{\rm e}}V}}, 
\eqno({\rm E}.13)
$$
where  $E_i\,(i =1,2,{\rm D},\nu_{\rm e})$ 
and $E^{\prime}_{\nu_{\rm e}}$ are the energies of the neutron, the proton, the deuteron, the initial and the final neutrino,
$V$ is the normalization volume.

Now we should take the r.h.s. of Eq.~({\rm D}.12) between the wave functions of the initial $|{\rm D}(k_{\rm D})
\nu_{\rm e}(k_{\nu_{\rm e}})>$ and the final 
$<p(p_2)n(p_1)\nu_{\rm e}(k^{\prime}_{\rm e})|$ states.  This gives
$$
(2\pi)^4\delta^{(4)}( p_2 + p_1 + k^{\prime}_{\nu_{\rm e}} 
- k_{\rm D} - k_{\nu_{\rm e}})\,\frac{{\cal M}(\nu_{\rm e} +
{\rm D} \to \nu_{\rm e}+ {\rm n} + {\rm p})}{\displaystyle \sqrt{2E_1V\,
2E_2V\,2E^{\prime}_{\nu_{\rm e}}V\,
2E_{\rm D}V\,2E_{\nu_{\rm e}}V}}=
$$
$$
= i\,g_{\rm A}G_{\rm \pi np}\frac{G_{\rm F}}{\sqrt{2}} 
\frac{g_{\rm V}}{16\pi^2}\int d^4x_1 
\int \frac{d^4x_2 d^4k_2}{(2\pi)^4}\frac{d^4x_3 d^4k_3}{(2\pi)^4}\,e^{\displaystyle i k_2\cdot (x_2-x_1)}e^{\displaystyle i k_3\cdot (x_3-x_1)}
$$
$$
\times\,\int d^3\rho \,U(\rho)\,
<p(p_2)n(p_1)\nu_{\rm e}(k^{\prime}_{\nu_{\rm e}})|{\rm T}([\bar{n}(t_1,\vec{x}_1 + \frac{1}{2}\,\vec{\rho}\,) \gamma_{\alpha}\gamma^5 p^c (t_1,\vec{x}_1 - \frac{1}{2}\,\vec{\rho}\,)
$$
$$
+ \bar{p}(t_1,\vec{x}_1 + \frac{1}{2}\,\vec{\rho}\,) \gamma_{\alpha}\gamma^5 n^c (t_1,\vec{x}_1 - \frac{1}{2}\,\vec{\rho}\,)] 
$$
$$
D_{\mu}(x_2)\,[\bar{\psi}_{\nu_{\rm e}}(x_3)
\gamma_{\nu}(1 - \gamma^5) 
\psi_{\nu_{\rm e}}(x_3)])|
{\rm D}(k_{\rm D})\nu_{\rm e}(k_{\nu_{\rm e}})>
$$
$$
\times\,\int\frac{d^4k_1}{\pi^2i}\,
e^{\displaystyle - i\vec{q}\cdot \vec{\rho}}\,
{\rm tr}\Bigg\{\gamma^{\alpha}\gamma^5
\frac{1}{M_{\rm N} - \hat{k}_1 - \hat{k}_2}
\gamma^{\mu}\frac{1}{M_{\rm N} - \hat{k}_1}
\gamma^{\nu}\gamma^5 
\frac{1}{M_{\rm N} - \hat{k}_1 
+ \hat{k}_3}\Bigg\}
$$
$$
+ i\,g_{\rm A}G_{\rm \pi np}
\frac{G_{\rm F}}{\sqrt{2}}\frac{g_{\rm V}}{16\pi^2}
\int d^4x_1 \int \frac{d^4x_2 d^4k_2}{(2\pi)^4}
\frac{d^4x_3 d^4k_3}{(2\pi)^4}\,
e^{\displaystyle i k_2\cdot (x_2-x_1)}
e^{\displaystyle i k_3\cdot (x_3-x_1)}
$$
$$
\times\,\int d^3\rho \,U(\rho)\,
<p(p_2)n(p_1)\nu_{\rm e}(k^{\prime}_{\nu_{\rm e}})|
{\rm T}([\bar{n}(t_1,\vec{x}_1 + \frac{1}{2}\,\vec{\rho}\,) 
\gamma^5 p^c (t_1,\vec{x}_1 - \frac{1}{2}\,\vec{\rho}\,) 
$$
$$
+ \bar{p}(t_1,\vec{x}_1 + \frac{1}{2}\,\vec{\rho}\,)
\gamma^5 n^c (t_1,\vec{x}_1 - 
\frac{1}{2}\,\vec{\rho}\,)] 
$$
$$
D_{\mu}(x_2)\,[\bar{\psi}_{\nu_{\rm e}}(x_3)
\gamma_{\nu}(1 - \gamma^5) \psi_{\nu_{\rm e}}(x_3)])|
{\rm D}(k_{\rm D})\nu_{\rm e}(k_{\nu_{\rm e}})>
$$
$$
\times\,\int\frac{d^4k_1}{\pi^2i}\,
e^{\displaystyle - i\vec{q}\cdot \vec{\rho}}\,
{\rm tr}\Bigg\{\gamma^5
\frac{1}{M_{\rm N} - \hat{k}_1 - \hat{k}_2}
\gamma^{\mu}\frac{1}{M_{\rm N} - \hat{k}_1} 
\gamma^{\nu}\gamma^5 
\frac{1}{M_{\rm N} - \hat{k}_1 
+ \hat{k}_3}\Bigg\}.\eqno({\rm E}.14)
$$
Between the initial $|{\rm D}(k_{\rm D})\nu_{\rm e}
(k_{\nu_{\rm e}})>$ and the final 
$<p(p_2)n(p_1)\nu_{\rm e}
(k^{\prime}_{\nu_{\rm e}})|$ states the matrix elements
in Eq.~({\rm E}.13) are defined (see Eq.~({\rm C}.23)
and Eqs.~({\rm C}.25) -- ({\rm C}.34)):
$$
<p(p_2)n(p_1)\nu_{\rm e}
(k^{\prime}_{\nu_{\rm e}})|
{\rm T}([\bar{n}(t_1,\vec{x}_1 + 
\frac{1}{2}\,\vec{\rho}\,) 
\gamma_{\alpha}\gamma^5 
p^c (t_1,\vec{x}_1 - \frac{1}{2}\,\vec{\rho}\,) 
$$
$$
+ \bar{p}(t_1,\vec{x}_1 + 
\frac{1}{2}\,\vec{\rho}\,) 
\gamma_{\alpha}\gamma^5 
n^c (t_1,\vec{x}_1 - 
\frac{1}{2}\,\vec{\rho}\,)] 
$$
$$
D_{\mu}(x_2)\,[\bar{\psi}_{\nu_{\rm e}}(x_3)
\gamma_{\nu}(1 - \gamma^5) 
\psi_{\nu_{\rm e}}(x_3)])|
{\rm D}(k_{\rm D})\nu_{\rm e}(k_{\nu_{\rm e}})> = 
$$
$$
=[\bar{u}(p_1)\gamma_{\alpha}\gamma^5 u^c(p_2)][\bar{u}(k^{\prime}_{\nu_{\rm e}})\gamma_{\nu}
(1-\gamma^5)u(k_{\nu_{\rm e}})]\,
$$
$$
\times\,e_{\mu}(k_{\rm D})\,\times\,2\times\,
\psi_{\rm np}(\rho)\,\frac{\displaystyle 
e^{\displaystyle i(p_1+p_2)\cdot x_1}\,
e^{\displaystyle -ik_{\rm D}\cdot x_2}\,
e^{\displaystyle - ik_{{\ell}}\cdot x_3}}{\displaystyle \sqrt{2E_1V\,2E_2V\,
2E^{\prime}_{\nu_{\rm e}}V\,
2E_{\rm D}V\,2E_{\nu_{\rm e}}V}},
$$
$$
<p(p_2)n(p_1)\nu_{\rm e}
(k^{\prime}_{\nu_{\rm e}})|
{\rm T}([\bar{n}(t_1,\vec{x}_1 + 
\frac{1}{2}\,\vec{\rho}\,) \gamma^5 
p^c (t_1,\vec{x}_1 - 
\frac{1}{2}\,\vec{\rho}\,) 
$$
$$
+ \bar{p}(t_1,\vec{x}_1 + \frac{1}{2}\,\vec{\rho}\,) 
\gamma^5 n^c (t_1,\vec{x}_1 - 
\frac{1}{2}\,\vec{\rho}\,)] 
$$
$$
D_{\mu}(x_2)\,[\bar{\psi}_{\nu_{\rm e}}
(x_3)\gamma_{\nu}(1 - \gamma^5) 
\psi_{\nu_{\rm e}}(x_3)])|{\rm D}(k_{\rm D})
\nu_{\rm e}(k_{\nu_{\rm e}})> = 
$$
$$
=[\bar{u}(p_1)\gamma^5 u^c(p_2)]
[\bar{u}(k^{\prime}_{\nu_{\rm e}})\gamma_{\nu}
(1-\gamma^5)u(k_{\nu_{\rm e}})]\,
$$
$$
\times\,e_{\mu}(k_{\rm D})\,\times\,2\times\,
\psi_{\rm np}(\rho)\,\frac{\displaystyle 
e^{\displaystyle i(p_1+p_2)\cdot x_1}\,
e^{\displaystyle -ik_{\rm D}\cdot x_2}\,
e^{\displaystyle - ik_{{\ell}}\cdot x_3}}{\displaystyle \sqrt{2E_1V\,2E_2V\,2E^{\prime}_{\nu_{\rm e}}V\,
2E_{\rm D}V\,2E_{\nu_{\rm e}}V}}.
\eqno({\rm E}.15)
$$
where $k_{\ell} = k^{\prime}_{\nu_{\rm e}} 
- k_{\nu_{\rm e}}$ and $\psi_{\rm np}(\rho)$ is the wave function of the relative movement of the neutron and the proton normalized per unit density [8]. It is given by Eq.~(\ref{label9.3}). For the computation of the matrix elements Eq.~({\rm D}.15) we have used the wave function of the neutron and the proton $<p(p_2)n(p_1)|$ determined in terms of the operators of the annihilation in the standard form [31]
$$
<p(p_2)n(p_1)| =<0|a_{\rm p}
(\vec{p}_2,\sigma_2) a_{\rm n}
(\vec{p}_1,\sigma_1),\eqno({\rm E}.16)
$$
where $a_{\rm p}(\vec{p}_2,\sigma_2)$ and
$a_{\rm n}(\vec{p}_1,\sigma_1)$ are the operators of the annihilation of the proton and the neutron.

Inserting the matrix elements  Eq.~({\rm E}.15) to the r.h.s. of  Eq.~({\rm E}.14) we obtain the matrix element of the 
$\nu_{\rm e}$ + D $\to$ $\nu_{\rm e}$ + n + p process in the following form
$$
(2\pi)^4\delta^{(4)}(p_2 + p_1 + 
k^{\prime}_{\nu_{\rm e}} - k_{\rm D} - 
k_{\nu_{\rm e}})\,i{\cal M}(\nu_{\rm e} + 
{\rm D}\to \nu_{\rm e} + {\rm n} + {\rm p}) = 
$$
$$
= -\,g_{\rm A}G_{\rm \pi np}\frac{G_{\rm F}}{\sqrt{2}} 
\frac{g_{\rm V}}{8\pi^2}[\bar{u}(p_1)\gamma_{\alpha}\gamma^5 u^c(p_2)][\bar{u}(k^{\prime}_{\nu_{\rm e}})\gamma_{\nu}(1-\gamma^5)u(k_{\nu_{\rm e}})]\,
e_{\mu}(k_{\rm D})\int d^4x_1 
$$
$$
\times \int \frac{d^4x_2 d^4k_2}{(2\pi)^4}
\frac{d^4x_3 d^4k_3}{(2\pi)^4}\,
e^{\displaystyle i (p_2 + p_1 - k_2 - k_3)\cdot x_1}\,
e^{\displaystyle i(k_2 - k_{\rm D})\cdot x_2}\,
e^{\displaystyle i(k_3 - k_{\ell})\cdot x_3}
$$
$$
\times\int d^3\rho \,U(\rho)\,\psi_{\rm np}(\rho) \int\frac{d^4k_1}{\pi^2i}
e^{\displaystyle - i\vec{q}\cdot \vec{\rho}} {\rm tr}\Bigg\{\gamma^{\alpha}\gamma^5
\frac{1}{M_{\rm N} - \hat{k}_1 - \hat{k}_2} 
\gamma^{\mu} \frac{1}{M_{\rm N} - \hat{k}_1}
\gamma^{\nu}\gamma^5 \frac{1}{M_{\rm N} - 
\hat{k}_1 + \hat{k}_3}\Bigg\}
$$
$$
-\, g_{\rm A}G_{\rm \pi np}\frac{G_{\rm F}}{\sqrt{2}} 
\frac{g_{\rm V}}{8\pi^2}[\bar{u}(p_1)\gamma^5 u^c(p_2)][\bar{u}(k^{\prime}_{\nu_{\rm e}}) 
\gamma_{\nu} (1-\gamma^5) u(k_{\nu_{\rm e}})] 
\, e_{\mu}(k_{\rm D}) \int d^4x_1 
$$
$$
\times \int \frac{d^4x_2 d^4k_2}{(2\pi)^4}
\frac{d^4x_3 d^4k_3}{(2\pi)^4}\,
e^{\displaystyle i (k_2 + k_3 - p_1 - p_2)\cdot x_1}
\,e^{\displaystyle i(k_2 - k_{\rm D})\cdot x_2}\,
e^{\displaystyle i(k_3 - k_{\ell})\cdot x_3}
$$
$$
\times\int d^3\rho\,U(\rho) \,\psi_{\rm np}(\rho) \int\frac{d^4k_1}{\pi^2i}
e^{\displaystyle -i\vec{q}\cdot \vec{\rho}}
{\rm tr}\Bigg\{\gamma^5
\frac{1}{M_{\rm N} - \hat{k}_1 - \hat{k}_2}
\gamma^{\mu} \frac{1}{M_{\rm N} - \hat{k}_1}
\gamma^{\nu}\gamma^5 \frac{1}{M_{\rm N} - \hat{k}_1 + \hat{k}_3}\Bigg\}.\eqno({\rm E}.17)
$$
Integrating over $x_1$, $x_2$, $x_3$, 
$k_2$ and $k_3$ we obtain in the r.h.s. 
of Eq.~({\rm E}.17) the $\delta$--function describing 
the 4--momentum conservation. Then, the matrix element of the  $\nu_{\rm e}$ + D $\to$ $\nu_{\rm e}$ + n + p process becomes equal
$$
i{\cal M}(\nu_{e} + {\rm D} \to \nu_{e} + 
{\rm n} + {\rm p}) = -\, g_{\rm A}
G_{\rm \pi np}\frac{G_{\rm F}}{\sqrt{2}}
\frac{g_{\rm V}}{8\pi^2}
$$
$$
\times\,[\bar{u}(p_1)\gamma_{\alpha}\gamma^5 u^c(p_2)][\bar{u}(k^{\prime}_{\nu_{\rm e}})\gamma_{\nu}(1-\gamma^5)u(k_{\nu_{\rm e}})]
\,e_{\mu}(k_{\rm D})\int d^3\rho\,  U(\rho) \,
\psi_{\rm np}(\rho) 
$$
$$
\times\,\int\frac{d^4k}{\pi^2i} 
e^{\displaystyle -i\vec{q}\cdot \vec{\rho}}\,
{\rm tr}\Bigg\{\gamma^{\alpha}\gamma^5
\frac{1}{M_{\rm N} - \hat{k} - 
\hat{k}_{\rm D}}
\gamma^{\mu} \frac{1}{M_{\rm N} - \hat{k}}
\gamma^{\nu}\gamma^5
\frac{1}{M_{\rm N} - \hat{k} + 
\hat{k}_{\ell}}\Bigg\} - g_{\rm A}G_{\rm \pi np}
$$
$$
\times\,\frac{G_{\rm F}}{\sqrt{2}} 
\frac{g_{\rm V}}{8\pi^2}
[\bar{u}(p_1)\gamma^5 u^c(p_2)]
[\bar{u}(k^{\prime}_{\nu_{\rm e}})
\gamma_{\nu}(1-\gamma^5) u(k_{\nu_{\rm e}})]
\,e_{\mu}(k_{\rm D})\int d^3\rho\,  U(\rho) \,
\psi_{\rm np}
(\rho)
$$
$$
\times \int\frac{d^4k}{\pi^2i}
e^{\displaystyle -i\vec{q}\cdot \vec{\rho}}
{\rm tr}\Bigg\{\gamma^5
\frac{1}{M_{\rm N} - \hat{k} -
\hat{k}_{\rm D}}\gamma^{\mu}
\frac{1}{M_{\rm N} - \hat{k}} 
\gamma^{\nu}\gamma^5 
\frac{1}{M_{\rm N} - \hat{k}
+ \hat{k}_{\ell}}\Bigg\},\eqno({\rm E}.18)
$$
where $\vec{q} = \vec{k} + (\vec{k}_{\ell}
- \vec{k}_{\rm D})/2$.

Then we represent as usually the matrix element 
Eq.~({\rm E}.18) in terms of the structure functions
$\bar{{\cal J}}^{\alpha\mu\nu}_{\rm np}
(k_{\rm D}, k_{\ell})$ and 
$\bar{{\cal J}}^{\mu\nu}_{\rm np}
(k_{\rm D}, k_{\ell})$:
$$
i{\cal M}(\nu_{\rm e} + {\rm D} \to 
\nu_{\rm e} + {\rm n} + {\rm p}) =
$$
$$
= g_{\rm A}G_{\rm \pi np}\frac{G_{\rm F}}{\sqrt{2}} 
\frac{g_{\rm V}}{8\pi^2}
[\bar{u}(p_1)\gamma_{\alpha}\gamma^5 u^c(p_2)][\bar{u}(k^{\prime}_{\nu_{\rm e}})\gamma_{\nu}(1-\gamma^5)u(k_{\nu_{\rm e}})]
\,e_{\mu}(k_{\rm D})\,
\bar{{\cal J}}^{\alpha\mu\nu}_{\rm np}
(k_{\rm D}, k_{\ell})
$$
$$
+ g_{\rm A}G_{\rm \pi np}\frac{G_{\rm F}}{\sqrt{2}}
\frac{g_{\rm V}}{4\pi^2}[\bar{u}(p_1)\gamma^5 u^c(p_2)][\bar{u}(k^{\prime}_{\nu_{\rm e}})\gamma_{\nu}(1-\gamma^5)u(k_{\nu_{\rm e}})]
\,e_{\mu}(k_{\rm D}) 
\bar{{\cal J}}^{\mu\nu}_{\rm np}(k_{\rm D},
k_{\ell}),\eqno({\rm E}.19)
$$
where the structure functions 
$\bar{{\cal J}}^{\alpha\mu\nu}_{\rm np}
(k_{\rm D}, k_{\ell})$ and 
$\bar{{\cal J}}^{\mu\nu}_{\rm np}
(k_{\rm D}, k_{\ell})$ are defined as
$$
\bar{{\cal J}}^{\alpha\mu\nu}_{\rm np}
(k_{\rm D}, k_{\ell}) = -\int d^3\rho\, 
U(\rho)\,\psi_{\rm np}(\rho)
\int\frac{d^4k}{\pi^2i} 
e^{\displaystyle  - i\vec{q}\cdot \vec{\rho}}
$$
$$
\times\,
{\rm tr} 
\Bigg\{\gamma^{\alpha}
\gamma^5
\frac{1}{M_{\rm N} - \hat{k} - 
\hat{k}_{\rm D}} 
\gamma^{\mu}
\frac{1}{M_{\rm N} - \hat{k}}
\gamma^{\nu}\gamma^5 
\frac{1}{M_{\rm N} - \hat{k} 
+ \hat{k}_{\ell}}\Bigg\},
$$
$$
\bar{{\cal J}}^{\mu\nu}_{\rm np}
(k_{\rm D}, k_{\ell}) = - \int d^3\rho\,  
U(\rho)\,\psi_{\rm np}(\rho)
\int\frac{d^4k}{\pi^2i}
e^{\displaystyle - i\vec{q}\cdot \vec{\rho}}
$$
$$
\times\,
{\rm tr} 
\Bigg\{\gamma^5
\frac{1}{M_{\rm N} - \hat{k} - 
\hat{k}_{\rm D}} \gamma^{\mu}
\frac{1}{M_{\rm N} - \hat{k}} 
\gamma^{\nu}\gamma^5
\frac{1}{M_{\rm N} - \hat{k} 
+ \hat{k}_{\ell}}\Bigg\},
\eqno({\rm E}.20)
$$
Since the energy  of 
the incident neutrino $E_{\nu_{\rm e}}$ 
ranges the region  $E_{\rm th} \le  E_{\nu_{\rm e}} 
\le  10\,{\rm MeV}$ and the deuteron in the rest 
frame $k^{\mu}_{\rm D} =
(k^0_{\rm D}, \vec{0}\,)$, we would calculate the structure functions setting $k^{\mu}_{\ell} = 
\vec{k}_{\rm D} = 0$:
$$
\bar{{\cal J}}^{\alpha\mu\nu}_{\rm np}
(k_{\rm D}, k_{\ell}) = -\,4\pi 
\int\limits^{\infty}_0 d\rho\,\rho^2\,U(\rho)\,
\psi_{\rm np}
(\rho)\int\frac{d^4k}{\pi^2i}\, \frac{\sin|\vec{k}\,|\rho}{|\vec{k}\,|\rho}
$$
$$
\times\,
{\rm tr} 
\Bigg\{\gamma^{\alpha}
\gamma^5
\frac{1}{M_{\rm N} - \hat{k}  - 
\hat{k}_{\rm D}}
\gamma^{\nu}\gamma^5 
\frac{1}{M_{\rm N} - \hat{k}}
\gamma^{\mu} 
\frac{1}{M_{\rm N} - \hat{k}}\Bigg\},
$$
$$
\bar{{\cal J}}^{\mu\nu}_{\rm np}
(k_{\rm D}, k_{\ell}) = - \,4\pi 
\int\limits^{\infty}_0 d\rho\,\rho^2\,U(\rho)\,
\psi_{\rm np}
(\rho)
\int\frac{d^4k}{\pi^2i}\,
\frac{\sin|\vec{k}\,|\rho}{|\vec{k}\,|\rho}
$$
$$
\times\,
{\rm tr}
\Bigg\{\gamma^5
\frac{1}{M_{\rm N} - \hat{k} -
\hat{k}_{\rm D}} 
\gamma^{\nu}\gamma^5
\frac{1}{M_{\rm N} - \hat{k}}
\gamma^{\mu} \frac{1}{M_{\rm N} 
- \hat{k}}\Bigg\}.
\eqno({\rm E}.21)
$$
The computation of the momentum integrals in the structure functions Eq.~({\rm D}21) is analogous to the momentum integrals of the structure functions 
${\cal J}^{\alpha\mu\nu}(k_{\rm D}, k_{\ell})$ and  
${\cal J}^{\mu\nu}(k_{\rm D}, k_{\ell})$ 
(see Appendix C). Then, due to the low--energy reduction
$$
[\bar{u}(p_1)\gamma_{\alpha}\gamma^5 u^c(p_2)] = 
g_{\alpha 0}[\bar{u}(p_1)\gamma^0\gamma^5 u^c(p_2)] = 
g_{\alpha 0}[\bar{u}(p_1)\gamma^5 u^c(p_2)],
$$
$$
[\bar{u}(p_1)\gamma_{\alpha}
\gamma^5 u^c(p_2)] \to  
g_{\alpha 0}[\bar{u}(p_1)
\gamma^5 u^c(p_2)],
\eqno({\rm E}.22)
$$
where we have used the relation 
$\bar{u}(p_1)\gamma^0 = \bar{u}(p_1)$, the matrix element Eq.~({\rm E}.19) can be written in the form
$$
i{\cal M}(\nu_{\rm e} + {\rm D} \to 
\nu_{\rm e} + {\rm n} + {\rm p}) =
$$
$$
=g_{\rm A}G_{\rm \pi np}\frac{G_{\rm F}}{\sqrt{2}} 
\frac{g_{\rm V}}{2\pi^2}[\bar{u}(p_1)\gamma^5 u^c(p_2)]
[\bar{u}(k_{\rm e^-})\gamma_{\nu}
(1-\gamma^5)u(k_{\nu_{\rm e}})]\,e_{\mu}(k_{\rm D})\,
$$
$$
\times\,[\bar{{\cal J}}^{0\mu\nu}_{\rm np}
(k_{\rm D}, k_{\ell}) + 
\bar{{\cal J}}^{\mu\nu}_{\rm np}
(k_{\rm D}, k_{\ell})].\eqno({\rm E}.23)
$$
For the computation of the structure function defining the amplitude of the process $\nu_{\rm e}$ + D $\to$ $\nu_{\rm e}$ + n + p given by Eq.~({\rm E}.23) we use the wave function $\psi_{\rm np}(\rho)$ given by Eq.~(\ref{label9.3}) and set
$$
\bar{{\cal J}}^{0\mu\nu}_{\rm np}
(k_{\rm D}, k_{\ell}) + 
\bar{{\cal J}}^{\mu\nu}_{\rm np}(k_{\rm D}, k_{\ell}) = 
e^{\displaystyle
i\delta_{\rm np}(k)}\frac{\displaystyle 
\sin\delta_{\rm
np}(k)}{\displaystyle a_{\rm np} k }\,
6\,M_{\rm N}\,g^{\mu\nu}\,
{\cal F}_{\rm np\nu_{\rm e}}\,
v_{\rm np}(0).\eqno({\rm E}.24)
$$
The factor ${\cal F}_{\rm np\nu_{\rm e}}$ amounts to 
(see Eq.~({\rm D}.34)):
$$
{\cal F}_{\rm np\nu_{\rm e}} = -a_{\rm np}\,
\frac{44}{27}\,
\Bigg[\frac{M^2_{\pi}}{\sqrt{M^2_{\rm N} - M^2_{\pi}}}\,
{\rm arctg}\frac{\sqrt{M^2_{\rm N} - M^2_{\pi}}}{M_{\pi}} - \frac{8}{11}\,\frac{M^3_{\pi}}{M^2_{\rm N} - M^2_{\pi}}
$$
$$
+ \frac{8}{11}\,\frac{M^4_{\pi}}{(M^2_{\rm N} - M^2_{\pi})^{3/2}}\,{\rm arctg}\frac{\sqrt{M^2_{\rm N} - M^2_{\pi}}}{M_{\pi}}\Bigg].\eqno({\rm E}.25)
$$
The amplitude of the process $\nu_{\rm e}$ + D $\to$ $\nu_{\rm e}$ + n + p is then given by
$$
i{\cal M}(\nu_{\rm e} + {\rm D} \to 
\nu_{\rm e} + {\rm n} + {\rm p}) =
g_{\rm A}G_{\rm \pi np}M_{\rm N}\frac{G_{\rm F}}{\sqrt{2}} \frac{3g_{\rm V}}{4\pi^2}\,\times\,
{\cal F}_{\rm np\nu_{\rm e}}\,v_{\rm np}(0)
$$
$$
\times\,[\bar{u}(p_1)\gamma^5 u^c(p_2)][\bar{u}(k^{\prime}_{\nu_{\rm e}}) \gamma^{\mu} (1-\gamma^5) u(k_{\nu_{\rm e}})]\,e_{\mu}(k_{\rm D})\, e^{\displaystyle i\delta_{\rm np}(k)} \frac{\displaystyle \sin\delta_{\rm
np}(k)}{\displaystyle a_{\rm np} k}.\eqno({\rm E}.26)
$$
This completes the computation of the matrix element of the process of the  $\nu_{\rm e}$ + D $\to$ $\nu_{\rm e}$ + n + p. 

\section*{Appendix F. Computation of the matrix element of the neutron--proton radiative capture}

In this Appendix we give a detailed computation of the matrix element of the neutron--proton radiative capture n + p $\to$ D + $\gamma$ for thermal neutrons caused by  pure M1 transition. In the RFMD the matrix element of the neutron--proton radiative capture is determined by the interactions
$$
{\cal L}_{\rm npD}(x) - ig_{\rm V}\,[\bar{p^c}(x)\gamma^{\mu} n(x) - \bar{n^c}(x)\gamma^{\mu}p(x)]\,D^{\dagger}_{\mu}(x),
$$
$$
{\cal L}^{\rm np \to np}_{\rm eff}(x) =  G_{\rm \pi
np}\,\int d^3\rho\,U(\rho)
$$
$$
\times\,\{[\bar{n}(t,\vec{x} + \frac{1}{2}\vec{\rho}\,)\gamma_{\mu}
\gamma^5 p^c(t,\vec{x} - \frac{1}{2}\vec{\rho}\,)]
[\bar{p^c}(t,\vec{x} + \frac{1}{2}\vec{\rho}\,)\gamma^{\mu}\gamma^5
n(t,\vec{x} - \frac{1}{2}\vec{\rho}\,)]
$$
$$
+ (\gamma_{\mu}\gamma^5 \otimes \gamma^{\mu}\gamma^5 \to \gamma^5 \otimes
\gamma^5)\},
$$
$$
{\cal L}_{\rm pp\gamma}(x) = -e\,\bar{p}(x)\gamma^{\nu}p(x)\,A_{\nu}(x) 
+ ie\,\frac{\kappa_{\rm p}}{4 M_{\rm N}}\,\bar{p}(x)\sigma^{\mu\nu}p(x)
\,F_{\mu\nu}(x),
$$
$$
{\cal L}_{\rm nn\gamma}(x) =  ie\,
\frac{\kappa_{\rm n}}{4 M_{\rm N}}\,\bar{p}(x)
\sigma^{\mu\nu}p(x)
\,F_{\mu\nu}(x),\eqno({\rm F}.1)
$$
where $F_{\mu\nu}(x) = \partial_{\mu}A_{\nu}(x) - \partial_{\nu}A_{\mu}(x)$ and $A_{\mu}(x)$ are the electromagnetic field strength and the electromagnetic potential, respectively, and $\sigma^{\mu\nu} = \frac{1}{2}\,[\gamma^{\mu},\gamma^{\nu}]$. Then, $\kappa_{\rm p} = 1.793$ and $\kappa_{\rm n} = -1.913$ are the anomalous magnetic dipole moments of the proton and the neutron measured in nuclear magnetons $\mu_{\rm N} = e/2 M_{\rm N}$, where $e$ is the proton electric charge. The total magnetic dipole moments of the proton and the neutron amount to $\mu_{\rm p} = 1 + \kappa_{\rm p} = 2.793$ and $\mu_{\rm n} = \kappa_{\rm n} = -1.913$, respectively.

The S matrix element ${\rm S}^{(3)}_{\rm np \to D\gamma}$ responsible for the transition n + p $\to$ D + $\gamma$ can be obtained by analogy with the S matrix element ${\rm S}^{(3)}_{\rm pp \to DW^+}$ (see Eq.~({\rm C}.8)) and reads
$$
{\rm S}^{(3)}_{\rm np \to D\gamma} =  -i \int d^4x_1 d^4x_2 d^4x_3\,{\rm T}({\cal L}^{\rm np\to np}_{\rm eff}(x_1)
{\cal L}_{\rm npD}(x_2){\cal L}_{\rm pp\gamma}(x_3))
$$
$$
-i \int d^4x_1 d^4x_2 d^4x_3\,{\rm T}(
{\cal L}^{\rm np\to np}_{\rm eff}(x_1){\cal L}_{\rm npD}(x_2)
{\cal L}_{\rm nn\gamma}(x_3)).\eqno({\rm F}.2)
$$
Then, for the derivation of the effective Lagrangian 
${\cal L}_{\rm np \to D\gamma}(x)$ containing only the fields of the initial and the final particles we have to make all necessary contractions of the operators of the neutron and the proton fields. Symbolically the result of the contractions we denote by brackets and get 
$$
<{\rm S}^{(3)}_{\rm np\to D\gamma}> = -i \int d^4x_1 d^4x_2 d^4x_3\,<{\rm T}({\cal L}^{\rm np\to np}_{\rm eff}(x_1)
{\cal L}_{\rm npD}(x_2){\cal L}_{\rm pp\gamma}(x_3))>
$$
$$
 -i \int d^4x_1 d^4x_2 d^4x_3\,<{\rm T}(
 {\cal L}^{\rm np\to np}_{\rm eff}(x_1)
 {\cal L}_{\rm npD}(x_2)
 {\cal L}_{\rm nn\gamma}(x_3))>.\eqno({\rm F}.3)
$$
The effective Lagrangian 
${\cal L}_{\rm np \to D\gamma}(x)$ related to the S matrix element $<{\rm S}^{(3)}_{\rm np\to D\gamma}>$ is defined as
$$
<{\rm S}^{(3)}_{\rm np\to D\gamma}> = 
i\int d^4x\,{\cal L}_{\rm np\to D\gamma}(x) = 
$$
$$
= -i \int d^4x_1 d^4x_2 d^4x_3\,<{\rm T}(
{\cal L}^{\rm np\to np}_{\rm eff}(x_1)
{\cal L}_{\rm npD}(x_2){\cal L}_{\rm pp\gamma}(x_3))>
$$
$$
-i \int d^4x_1 d^4x_2 d^4x_3\,<{\rm T}(
{\cal L}^{\rm np\to np}_{\rm eff}(x_1)
{\cal L}_{\rm npD}(x_2)
{\cal L}_{\rm nn\gamma}(x_3))>.\eqno({\rm F}.4)
$$
In terms of the operators of the interacting fields the effective Lagrangian  ${\cal L}_{\rm np\to D\gamma}(x)$ reads
$$
\int d^4x\,{\cal L}_{\rm np\to D\gamma}(x) =  - \int d^4x_1 d^4x_2 d^4x_3\,<{\rm T}({\cal L}^{\rm np\to np}_{\rm eff}(x_1)
{\cal L}_{\rm npD}(x_2){\cal L}_{\rm pp\gamma}(x_3))>
$$
$$
- \int d^4x_1 d^4x_2 d^4x_3\,<{\rm T}(
{\cal L}^{\rm np\to np}_{\rm eff}(x_1)
{\cal L}_{\rm npD}(x_2){\cal L}_{\rm nn\gamma}(x_3))>
$$
$$
= G_{\rm \pi np}\,\times\,(-ig_{\rm V})\,\times \,
(-e)\int d^4x_1 d^4x_2 d^4x_3\,\int d^3\rho\,U(\rho)\,
$$
$$
\times\,{\rm T}([\bar{p^c}(t_1,\vec{x}_1 + \frac{1}{2}\,\vec{\rho}\,)\,\gamma_{\alpha}\gamma^5 
n(t_1,\vec{x}_1 - \frac{1}{2}\,\vec{\rho}\,)]\,
D^{\dagger}_{\mu}(x_2)\,A_{\nu}(x_3))
$$
$$
\times <0|{\rm T}([\bar{n}(t_1,\vec{x}_1 + \frac{1}{2}\,\vec{\rho}\,)\,\gamma^{\alpha}\gamma^5 
p^c (t_1,\vec{x}_1 - \frac{1}{2}\,\vec{\rho}\,)][\bar{p^c}(x_2)\gamma^{\mu}n(x_2) - \bar{n^c}(x_2)\gamma^{\mu}p(x_2)]\,
$$
$$[\bar{p}(x_3)\gamma^{\nu}p(x_3)])|0>
$$
$$
+ G_{\rm \pi np}\,\times\,(-ig_{\rm V})\,\times 
\,(-e)\int d^4x_1 d^4x_2 d^4x_3\,\int d^3\rho\,U(\rho)\,
$$
$$
\times\,{\rm T}([\bar{p^c}(t_1,\vec{x}_1 + \frac{1}{2}\,\vec{\rho}\,)\,\gamma^5 n(t_1,\vec{x}_1 - \frac{1}{2}\,\vec{\rho}\,)]
\,D^{\dagger}_{\mu}(x_2)
\,A_{\nu}(x_3))
$$
$$
\times <0|{\rm T}([\bar{n}(t_1,\vec{x}_1 + \frac{1}{2}\,\vec{\rho}\,)\,\gamma^5 p^c (t_1,\vec{x}_1 - \frac{1}{2}\,\vec{\rho}\,)][\bar{p^c}(x_2)\gamma^{\mu}n(x_2) - \bar{n^c}(x_2)\gamma^{\mu}p(x_2)]\,
$$
$$
\times\,[\bar{p}(x_3)\gamma^{\nu}p(x_3)])|0> 
$$
$$
+ G_{\rm \pi np}\,\times\,(-ig_{\rm V})\,\times \,i\,e\,\frac{\kappa_{\rm p}}{4 M_{\rm N}}\int d^4x_1 d^4x_2 d^4x_3\,\int d^3\rho\,U(\rho)\,
$$
$$
\times\,{\rm T}([\bar{p^c}(t_1,\vec{x}_1 + \frac{1}{2}\,\vec{\rho}\,)\,\gamma_{\alpha}
\gamma^5 n(t_1,\vec{x}_1 - \frac{1}{2}\,\vec{\rho}\,)]\,D^{\dagger}_{\mu}(x_2)\,
F_{\nu\beta}(x_3))
$$
$$
\times <0|{\rm T}([\bar{n}(t_1,\vec{x}_1 + \frac{1}{2}\,\vec{\rho}\,)\,\gamma^{\alpha}
\gamma^5 p^c (t_1,\vec{x}_1 - \frac{1}{2}\,\vec{\rho}\,)][\bar{p^c}(x_2)
\gamma^{\mu}n(x_2) - 
\bar{n^c}(x_2)\gamma^{\mu}p(x_2)]\,
$$
$$
\times\,[\bar{p}(x_3)\sigma^{\nu\beta} p(x_3)])|0> 
$$
$$
+ G_{\rm \pi np}\,\times\,(-ig_{\rm V})\,\times \,i\,e\,\frac{\kappa_{\rm p}}{4 M_{\rm N}}\int d^4x_1 d^4x_2 d^4x_3\,\int d^3\rho\,U(\rho)\,
$$
$$
\times\,{\rm T}([\bar{p^c}(t_1,\vec{x}_1 + \frac{1}{2}\,\vec{\rho}\,)\,\gamma^5 n(t_1,\vec{x}_1 - \frac{1}{2}\,\vec{\rho}\,)]\,
D^{\dagger}_{\mu}(x_2)\,F_{\nu\beta}(x_3))
$$
$$
\times <0|{\rm T}([\bar{n}(t_1,\vec{x}_1 + \frac{1}{2}\,\vec{\rho}\,)\,\gamma^5 
p^c (t_1,\vec{x}_1 - \frac{1}{2}\,\vec{\rho}\,)]
[\bar{p^c}(x_2)\gamma^{\mu}n(x_2) - \bar{n^c}(x_2)\gamma^{\mu}p(x_2)]\,
$$
$$
\times\,[\bar{p}(x_3)\sigma^{\nu\beta} p(x_3)])|0> 
$$
$$
+ G_{\rm \pi np}\,\times\,(-ig_{\rm V})\,\times \,i\,e\,\frac{\kappa_{\rm n}}{4 M_{\rm N}}
\int d^4x_1 d^4x_2 d^4x_3\,\int d^3\rho\,U(\rho)\,
$$
$$
\times\,{\rm T}([\bar{p^c}(t_1,\vec{x}_1 + \frac{1}{2}\,\vec{\rho}\,)\,\gamma_{\alpha}\gamma^5 
n(t_1,\vec{x}_1 - \frac{1}{2}\,\vec{\rho}\,)]\,
D^{\dagger}_{\mu}(x_2)\,F_{\nu\beta}(x_3))
$$
$$
\times <0|{\rm T}([\bar{n}(t_1,\vec{x}_1 + \frac{1}{2}\,\vec{\rho}\,)\,
\gamma^{\alpha}\gamma^5 p^c (t_1,\vec{x}_1 - \frac{1}{2}\,\vec{\rho}\,)][\bar{p^c}(x_2)
\gamma^{\mu}n(x_2) - \bar{n^c}(x_2)
\gamma^{\mu}p(x_2)]\,
$$
$$
\times\,[\bar{n}(x_3)\sigma^{\nu\beta} n(x_3)])|0> 
$$
$$
+ G_{\rm \pi np}\,\times\,(-ig_{\rm V})\,\times \,i\,e\,\frac{\kappa_{\rm n}}{4 M_{\rm N}}
\int d^4x_1 d^4x_2 d^4x_3\,
\int d^3\rho\,U(\rho)\,
$$
$$
\times\,{\rm T}([\bar{p^c}(t_1,\vec{x}_1 + \frac{1}{2}\,\vec{\rho}\,)\,\gamma^5 
n(t_1,\vec{x}_1 - \frac{1}{2}\,\vec{\rho}\,)]\,
D^{\dagger}_{\mu}(x_2)\,F_{\nu\beta}(x_3))
$$
$$
\times <0|{\rm T}([\bar{n}(t_1,\vec{x}_1 + \frac{1}{2}\,\vec{\rho}\,)\,
\gamma^5 p^c (t_1,\vec{x}_1 - \frac{1}{2}\,\vec{\rho}\,)]
[\bar{p^c}(x_2)\gamma^{\mu}n(x_2) - \bar{n^c}(x_2)\gamma^{\mu}p(x_2)]\,
$$
$$
\times\,[\bar{n}(x_3)
 \sigma^{\nu\beta} n(x_3)])|0>.\eqno({\rm F}.5)
$$
Before we have made the necessary contractions we suggest to rewrite the r.h.s. of Eq.~({\rm F}.5) in the more convenient form
$$
\int d^4x\,{\cal L}_{\rm np\to D\gamma}(x) =  - \int d^4x_1 d^4x_2 d^4x_3\,<{\rm T}
({\cal L}^{\rm np\to np}_{\rm eff}(x_1)
{\cal L}_{\rm npD}(x_2){\cal L}_{\rm pp\gamma}(x_3))>
$$
$$
- \int d^4x_1 d^4x_2 d^4x_3\,<{\rm T}
({\cal L}^{\rm np\to np}_{\rm eff}(x_1)
{\cal L}_{\rm npD}(x_2){\cal L}_{\rm nn\gamma}(x_3))>
$$
$$
= G_{\rm \pi np}\,\times\,2ig_{\rm V}\,\times \,
(-e)\int d^4x_1 d^4x_2 d^4x_3\,\int d^3\rho\,U(\rho)\,
$$
$$
\times\,{\rm T}([\bar{p^c}(t_1,\vec{x}_1 + \frac{1}{2}\,\vec{\rho}\,)\,
\gamma_{\alpha}\gamma^5 n(t_1,\vec{x}_1 - \frac{1}{2}\,\vec{\rho}\,)]\,
D^{\dagger}_{\mu}(x_2)\,A_{\nu}(x_3))
$$
$$
\times <0|{\rm T}([\bar{p}(t_1,\vec{x}_1 - \frac{1}{2}\,\vec{\rho}\,)\,\gamma^{\alpha}
\gamma^5 n^c (t_1,\vec{x}_1 + \frac{1}{2}\,\vec{\rho}\,)][ \bar{n^c}(x_2)\gamma^{\mu}p(x_2)][\bar{p}(x_3)
\gamma^{\nu}p(x_3)])|0>
$$
$$
+ G_{\rm \pi np}\,\times\,2ig_{\rm V}\,\times \,
(-e)\int d^4x_1 d^4x_2 d^4x_3\,\int d^3\rho\,U(\rho)\,
$$
$$
\times\,{\rm T}([\bar{p^c}(t_1,\vec{x}_1 + \frac{1}{2}\,\vec{\rho}\,)\,\gamma^5 n(t_1,\vec{x}_1 - \frac{1}{2}\,\vec{\rho}\,)]\,
D^{\dagger}_{\mu}(x_2)\,A_{\nu}(x_3))
$$
$$
\times <0|{\rm T}([\bar{p}(t_1,\vec{x}_1 - \frac{1}{2}\,\vec{\rho}\,)\,\gamma^5
n^c (t_1,\vec{x}_1 + \frac{1}{2}\,\vec{\rho}\,)]
[\bar{n^c}(x_2)
\gamma^{\mu}p(x_2)][\bar{p}(x_3)
\gamma^{\nu}p(x_3)])|0> 
$$
$$
+ G_{\rm \pi np}\,\times\,2ig_{\rm V}\,\times \,i\,e\,\frac{\kappa_{\rm p}}{4 M_{\rm N}}
\int d^4x_1 d^4x_2 d^4x_3\,\int d^3\rho\,U(\rho)\,
$$
$$
\times\,{\rm T}([\bar{p^c}(t_1,\vec{x}_1 + \frac{1}{2}\,\vec{\rho}\,)\,\gamma_{\alpha}
\gamma^5 n(t_1,\vec{x}_1 - \frac{1}{2}\,\vec{\rho}\,)]\,
D^{\dagger}_{\mu}(x_2)\,F_{\nu\beta}(x_3))
$$
$$
\times <0|{\rm T}([\bar{p}(t_1,\vec{x}_1 - \frac{1}{2}\,\vec{\rho}\,)\,\gamma^{\alpha}
\gamma^5 n^c (t_1,\vec{x}_1 + \frac{1}{2}\,\vec{\rho}\,)][\bar{n^c}(x_2)
\gamma^{\mu}p(x_2)][\bar{p}(x_3)
\sigma^{\nu\beta} p(x_3)])|0> 
$$
$$
+ G_{\rm \pi np}\,\times\,2ig_{\rm V}\,\times \,i\,e\,\frac{\kappa_{\rm p}}{4 M_{\rm N}}
\int d^4x_1 d^4x_2 d^4x_3\,
\int d^3\rho\,U(\rho)\,
$$
$$
\times\,{\rm T}([\bar{p^c}(t_1,\vec{x}_1 + \frac{1}{2}\,\vec{\rho}\,)\,\gamma^5 n(t_1,\vec{x}_1 - \frac{1}{2}\,\vec{\rho}\,)]\,
D^{\dagger}_{\mu}(x_2)\,F_{\nu\beta}(x_3))
$$
$$
\times <0|{\rm T}([\bar{p}(t_1,\vec{x}_1 - \frac{1}{2}\,\vec{\rho}\,)\,\gamma^5 n^c (t_1,\vec{x}_1 + \frac{1}{2}\,\vec{\rho}\,)][\bar{n^c}(x_2)
\gamma^{\mu}p(x_2)][\bar{p}(x_3)
\sigma^{\nu\beta} p(x_3)])|0> 
$$
$$
+ G_{\rm \pi np}\,\times\,(-2ig_{\rm V})\,\times \,i\,e\,\frac{\kappa_{\rm n}}{4 M_{\rm N}}
\int d^4x_1 d^4x_2 d^4x_3\,\int d^3\rho\,U(\rho)\,
$$
$$
\times\,{\rm T}([\bar{p^c}(t_1,\vec{x}_1 - \frac{1}{2}\,\vec{\rho}\,)\,\gamma_{\alpha}
\gamma^5 n(t_1,\vec{x}_1 + \frac{1}{2}\,\vec{\rho}\,)]\,
D^{\dagger}_{\mu}(x_2)\,F_{\nu\beta}(x_3))
$$
$$
\times <0|{\rm T}([\bar{n}(t_1,\vec{x}_1 - \frac{1}{2}\,\vec{\rho}\,)\,\gamma^{\alpha}
\gamma^5 p^c (t_1,\vec{x}_1 + \frac{1}{2}\,\vec{\rho}\,)] [\bar{p^c}(x_2)\gamma^{\mu}n(x_2)]
[\bar{n}(x_3)\sigma^{\nu\beta} n(x_3)])|0> 
$$
$$
+ G_{\rm \pi np}\,\times\,(-2ig_{\rm V})\,\times \,i\,e\,\frac{\kappa_{\rm n}}{4 M_{\rm N}}
\int d^4x_1 d^4x_2 d^4x_3\,\int d^3\rho\,U(\rho)\,
$$
$$
\times\,{\rm T}([\bar{p^c}(t_1,\vec{x}_1 - \frac{1}{2}\,\vec{\rho}\,)\,
\gamma^5 n(t_1,\vec{x}_1 + \frac{1}{2}\,\vec{\rho}\,)]\,
D^{\dagger}_{\mu}(x_2)\,F_{\nu\beta}(x_3))
$$
$$
\times <0|{\rm T}([\bar{n}(t_1,\vec{x}_1 - \frac{1}{2}\,\vec{\rho}\,)\,
\gamma^5 p^c (t_1,\vec{x}_1 + \frac{1}{2}\,\vec{\rho}\,)]
[\bar{p^c}(x_2)\gamma^{\mu}n(x_2)]
[\bar{n}(x_3)
\sigma^{\nu\beta} n(x_3)])|0>.\eqno({\rm F}.6)
$$
Making all contractions we obtain
$$
\int d^4x\,{\cal L}_{\rm np\to D\gamma}(x) =
G_{\rm \pi np}\,\times\,2ig_{\rm V}\,
\times \,(-e)\int d^4x_1 d^4x_2 d^4x_3\,
\int d^3\rho\,U(\rho)\,
$$
$$
\times\,{\rm T}([\bar{p^c}(t_1,\vec{x}_1 + \frac{1}{2}\,\vec{\rho}\,)\,\gamma_{\alpha}
\gamma^5 n(t_1,\vec{x}_1 - \frac{1}{2}\,\vec{\rho}\,)]\,
D^{\dagger}_{\mu}(x_2)\,A_{\nu}(x_3))
$$
$$
\times \frac{1}{i}\,{\rm tr}\{S_F(t_3 - t_1, \vec{x}_3 - \vec{x}_1 +\frac{1}{2}\,\vec{\rho}\,) \gamma_{\alpha}
\gamma^5 S^c_F(t_1 - t_2,\vec{x}_1 - \vec{x}_2 + \frac{1}{2}\,\vec{\rho}\,) \gamma^{\mu} 
S_F(x_2 - x_3)\gamma^{\nu}\}
$$
$$
+ G_{\rm \pi np}\,\times\,2ig_{\rm V}\,\times 
\,(-e)\int d^4x_1 d^4x_2 d^4x_3\,\int d^3\rho\,U(\rho)\,
$$
$$
\times\,{\rm T}([\bar{p^c}(t_1,\vec{x}_1 + \frac{1}{2}\,\vec{\rho}\,)\,\gamma^5 
n(t_1,\vec{x}_1 - \frac{1}{2}\,\vec{\rho}\,)]\,
D^{\dagger}_{\mu}(x_2)\,A_{\nu}(x_3))
$$
$$
\times \frac{1}{i}\,{\rm tr}\{S_F(t_3 - t_1, \vec{x}_3 - \vec{x}_1 +\frac{1}{2}\,\vec{\rho}\,)\gamma^5
S^c_F(t_1 - t_2,\vec{x}_1 - \vec{x}_2 + \frac{1}{2}\,\vec{\rho}\,) \gamma^{\mu} S_F(x_2 - x_3)\gamma^{\nu}\}
$$
$$
+ G_{\rm \pi np}\,\times\,2ig_{\rm V}\,\times \,i\,e\,\frac{\kappa_{\rm p}}{4 M_{\rm N}}
\int d^4x_1 d^4x_2 d^4x_3\,\int d^3\rho\,U(\rho)\,
$$
$$
\times\,{\rm T}([\bar{p^c}(t_1,\vec{x}_1 + \frac{1}{2}\,\vec{\rho}\,)\,\gamma_{\alpha}\gamma^5 
n(t_1,\vec{x}_1 - \frac{1}{2}\,\vec{\rho}\,)]\,D^{\dagger}_{\mu}(x_2)
\,F_{\nu\beta}(x_3))
$$
$$
\times \frac{1}{i}\,{\rm tr}\{S_F(t_3 - t_1, \vec{x}_3 - \vec{x}_1 +\frac{1}{2}\,\vec{\rho}\,) \gamma_{\alpha}\gamma^5 
S^c_F(t_1 - t_2,\vec{x}_1 - \vec{x}_2 + \frac{1}{2}\,\vec{\rho}\,) \gamma^{\mu} S_F(x_2 - x_3)\sigma^{\nu\beta}\}
$$
$$
+ G_{\rm \pi np}\,\times\,2ig_{\rm V}\,\times \,i\,e\,\frac{\kappa_{\rm p}}{4 M_{\rm N}}\int d^4x_1 d^4x_2 d^4x_3\,\int d^3\rho\,U(\rho)\,
$$
$$
\times\,{\rm T}([\bar{p^c}(t_1,\vec{x}_1 + \frac{1}{2}\,\vec{\rho}\,)\,\gamma^5 n(t_1,\vec{x}_1 - \frac{1}{2}\,\vec{\rho}\,)]\,
D^{\dagger}_{\mu}(x_2)\,F_{\nu\beta}(x_3))
$$
$$
\times \frac{1}{i}\,{\rm tr}\{
S_F(t_3 - t_1, \vec{x}_3 - \vec{x}_1 +\frac{1}{2}\,\vec{\rho}\,) \gamma^5 S^c_F(t_1 - t_2,\vec{x}_1 - \vec{x}_2 + \frac{1}{2}\,\vec{\rho}\,) 
\gamma^{\mu} S_F(x_2 - x_3)\sigma^{\nu\beta}\}
$$
$$
+ G_{\rm \pi np}\,\times\,(-2ig_{\rm V})\,\times \,i\,e\,\frac{\kappa_{\rm n}}{4 M_{\rm N}}
\int d^4x_1 d^4x_2 d^4x_3\,\int d^3\rho\,U(\rho)\,
$$
$$
\times\,{\rm T}([\bar{p^c}(t_1,\vec{x}_1 - \frac{1}{2}\,\vec{\rho}\,)\,\gamma_{\alpha}\gamma^5 n(t_1,\vec{x}_1 + \frac{1}{2}\,\vec{\rho}\,)]\,
D^{\dagger}_{\mu}(x_2)\,F_{\nu\beta}(x_3))
$$
$$
\times \frac{1}{i}\,{\rm tr}\{S_F(t_3 - t_1, \vec{x}_3 - \vec{x}_1 +\frac{1}{2}\,\vec{\rho}\,) 
\gamma_{\alpha}\gamma^5 S^c_F(t_1 - t_2,\vec{x}_1 - \vec{x}_2 + \frac{1}{2}\,\vec{\rho}\,) 
\gamma^{\mu} S_F(x_2 - x_3)
\sigma^{\nu\beta}\}
$$
$$
+ G_{\rm \pi np}\,\times\,(-2ig_{\rm V})\,\times \,i\,e\,\frac{\kappa_{\rm n}}{4 M_{\rm N}}
\int d^4x_1 d^4x_2 d^4x_3\,
\int d^3\rho\,U(\rho)\,
$$
$$
\times\,{\rm T}([\bar{p^c}(t_1,\vec{x}_1 - \frac{1}{2}\,\vec{\rho}\,)\,\gamma^5 n(t_1,\vec{x}_1 + \frac{1}{2}\,\vec{\rho}\,)]\,
D^{\dagger}_{\mu}(x_2)\,F_{\nu\beta}(x_3))
$$
$$
\times \frac{1}{i}\,{\rm tr}\{S_F(t_3 - t_1, \vec{x}_3 - \vec{x}_1 +\frac{1}{2}\,\vec{\rho}\,) \gamma^5
S^c_F(t_1 - t_2,\vec{x}_1 - \vec{x}_2 + \frac{1}{2}\,\vec{\rho}\,) \gamma^{\mu} S_F(x_2 - x_3)\sigma^{\nu\beta}\}.
\eqno({\rm F}.7)
$$
Since the neutron and the proton are in the 
${^1}{\rm S}_0$--state we can sum up the contributions and get
$$
\int d^4x\,{\cal L}_{\rm np\to D\gamma}(x) =
G_{\rm \pi np}\,\times\,2ig_{\rm V}\,\times 
\,(-e)\int d^4x_1 d^4x_2 d^4x_3\,
\int d^3\rho\,U(\rho)\,
$$
$$
\times\,{\rm T}([\bar{p^c}(t_1,\vec{x}_1 + \frac{1}{2}\,\vec{\rho}\,)\,\gamma_{\alpha}
\gamma^5 n(t_1,\vec{x}_1 - \frac{1}{2}\,\vec{\rho}\,)]\,
D^{\dagger}_{\mu}(x_2)\,A_{\nu}(x_3))
$$
$$
\times \frac{1}{i}\,{\rm tr}\{S_F(t_3 - t_1, \vec{x}_3 - \vec{x}_1 +\frac{1}{2}\,\vec{\rho}\,) \gamma_{\alpha}
\gamma^5 S^c_F(t_1 - t_2,\vec{x}_1 - \vec{x}_2 + \frac{1}{2}\,\vec{\rho}\,) \gamma^{\mu} 
S_F(x_2 - x_3)\gamma^{\nu}\}
$$
$$
+ G_{\rm \pi np}\,\times\,2ig_{\rm V}\,
\times \,(-e)\int d^4x_1 d^4x_2 d^4x_3\,
\int d^3\rho\,U(\rho)\,
$$
$$
\times\,{\rm T}([\bar{p^c}(t_1,\vec{x}_1 + \frac{1}{2}\,\vec{\rho}\,)\,\gamma^5 n(t_1,\vec{x}_1 - \frac{1}{2}\,\vec{\rho}\,)]\,
D^{\dagger}_{\mu}(x_2)\,A_{\nu}(x_3))
$$
$$
\times \frac{1}{i}\,{\rm tr}\{S_F(t_3 - t_1, \vec{x}_3 - \vec{x}_1 +\frac{1}{2}\,\vec{\rho}\,)
\gamma^5 S^c_F(t_1 - t_2,\vec{x}_1 - \vec{x}_2 + \frac{1}{2}\,\vec{\rho}\,) \gamma^{\mu}
S_F(x_2 - x_3)\gamma^{\nu}\}
$$
$$
+ G_{\rm \pi np}\,\times\,2ig_{\rm V}\,\times \,i\,e\,
\frac{\kappa_{\rm p} - \kappa_{\rm n}}{4 M_{\rm N}}
\int d^4x_1 d^4x_2 d^4x_3\,\int d^3\rho\,U(\rho)\,
$$
$$
\times\,{\rm T}([\bar{p^c}(t_1,\vec{x}_1 + \frac{1}{2}\,\vec{\rho}\,)\,\gamma_{\alpha}
\gamma^5 n(t_1,\vec{x}_1 - \frac{1}{2}\,\vec{\rho}\,)]\,
D^{\dagger}_{\mu}(x_2)\,F_{\nu\beta}(x_3))
$$
$$
\times \frac{1}{i}\,{\rm tr}\{S_F(t_3 - t_1, \vec{x}_3 - \vec{x}_1 +\frac{1}{2}\,\vec{\rho}\,) \gamma_{\alpha}\gamma^5 S^c_F(t_1 - t_2,\vec{x}_1 - \vec{x}_2 + \frac{1}{2}\,\vec{\rho}\,) \gamma^{\mu} S_F(x_2 - x_3)\sigma^{\nu\beta}\}
$$
$$
+ G_{\rm \pi np}\,\times\,2ig_{\rm V}\,\times \,i\,e\,
\frac{\kappa_{\rm p} - \kappa_{\rm n}}{4 M_{\rm N}}
\int d^4x_1 d^4x_2 d^4x_3\,\int d^3\rho\,U(\rho)\,
$$
$$
\times\,{\rm T}([\bar{p^c}(t_1,\vec{x}_1 + \frac{1}{2}\,\vec{\rho}\,)\,\gamma^5 n(t_1,\vec{x}_1 - \frac{1}{2}\,\vec{\rho}\,)]\,
D^{\dagger}_{\mu}(x_2)\,F_{\nu\beta}(x_3))
$$
$$
\times \frac{1}{i}\,{\rm tr}\{S_F(t_3 - t_1, \vec{x}_3 - \vec{x}_1 +\frac{1}{2}\,\vec{\rho}\,) \gamma^5 S^c_F(t_1 - t_2,\vec{x}_1 - \vec{x}_2 + \frac{1}{2}\,\vec{\rho}\,) \gamma^{\mu} S_F(x_2 - x_3)\sigma^{\nu\beta}\}.\eqno({\rm F}.8)
$$
In the momentum representation of the nucleon Green functions we define the effective Lagrangian ${\cal L}_{\rm np\to D\gamma}(x)$ as follows:
$$
\int d^4x\,{\cal L}_{\rm np\to D\gamma}(x) =
$$
$$
=- i\,e\,G_{\rm \pi np}\,\frac{g_{\rm V}}{8\pi^2}
\int d^4x_1 \int \frac{d^4x_2 d^4k_2}{(2\pi)^4} 
\frac{d^4x_3 d^4k_3}{(2\pi)^4}\,
e^{\displaystyle  - i k_2\cdot (x_2 - x_1)}
e^{\displaystyle  - i k_3\cdot (x_3 - x_1)}
$$
$$
\times\,\int d^3\rho \,U(\rho)\,{\rm T}
([\bar{p^c}(t_1,\vec{x}_1 + \frac{1}{2}\,\vec{\rho}\,)\,
\gamma_{\alpha}\gamma^5 n(t_1,\vec{x}_1 - \frac{1}{2}\,\vec{\rho}\,)]\,
D^{\dagger}_{\mu}(x_2)\,A_{\nu}(x_3))
$$
$$
\times\,\int\frac{d^4k_1}{\pi^2i}\,
e^{\displaystyle - i\vec{q}\cdot \vec{\rho}}\,
{\rm tr}\Bigg\{\gamma^{\alpha}\gamma^5
\frac{1}{M_{\rm N} - \hat{k}_1 + \hat{k}_2}
\gamma^{\mu}\frac{1}{M_{\rm N} - \hat{k}_1}
\gamma^{\nu}\frac{1}{M_{\rm N} - 
\hat{k}_1 - \hat{k}_3}\Bigg\}
$$
$$
-i\,e\,G_{\rm \pi np}\,\frac{g_{\rm V}}{8\pi^2}
\int d^4x_1 \int \frac{d^4x_2 d^4k_2}{(2\pi)^4}
\frac{d^4x_3 d^4k_3}{(2\pi)^4}\,
e^{\displaystyle  - i k_2\cdot (x_2 - x_1)}
e^{\displaystyle  - i k_3\cdot (x_3 - x_1)}
$$
$$
\times\,\int d^3\rho \,U(\rho)\,{\rm T}([\bar{p^c}(t_1,\vec{x}_1 + \frac{1}{2}\,\vec{\rho}\,)\,\gamma^5 n(t_1,\vec{x}_1 - \frac{1}{2}\,\vec{\rho}\,)]\,
D^{\dagger}_{\mu}(x_2)\,A_{\nu}(x_3))
$$
$$
\times\,\int\frac{d^4k_1}{\pi^2i}\,
e^{\displaystyle - i\vec{q}\cdot \vec{\rho}}\,
{\rm tr}\Bigg\{\gamma^5
\frac{1}{M_{\rm N} - \hat{k}_1 + \hat{k}_2}
\gamma^{\mu}\frac{1}{M_{\rm N} - \hat{k}_1}
\gamma^{\nu}\frac{1}{M_{\rm N} -
\hat{k}_1 - \hat{k}_3}\Bigg\}
$$
$$
-e\, G_{\rm \pi np}\,
\frac{\kappa_{\rm p}- \kappa_{\rm n}}{4 M_{\rm N}}
\frac{g_{\rm V}}{8\pi^2}\int d^4x_1 \int \frac{d^4x_2 d^4k_2}{(2\pi)^4} \frac{d^4x_3 d^4k_3}{(2\pi)^4}
\,e^{\displaystyle  - i k_2\cdot (x_2 - x_1)}
e^{\displaystyle  - i k_3\cdot (x_3 - x_1)}
$$
$$
\times\,\int d^3\rho \,U(\rho)\,{\rm T}([\bar{p^c}(t_1,\vec{x}_1 + \frac{1}{2}\,\vec{\rho}\,)\,\gamma_{\alpha}\gamma^5 n(t_1,\vec{x}_1 - \frac{1}{2}\,\vec{\rho}\,)]\,
D^{\dagger}_{\mu}(x_2)\,F_{\nu\beta}(x_3))
$$
$$
\times\,\int\frac{d^4k_1}{\pi^2i}\,e^{\displaystyle - i\vec{q}\cdot \vec{\rho}}\,{\rm tr}\Bigg\{\gamma^{\alpha}
\gamma^5\frac{1}{M_{\rm N} - \hat{k}_1 + \hat{k}_2}\gamma^{\mu}\frac{1}{M_{\rm N} - 
\hat{k}_1}\sigma^{\nu\beta}
\frac{1}{M_{\rm N} - \hat{k}_1 - \hat{k}_3}\Bigg\}
$$
$$
-e\, G_{\rm \pi np}\,
\frac{\kappa_{\rm p}- \kappa_{\rm n}}{4 M_{\rm N}}
\frac{g_{\rm V}}{8\pi^2}\int d^4x_1 \int \frac{d^4x_2 d^4k_2}{(2\pi)^4} \frac{d^4x_3 d^4k_3}{(2\pi)^4}\,
e^{\displaystyle  - i k_2\cdot (x_2 - x_1)}
e^{\displaystyle  - i k_3\cdot (x_3 - x_1)}
$$
$$
\times\,\int d^3\rho \,U(\rho)\,{\rm T}([\bar{p^c}(t_1,\vec{x}_1 + \frac{1}{2}\,\vec{\rho}\,)\,\gamma^5 n(t_1,\vec{x}_1 - \frac{1}{2}\,\vec{\rho}\,)]\,
D^{\dagger}_{\mu}(x_2)\,F_{\nu\beta}(x_3))
$$
$$
\times\,\int\frac{d^4k_1}{\pi^2i}\,
e^{\displaystyle - i\vec{q}\cdot \vec{\rho}}\,{\rm tr}\Bigg\{\gamma^5\frac{1}{M_{\rm N} - \hat{k}_1 + \hat{k}_2}\gamma^{\mu}\frac{1}{M_{\rm N} - \hat{k}_1}\sigma^{\nu\beta}\frac{1}{M_{\rm N} - \hat{k}_1 - \hat{k}_3}\Bigg\},\eqno({\rm F}.9)
$$
where $\vec{q} = \vec{k}_1 + (\vec{k}_3 - \vec{k}_2)/2$.

The matrix element of the neutron--proton radiative capture n + p $\to$ D + $\gamma$ we define by a usual way as
$$
\int d^4x<D(k_{\rm D})\gamma(k)|{\cal L}_{\rm np\to D\gamma}(x)|n(p_1)p(p_2)> = 
$$
$$
=(2\pi)^4\delta^{(4)}( k_{\rm D} + k - p_1 - p_2)\,\frac{{\cal M}({\rm n} + {\rm p}\to {\rm D} + \gamma)}{\displaystyle
\sqrt{2E_1V\,2E_2V\,
2E_{\rm D}V\,2\omega V}},
\eqno({\rm F}.10)
$$
where  $E_i\,(i =1,2,{\rm D})$ and $\omega$ are the energies of the neutron, the proton, the deuteron and the photon, $V$ is the normalization volume.

Now we should take the r.h.s. of Eq.~({\rm F}.9) between the wave functions of the initial $|n(p_1) p(p_2)>$ and the final $<D(k_{\rm D})\gamma(k)|$ states.  This gives
$$
(2\pi)^4\delta^{(4)}( k_{\rm D} + k - p_1 - p_2)\,\frac{{\cal M}({\rm n} + {\rm p}\to {\rm D} + \gamma)}{\displaystyle 
\sqrt{2E_1V\,2E_2V\,
2E_{\rm D}V\,2\omega V}}
$$
$$
=-i\,e\,G_{\rm \pi np}\,\frac{g_{\rm V}}{8\pi^2}\int d^4x_1 \int \frac{d^4x_2 d^4k_2}{(2\pi)^4} \frac{d^4x_3 d^4k_3}{(2\pi)^4}\,e^{\displaystyle  - i k_2\cdot (x_2 - x_1)}e^{\displaystyle  - i k_3\cdot (x_3 - x_1)}
$$
$$
\times\,\int d^3\rho \,U(\rho)\,
\int\frac{d^4k_1}{\pi^2i}\,e^{\displaystyle - i\vec{q}\cdot \vec{\rho}}\,{\rm tr}\Bigg\{\gamma^{\alpha}\gamma^5\frac{1}{M_{\rm N} - \hat{k}_1 + \hat{k}_2}\gamma^{\mu}\frac{1}{M_{\rm N} - \hat{k}_1}\gamma^{\nu}\frac{1}{M_{\rm N} - \hat{k}_1 - \hat{k}_3}\Bigg\}
$$
$$
\times <D(k_{\rm D})\gamma(k)|{\rm T}([\bar{p^c}(t_1,\vec{x}_1 + \frac{1}{2}\,\vec{\rho}\,)\,\gamma_{\alpha}\gamma^5 n(t_1,\vec{x}_1 - \frac{1}{2}\,\vec{\rho}\,)]\,
D^{\dagger}_{\mu}(x_2)\,
A_{\nu}(x_3))|n(p_1) p(p_2)>
$$
$$
-i\,e\,G_{\rm \pi np}\,\frac{g_{\rm V}}{8\pi^2}\int d^4x_1 \int \frac{d^4x_2 d^4k_2}{(2\pi)^4} \frac{d^4x_3 d^4k_3}{(2\pi)^4}\,e^{\displaystyle  - i k_2\cdot (x_2 - x_1)}e^{\displaystyle  - i k_3\cdot (x_3 - x_1)}
$$
$$
\times\,\int d^3\rho \,U(\rho)\,\int\frac{d^4k_1}{\pi^2i}\,e^{\displaystyle - i\vec{q}\cdot \vec{\rho}}\,{\rm tr}\Bigg\{\gamma^5\frac{1}{M_{\rm N} - \hat{k}_1 + \hat{k}_2}\gamma^{\mu}\frac{1}{M_{\rm N} - \hat{k}_1}\gamma^{\nu}\frac{1}{M_{\rm N} - \hat{k}_1 - \hat{k}_3}\Bigg\}
$$
$$
\times <D(k_{\rm D})\gamma(k)|{\rm T}([\bar{p^c}(t_1,\vec{x}_1 + \frac{1}{2}\,\vec{\rho}\,)\,\gamma^5 n(t_1,\vec{x}_1 - \frac{1}{2}\,\vec{\rho}\,)]\,
D^{\dagger}_{\mu}(x_2)\,
A_{\nu}(x_3))|n(p_1) p(p_2)>
$$
$$
-e\, G_{\rm \pi np}\,\frac{\kappa_{\rm p}- \kappa_{\rm n}}{4 M_{\rm N}}\frac{g_{\rm V}}{8\pi^2}\int d^4x_1 \int \frac{d^4x_2 d^4k_2}{(2\pi)^4} \frac{d^4x_3 d^4k_3}{(2\pi)^4}\,e^{\displaystyle  - i k_2\cdot (x_2 - x_1)}e^{\displaystyle  - i k_3\cdot (x_3 - x_1)}
$$
$$
\times\,\int d^3\rho \,U(\rho)\,\int\frac{d^4k_1}{\pi^2i}\,e^{\displaystyle - i\vec{q}\cdot \vec{\rho}}\,{\rm tr}\Bigg\{\gamma^{\alpha}\gamma^5\frac{1}{M_{\rm N} - \hat{k}_1 + \hat{k}_2}\gamma^{\mu}\frac{1}{M_{\rm N} - \hat{k}_1}\sigma^{\nu\beta}\frac{1}{M_{\rm N} - \hat{k}_1 - \hat{k}_3}\Bigg\}
$$
$$
\times <D(k_{\rm D})\gamma(k)|{\rm T}([\bar{p^c}(t_1,\vec{x}_1 + \frac{1}{2}\,\vec{\rho}\,)\,\gamma_{\alpha}\gamma^5 n(t_1,\vec{x}_1 - \frac{1}{2}\,\vec{\rho}\,)]\,D^{\dagger}_{\mu}(x_2)\,F_{\nu\beta}(x_3))|n(p_1) p(p_2)>
$$
$$ 
-e\, G_{\rm \pi np}\,\frac{\kappa_{\rm p} - \kappa_{\rm n}}{4 M_{\rm N}}\frac{g_{\rm V}}{8\pi^2}\int d^4x_1 \int \frac{d^4x_2 d^4k_2}{(2\pi)^4} \frac{d^4x_3 d^4k_3}{(2\pi)^4}\,e^{\displaystyle  - i k_2\cdot (x_2 - x_1)}e^{\displaystyle  - i k_3\cdot (x_3 - x_1)}
$$
$$
\times\,\int d^3\rho \,U(\rho)\,\int\frac{d^4k_1}{\pi^2i}\,e^{\displaystyle - i\vec{q}\cdot \vec{\rho}}\,{\rm tr}\Bigg\{\gamma^5\frac{1}{M_{\rm N} - \hat{k}_1 + \hat{k}_2}\gamma^{\mu}\frac{1}{M_{\rm N} - \hat{k}_1}\sigma^{\nu\beta}\frac{1}{M_{\rm N} - \hat{k}_1 - \hat{k}_3}\Bigg\}
$$
$$
\times <D(k_{\rm D})\gamma(k)|{\rm T}([\bar{p^c}(t_1,\vec{x}_1 + \frac{1}{2}\,\vec{\rho}\,)\,\gamma^5 n(t_1,\vec{x}_1 - \frac{1}{2}\,\vec{\rho}\,)]\,D^{\dagger}_{\mu}(x_2)\,F_{\nu\beta}(x_3))|n(p_1) p(p_2)>.\eqno({\rm F}.11)
$$
Between the initial $|n(p_1) p(p_2)>$ and the final $<D(k_{\rm D})\gamma(k)|$ states the matrix elements in Eq.~({\rm F}.11) are defined (see Eq.~({\rm C}.23), Eqs.~({\rm C}.25) -- ({\rm C}.34) and Eq.~({\rm E}.15)):
$$
<D(k_{\rm D})\gamma(k)|{\rm T}([\bar{p^c}(t_1,\vec{x}_1 + \frac{1}{2}\,\vec{\rho}\,)\,\gamma_{\alpha}\gamma^5 
n(t_1,\vec{x}_1 - \frac{1}{2}\,\vec{\rho}\,)]\,
D^{\dagger}_{\mu}(x_2)
\,A_{\nu}(x_3))|n(p_1) p(p_2)>
$$
$$
=[\bar{u^c}(p_2)\gamma_{\alpha}\gamma^5 u(p_1)]
\,e^*_{\mu}(k_{\rm D})\,e^*_{\nu}(k)\,\psi_{\rm np}(\rho)\,\frac{\displaystyle
e^{\displaystyle -i(p_1+p_2)\cdot x_1}\,
e^{\displaystyle ik_{\rm D}\cdot x_2}\,
e^{\displaystyle ik\cdot x_3}}{\displaystyle 
\sqrt{2E_1V\,2E_2V\,
2E_{\rm D}V\,2\omega V}},
$$
$$
<D(k_{\rm D})\gamma(k)|{\rm T}([\bar{p^c}
(t_1,\vec{x}_1 + \frac{1}{2}\,\vec{\rho}\,)\,
\gamma^5 n(t_1,\vec{x}_1 - \frac{1}{2}\,\vec{\rho}\,)]\,
D^{\dagger}_{\mu}(x_2)\,A_{\nu}(x_3))|n(p_1) p(p_2)>
$$
$$
=[\bar{u^c}(p_2)u(p_1)]\,e^*_{\mu}(k_{\rm D})\,
e^*_{\nu}(k)\,\psi_{\rm np}(\rho)\,
\frac{\displaystyle e^{\displaystyle 
-i(p_1+p_2)\cdot x_1}\,e^{\displaystyle ik_{\rm D}\cdot x_2}\,e^{\displaystyle  ik\cdot x_3}}{\displaystyle \sqrt{2E_1V\,2E_2V\,
2E_{\rm D}V\,2\omega V}},
$$
$$
<D(k_{\rm D})\gamma(k)|{\rm T}([\bar{p^c}
(t_1,\vec{x}_1 + \frac{1}{2}\,\vec{\rho}\,)\,
\gamma_{\alpha}\gamma^5 n(t_1,\vec{x}_1 - \frac{1}{2}\,\vec{\rho}\,)]\,D^{\dagger}_{\mu}(x_2)
\,F_{\nu\beta}(x_3))|n(p_1) p(p_2)>
$$
$$
=[\bar{u^c}(p_2)\gamma_{\alpha}\gamma^5 u(p_1)]
\,e^*_{\mu}(k_{\rm D})\,i\,(k_{\nu}e^*_{\beta}(k) - k_{\beta}e^*_{\nu}(k))
$$
$$
\times\,\psi_{\rm np}(\rho)\,
\frac{\displaystyle e^{\displaystyle 
-i(p_1+p_2)\cdot x_1}\,e^{\displaystyle ik_{\rm D}\cdot x_2}\,e^{\displaystyle ik\cdot x_3}}{\displaystyle \sqrt{2E_1V\,2E_2V\,
2E_{\rm D}V\,2\omega V}}
$$
$$
<D(k_{\rm D})\gamma(k)|{\rm T}([\bar{p^c}(t_1,\vec{x}_1 + \frac{1}{2}\,\vec{\rho}\,)\,\gamma^5 n(t_1,\vec{x}_1 - \frac{1}{2}\,\vec{\rho}\,)]\,
D^{\dagger}_{\mu}(x_2)\,
F_{\nu\beta}(x_3))|n(p_1) p(p_2)>
$$
$$
=[\bar{u^c}(p_2) \gamma^5 u(p_1)]
\,e^*_{\mu}(k_{\rm D})\,i\,(k_{\nu}e^*_{\beta}(k) - k_{\beta}e^*_{\nu}(k))
$$
$$
\times\,\psi_{\rm np}(\rho)\,
\frac{\displaystyle e^{\displaystyle 
-i(p_1+p_2)\cdot x_1}\,e^{\displaystyle ik_{\rm D}\cdot x_2}\,e^{\displaystyle ik\cdot x_3}}{\displaystyle \sqrt{2E_1V\,2E_2V\,2E_{\rm D}V\,
2\omega V}},\eqno({\rm F}.12)
$$
where $k_{\rm D}$ and $k$ are the 4--momenta of the deuteron and the photon, then $\psi_{\rm np}(\rho)$ is the wave function of the relative movement of the neutron and the proton normalized per unit density [8]. It is given by Eq.~(\ref{label9.3}). For the computation of the matrix elements Eq.~({\rm F}.12) we have used the wave function of the neutron and the proton $|n(p_1)p(p_2)>$ determined by the operators of annihilation in the standard form [31]
$$
|n(p_1)p(p_2)>  = a^{\dagger}_{\rm n}(\vec{p}_1,\sigma_1)a^{\dagger}_{\rm p}(\vec{p}_2,\sigma_2)|0>,\eqno({\rm F}.13)
$$
where $a^{\dagger}_{\rm n}(\vec{p}_1,\sigma_1)$ and $a^{\dagger}_{\rm p}(\vec{p}_2,\sigma_2)$ are the operators of the creation of the neutron and the proton.

Substituting Eq.~({\rm F}.12) in Eq.~({\rm F}.11) we obtain the matrix element of the neutron--proton radiative capture in the following form
$$
(2\pi)^4\delta^{(4)}(k_{\rm D} + k - p_2 - p_1)\,{\cal M}({\rm n} + {\rm p} \to {\rm D} + \gamma) = 
$$
$$
= -i\,e\,G_{\rm \pi np}\,\frac{g_{\rm V}}{8\pi^2}\,[\bar{u^c}(p_2)\gamma_{\alpha}\gamma^5 u(p_1)]\,e^*_{\mu}(k_{\rm D})\,e^*_{\nu}(k)
\int d^4x_1 
$$
$$
\times \int \frac{d^4x_2 d^4k_2}{(2\pi)^4}
\frac{d^4x_3 d^4k_3}{(2\pi)^4}\,
e^{\displaystyle i (k_2 + k_3 - p_2 - p_1)\cdot x_1}
\,e^{\displaystyle i(k_{\rm D} - k_2)\cdot x_2}\,
e^{\displaystyle i(k - k_3)\cdot x_3}
$$
$$
\times\int d^3\rho \,U(\rho)\,\psi_{\rm np}(\rho) \int\frac{d^4k_1}{\pi^2i} 
e^{\displaystyle - i\vec{q}\cdot \vec{\rho}} 
{\rm tr}\Bigg\{\gamma^{\alpha}
\gamma^5\frac{1}{M_{\rm N} - \hat{k}_1 + \hat{k}_2} 
\gamma^{\mu} \frac{1}{M_{\rm N} - \hat{k}_1} \gamma^{\nu}\frac{1}{M_{\rm N} - \hat{k}_1 
- \hat{k}_3}\Bigg\}
$$
$$
-i\,e\,G_{\rm \pi np}\,\frac{g_{\rm V}}{8\pi^2}\,[\bar{u^c}(p_2)\gamma^5 u(p_1)]
\,e^*_{\mu}(k_{\rm D})\,e^*_{\nu}(k)\int d^4x_1 
$$
$$
\times \int \frac{d^4x_2 d^4k_2}{(2\pi)^4}
\frac{d^4x_3 d^4k_3}{(2\pi)^4}\,e^{\displaystyle
i (k_2 + k_3 - p_2 - p_1)\cdot x_1}\,e^{\displaystyle
i(k_{\rm D} - k_2)\cdot x_2}\,
e^{\displaystyle i(k - k_3)\cdot x_3}
$$
$$
\times\int d^3\rho \,U(\rho)\,
\psi_{\rm np}(\rho) \int\frac{d^4k_1}{\pi^2i} 
e^{\displaystyle - i\vec{q}\cdot \vec{\rho}} 
{\rm tr}\Bigg\{\gamma^5
\frac{1}{M_{\rm N} - \hat{k}_1 + \hat{k}_2}
\gamma^{\mu} \frac{1}{M_{\rm N} - \hat{k}_1} \gamma^{\nu}\frac{1}{M_{\rm N} - \hat{k}_1 
- \hat{k}_3}\Bigg\}
$$
$$
-i\,e\, G_{\rm \pi np}\,
\frac{\kappa_{\rm p}- \kappa_{\rm n}}{2 M_{\rm N}}
\frac{g_{\rm V}}{8\pi^2}\,
[\bar{u^c}(p_2)\gamma_{\alpha}
\gamma^5 u(p_1)]\,e^*_{\mu}(k_{\rm D})
\,k_{\nu}\,e^*_{\beta}(k)\int d^4x_1 
$$
$$
\times \int \frac{d^4x_2 d^4k_2}{(2\pi)^4}
\frac{d^4x_3 d^4k_3}{(2\pi)^4}\,
e^{\displaystyle i (k_2 + k_3 - p_2 - p_1)\cdot x_1}
\,e^{\displaystyle i(k_{\rm D} - k_2)\cdot x_2}
\,e^{\displaystyle i(k - k_3)\cdot x_3}
$$
$$
\times\int d^3\rho \,U(\rho)\,
\psi_{\rm np}(\rho) \int\frac{d^4k_1}{\pi^2i} 
e^{\displaystyle - i\vec{q}\cdot \vec{\rho}} 
{\rm tr}\Bigg\{\gamma^{\alpha}\gamma^5
\frac{1}{M_{\rm N} - \hat{k}_1 + \hat{k}_2}
\gamma^{\mu} \frac{1}{M_{\rm N} - \hat{k}_1} \sigma^{\nu\beta}\frac{1}{M_{\rm N} - 
\hat{k}_1 - \hat{k}_3}\Bigg\}
$$
$$
-i\,e\, G_{\rm \pi np}\,
\frac{\kappa_{\rm p}- \kappa_{\rm n}}{2 M_{\rm N}}
\frac{g_{\rm V}}{8\pi^2}\,[\bar{u^c}(p_2)\gamma^5 u(p_1)]\,e^*_{\mu}(k_{\rm D})\,k_{\nu}\,
e^*_{\beta}(k)\int d^4x_1 
$$
$$
\times \int \frac{d^4x_2 d^4k_2}{(2\pi)^4}
\frac{d^4x_3 d^4k_3}{(2\pi)^4}\,e^{\displaystyle
i (k_2 + k_3 - p_2 - p_1)\cdot x_1}\,
e^{\displaystyle i(k_{\rm D} - k_2)\cdot x_2}\,
e^{\displaystyle i(k - k_3)\cdot x_3}
$$
$$
\times\int d^3\rho \,U(\rho)\,\psi_{\rm np}(\rho) \int\frac{d^4k_1}{\pi^2i} e^{\displaystyle 
- i\vec{q}\cdot \vec{\rho}} {\rm tr}\Bigg\{
\gamma^5\frac{1}{M_{\rm N} - \hat{k}_1 + \hat{k}_2} 
\gamma^{\mu} \frac{1}{M_{\rm N} - \hat{k}_1} \sigma^{\nu\beta}\frac{1}{M_{\rm N} - \hat{k}_1 - \hat{k}_3}\Bigg\}.\eqno({\rm F}.14)
$$
Integrating over $x_1$, $x_2$, $x_3$,
$k_2$ and $k_3$ we obtain in the r.h.s. 
of Eq.~({\rm F}.14) the $\delta$--function describing the 4--momentum conservation. Then, the matrix element of the  neutron--proton radiative capture becomes equal
$$
{\cal M}({\rm n} + {\rm p} \to {\rm D} + \gamma) =  
-i\,e\,G_{\rm \pi np}\,\frac{g_{\rm V}}{8\pi^2}\,[\bar{u^c}(p_2)\gamma_{\alpha}\gamma^5 u(p_1)]\,e^*_{\mu}(k_{\rm D})\,e^*_{\nu}(k)
$$
$$
\times\int d^3\rho \,U(\rho)\,\psi_{\rm np}(\rho) \int\frac{d^4k_1}{\pi^2i} e^{\displaystyle - i\vec{q}\cdot \vec{\rho}} {\rm tr}\Bigg\{\gamma^{\alpha}\gamma^5\frac{1}{M_{\rm N} - \hat{k}_1 + \hat{k}_{\rm D}} \gamma^{\mu} \frac{1}{M_{\rm N} - \hat{k}_1} \gamma^{\nu}\frac{1}{M_{\rm N} - \hat{k}_1 - \hat{k}}\Bigg\}
$$
$$
 -i\,e\,G_{\rm \pi np}\,\frac{g_{\rm V}}{8\pi^2}\,[\bar{u^c}(p_2)\gamma^5 u(p_1)]\,
 e^*_{\mu}(k_{\rm D})\,e^*_{\nu}(k)
$$
$$
\times\int d^3\rho \,U(\rho)\,\psi_{\rm np}(\rho) \int\frac{d^4k_1}{\pi^2i} e^{\displaystyle 
- i\vec{q}\cdot \vec{\rho}} {\rm tr}\Bigg\{\gamma^5
\frac{1}{M_{\rm N} - \hat{k}_1 + \hat{k}_{\rm D}} 
\gamma^{\mu} \frac{1}{M_{\rm N} - \hat{k}_1} \gamma^{\nu}\frac{1}{M_{\rm N} - 
\hat{k}_1 - \hat{k}}\Bigg\}
$$
$$
-i\,e\, G_{\rm \pi np}\,\frac{\kappa_{\rm p}- 
\kappa_{\rm n}}{2 M_{\rm N}}\frac{g_{\rm V}}{8\pi^2}\,[\bar{u^c}(p_2)\gamma_{\alpha}\gamma^5 u(p_1)]\,e^*_{\mu}(k_{\rm D})\,k_{\nu}\,e^*_{\beta}(k)
$$
$$
\times\int d^3\rho \,U(\rho)\,\psi_{\rm np}(\rho) \int\frac{d^4k_1}{\pi^2i} e^{\displaystyle 
- i\vec{q}\cdot \vec{\rho}} 
{\rm tr}\Bigg\{\gamma^{\alpha}\gamma^5
\frac{1}{M_{\rm N} - \hat{k}_1 + \hat{k}_{\rm D}}
\gamma^{\mu} \frac{1}{M_{\rm N} - \hat{k}_1} \sigma^{\nu\beta}\frac{1}{M_{\rm N} -
\hat{k}_1 - \hat{k}}\Bigg\}
$$
$$
-i\,e\, G_{\rm \pi np}\,\frac{\kappa_{\rm p} -
\kappa_{\rm n}}{2 M_{\rm N}}\frac{g_{\rm V}}{8\pi^2}\,
[\bar{u^c}(p_2)
\gamma^5 u(p_1)]\,e^*_{\mu}(k_{\rm D})\,k_{\nu}
\,e^*_{\beta}(k)
$$
$$
\times\int d^3\rho \,U(\rho)\,\psi_{\rm np}(\rho) \int\frac{d^4k_1}{\pi^2i} e^{\displaystyle
- i\vec{q}\cdot \vec{\rho}} {\rm tr}\Bigg\{\gamma^5
\frac{1}{M_{\rm N} - \hat{k}_1 + \hat{k}_{\rm D}} 
\gamma^{\mu} \frac{1}{M_{\rm N} - \hat{k}_1} \sigma^{\nu\beta}\frac{1}{M_{\rm N} - \hat{k}_1 - \hat{k}}\Bigg\},\eqno({\rm F}.15)
$$
where $\vec{q} = \vec{k}_1 +(\vec{k} - 
\vec{k}_{\rm D})/2$.

As usually we represent the matrix element 
Eq.~({\rm F}.15) in terms of the structure functions
$$
{\cal M}({\rm n} + {\rm p} \to {\rm D} + \gamma) = 
$$
$$
-i\,e\,G_{\rm \pi np}\,\frac{g_{\rm V}}{8\pi^2}\,[\bar{u^c}(p_2)\gamma_{\alpha}\gamma^5 u(p_1)]\,e^*_{\mu}(k_{\rm D})\,e^*_{\nu}(k)\,
{\cal J}^{\alpha\mu\nu}_5(k_{\rm D},k;Q)
$$
$$
 -i\,e\,G_{\rm \pi np}\,\frac{g_{\rm V}}{8\pi^2}\,[\bar{u^c}(p_2)\gamma^5 u(p_1)]\,
 e^*_{\mu}(k_{\rm D})\,e^*_{\nu}(k)\,
 {\cal J}^{\mu\nu}_5(k_{\rm D},k;Q)
$$
$$
-i\,e\, G_{\rm \pi np}\,\frac{\kappa_{\rm p}- 
\kappa_{\rm n}}{2 M_{\rm N}}\frac{g_{\rm V}}{8\pi^2}\,[\bar{u^c}(p_2)\gamma_{\alpha}
\gamma^5 u(p_1)]\,e^*_{\mu}(k_{\rm D})\,k_{\nu}\,e^*_{\beta}(k)\,\bar{{\cal J}}^{\alpha\mu\nu\beta}_5(k_{\rm D},k;Q)
$$
$$
-i\,e\, G_{\rm \pi np}\,\frac{\kappa_{\rm p}- 
\kappa_{\rm n}}{2 M_{\rm N}}\frac{g_{\rm V}}{8\pi^2}\,[\bar{u^c}(p_2)\gamma^5 u(p_1)]\,
e^*_{\mu}(k_{\rm D})\,k_{\nu}\,e^*_{\beta}(k)\,
\bar{{\cal J}}^{\mu\nu\beta}_5
(k_{\rm D},k;Q),\eqno({\rm F}.16)
$$
where we have denoted
$$
{\cal J}^{\alpha\mu\nu}_5(k_{\rm D},k;Q) =\int d^3\rho \,U(\rho)\,\psi_{\rm np}(\rho) 
$$
$$
\times\int\frac{d^4k_1}{\pi^2i} e^{\displaystyle 
- i\vec{q}\cdot \vec{\rho}} 
{\rm tr}\Bigg\{\gamma^{\alpha}\gamma^5
\frac{1}{M_{\rm N} - \hat{k}_1 - 
\hat{Q} + \hat{k}_{\rm D}} \gamma^{\mu} 
\frac{1}{M_{\rm N} - \hat{k}_1 - \hat{Q}} 
\gamma^{\nu}\frac{1}{M_{\rm N} - \hat{k}_1 - 
\hat{Q} - \hat{k}}\Bigg\},
$$
$$
{\cal J}^{\mu\nu}_5(k_{\rm D},k;Q) =\int d^3\rho 
\,U(\rho)\,\psi_{\rm np}(\rho)  
$$
$$
\times\int\frac{d^4k_1}{\pi^2i} e^{\displaystyle 
- i\vec{q}\cdot \vec{\rho}} {\rm tr}\Bigg\{\gamma^5
\frac{1}{M_{\rm N} - \hat{k}_1 - \hat{Q} + \hat{k}_{\rm D}} \gamma^{\mu} \frac{1}{M_{\rm N} - \hat{k}_1 - \hat{Q}} \gamma^{\nu}\frac{1}{M_{\rm N} - \hat{k}_1 - \hat{Q} - \hat{k}}\Bigg\},
$$
$$
\bar{{\cal J}}^{\alpha\mu\nu\beta}_5
(k_{\rm D},k;Q) =\int d^3\rho \,U(\rho)\,
\psi_{\rm np}(\rho) 
$$
$$
\times\int\frac{d^4k_1}{\pi^2i} e^{\displaystyle - i\vec{q}\cdot \vec{\rho}} {\rm tr}\Bigg\{\gamma^{\alpha}\gamma^5
\frac{1}{M_{\rm N} - \hat{k}_1 - \hat{Q} 
+ \hat{k}_{\rm D}} \gamma^{\mu} 
\frac{1}{M_{\rm N} - \hat{k}_1 - \hat{Q}} \sigma^{\nu\beta}\frac{1}{M_{\rm N} - 
\hat{k}_1 - \hat{Q} - \hat{k}}\Bigg\},
$$
$$
\bar{{\cal J}}^{\mu\nu\beta}_5(k_{\rm D},k;Q) 
=\int d^3\rho \,U(\rho)\,\psi_{\rm np}(\rho)  
$$
$$
\times\int\frac{d^4k_1}{\pi^2i} e^{\displaystyle 
- i\vec{q}\cdot \vec{\rho}} {\rm tr}\Bigg\{\gamma^5
\frac{1}{M_{\rm N} - \hat{k}_1 - \hat{Q} + \hat{k}_{\rm D}} \gamma^{\mu} \frac{1}{M_{\rm N} - \hat{k}_1 - \hat{Q}} \sigma^{\nu\beta}
\frac{1}{M_{\rm N} - \hat{k}_1 - 
\hat{Q} - \hat{k}}\Bigg\}.\eqno({\rm F}.17)
$$
Here $Q = a\,k_{\rm D} + b\,k$ is an arbitrary shift of a virtual momentum, where $a$ and $b$ are arbitrary parameters.

First, let us restore our result for the matrix element of the neutron--proton radiative capture obtained for the $\delta^{(3)}(\vec{\rho}\,)$--potential [2,4]. In the case of the local four--nucleon interaction the structure functions are defined as follows:
$$
{\cal J}^{\alpha\mu\nu}_5(k_{\rm D},k;Q) =
$$
$$
\times\int\frac{d^4k_1}{\pi^2i} \,{\rm tr}\Bigg\{\gamma^{\alpha}\gamma^5
\frac{1}{M_{\rm N} - \hat{k}_1 - 
\hat{Q} + \hat{k}_{\rm D}} \gamma^{\mu} 
\frac{1}{M_{\rm N} - \hat{k}_1 - \hat{Q}} 
\gamma^{\nu}\frac{1}{M_{\rm N} - \hat{k}_1 
- \hat{Q} - \hat{k}}\Bigg\},
$$
$$
{\cal J}^{\mu\nu}_5(k_{\rm D},k;Q) = 
$$
$$
\times\int\frac{d^4k_1}{\pi^2i}\,
{\rm tr} \Bigg\{\gamma^5 
\frac{1}{M_{\rm N} - \hat{k}_1 - 
\hat{Q} + \hat{k}_{\rm D}} \gamma^{\mu} 
\frac{1}{M_{\rm N} - \hat{k}_1 - \hat{Q}} 
\gamma^{\nu}\frac{1}{M_{\rm N} - \hat{k}_1 
- \hat{Q} - \hat{k}}\Bigg\},
$$
$$
\bar{{\cal J}}^{\alpha\mu\nu\beta}_5(k_{\rm D},k;Q) =
$$
$$
\times\int\frac{d^4k_1}{\pi^2i} \,{\rm tr}\Bigg\{\gamma^{\alpha}\gamma^5
\frac{1}{M_{\rm N} - \hat{k}_1 - 
\hat{Q} + \hat{k}_{\rm D}} \gamma^{\mu}
\frac{1}{M_{\rm N} - \hat{k}_1 - \hat{Q}} \sigma^{\nu\beta}\frac{1}{M_{\rm N} - 
\hat{k}_1 - \hat{Q} - \hat{k}}\Bigg\},
$$
$$
\bar{{\cal J}}^{\mu\nu\beta}_5(k_{\rm D},k;Q) =
$$
$$
\times\int\frac{d^4k_1}{\pi^2i} \,{\rm tr}\Bigg\{\gamma^5
\frac{1}{M_{\rm N} - \hat{k}_1 - \hat{Q} + \hat{k}_{\rm D}} \gamma^{\mu} \frac{1}{M_{\rm N} - \hat{k}_1 - \hat{Q}} \sigma^{\nu\beta}\frac{1}{M_{\rm N} - \hat{k}_1 
- \hat{Q} - \hat{k}}\Bigg\}.\eqno({\rm F}.18)
$$
As has been shown in [2] the structure functions ${\cal J}^{\mu\nu}_5(k_{\rm D},k;Q)$ and $\bar{{\cal J}}^{\alpha\mu\nu\beta}_5(k_{\rm D},k;Q)$ are finite and unambiguously defined [2]
$$
{\cal J}^{\mu\nu}_5(k_{\rm D},k;Q) = \frac{2i}{M_{\rm N}}\,\varepsilon^{\mu\nu\alpha\beta}k_{\rm D \alpha} k_{\beta},
$$
$$
\bar{{\cal J}}^{\alpha\mu\nu\beta}_5(k_{\rm D},k;Q) = 
2M_{\rm N}\,i\,
\varepsilon^{\alpha\mu\nu\beta},
\eqno({\rm F}.19)
$$
where $\varepsilon^{0123} = 1$. In turn the structure 
functions ${\cal J}^{\alpha\mu\nu}_5
(k_{\rm D},k;Q)$ and $\bar{{\cal J}}^{\mu\nu\beta}_5
(k_{\rm D},k;Q)$ are defined ambiguously with respect to the shift of the virtual momentum $k_1 \to k_1 + Q$. Following the procedure [25] we obtain [2]
$$
{\cal J}^{\alpha\mu\nu}_5(k_{\rm D},k;Q) = 2i\varepsilon^{\alpha\mu\nu\beta}Q_{\beta} = 2i\varepsilon^{\alpha\mu\nu\beta}(a\,k_{\rm D} + b\,k)_{\beta},
$$
$$
\bar{{\cal J}}^{\mu\nu\beta}_5(k_{\rm D},k;Q) = - 2i\,\varepsilon^{\mu\nu\beta\alpha}Q_{\alpha} = - 2i\,\varepsilon^{\mu\nu\beta\alpha}(a\,k_{\rm D} + b\,k)_{\alpha}.\eqno({\rm F}.20)
$$
Since the shifts along $k_{\rm D}$ for ${\cal J}^{\alpha\mu\nu}_5(k_{\rm D},k;Q)$ violate the electromagnetic gauge invariance, we should set $a = 0$. Then, for $\bar{{\cal J}}^{\mu\nu\beta}_5(k_{\rm D},k;Q)$ the term proportional to the photon momentum does not contribute to the matrix element of the neutron--proton radiative capture. Therefore, we can set $b = 0$ too. Thus, the structure functions Eq.~({\rm F}.19) reduce themselves to the form
$$
{\cal J}^{\alpha\mu\nu}_5(k_{\rm D},k;Q) =2\,b\,i\varepsilon^{\alpha\mu\nu\beta}\,k_{\beta},
$$
$$
\bar{{\cal J}}^{\mu\nu\beta}_5(k_{\rm D},k;Q) = - 2\,a\,i\varepsilon^{\mu\nu\beta\alpha}\,k_{\rm D \alpha}.\eqno({\rm F}.21)
$$
For the structure functions Eq.~({\rm F}.19) and Eq.~({\rm F}.21) the matrix element of the neutron proton radiative capture is defined [2]: 
$$
{\cal M}({\rm n + p\to D + \gamma}) =
$$
$$
= (1 - a\,(\kappa_{\rm p} - \kappa_{\rm n}))\,\frac{e}{2 M_{\rm N}}\,\frac{g_{\rm V}}{4\pi^2}\,G_{\rm \pi np}\, \varepsilon^{\alpha\beta\mu\nu} k_{\alpha} e^*_{\beta}(k) e^*_{\mu}(k_{\rm D}) \,2\, k_{\rm D \nu} [\bar{u^c}(p_2)\gamma^5 u(p_1)] -
$$
$$
- (\kappa_{\rm p} - \kappa_{\rm n} - 2\,b) \,\frac{e}{2 M_{\rm N}}\,\frac{g_{\rm V}}{4\pi^2}\,G_{\rm \pi np}\, \varepsilon^{\alpha\beta\mu\nu} k_{\alpha} e^*_{\beta}(k) e^*_{\mu}(k_{\rm D})\,M_{\rm N}\,[\bar{u^c}(p_2)\gamma_{\nu} \gamma^5 u(p_1)].\eqno({\rm F}.22)
$$
As the neutron--proton radiative capture n + p $\to$ D + $\gamma$ is a magnetic dipole M1 transition, i.e., ${^1}{\rm S}_0 \to {^3}{\rm S}_1$, the matrix element of the transition should be proportional to the difference of the total magnetic dipole moments of the proton and the neutron $(\mu_{\rm p} - \mu_{\rm n})$.

It is reasonable to satisfy this constraint adjusting only ambiguously defined contributions. Thereby, setting $a = - 1$ and $b = - 1/2$ we get
$$
{\cal M}({\rm n + p\to D + \gamma}) = (\mu_{\rm p} - \mu_{\rm n}) \,\frac{e}{2 M_{\rm N}}\,\frac{g_{\rm V}}{4\pi^2}\,G_{\rm \pi np}\, \varepsilon^{\alpha\beta\mu\nu} k_{\alpha}\, e^*_{\beta}(k)\, e^*_{\mu}(k_{\rm D}) 
$$
$$
\times \,[\bar{u^c}(p_2)(2\, k_{\rm D \nu} - M_{\rm N}\gamma_{\nu})\gamma^5 u(p_1)].\eqno({\rm F}.23)
$$
The matrix element Eq.~({\rm F}.23) coincides completely with the matrix element Eq.~(5.15) of Ref.~[2].

In the low--energy limit when 
$$
[\bar{u^c}(p_2)\gamma_{\nu}\gamma^5 u(p_1)] \to - g_{\nu 0}\,[\bar{u^c}(p_2)\gamma^5 u(p_1)]
$$
and  $k_{\rm D \nu} \to g_{\nu 0}\,2 M_{\rm N}$ the matrix element Eq.~({\rm F}.23) acquires the form
$$
{\cal M}({\rm n + p\to D + \gamma}) = e\,(\mu_{\rm p} - \mu_{\rm n}) \,\frac{5 g_{\rm V}}{8\pi^2}\,G_{\rm \pi np}\,(\vec{k}\times \vec{e}^{\,*}(\vec{k}\,))\cdot \vec{e}^{\,*}(\vec{k}_{\rm D}) \,[\bar{u^c}(p_2)\gamma^5 u(p_1)].\eqno({\rm F}.24)
$$
Now let us proceed to the computation of the structure functions defined by the four--nucleon interaction Eq.~(\ref{label1.7}) with the Yukawa potential $U(\rho)$ given by Eq.~(\ref{label1.4}). Following the prescription of the RFMD we should expand the integrand of the structure functions in powers of $k_{\rm D}$ and $k$ and keep only leading contributions.

\noindent {\bf The calculation of the structure function ${\cal J}^{\mu\nu}_5(k_{\rm D},k;Q)$}. After the algebraical manipulations with the Dirac matrices we arrive at the expression
$$
{\cal J}^{\mu\nu}_5(k_{\rm D},k;Q) = - \int d^3\rho \,U(\rho)\,\psi_{\rm np}(\rho) \int\frac{d^4k_1}{\pi^2i} e^{\displaystyle - i\vec{q}\cdot \vec{\rho}}  
$$
$$
\times\,\frac{M_{\rm N}\,{\rm tr}\{\gamma^5 \hat{k}_{\rm D}\gamma^{\mu}\gamma^{\nu}\hat{k}\}}{[M^2_{\rm N} - (k_1 + Q - k_{\rm D})^2 -i0][M^2_{\rm N} - (k_1 + Q )^2 -i0][M^2_{\rm N} - (k_1 + Q + k)^2 -i0]}.\eqno({\rm F}.25) 
$$
It is seen that  the leading term of the structure function is given by the numerator which does not depend on both the virtual momentum $k_1$ and the shift $Q$. Therefore, taking away the trace over Dirac matrices we should calculate the residual integral at $k_{\rm D} = k = 0$:
$$
{\cal J}^{\mu\nu}_5(k_{\rm D},k;Q) = - {\rm tr}\{\gamma^5 \hat{k}_{\rm D}\gamma^{\mu}\gamma^{\nu}\hat{k}\}\int d^3\rho \,U(\rho)\,\psi_{\rm np}(\rho)\int\frac{d^4k_1}{\pi^2i} e^{\displaystyle - i\vec{k}_1\cdot \vec{\rho}}\,\frac{M_{\rm N}\,}{[M^2_{\rm N} - k^2_1 -i0]^3}.\eqno({\rm F}.26) 
$$
The computation of the trace is trivial and reads
$$
{\rm tr}\{\gamma^5 \hat{k}_{\rm D}\gamma^{\mu}\gamma^{\nu}\hat{k}\} = -  4\,i\,\varepsilon^{\mu\nu\alpha\beta}k_{\rm D \alpha}k_{\beta}.\eqno({\rm F}.27) 
$$
The residual integral over $k_1$ can be expressed in terms of the McDonald function:
$$
\int\frac{d^4k_1}{\pi^2i} e^{\displaystyle - i\vec{k}_1\cdot \vec{\rho}}\,\frac{M_{\rm N}}{[M^2_{\rm N} - k^2_1 -i0]^3} =  \int\frac{d^3k_1}{\pi^2i} e^{\displaystyle - i\vec{k}_1\cdot \vec{\rho}}\,\int\limits^{\infty}_{-\infty}\frac{dk_{10}M_{\rm N}\,}{[E^2_{\vec{k}_1} - k^2_{10} -i0]^3}=
$$
$$
=\frac{3\pi}{8}\int\frac{d^3k_1}{\pi^2} e^{\displaystyle - i\vec{k}_1\cdot \vec{\rho}}\,\frac{M_{\rm N}}{E^5_{\vec{k}_1}}= \frac{ M_{\rm N}}{2}\frac{3}{\rho}\int\limits^{\infty}_0\frac{\displaystyle d|\vec{k}_1||\vec{k}_1|\sin |\vec{k}_1|\rho}{\displaystyle (M^2_{\rm N} + \vec{k}^{\,2}_1)^{5/2}} = \frac{M_{\rm N}}{2}\int\limits^{\infty}_0\frac{\displaystyle d|\vec{k}_1|\cos |\vec{k}_1|\rho}{\displaystyle (M^2_{\rm N} + \vec{k}^{\,2}_1)^{3/2}}=
$$
$$
=\frac{1}{2}\,\rho\,K_1(M_{\rm N}\rho),
$$
$$
\int\frac{d^4k_1}{\pi^2i} e^{\displaystyle - i\vec{k}_1\cdot \vec{\rho}}\,\frac{M_{\rm N}}{[M^2_{\rm N} - k^2_1 -i0]^3} = \frac{1}{2}\,\rho\,K_1(M_{\rm N}\rho).\eqno({\rm F}.28) 
$$
Thus, the structure function ${\cal J}^{\mu\nu}_5(k_{\rm D},k;Q)$ is given by
$$
{\cal J}^{\mu\nu}_5(k_{\rm D},k;Q) = \frac{2i}{M_{\rm N}}\,\varepsilon^{\mu\nu\alpha\beta}k_{\rm D \alpha}k_{\beta}\,\int d^3\rho\,U(\rho)\,\psi_{\rm np}(\rho)\,M_{\rm N}\rho\,K_1(M_{\rm N}\rho)
$$
$$
= \frac{2i}{M_{\rm N}}\,\varepsilon^{\mu\nu\alpha\beta}k_{\rm D \alpha}k_{\beta}\,v_{\rm np}(0)\,a_{\rm np}\int \frac{d^3\rho}{\rho}\,U(\rho)\,M_{\rm N}\rho\,K_1(M_{\rm N}\rho) \times\,e^{\displaystyle i\delta_{\rm np}(K)}\,\frac{\sin\delta_{\rm np}(K)}{a_{\rm np} K},\eqno({\rm F}.29) 
$$
where we have inserted the wave function $\psi_{\rm np}(\rho)$ in the form of Eq.~(\ref{label9.3}) and $K$ is the relative 3--momentum of the neutron and the proton.

\noindent {\bf The calculation of the structure function $\bar{{\cal J}}^{\mu\nu\beta}_5(k_{\rm D},k;Q)$}. Emphasize, since the structure function $\bar{{\cal J}}^{\mu\nu\beta}_5(k_{\rm D},k;Q)$ enters to the matrix element of the transition n + p $\to$ D + $\gamma$ multiplied by the photon momentum $k_{\nu}$, therefore, in the integrand we should set $k=0$  and then expand in powers of $k_{\rm D}$ keeping only linear  terms. Recall, that we consider the neutron--proton radiative capture for thermal neutrons, therefore, the deuteron is almost at the rest, i.e., $k^{\mu}_{\rm D} = (k^0_{\rm D}, \vec{0}\,)$. For the computation of the momentum integral we should assume that $k^0_{\rm D} \ll M_{\rm N}$ and continue the final result on--mass shell of the deuteron (see Appendix C). At $k=0$ a 4--vector $Q$ is proportional to $k_{\rm D}$ only, i.e., $Q = a\,k_{\rm D}$.

For the computation of the momentum integral in $\bar{{\cal J}}^{\mu\nu\beta}_5(k_{\rm D},k;Q)$ we should take into account that the indices $\mu, \nu$ and $\beta$ should be spatial. Then, the computation runs as follows:
$$
\int\frac{d^4k_1}{\pi^2i} e^{\displaystyle - i\vec{k}_1\cdot \vec{\rho}}\,{\rm tr}\Bigg\{\gamma^5\frac{1}{M_{\rm N} - \hat{k}_1 - (a -1)\hat{k}_{\rm D}} \gamma^{\mu} \frac{1}{M_{\rm N} - \hat{k}_1 - a\,\hat{k}_{\rm D}} \sigma^{\nu\beta}\frac{1}{M_{\rm N} - \hat{k}_1 - a\,\hat{k}_{\rm D}}\Bigg\}=
$$
$$
=\int\frac{d^4k_1}{\pi^2i} e^{\displaystyle - i\vec{k}_1\cdot \vec{\rho}}\,\frac{1}{\displaystyle [E^2_{\vec{k}_1} - k^2_{10} - 2\,(a-1)\, k_{10} k^0_{\rm D} - i0][E^2_{\vec{k}_1} - k^2_{10} - 2\,a\,k_{10} k^0_{\rm D} - i0]^2}
$$
$$
\times\,
{\rm tr}
 \{\gamma^5 (M_{\rm N} + \hat{k}_1 + 
 (a-1)\,\hat{k}_{\rm D}) 
 \gamma^{\mu}(M_{\rm N} + \hat{k}_1 + 
 a \hat{k}_{\rm D}) \sigma^{\nu\beta}
 (M_{\rm N} + \hat{k}_1 + a \hat{k}_{\rm D})\}
$$
$$
=\int\frac{d^4k_1}{\pi^2i} e^{\displaystyle - i\vec{k}_1\cdot \vec{\rho}}\,\frac{1}{\displaystyle [E^2_{\vec{k}_1} - k^2_{10} - i0]^3}\,\Bigg\{1 + (6\,a - 2)\,\frac{k_{10}k^0_{\rm D}}{[E^2_{\vec{k}_1} - k^2_{10} - i0]} + \ldots \Bigg\}
$$
$$
\times\,
{\rm tr}\{\gamma^5 (M_{\rm N} + 
\hat{k}_1 + (a-1)\,\hat{k}_{\rm D}) \gamma^{\mu}
(M_{\rm N} + \hat{k}_1 + a \hat{k}_{\rm D}) 
 \sigma^{\nu\beta} (M_{\rm N} + \hat{k}_1 + a \hat{k}_{\rm D})\}.\eqno({\rm F}.30)
$$
Now we should bring up the trace over Dirac matrices to the more convenient form
$$
{\rm tr}\{\gamma^5 (M_{\rm N} + \hat{k}_1 + (a-1)\,\hat{k}_{\rm D}) \gamma^{\mu}(M_{\rm N} + \hat{k}_1 + a \hat{k}_{\rm D}) \sigma^{\nu\beta} (M_{\rm N} + \hat{k}_1 + a \hat{k}_{\rm D})\}=
$$
$$
={\rm tr}\{(M_{\rm N} + \hat{k}_1 + a \hat{k}_{\rm D})\gamma^5 (M_{\rm N} + \hat{k}_1 + (a-1)\,\hat{k}_{\rm D}) \gamma^{\mu}(M_{\rm N} + \hat{k}_1 + a \hat{k}_{\rm D}) \sigma^{\nu\beta}\}=
$$
$$
={\rm tr}\{\gamma^5 [(M^2_{\rm N} - (k_1 + a k_{\rm D})^2) - (M_{\rm N} - \hat{k}_1 - a \hat{k}_{\rm D}) \hat{k}_{\rm D}]\gamma^{\mu}(M_{\rm N} + \hat{k}_1 + a \hat{k}_{\rm D}) \sigma^{\nu\beta}\}=
$$
$$
={\rm tr}\{\gamma^5 [(M^2_{\rm N} - k^2_1) - 2a k_1\cdot k_{\rm D}- (M_{\rm N} - \hat{k}_1) \hat{k}_{\rm D}]\gamma^{\mu}(M_{\rm N} + \hat{k}_1 + a \hat{k}_{\rm D}) \sigma^{\nu\beta}\}=
$$
$$
= (M^2_{\rm N} - k^2_1){\rm tr}\{\gamma^5 \gamma^{\mu}\hat{k}_1\sigma^{\nu\beta}\} - 2a k_1\cdot k_{\rm D}{\rm tr}\{\gamma^5 \gamma^{\mu}\hat{k}_1\sigma^{\nu\beta}\} + a(M^2_{\rm N} - k^2_1){\rm tr}\{\gamma^5 \gamma^{\mu}\hat{k}_{\rm D}\sigma^{\nu\beta}\}
$$
$$
- M^2_{\rm N}{\rm tr}\{\gamma^5 \hat{k}_{\rm D}\gamma^{\mu}\sigma^{\nu\beta}\} + {\rm tr}\{\gamma^5 \hat{k}_1\hat{k}_{\rm D}\gamma^{\mu}\hat{k}_1\sigma^{\nu\beta}\}.\eqno({\rm F}.31)
$$
Substituting Eq.~({\rm F}.31) in Eq.~({\rm F}.30) and keeping only linear  terms in power momentum expansion we get
$$
\int\frac{d^4k_1}{\pi^2i} e^{\displaystyle - i\vec{k}_1\cdot \vec{\rho}}\,{\rm tr}\Bigg\{\gamma^5\frac{1}{M_{\rm N} - \hat{k}_1 - (a -1)\hat{k}_{\rm D}} \gamma^{\mu} \frac{1}{M_{\rm N} - \hat{k}_1 - a\,\hat{k}_{\rm D}} \sigma^{\nu\beta}\frac{1}{M_{\rm N} - \hat{k}_1 - a\,\hat{k}_{\rm D}}\Bigg\}=
$$
$$
=(4a-2)\,{\rm tr}\{\gamma^5 \gamma^{\mu}\hat{k}_{\rm D}\sigma^{\nu\beta}\} \int\frac{d^4k_1}{\pi^2i} e^{\displaystyle - i\vec{k}_1\cdot \vec{\rho}}\,\frac{k^2_{10}}{\displaystyle[E^2_{\vec{k}_1} - k^2_{10} - i0]^3} 
$$
$$
+ a\,{\rm tr}\{\gamma^5 \gamma^{\mu}\hat{k}_{\rm D}\sigma^{\nu\beta}\} \int\frac{d^4k_1}{\pi^2i} e^{\displaystyle - i\vec{k}_1\cdot \vec{\rho}}\,\frac{1}{\displaystyle[E^2_{\vec{k}_1} - k^2_{10} - i0]^2}
$$
$$
+ {\rm tr}\{\gamma^5 \gamma^{\mu}\hat{k}_{\rm D}\sigma^{\nu\beta}\}\int\frac{d^4k_1}{\pi^2i} e^{\displaystyle - i\vec{k}_1\cdot \vec{\rho}}\,\frac{M^2_{\rm N}}{\displaystyle[E^2_{\vec{k}_1} - k^2_{10} - i0]^3}
$$
$$
+ \int\frac{d^4k_1}{\pi^2i} e^{\displaystyle - i\vec{k}_1\cdot \vec{\rho}}\,\frac{{\rm tr}\{\gamma^5 \hat{k}_{\rm D}(\gamma^0 k_{10} + \vec{\gamma}\cdot\vec{k}_1)\gamma^{\mu}(\gamma^0 k_{10} - \vec{\gamma}\cdot\vec{k}_1) \sigma^{\nu\beta}\}}{\displaystyle[E^2_{\vec{k}_1} - k^2_{10} - i0]^3}
$$
$$
=(4a-2)\,{\rm tr}\{\gamma^5 \gamma^{\mu}\hat{k}_{\rm D}\sigma^{\nu\beta}\} \int\frac{d^4k_1}{\pi^2i} e^{\displaystyle - i\vec{k}_1\cdot \vec{\rho}}\,\frac{k^2_{10}}{\displaystyle[E^2_{\vec{k}_1} - k^2_{10} - i0]^3} 
$$
$$
+ a\,{\rm tr}\{\gamma^5 \gamma^{\mu}\hat{k}_{\rm D}\sigma^{\nu\beta}\} \int\frac{d^4k_1}{\pi^2i} e^{\displaystyle - i\vec{k}_1\cdot \vec{\rho}}\,\frac{1}{\displaystyle[E^2_{\vec{k}_1} - k^2_{10} - i0]^2}
$$
$$
+ {\rm tr}\{\gamma^5 \gamma^{\mu}\hat{k}_{\rm D}\sigma^{\nu\beta}\}\int\frac{d^4k_1}{\pi^2i} e^{\displaystyle - i\vec{k}_1\cdot \vec{\rho}}\,\frac{M^2_{\rm N}}{\displaystyle[E^2_{\vec{k}_1} - k^2_{10} - i0]^3}
$$
$$
+ {\rm tr}\{\gamma^5 \gamma^{\mu}\hat{k}_{\rm D}\sigma^{\nu\beta}\}\int\frac{d^4k_1}{\pi^2i} e^{\displaystyle - i\vec{k}_1\cdot \vec{\rho}}\,\frac{\displaystyle k^2_{10} + \frac{1}{3}\,\vec{k}^{\,2}_1}{\displaystyle[E^2_{\vec{k}_1} - k^2_{10} - i0]^3}=
$$
$$
=\Bigg(4a-\frac{2}{3}\Bigg)\,{\rm tr}\{\gamma^5 \gamma^{\mu}\hat{k}_{\rm D}\sigma^{\nu\beta}\} \int\frac{d^4k_1}{\pi^2i} e^{\displaystyle - i\vec{k}_1\cdot \vec{\rho}}\,\frac{k^2_{10}}{\displaystyle[E^2_{\vec{k}_1} - k^2_{10} - i0]^3} 
$$
$$
+ \Bigg(a + \frac{1}{3}\Bigg)\,{\rm tr}\{\gamma^5 \gamma^{\mu}\hat{k}_{\rm D}\sigma^{\nu\beta}\} \int\frac{d^4k_1}{\pi^2i} e^{\displaystyle - i\vec{k}_1\cdot \vec{\rho}}\,\frac{1}{\displaystyle[E^2_{\vec{k}_1} - k^2_{10} - i0]^2}
$$
$$
+ \frac{2}{3}\,{\rm tr}\{\gamma^5 \gamma^{\mu}\hat{k}_{\rm D}\sigma^{\nu\beta}\}\int\frac{d^4k_1}{\pi^2i} e^{\displaystyle - i\vec{k}_1\cdot \vec{\rho}}\,\frac{M^2_{\rm N}}{\displaystyle[E^2_{\vec{k}_1} - k^2_{10} - i0]^3}.\eqno({\rm F}.32)
$$
We have used here that $\hat{k}_{\rm D} = \gamma_0 k^0_{\rm D}$ and the relation $\vec{\gamma}\gamma^{\mu}\cdot\vec{\gamma} = \gamma^{\mu}$ valid for $\mu =1,2,3$. We have carried out the integration over the directions of $\vec{k}_1$ as if the integrand has been a spherically symmetric. It is true, since the factor $e^{\textstyle - i\vec{k}_1\cdot \vec{\rho}}$ can be taken as a spherically symmetric if to keep in mind that the integration over $\rho$ is spherically symmetric and should reduce this exponential to the form $\sin(|\vec{k}_1|\rho)/|\vec{k}_1|\rho$.

Integrating over $k_{10}$ we leave with the integral over $\vec{k}_1$:
$$
\int\frac{d^4k_1}{\pi^2i} e^{\displaystyle - i\vec{k}_1\cdot \vec{\rho}}\,{\rm tr}\Bigg\{\gamma^5\frac{1}{M_{\rm N} - \hat{k}_1 - (a -1)\hat{k}_{\rm D}} \gamma^{\mu} \frac{1}{M_{\rm N} - \hat{k}_1 - a\,\hat{k}_{\rm D}} \sigma^{\nu\beta}\frac{1}{M_{\rm N} - \hat{k}_1 - a\,\hat{k}_{\rm D}}\Bigg\}=
$$
$$
= \frac{1}{4}\,{\rm tr}\{\gamma^5 \gamma^{\mu}\hat{k}_{\rm D}\sigma^{\nu\beta}\}\int\frac{d^3k_1}{\pi} e^{\displaystyle - i\vec{k}_1\cdot \vec{\rho}}\,\frac{1}{E^3_{\vec{k}_1}}.\eqno({\rm F}.33)
$$
The integral over $\vec{k}_1$ can be expressed in terms of the McDonald function:
$$
\int\frac{d^3k_1}{\pi} e^{\displaystyle - i\vec{k}_1\cdot \vec{\rho}}\,\frac{1}{E^3_{\vec{k}_1}} = \frac{4}{\rho} \int\limits^{\infty}_0 \frac{d|\vec{k}_1||\vec{k}_1| \sin|\vec{k}_1|\rho}{(M^2_{\rm N} + \vec{k}^{\,2}_1)^{3/2}}= 
$$
$$
=4\int\limits^{\infty}_0 \frac{d|\vec{k}_1|\cos|\vec{k}_1|\rho}{(M^2_{\rm N} + \vec{k}^{\,2}_1)^{3/2}}= 4\,K_0(M_{\rm N}\rho).\eqno({\rm F}.34)
$$
This yields the structure function $\bar{\cal J}^{\mu\nu\beta}(k_{\rm D}, k; Q)$ in the form
$$
\bar{\cal J}^{\mu\nu\beta}(k_{\rm D}, k; Q) = -\, 4\,i\,\varepsilon^{\mu\nu\beta\alpha}k_{\rm D\alpha}\int d^3\rho\,U(\rho)\,\psi_{\rm np}(\rho)\,K_0(M_{\rm N}\rho)
$$
$$
= -\,4i\,\varepsilon^{\mu\nu\beta\alpha}k_{\rm D\alpha}\,v_{\rm np}(0)\,a_{\rm np}\int \frac{d^3\rho}{\rho}\,U(\rho)\,K_0(M_{\rm N}\rho) \times\,\,e^{\displaystyle i\delta_{\rm np}(K)}\,\frac{\sin\delta_{\rm np}(K)}{a_{\rm np} K},\eqno({\rm F}.35) 
$$
\noindent {\bf The calculation of the structure function $\bar{\cal J}^{\alpha\mu\nu\beta}(k_{\rm D},k;Q)$}. At leading order in the momentum expansion it does not depend on $Q$ and defined as
$$
\bar{{\cal J}}^{\alpha\mu\nu\beta}_5(k_{\rm D},k;Q) =\int d^3\rho \,U(\rho)\,\psi_{\rm np}(\rho)
$$
$$
\times \int\frac{d^4k_1}{\pi^2i} e^{\displaystyle - i\vec{k}_1\cdot \vec{\rho}}\,{\rm tr}\Bigg\{\gamma^{\alpha}\gamma^5\frac{1}{M_{\rm N} - \hat{k}_1} \gamma^{\mu} \frac{1}{M_{\rm N} - \hat{k}_1} \sigma^{\nu\beta}\frac{1}{M_{\rm N} - \hat{k}_1}\Bigg\}.
$$
The computation of the momentum integral runs as follows:
$$
\int\frac{d^4k_1}{\pi^2i} e^{\displaystyle - i\vec{k}_1\cdot \vec{\rho}}\,{\rm tr}\Bigg\{\gamma^{\alpha}\gamma^5\frac{1}{M_{\rm N} - \hat{k}_1} \gamma^{\mu} \frac{1}{M_{\rm N} - \hat{k}_1} \sigma^{\nu\beta}\frac{1}{M_{\rm N} - \hat{k}_1}\Bigg\}=
$$
$$
=-\int\frac{d^4k_1}{\pi^2i} e^{\displaystyle - i\vec{k}_1\cdot \vec{\rho}}\,\frac{{\rm tr}\{\gamma^5(M_{\rm N} - \hat{k}_1) \gamma^{\alpha}(M_{\rm N} + \hat{k}_1) \gamma^{\mu}(M_{\rm N} + \hat{k}_1) \sigma^{\nu\beta}\}}{[M^2_{\rm N} - k^2_1 - i0]^3}=
$$
$$
=-\int\frac{d^4k_1}{\pi^2i} e^{\displaystyle - i\vec{k}_1\cdot \vec{\rho}}\,\frac{{\rm tr}\{\gamma^5(M_{\rm N} - \hat{k}_1) \gamma^{\alpha}(M^2_{\rm N}\gamma^{\mu} + 2M_{\rm N}k^{\mu}_1 + \hat{k}_1\gamma^{\mu}\hat{k}_1) \sigma^{\nu\beta}\}}{[M^2_{\rm N} - k^2_1 - i0]^3}=
$$
$$
=-\int\frac{d^4k_1}{\pi^2i} e^{\displaystyle - i\vec{k}_1\cdot \vec{\rho}}\,\frac{{\rm tr}\{\gamma^5(M^3_{\rm N}\gamma^{\alpha}\gamma^{\mu} - 2M_{\rm N}k^{\mu}_1\hat{k}_1\gamma^{\alpha} + M_{\rm N}\gamma^{\alpha}\hat{k}_1\gamma^{\mu}\hat{k}_1) \sigma^{\nu\beta}\}}{[M^2_{\rm N} - k^2_1 - i0]^3}.\eqno({\rm F}.36)
$$
Integrating over directions of $\vec{k}_1$ and using the relations
$$
k^{\mu}_1\hat{k}_1\gamma^{\alpha} \to - \frac{1}{3}\,\vec{k}^{\,2}_1\gamma^{\mu}\gamma^{\alpha},
$$
$$
\gamma^{\alpha}\hat{k}_1 \gamma^{\mu}\hat{k}_1 \to \Bigg( - k^2_{10} + \frac{1}{3}\,\vec{k}^{\,2}_1\Bigg)\,\gamma^{\alpha}\gamma^{\mu}\eqno({\rm F}.37)
$$
valid for $\mu = 1,2,3$, we arrive at the expression
$$
\int\frac{d^4k_1}{\pi^2i} e^{\displaystyle - i\vec{k}_1\cdot \vec{\rho}}\,{\rm tr}\Bigg\{\gamma^{\alpha}\gamma^5\frac{1}{M_{\rm N} - \hat{k}_1} \gamma^{\mu} \frac{1}{M_{\rm N} - \hat{k}_1} \sigma^{\nu\beta}\frac{1}{M_{\rm N} - \hat{k}_1}\Bigg\}=
$$
$$
= - {\rm tr} \{\gamma^5\gamma^{\alpha}\gamma^{\mu}\sigma^{\nu\beta}\} M_{\rm N} \int\frac{d^4k_1}{\pi^2i} e^{\displaystyle - i\vec{k}_1\cdot \vec{\rho}}\,\frac{\displaystyle M^2_{\rm N} - \frac{1}{3}\,\vec{k}^{\,2}_1 - k^2_{10}}{[M^2_{\rm N} - k^2_1 - i0]^3}
$$
$$
= - {\rm tr}\{\gamma^5\gamma^{\alpha}\gamma^{\mu}\sigma^{\nu\beta}\} M_{\rm N} \int\frac{d^4k_1}{\pi^2i} e^{\displaystyle - i\vec{k}_1\cdot \vec{\rho}}\,\frac{\displaystyle \frac{4}{3}\,M^2_{\rm N} - \frac{1}{3}\,E^2_{\vec{k}_1} - k^2_{10}}{\displaystyle [E^2_{\vec{k}_1} - k^2_{10} - i0]^3}.\eqno({\rm F}.38)
$$
Then, we integrate over $k_{10}$ and get
$$
\int\frac{d^4k_1}{\pi^2i} e^{\displaystyle - i\vec{k}_1\cdot \vec{\rho}}\,{\rm tr}\Bigg\{\gamma^{\alpha}\gamma^5\frac{1}{M_{\rm N} - \hat{k}_1} \gamma^{\mu} \frac{1}{M_{\rm N} - \hat{k}_1} \sigma^{\nu\beta}\frac{1}{M_{\rm N} - \hat{k}_1}\Bigg\}=
$$
$$
= - \frac{1}{2}\,{\rm tr}\{\gamma^5\gamma^{\alpha}\gamma^{\mu}\sigma^{\nu\beta}\} M_{\rm N} \int\frac{d^3k_1}{\pi} e^{\displaystyle - i\vec{k}_1\cdot \vec{\rho}}\,\frac{\displaystyle M^2_{\rm N}}{\displaystyle E^5_{\vec{k}_1}}= 
$$
$$
=2\,i\,\varepsilon^{\alpha\mu\nu\beta} M_{\rm N}\int\frac{d^3k_1}{\pi} e^{\displaystyle - i\vec{k}_1\cdot \vec{\rho}}\,\frac{\displaystyle M^2_{\rm N}}{\displaystyle E^5_{\vec{k}_1}}.\eqno({\rm F}.39)
$$
The integral over $\vec{k}_1$ can be expressed in terms of the McDonald function:
$$
\int\frac{d^3k_1}{\pi} e^{\displaystyle - i\vec{k}_1\cdot \vec{\rho}}\,\frac{\displaystyle M^2_{\rm N}}{\displaystyle E^5_{\vec{k}_1}}= M^2_{\rm N}\,\frac{4}{\rho}\int\limits^{\infty}_0 \frac{d|\vec{k}_1||\vec{k}_1|\sin|\vec{k}_1|\rho}{(M^2_{\rm N} + \vec{k}^{\,2}_1)^{5/2}}=
$$
$$
=\frac{4}{3}\,M^2_{\rm N}\int\limits^{\infty}_0 \frac{d|\vec{k}_1|\cos|\vec{k}_1|\rho}{(M^2_{\rm N} + \vec{k}^{\,2}_1)^{5/2}}= \frac{4}{3}\,M_{\rm N}\rho\,K_1(M_{\rm N}\rho).\eqno({\rm F}.40)
$$
This gives the structure function $\bar{{\cal J}}^{\alpha\mu\nu\beta}_5(k_{\rm D},k;Q)$ in the form
$$
\bar{{\cal J}}^{\alpha\mu\nu\beta}_5(k_{\rm D},k;Q) = \frac{8}{3}\,M_{\rm N}\,i\,\varepsilon^{\alpha\mu\nu\beta}\int d^3\rho \,U(\rho)\,\psi_{\rm np}(\rho)\,M_{\rm N}\rho\,K_1(M_{\rm N}\rho)
$$
$$
= \frac{8}{3}\,M_{\rm N}\,i\,\varepsilon^{\alpha\mu\nu\beta}\,v_{\rm np}(0)\,a_{\rm np}\int \frac{d^3\rho}{\rho}\,U(\rho)\,M_{\rm N}\rho\,K_1(M_{\rm N}\rho)\times\,e^{\displaystyle i\delta_{\rm np}(K)}\,\frac{\sin\delta_{\rm np}(K)}{a_{\rm np} K},\eqno({\rm F}.41) 
$$
\noindent {\bf The calculation of the structure function ${\cal J}^{\alpha\mu\nu}_5(k_{\rm D},k;Q)$}. The computation of the momentum integral defining the structure function ${\cal J}^{\alpha\mu\nu}_5(k_{\rm D},k;Q)$ we perform as usually:
$$
\int\frac{d^4k_1}{\pi^2i} e^{\displaystyle - i\vec{q}\cdot \vec{\rho}} {\rm tr}\Bigg\{\gamma^{\alpha}\gamma^5\frac{1}{M_{\rm N} - \hat{k}_1 - \hat{Q} + \hat{k}_{\rm D}} \gamma^{\mu} \frac{1}{M_{\rm N} - \hat{k}_1 - \hat{Q}} \gamma^{\nu}\frac{1}{M_{\rm N} - \hat{k}_1 - \hat{Q} - \hat{k}}\Bigg\}=
$$
$$
\int\frac{d^4k_1}{\pi^2i} e^{\displaystyle - i\vec{q}\cdot \vec{\rho}} \,{\rm tr}\{\gamma^{\alpha}\gamma^5(M_{\rm N} + \hat{k}_1 + \hat{Q} - \hat{k}_{\rm D})\gamma^{\mu}(M_{\rm N} + \hat{k}_1 + \hat{Q}) \gamma^{\nu}(M_{\rm N} + \hat{k}_1 + \hat{Q} + \hat{k})\}
$$
$$
\times\frac{1}{[M^2_{\rm N} - (k_1 + Q -k_{\rm D})^2 - i0][M^2_{\rm N} - (k_1 + Q )^2 - i0][M^2_{\rm N} - (k_1 + Q + k)^2 - i0]}=
$$
$$
=\int\frac{d^4k_1}{\pi^2i} e^{\displaystyle - i\vec{k}_1\cdot \vec{\rho}} \,{\rm tr}\{\gamma^{\alpha}\gamma^5
(M_{\rm N} + \hat{k}_1 + \hat{Q} - \hat{k}_{\rm D})
\gamma^{\mu}(M_{\rm N} + \hat{k}_1 + \hat{Q}) 
\gamma^{\nu}(M_{\rm N} + \hat{k}_1 + 
\hat{Q} + \hat{k})\}
$$
$$
\times\frac{1}{[M^2_{\rm N} - k^2_1 - i0]^3}\,
\Bigg\{1 + \frac{2 k_1\cdot (3Q - k_{\rm D} 
+ k)}{[M^2_{\rm N} - k^2_1 - i0]} + 
\ldots \Bigg\}.\eqno({\rm F}.42)
$$
The trace over Dirac matrices we transform as follows:
$$
{\rm tr}\{\gamma^{\alpha}\gamma^5
(M_{\rm N} + \hat{k}_1 + \hat{Q} - \hat{k}_{\rm D})
\gamma^{\mu}(M_{\rm N} + \hat{k}_1 + \hat{Q}) 
\gamma^{\nu}(M_{\rm N} + \hat{k}_1 + 
\hat{Q} + \hat{k})\}=
$$
$$
=-{\rm tr}\{\gamma^5 \gamma^{\alpha} 
(M_{\rm N} + \hat{k}_1 + \hat{Q} - 
\hat{k}_{\rm D})[M^2_{\rm N}\gamma^{\mu}
\gamma^{\nu} + M_{\rm N}\gamma^{\mu}(\hat{k}_1 
+ \hat{Q})\gamma^{\nu} + M_{\rm N}\gamma^{\mu}
\gamma^{\nu}(\hat{k}_1 + \hat{Q} + \hat{k}) 
$$
$$
+ \gamma^{\mu}(\hat{k}_1 + \hat{Q}) \gamma^{\nu}
(\hat{k}_1 + \hat{Q} + \hat{k})]\}=
$$
$$
=-{\rm tr}\{M^2\gamma^5\gamma^{\alpha}(\hat{k}_1 + \hat{Q} - \hat{k}_{\rm D})\gamma^{\mu}\gamma^{\nu} + M^2_{\rm N}
\gamma^5 \gamma^{\alpha}\gamma^{\mu}(\hat{k}_1 + \hat{Q})
\gamma^{\nu} + M^2_{\rm N}\gamma^5 \gamma^{\alpha}\gamma^{\mu}\gamma^{\nu}(\hat{k}_1 + 
\hat{Q} + \hat{k}) 
$$
$$
+ \gamma^5 \gamma^{\alpha}(\hat{k}_1 + \hat{Q} - 
\hat{k}_{\rm D}) \gamma^{\mu}(\hat{k}_1 + \hat{Q}) \gamma^{\nu}(\hat{k}_1 + \hat{Q} + \hat{k})\}= 
$$
$$
=-{\rm tr}\{M^2_{\rm N}\gamma^5 \gamma^{\alpha}\gamma^{\mu}\gamma^{\nu}(\hat{k}_1 + 
\hat{Q} - \hat{k}_{\rm D} + \hat{k}) + \gamma^5
\gamma^{\alpha}(\hat{Q} - \hat{k}_{\rm D})\gamma^{\mu}\hat{k}_1\gamma^{\nu}\hat{k}_1 - 
\gamma^5 \hat{k}_1 \gamma^{\alpha} \hat{k}_1 
\gamma^{\mu} \hat{Q} \gamma^{\nu}
$$
$$
+ \gamma^5 \gamma^{\alpha} \hat{k}_1 \gamma^{\mu} 
\hat{k}_1 \gamma^{\nu} (\hat{Q} + \hat{k}) + \gamma^5
\gamma^{\alpha} \hat{k}_1 \gamma^{\mu} \hat{k}_1 
\gamma^{\nu} \hat{k}_1\}.\eqno({\rm F}.43)
$$
Substituting Eq.~({\rm F}.43) in Eq.~({\rm F}.42) we obtain
$$
\int\frac{d^4k_1}{\pi^2i} 
e^{\displaystyle - i\vec{k}_1\cdot \vec{\rho}} {\rm tr}\Bigg\{\gamma^{\alpha}\gamma^5
\frac{1}{M_{\rm N} - \hat{k}_1 - \hat{Q} + 
\hat{k}_{\rm D}} \gamma^{\mu} 
\frac{1}{M_{\rm N} - \hat{k}_1 - \hat{Q}}
\gamma^{\nu}\frac{1}{M_{\rm N} - \hat{k}_1 
- \hat{Q} - \hat{k}}\Bigg\}=
$$
$$
= - {\rm tr}\{\gamma^5\gamma^{\alpha}\gamma^{\mu}
\gamma^{\nu}(\hat{Q} - \hat{k}_{\rm D} + \hat{k})\}\int\frac{d^4k_1}{\pi^2i} 
e^{\displaystyle - i\vec{k}_1\cdot \vec{\rho}}\,
\frac{M^2_{\rm N}}{[M^2_{\rm N} - k^2_1 - i0]^3}
$$
$$
- M^2_{\rm N}\int\frac{d^4k_1}{\pi^2i} 
e^{\displaystyle - i\vec{k}_1\cdot \vec{\rho}}\,
\frac{{\rm tr}\{\gamma^5\gamma^{\alpha}\gamma^{\mu}
\gamma^{\nu}\hat{k}_1\}}{[M^2_{\rm N} - k^2_1 - i0]^4}\,2\,k_1\cdot(3\,Q - k_{\rm D} + k)
$$
$$
-\int\frac{d^4k_1}{\pi^2i} 
e^{\displaystyle - i\vec{k}_1\cdot \vec{\rho}}\,
\frac{1}{[M^2_{\rm N} - k^2_1 - i0]^3}\,
$$
$$
\times\,
{\rm tr}\{\gamma^5\gamma^{\alpha}
(\hat{Q} - \hat{k}_{\rm D})\gamma^{\mu}\hat{k}_1
\gamma^{\nu}\hat{k}_1 - \gamma^5 \hat{k}_1 
\gamma^{\alpha} \hat{k}_1 \gamma^{\mu} \hat{Q} 
\gamma^{\nu}
+ \gamma^5 \gamma^{\alpha} \hat{k}_1 \gamma^{\mu} 
\hat{k}_1 \gamma^{\nu} (\hat{Q} + \hat{k})\}
$$
$$
-\int\frac{d^4k_1}{\pi^2i} 
e^{\displaystyle - i\vec{k}_1\cdot \vec{\rho}}\,
\frac{{\rm tr}\{\gamma^5 \gamma^{\alpha} \hat{k}_1
\gamma^{\mu} \hat{k}_1 \gamma^{\nu} 
\hat{k}_1\}}{[M^2_{\rm N} - k^2_1 
- i0]^4}\,\,2\,k_1\cdot(3\,Q - 
k_{\rm D} + k).\eqno({\rm F}.44)
$$
Emphasize that in the low--energy limit the main contribution comes from the component with $\alpha=0$ and $\mu,\nu$ spatial. Therefore, due to $\hat{k}_{\rm D} = 
\gamma_0 k^0_{\rm D}$  the structure function ${\cal J}^{\alpha\mu\nu}_5
(k_{\rm D},k;Q)$ should be proportional to the photon momentum $k$. Thus, in the low--energy limit the momentum integral
Eq.~({\rm F}.44) can be transformed as follows:
$$
\int\frac{d^4k_1}{\pi^2i} 
e^{\displaystyle - i\vec{k}_1\cdot \vec{\rho}} 
{\rm tr}\Bigg\{\gamma^{\alpha}\gamma^5
\frac{1}{M_{\rm N} - \hat{k}_1 - \hat{Q} + 
\hat{k}_{\rm D}} \gamma^{\mu} \frac{1}{M_{\rm N} - 
\hat{k}_1 - \hat{Q}} \gamma^{\nu}
\frac{1}{M_{\rm N} - \hat{k}_1 - \hat{Q} - 
\hat{k}}\Bigg\}=
$$
$$
= - (b + 1)\,{\rm tr}\{\gamma^5 \gamma^{\alpha} 
\gamma^{\mu} \gamma^{\nu} \hat{k}\} 
\int\frac{d^4k_1}{\pi^2i} 
e^{\displaystyle - i\vec{k}_1\cdot \vec{\rho}}\,
\frac{M^2_{\rm N}}{[M^2_{\rm N} - k^2_1 - i0]^3}
$$
$$
- 2\,(3b+1)\,M^2_{\rm N}\int\frac{d^4k_1}{\pi^2i} 
e^{\displaystyle - i\vec{k}_1\cdot \vec{\rho}}\,
\frac{{\rm tr}\{\gamma^5 \gamma^{\alpha} \gamma^{\mu} \gamma^{\nu}\hat{k}_1\}}
{[M^2_{\rm N} - k^2_1 - i0]^4}\,
(k_1\cdot k)
$$
$$
-(3b+1)\int\frac{d^4k_1}{\pi^2i} 
e^{\displaystyle - i\vec{k}_1\cdot \vec{\rho}}\,
\frac{{\rm tr}\{\gamma^5 \gamma^{\alpha} \hat{k}_1 
\gamma^{\mu} \hat{k}_1 \gamma^{\nu} 
\hat{k}\}}{[M^2_{\rm N} - k^2_1 - i0]^3}
$$
$$
- 2(3b+1)\int\frac{d^4k_1}{\pi^2i} 
e^{\displaystyle - i\vec{k}_1\cdot \vec{\rho}}\,\frac{{\rm tr}\{\gamma^5 \gamma^{\alpha} \hat{k}_1 
\gamma^{\mu} \hat{k}_1 \gamma^{\nu}
\hat{k}_1\}}{[M^2_{\rm N} - k^2_1 - i0]^4}\,
(k_1\cdot k).\eqno({\rm F}.45)
$$
Now let us integrate over directions of $\vec{k}_1$:
$$
{\rm tr}\{\gamma^5 \gamma^{\alpha} \gamma^{\mu} \gamma^{\nu}\hat{k}_1\}\,(k_1\cdot k) \to - \frac{1}{3}\,\vec{k}^{\,2}_1\,{\rm tr}\{\gamma^5 
\gamma^{\alpha} \gamma^{\mu} \gamma^{\nu}\hat{k}\},
$$
$$
{\rm tr}\{\gamma^5 \gamma^{\alpha} \hat{k}_1 \gamma^{\mu} 
\hat{k}_1 \gamma^{\nu} \hat{k}\} \to \Bigg(- k^2_{10} + \frac{1}{3}\,\vec{k}^{\,2}_1\Bigg)\,{\rm tr}\{\gamma^5 
\gamma^{\alpha} \gamma^{\mu} \gamma^{\nu}\hat{k}\},
$$
$$
{\rm tr}\{\gamma^5 \gamma^{\alpha} \hat{k}_1 
\gamma^{\mu} \hat{k}_1 \gamma^{\nu} \hat{k}_1\}\,
(k_1\cdot k) = k^2_1\,{\rm tr}\{\gamma^5 
\gamma^{\alpha} \gamma^{\mu} \hat{k}_1 \gamma^{\nu}\}\,
(k_1\cdot k) =
$$
$$
\to \frac{1}{3}\,\vec{k}^{\,2}_1\,(k^2_{10} - 
\vec{k}^{\,2}_1)\,{\rm tr}\{\gamma^5 
\gamma^{\alpha} \gamma^{\mu} \gamma^{\nu}\hat{k}\}.
\eqno({\rm F}.46)
$$
Substituting Eq.~({\rm F}.46) in Eq.~({\rm F}.45) we arrive at the expression
$$
\int\frac{d^4k_1}{\pi^2i} 
e^{\displaystyle - i\vec{q}\cdot \vec{\rho}} {\rm tr}\Bigg\{\gamma^{\alpha}\gamma^5
\frac{1}{M_{\rm N} - \hat{k}_1 - \hat{Q} + \hat{k}_{\rm D}} \gamma^{\mu} \frac{1}{M_{\rm N} - \hat{k}_1 - \hat{Q}} \gamma^{\nu}\frac{1}{M_{\rm N} - \hat{k}_1 - \hat{Q} - \hat{k}}\Bigg\}=
$$
$$
={\rm tr}\{\gamma^5 \gamma^{\alpha} \gamma^{\mu} \gamma^{\nu} \hat{k}\}\,\Bigg\{ (- b - 1)\,\int\frac{d^4k_1}{\pi^2i} e^{\displaystyle - i\vec{k}_1\cdot \vec{\rho}}\,
\frac{M^2_{\rm N}}{\displaystyle  
[E^2_{\vec{k}_1} - k^2_{10} - i0]^3}
$$
$$
+\Bigg(2b+\frac{2}{3}\Bigg)\,M^2_{\rm N}
\int\frac{d^4k_1}{\pi^2i} e^{\displaystyle
- i\vec{k}_1\cdot \vec{\rho}}\, \frac{\displaystyle \vec{k}^{\,2}_1}{\displaystyle 
[E^2_{\vec{k}_1} - k^2_{10} - i0]^4}\,
$$
$$
+(-3\,b - 1)\int\frac{d^4k_1}{\pi^2i} e^{\displaystyle - i\vec{k}_1\cdot \vec{\rho}} \, \frac{\displaystyle -k^2_{10} + \frac{1}{3}\, \vec{k}^{\,2}_1}{\displaystyle [E^2_{\vec{k}_1} - k^2_{10} - i0]^3}
$$
$$
+\Bigg(- 2b - \frac{2}{3}\Bigg) \int\frac{d^4k_1}{\pi^2i} e^{\displaystyle - i\vec{k}_1\cdot \vec{\rho}} \, \frac{\displaystyle \vec{k}^{\,2}_1 (k^2_{10} - \vec{k}^{\,2}_1)}{\displaystyle [E^2_{\vec{k}_1} - k^2_{10} - i0]^4}\,\Bigg\}.\eqno({\rm F}.47)
$$
Integrating over $k_{10}$ we get
$$
\int\frac{d^4k_1}{\pi^2i} e^{\displaystyle - i\vec{q}\cdot \vec{\rho}} {\rm tr}\Bigg\{\gamma^{\alpha}\gamma^5\frac{1}{M_{\rm N} - \hat{k}_1 - \hat{Q} + \hat{k}_{\rm D}} \gamma^{\mu} \frac{1}{M_{\rm N} - \hat{k}_1 - \hat{Q}} \gamma^{\nu}\frac{1}{M_{\rm N} - \hat{k}_1 - \hat{Q} - \hat{k}}\Bigg\}=
$$
$$
={\rm tr}\{\gamma^5 \gamma^{\alpha} \gamma^{\mu} \gamma^{\nu} \hat{k}\}\,\Bigg\{ \Bigg(- \frac{3}{8}\,b - \frac{3}{8}\Bigg) \int\frac{d^3k_1}{\pi} e^{\displaystyle - i\vec{k}_1\cdot \vec{\rho}}\,\frac{M^2_{\rm N}}{\displaystyle   E^5_{\vec{k}_1}}
$$
$$
+\Bigg(\frac{5}{8}b+\frac{5}{24}\Bigg) \int\frac{d^4k_1}{\pi^2i} e^{\displaystyle - i\vec{k}_1\cdot \vec{\rho}}\, \frac{\displaystyle M^2_{\rm N}\vec{k}^{\,2}_1}{\displaystyle  E^7_{\vec{k}_1}}\,
$$
$$
+\Bigg(- \frac{3}{8}\,b - \frac{1}{8}\Bigg)\int\frac{d^3k_1}{\pi} e^{\displaystyle - i\vec{k}_1\cdot \vec{\rho}} \, \frac{\displaystyle E^2_{\vec{k}_1} + \vec{k}^{\,2}_1}{\displaystyle  E^5_{\vec{k}_1}}
$$
$$
+\Bigg(\frac{1}{4}\,b + \frac{1}{24}\Bigg) \int\frac{d^3k_1}{\pi} e^{\displaystyle - i\vec{k}_1\cdot \vec{\rho}} \, \frac{\displaystyle \vec{k}^{\,2}_1 (E^2_{\vec{k}_1} + 3\,\vec{k}^{\,2}_1)}{\displaystyle  E^7_{\vec{k}_1}}\,\Bigg\}=
$$
$$
={\rm tr}\{\gamma^5 \gamma^{\alpha} \gamma^{\mu} \gamma^{\nu} \hat{k}\}\,\Bigg\{ \Bigg(\frac{1}{4}\,b - \frac{1}{12}\Bigg)\int\frac{d^3k_1}{\pi} e^{\displaystyle - i\vec{k}_1\cdot \vec{\rho}}\,\frac{1}{\displaystyle   E^3_{\vec{k}_1}}
$$
$$
+ \Bigg(-\frac{9}{8}\,b - \frac{1}{32}\Bigg)\int\frac{d^3k_1}{\pi} e^{\displaystyle - i\vec{k}_1\cdot \vec{\rho}}\,\frac{M^2_{\rm N}}{\displaystyle   E^5_{\vec{k}_1}}
+ \Bigg(\frac{1}{8}\,b - \frac{1}{12}\Bigg)\int\frac{d^3k_1}{\pi} e^{\displaystyle - i\vec{k}_1\cdot \vec{\rho}}\,\frac{M^4_{\rm N}}{\displaystyle   E^7_{\vec{k}_1}}\Bigg\}.\eqno({\rm F}.48)
$$
We have found the structure function ${\cal J}^{\alpha\mu\nu} (k_{\rm D}, k;Q)$ dependent on the arbitrary shift of the virtual momentum:
$$
{\cal J}^{\alpha\mu\nu} (k_{\rm D}, k;Q) = -\,i\,\varepsilon^{\mu\nu\alpha\beta}k_{\beta} \int d^3\rho\,U(\rho)\,\psi_{\rm np}(\rho)\,\Bigg\{ \Bigg(b - \frac{1}{3}\Bigg)\int\frac{d^3k_1}{\pi} e^{\displaystyle - i\vec{k}_1\cdot \vec{\rho}}\,\frac{1}{\displaystyle   E^3_{\vec{k}_1}}
$$
$$
+ \Bigg(-\frac{9}{2}\,b - \frac{1}{8}\Bigg)\int\frac{d^3k_1}{\pi} e^{\displaystyle - i\vec{k}_1\cdot \vec{\rho}}\,\frac{M^2_{\rm N}}{\displaystyle   E^5_{\vec{k}_1}}
+ \Bigg(\frac{1}{2}\,b - \frac{1}{3}\Bigg)\int\frac{d^3k_1}{\pi} e^{\displaystyle - i\vec{k}_1\cdot \vec{\rho}}\,\frac{M^4_{\rm N}}{\displaystyle   E^7_{\vec{k}_1}}\Bigg\}.\eqno({\rm F}.49)
$$
In the more convenient form the structure function ${\cal J}^{\alpha\mu\nu} (k_{\rm D}, k;Q)$ can be defined as follows:
$$
{\cal J}^{\alpha\mu\nu} (k_{\rm D}, k;Q) =  i \varepsilon^{\alpha\mu\nu\beta} k_{\beta} \times\,{\cal C}\times\,e^{\displaystyle i\delta_{\rm np}(K)}\frac{\sin \delta_{\rm np}(K)}{a_{\rm np} K},\eqno({\rm F}.50)
$$
where the arbitrary constant ${\cal C}$ contains all uncertainties caused by  a shift of a virtual momentum: $k_1 \to k_1 + b\,k$. 

In terms of the structure functions the amplitude Eq.~({\rm F}.16) reads
$$
{\cal M}({\rm n + p} \to {\rm D + \gamma}) = e^{\displaystyle i\delta_{\rm np}(K)}\frac{\sin \delta_{\rm np}(K)}{a_{\rm np} K}\,\times\,\frac{e}{2M_{\rm N}}\,\frac{g_{\rm V}}{4\pi^2}\,G_{\rm \pi np}
$$
$$
\times\,\Bigg\{v_{\rm np}\,a_{\rm np}\int \frac{d^3\rho}{\rho}\,U(\rho)\,[M_{\rm N}\rho\,K_1(M_{\rm N}\rho) - (\kappa_{\rm p} - \kappa_{\rm n})\,K_0(M_{\rm N}\rho)]
$$
$$
\times\,\varepsilon^{\alpha\beta\mu\nu}\,k_{\alpha}\,e^*_{\beta}(k) \,e^*_{\mu}(k_{\rm D})\, [\bar{u^c}(p_2) 2 k_{\rm D\nu}\gamma^5 u(p_1)]
$$
$$
-\Bigg[ -2 \,{\cal C} + \frac{2}{3}\,(\kappa_{\rm p} - \kappa_{\rm n})\,v_{\rm np}\,a_{\rm np}\int \frac{d^3\rho}{\rho}\,U(\rho)\,[M_{\rm N}\rho\,K_1(M_{\rm N}\rho)\Bigg]
$$
$$
\times\,\varepsilon^{\alpha\beta\mu\nu}\,k_{\alpha}\,e^*_{\beta}(k) \,e^*_{\mu}(k_{\rm D})\, [\bar{u^c}(p_2) M_{\rm N}\gamma_{\nu} \gamma^5 u(p_1)]\Bigg\}.\eqno({\rm F}.51)
$$
In the low--energy limit the matrix element Eq.~({\rm F}.51) reads
$$
{\cal M}({\rm n + p} \to {\rm D + \gamma}) = e^{\displaystyle i\delta_{\rm np}(K)}\frac{\sin \delta_{\rm np}(K)}{a_{\rm np} K}\,
$$
$$
\times\,e\,\frac{g_{\rm V}}{8\pi^2}\,G_{\rm \pi np}\,\,(\vec{k}\times \vec{e}^{\,*}(\vec{k}\,))\cdot \vec{e}^{\,*}(\vec{k}_{\rm D}) \,[\bar{u^c}(p_2)\gamma^5 u(p_1)]
$$
$$
\times\,\Bigg\{ -2 \,{\cal C} + v_{\rm np}\,a_{\rm np}\int \frac{d^3\rho}{\rho}\,U(\rho)\,[M_{\rm N}\rho\,\Bigg[\Bigg(\frac{2}{3}\,(\kappa_{\rm p} - \kappa_{\rm n}) - \Bigg)\, K_1(M_{\rm N}\rho) - (\kappa_{\rm p} - \kappa_{\rm n})\,K_0(M_{\rm N}\rho)\Bigg]\Bigg\}.\eqno({\rm F}.52)
$$
The expression in the curls is completely arbitrary due to arbitrariness of ${\cal C}$. Therefore, we suggest to denote
$$
{\cal M}({\rm n + p} \to {\rm D + \gamma}) = e^{\displaystyle i\delta_{\rm np}(K)}\frac{\sin \delta_{\rm np}(K)}{a_{\rm np} K}\,
$$
$$
\times\,{\cal M}_0\times\,e\,(\mu_{\rm p} - \mu_{\rm n})\,\frac{g_{\rm V}}{8\pi^2}\,G_{\rm \pi np}\,\,(\vec{k}\times \vec{e}^{\,*}(\vec{k}\,))\cdot \vec{e}^{\,*}(\vec{k}_{\rm D}) \,[\bar{u^c}(p_2)\gamma^5 u(p_1)],\eqno({\rm F}.53)
$$
where ${\cal M}_0$ is an arbitrary parameter which we can fix by the consideration which has been used for the derivation of the low--energy theorem Eq.~(\ref{label8.7}). Indeed, in the low--energy limit $K\to 0$, when the np system  becomes localized in the region of order of $O(1/K)$ which is much larger than the range of nuclear forces, the wave function of the relative movement of the neutron and the proton can be described by a plane wave, and the matrix element of the neutron--proton radiative capture should not depend on the shape and the range of the nuclear potential [6]. Thereby, in the low--energy limit  the matrix element of the neutron--proton radiative capture ${\cal M}({\rm n + p} \to {\rm D + \gamma})_{\rm s.p}$ calculated for the smeared potential $U(\rho)$ should coincide with the matrix element ${\cal M}({\rm n + p} \to {\rm D + \gamma})_{\rm \delta.p}$ calculated for the $\delta^{(3)}(\vec{\rho}\,)$--potential. This gives ${\cal M}_0 = 5$. As a result the matrix element of the neutron--proton radiative capture calculated in the generalized RFMD reads
$$
{\cal M}({\rm n + p} \to {\rm D + \gamma}) = e^{\displaystyle i\delta_{\rm np}(K)}\frac{\sin \delta_{\rm np}(K)}{a_{\rm np} K}\,
$$
$$
\times\,e\,(\mu_{\rm p} - \mu_{\rm n})\,\frac{5 g_{\rm V}}{8\pi^2}\,G_{\rm \pi np}\,\,(\vec{k}\times \vec{e}^{\,*}(\vec{k}\,))\cdot \vec{e}^{\,*}(\vec{k}_{\rm D}) \,[\bar{u^c}(p_2)\gamma^5 u(p_1)].\eqno({\rm F}.54)
$$
The cross section for the process n + p $\to$ D + $\gamma$ is calculated in Sect.$\,10$, where we also compare our result with experimental data and the PMA and the EFT approach. 

\section*{Appendix G. Computation of cross section for low--energy elastic pp scattering}

In this Appendix we give the detailed calculation of the cross section for the low--energy elastic pp scattering in the ${^1}{\rm S}_0$--state caused by the effective strong local four--proton interaction
$$
{\cal L}^{\rm pp \to pp}_{\rm eff}(x)_{\rm cont.} = - \frac{2\pi\,a_{\rm pp}}{M_{\rm N}}\,\int d^3\rho \,\delta^{(3)}(\vec{\rho}\,)\,
$$
$$
[\bar{p}(t,\vec{x} + 
\frac{1}{2}\,\vec{\rho}\,)\,\gamma^5
p^c (t,\vec{x} - \frac{1}{2}\,\vec{\rho}\,)]
\,[\bar{p^c}(t,\vec{x} + \frac{1}{2}\,\vec{\rho}\,)\,\gamma^5 p(t,\vec{x} - \frac{1}{2}\,\vec{\rho}\,)],\eqno({\rm G}.1)
$$
In the quantum field theory the amplitude of the pp scattering is defined
$$
\int d^4x\,<p(p^{\prime}_2) p(p^{\prime}_1)|{\cal L}^{\rm cont}_{\rm eff}(x)|p(p_1) p(p_2)> =
$$
$$
= (2\pi)^4\,\delta^{(4)}(p^{\prime}_1 + p^{\prime}_2 - p_1 - p_2)\,\frac{{\cal M}(p(p_1) + p(p_2) \to p(p^{\prime}_1) + p(p^{\prime}_2))}{\displaystyle \sqrt{2 E^{\prime}_1 V\,2 E^{\prime}_2 V\,2 E_1 V\,2 E_2 V}},\eqno({\rm G}.2)
$$
where $E_i\,(E^{\prime}_i)\,(i=1,2)$ are the energies of the protons in the initial(final) state.

Substituting the effective Lagrangian Eq.~({\rm G}.1) in the l.h.s. of Eq.~({\rm G}.2) we get
$$
\int d^4x\,<p(p^{\prime}_2) p(p^{\prime}_1)|{\cal L}^{\rm pp \to pp}_{\rm eff}(x)_{\rm cont.}|p(p_1) p(p_2)> =
$$
$$
= - \frac{2\pi\,a_{\rm pp}}{M_{\rm N}}\,\int d^4x \int d^3\rho \,\delta^{(3)}(\vec{\rho}\,)\,
<p(p^{\prime}_2) p(p^{\prime}_1)|
[\bar{p}(t,\vec{x} + \frac{1}{2}\,\vec{\rho}\,)\,
\gamma^5 p^c (t,\vec{x} - \frac{1}{2}\,\vec{\rho}\,)]
$$
$$
\times\,[\bar{p^c}(t,\vec{x} + \frac{1}{2}\,\vec{\rho}\,)\,\gamma^5 p(t,\vec{x} - \frac{1}{2}\,\vec{\rho}\,)]|p(p_1) p(p_2)>=
$$
$$
= - \frac{2\pi\,a_{\rm pp}}{M_{\rm N}}\,\int d^3\rho \,\delta^{(3)}(\vec{\rho}\,)\,
<p(p^{\prime}_2) p(p^{\prime}_1)|[\bar{p}
(t,\vec{x} + \frac{1}{2}\,\vec{\rho}\,)\,
\gamma^5 p^c
(t,\vec{x} - \frac{1}{2}\,\vec{\rho}\,)]|0>
$$
$$
\times\,<0|[\bar{p^c}(t,\vec{x} + \frac{1}{2}\,\vec{\rho}\,)\,
\gamma^5 p(t,\vec{x} - \frac{1}{2}\,\vec{\rho}\,)]|p(p_1) p(p_2)>.
\eqno({\rm G}.3)
$$
For the computation of the matrix elements in the r.h.s. of
Eq.~({\rm G}.3) we should write down the wave functions of 
the initial 
$|p(p_1) p(p_2)>$ and the final
$<p(p^{\prime}_2) 
p(p^{\prime}_1)|$ states. 

Since the protons are coupled in the ${^1}{\rm S}_0$--state and correlated in the initial state, we should take the wave function $|p(p_1) p(p_2)>$ in the symmetrized form Eq.~({\rm C}.26):
$$
|p(p_1) p(p_2)> =\frac{1}{\sqrt{2}} 
a^{\dagger}(\vec{p}_1,\sigma_1)\,
a^{\dagger}(\vec{p}_2,\sigma_2)|0>.
\eqno({\rm C}.26)
$$
The wave function of the final state 
$<p(p^{\prime}_2)
p(p^{\prime}_1)|$ 
we take in the form
$$
<p(p^{\prime}_2) p(p^{\prime}_1)| = <0|a(\vec{p}^{\,\prime}_2,\sigma^{\prime}_2)\, a(\vec{p}^{\,\prime}_1,\sigma^{\prime}_1).
\eqno({\rm G}.4)
$$
This wave function is antisymmetric under the permutations of the protons. The squared normalization factor $1/\sqrt{2}$ will be taken into account as usually for the computation of the phase volume.

For the proton field operators we use the plane--wave expansions Eq.~({\rm C}.27) and 
$$
\bar{p}(t,\vec{x} + \frac{1}{2}\,\vec{\rho}\,) = \sum_{\vec{q}^{\,\prime}_1,\alpha^{\prime}_1}\frac{1}{\displaystyle \sqrt{2E_{\vec{q}^{\,\prime}_1}V}}\Bigg[ a^{\dagger}(\vec{q}^{\,\prime}_1,\alpha^{\prime}_1)\,\bar{u}(q^{\prime}_1)\,e^{\displaystyle iE_{\vec{q}^{\,\prime}_1}t - i\vec{q}^{\,\prime}_1\cdot (\vec{x} + \vec{\rho}/2)}
$$
$$
\hspace{0.5in}+  
b(\vec{q}^{\,\prime}_1,\alpha^{\prime}_1)\,
\bar{v}(q^{\prime}_1)\,
e^{\displaystyle -iE_{\vec{q}^{\,\prime}_1}t + i\vec{q}^{\,\prime}_1\cdot (\vec{x} 
+ \vec{\rho}/2)}\Bigg],
$$
$$
p^c(t,\vec{x} - \frac{1}{2}\,\vec{\rho}\,) = \sum_{\vec{q}^{\,\prime}_2,\alpha^{\prime}_2}
\frac{1}{\displaystyle \sqrt{2E_{\vec{q}^{\,\prime}_2}V}}
\Bigg[ a^{\dagger}(\vec{q}^{\,\prime}_2,\alpha^{\prime}_2)\,
u^c(q^{\prime}_2)\,
e^{\displaystyle iE_{\vec{q}^{\,\prime}_2}t - i\vec{q}^{\,\prime}_2\cdot (\vec{x} - \vec{\rho}/2)}
$$
$$
 +  b(\vec{q}^{\,\prime}_2,\alpha^{\prime}_2)\,v^c(q^{\prime}_2)\,e^{\displaystyle -iE_{\vec{q}^{\,\prime}_2}t + i\vec{q}^{\,\prime}_2\cdot (\vec{x} - \vec{\rho}/2)}\Bigg],\eqno({\rm G}.5)
$$
Now we can compute the matrix elements in Eq.~({\rm G}.3). Keeping only the terms containing the operators of the creation and the annihilation of the protons and applying the anti--commutation relations Eq.~({\rm C}.30) we get
$$
<0|\bar{p^c}(t,\vec{x} + \frac{1}{2}\,\vec{\rho}\,)\,\gamma^5 p(t,\vec{x} - \frac{1}{2}\,\vec{\rho}\,)|p(p_1) p(p_2)> =
$$
$$
=\sum_{\vec{q}_1,\alpha_1}\sum_{\vec{q}_2,\alpha_2}\frac{1}{\displaystyle \sqrt{2E_{\vec{q}_1}V}}\frac{1}{\displaystyle \sqrt{2E_{\vec{q}_2}V}}\,e^{\displaystyle -i(q_1 + q_2)\cdot x + i(\vec{q}_1 - \vec{q}_2)\cdot \vec{\rho}/2}
$$
$$
\times\,[\bar{u^c}(q_1)\,\gamma^5\,u(q_2)]\,
\frac{1}{\sqrt{2}}
<0|a(\vec{q}_1,\alpha_1) 
a(\vec{q}_2,\alpha_2) 
a^{\dagger}(\vec{p}_1,\sigma_1)\,
a^{\dagger}(\vec{p}_2,\sigma_2)|0>=
$$
$$
=\sum_{\vec{q}_1,\alpha_1}
\sum_{\vec{q}_2,\alpha_2}
\frac{1}{\displaystyle 
\sqrt{2E_{\vec{q}_1}V}}\frac{1}{\displaystyle \sqrt{2E_{\vec{q}_2}V}}\,e^{\displaystyle 
-i(q_1 + q_2)\cdot x + i(\vec{q}_1 - 
\vec{q}_2)\cdot \vec{\rho}/2}
$$
$$
\times\,[\bar{u^c}(q_1)\,\gamma^5\,u(q_2)]\,
\frac{1}{\sqrt{2}}(-\delta_{\vec{q}_1\vec{p}_1}\,
\delta_{\alpha_1\sigma_1}\,
\delta_{\vec{q}_2\vec{p}_2}\,
\delta_{\alpha_2\sigma_2} + 
\delta_{\vec{q}_2\vec{p}_1}\,
\delta_{\alpha_2\sigma_1}\,
\delta_{\vec{q}_1\vec{p}_2}\,
\delta_{\alpha_1\sigma_2}) =
$$
$$
= \frac{\displaystyle e^{\displaystyle 
-i(p_1 + p_2)\cdot x}}{\displaystyle 
\sqrt{2E_1V\,2E_2V}}\,
$$
$$
\times\,\frac{1}{\sqrt{2}}\,\Bigg(- [\bar{u^c}(p_1)\,\gamma^5\,u(p_2)]\,
e^{\displaystyle 
i(\vec{p}_1 - \vec{p}_2)\cdot \vec{\rho}/2} +  [\bar{u^c}(p_2)\,\gamma^5\,u(p_1)]\,
e^{\displaystyle - 
i(\vec{p}_1 - \vec{p}_2)\cdot \vec{\rho}/2}\Bigg) = 
$$
$$
= \frac{\displaystyle 
e^{\displaystyle -i(p_1 + p_2)\cdot x}}{\displaystyle \sqrt{2E_1V\,2E_2V}}\,[\bar{u^c}(p_2)\,\gamma^5\,u(p_1)]\,
\left(\frac{\displaystyle e^{\displaystyle i(\vec{p}_1 - \vec{p}_2)\cdot \vec{\rho}/2} +  
e^{\displaystyle - 
i(\vec{p}_1 - \vec{p}_2)\cdot \vec{\rho}/2}}{\sqrt{2}}
\right).
$$
$$
= \frac{\displaystyle
e^{\displaystyle 
-i(p_1 + p_2)\cdot x}}{\displaystyle \sqrt{2E_1V\,2E_2V}}\,[\bar{u^c}(p_2)\,
\gamma^5\,u(p_1)]\,\left(\frac{\displaystyle 
e^{\displaystyle i \vec{k}\cdot \vec{\rho}} +  
e^{\displaystyle - i \vec{k}\cdot \vec{\rho}}}{\sqrt{2}}\right),\eqno({\rm G}.6)
$$
$$
<p(p^{\prime}_2) p(p^{\prime}_1)|
[\bar{p}(t,\vec{x} + \frac{1}{2}\,\vec{\rho}\,)\,
\gamma^5 p^c (t,\vec{x} - \frac{1}{2}\,\vec{\rho}\,)]|0>=
$$
$$
=\sum_{\vec{q}^{\,\prime}_1,\alpha^{\prime}_1}
\sum_{\vec{q}^{\,\prime}_2,\alpha^{\prime}_2}
\frac{1}{\displaystyle \sqrt{2E_{\vec{q}^{\,\prime}_1}V}}
\frac{1}{\displaystyle \sqrt{2E_{\vec{q}^{\,\prime}_2}V}}\,
e^{\displaystyle  i(q^{\prime}_1 + q^{\prime}_2)\cdot x - i(\vec{q}^{\,\prime}_1 - \vec{q}^{\,\prime}_2)
\cdot \vec{\rho}/2}
$$
$$
\times\,[\bar{u}(q^{\prime}_1)\,\gamma^5\,
u^c(q^{\prime}_2)]\,
(-\delta_{\vec{q}^{\,\prime}_1\vec{p}^{\,\prime}_1}\,
\delta_{\alpha^{\prime}_1\sigma^{\prime}_1}\,
\delta_{\vec{q}^{\,\prime}_2\vec{p}^{\,\prime}_2}\,
\delta_{\alpha^{\prime}_2\sigma^{\prime}_2} + \delta_{\vec{q}^{\,\prime}_2\vec{p}^{\,\prime}_1}\,
\delta_{\alpha^{\prime}_2\sigma^{\prime}_1}\,
\delta_{\vec{q}^{\,\prime}_1\vec{p}^{\,\prime}_2}\,
\delta_{\alpha^{\prime}_1\sigma^{\prime}_2})=
$$
$$
= \frac{\displaystyle e^{\displaystyle
i(p^{\prime}_1 + p^{\prime}_2)\cdot x}}{\displaystyle \sqrt{2E^{\prime}_1V\,2E^{\prime}_2V}}\,
$$
$$
\times\,\Bigg([- \bar{u}(p^{\prime}_1)\,\gamma^5\,
u(p^{\prime}_2)]\,
e^{\displaystyle - i(\vec{p}^{\,\prime}_1 - 
\vec{p}^{\,\prime}_2)\cdot \vec{\rho}/2} +  [\bar{u}(p^{\prime}_2)\,\gamma^5\,u^c(p^{\prime}_1)]\,
e^{\displaystyle  i(\vec{p}^{\,\prime}_1 - 
\vec{p}^{\,\prime}_2)\cdot \vec{\rho}/2}\Bigg) = 
$$
$$
= \frac{\displaystyle e^{\displaystyle i(p^{\prime}_1 + p^{\prime}_2)\cdot x}}{\displaystyle \sqrt{2E^{\prime}_1V\,2E^{\prime}_2V}}\,[\bar{u}(p^{\prime}_2)\,
\gamma^5\,u(p^{\prime}_1)]\,\left(\displaystyle e^{\displaystyle i \vec{k}^{\,\prime}\cdot \vec{\rho}} +  e^{\displaystyle - i \vec{k}^{\,\prime}\cdot \vec{\rho}}\right),\eqno({\rm G}.7)
$$
where $\vec{k} =(\vec{p}_1 - \vec{p}_2)/2$ and $\vec{k}^{\,\prime} =(\vec{p}^{\,\prime}_1 - \vec{p}^{\,\prime}_2)/2$ are the relative 3--momenta of the protons in the initial and the final state such as $|\vec{k}| = |\vec{k}^{\,\prime}\,| = k$. We have used too the relations: $[\bar{u^c}(p_1)\gamma^5 u(p_2)] = - [\bar{u^c}(p_2)\gamma^5 u(p_1)]$  and $[\bar{u}(p^{\prime}_1)\gamma^5 u^c(p^{\prime}_2)] = - [\bar{u}(p^{\prime}_2)\gamma^5 u^c(p^{\prime}_1)]$.

Expanding into spherical harmonics and keeping only the S--wave contributions we bring up the matrix elements Eq.~({\rm G}.6) and Eq.~({\rm G}.7) to the form
$$
<0|\bar{p^c}(t,\vec{x} + \frac{1}{2}\,\vec{\rho}\,)\,\gamma^5 p(t,\vec{x} - \frac{1}{2}\,\vec{\rho}\,)|p(p_1) p(p_2)> =
$$
$$
= \frac{\displaystyle e^{\displaystyle -i(p_1 + p_2)\cdot x}}{\displaystyle \sqrt{2E_1V\,2E_2V}}\,[\bar{u^c}(p_2)\,\gamma^5\,u(p_1)]\,
\sqrt{2}\,\frac{\sin k\rho}{k\rho},\eqno({\rm G}.8)
$$
$$
<p(p^{\prime}_2) p(p^{\prime}_1)|[\bar{p}
(t,\vec{x} + \frac{1}{2}\,\vec{\rho}\,)\,
\gamma^5 p^c 
(t,\vec{x} - \frac{1}{2}\,\vec{\rho}\,)]|0>=
$$
$$
= \frac{\displaystyle e^{\displaystyle i(p^{\prime}_1 + p^{\prime}_2)\cdot x}}{\displaystyle \sqrt{2E^{\prime}_1V\,2E^{\prime}_2V}}\,[\bar{u}(p^{\prime}_2)\gamma^5 u^c(p^{\prime}_1)]\,2\,\frac{\sin k\rho}{k\rho}.\eqno({\rm G}.9)
$$
Substituting Eq.~({\rm G}.8) and  Eq.~({\rm G}.9) in  Eq.~({\rm G}.3) we obtain
$$
\int d^4x\,<p(p^{\prime}_2) p(p^{\prime}_1)|
{\cal L}^{\rm cont}_{\rm eff}(x)|p(p_1) p(p_2)> =
$$
$$
= - \frac{4\sqrt{2}\pi}{M_{\rm N}}\,a_{\rm pp}
\int d^4x \frac{\displaystyle e^{\displaystyle 
i(p^{\prime}_1 + p^{\prime}_2 -
p_1 - p_2)\cdot x}}{\displaystyle \sqrt{2E^{\prime}_1V
\,2E^{\prime}_2V
\,2 E_1V\,2 E_2 V}}
$$
$$
\int d^3 \rho \,\delta^{(3)}(\vec{\rho}\,)\,
\frac{\sin^2 k\rho}{k^2\rho^2}\,[\bar{u}(p^{\prime}_2)
\gamma^5 u^c(p^{\prime}_1)]\,[\bar{u^c}(p_2)
\,\gamma^5\,u(p_1)]=
$$
$$
=(2\pi)^4\,\delta^{(4)}(p^{\prime}_1 + p^{\prime}_2 - p_1 - p_2)\,\frac{1}{\displaystyle 
\sqrt{2E^{\prime}_1V\,2E^{\prime}_2V\,
2 E_1V\,2 E_2 V}}
$$
$$
\times\,\Bigg\{- \frac{4\sqrt{2}\pi}{M_{\rm N}}\,
a_{\rm pp}\,[\bar{u}(p^{\prime}_2)\gamma^5 u^c(p^{\prime}_1)]\,[\bar{u^c}(p_2)\,
\gamma^5\,u(p_1)]\Bigg\}.\eqno({\rm G}.10)
$$
This gives
$$
{\cal M}(p(p_1) + p(p_2) \to p(p^{\prime}_1) + 
p(p^{\prime}_2)) = - \frac{4\sqrt{2}\pi}{M_{\rm N}}
\,a_{\rm pp}[\bar{u}(p^{\prime}_2)\gamma^5 u^c(p^{\prime}_1)]\,[\bar{u^c}(p_2)\,
\gamma^5\,u(p_1)].\eqno({\rm G}.11)
$$
The amplitude Eq.~({\rm G}.11) squared and averaged over the polarizations of the initial protons and summed over polarizations of the final protons reads
$$
\overline{|{\cal M}(p(p_1) + p(p_2) \to p(p^{\prime}_1) + p(p^{\prime}_2))|^2} = \frac{32\pi^2}{M^2_{\rm N}}
\,a^2_{\rm pp} 
$$
$$
\times\,\frac{1}{4}\,{\rm tr}\{(M_{\rm N} + \hat{p}^{\prime}_2)\gamma^5(M_{\rm N} - 
\hat{p}^{\prime}_1)\gamma^5\}
{\rm tr}\{(M_{\rm N} + \hat{p}_1)
\gamma^5(M_{\rm N} - \hat{p}_2)\gamma^5\} = 
\frac{32\pi^2}{M^2_{\rm N}}\,a^2_{\rm pp}\,s^2,
\eqno({\rm G}.12)
$$
where $s = (p_1 + p_2)^2 =(p^{\prime}_1 
+ p^{\prime}_2)^2$.

The cross section for the low--energy elastic pp scattering is defined
$$
\sigma({\rm pp}\to {\rm pp}) = \frac{1}{4 k \sqrt{s}}\int \overline{|{\cal M}(p(p_1) + p(p_2) 
\to p(p^{\prime}_1) + p(p^{\prime}_2))|^2}
$$
$$
\times\,\frac{1}{2}\,(2\pi)^4\,
\delta^{(4)}(p^{\prime}_1 + p^{\prime}_2 - p_1 -p_2) \frac{d^3p^{\prime}_1}{(2\pi)^2E^{\prime}_1}
\frac{d^3p^{\prime}_2}{(2\pi)^2E^{\prime}_2}=
$$
$$
=\frac{1}{4 k \sqrt{s}}\,\overline{|{\cal M}(p(p_1) + p(p_2) \to p(p^{\prime}_1) + p(p^{\prime}_2))|^2}\,\frac{1}{2}\,
\frac{k}{4\pi \sqrt{s}}= \pi\,a^2_{\rm pp}\,
\frac{s}{M^2_{\rm N}}.\eqno({\rm G}.13)
$$
In the low--energy limit $k\to 0$, when 
$s \to 4M^2_{\rm N}$, we get
$$
\sigma({\rm pp}\to {\rm pp}) = 4 \pi 
a^2_{\rm pp}.\eqno({\rm G}.14)
$$
By analogous way one can show that the cross section for the low--energy elastic np scattering (see Sect.~11) reads 
$\sigma({\rm np}\to {\rm np}) = 4 \pi a^2_{\rm np}$. 

\section*{Appendix H. Threshold behaviour of  amplitude and  cross section for $\bar{\nu}_{\rm e}$ + D $\to$ e$^+$ +  n + n}

In this Appendix by using the effective local four--nucleon interaction Eq.~(\ref{label1.1})
we adduce the calculation of the cross section for the
process $\bar{\nu}_{\rm e}$ + D $\to$ e$^+$ + n + n near threshold, when the relative momentum of the neutrons goes to zero.

The effective Lagrangian of the process 
$\bar{\nu}_{\rm e}$ + D $\to$ e$^+$ + n + n caused by the effective local four--nucleon interaction  interaction 
Eq.~(\ref{label1.1}) is given by [4]
$$
{\cal L}_{\rm \bar{\nu}_{\rm e}D \to e^+ nn }(x) = 
g_{\rm A}G_{\rm \pi nn}\frac{G_{\rm V}}{\sqrt{2}}
\frac{3g_{\rm V}}{8\pi^2}\,D_{\mu\nu}(x)\,[\bar{n}(x)
\gamma^{\mu}\gamma^5 n^c(x)]\,[\bar{\psi}_{\nu_{\rm e}}(x)\gamma^{\nu}(1 - \gamma^5) \psi_{\rm e}(x)]. 
\eqno ({\rm H}.1)
$$
The amplitude defined by the effective Lagrangian Eq.~({\rm H}.1) reads [4]
$$
\hspace{-3in}i{\cal M}(\bar{\nu}_{\rm e} + {\rm D} 
\to {\rm e}^+ + {\rm n} + {\rm n})_{\rm \delta. p.} =
$$
$$
= - g_{\rm A} M_{\rm N} \frac{G_{\rm
V}}{\sqrt{2}} \frac{3g_{\rm V}}{2\pi^2}\,
G_{\rm \pi NN} e_{\mu}(Q) [\bar{v}(k_{\bar{\nu}_{\rm
e}})\gamma^{\mu}(1-\gamma^5)
v(k_{{\rm e}^+})]\,[\bar{u}(p_1) \gamma^5 u^c(p_2)]. \eqno ({\rm H}.2)
$$
The amplitude Eq.~({\rm H}.2) squared, averaged over polarizations of the
deuteron and summed over polarizations of the final particles is defined by
$$
\overline{|{\cal M}(\bar{\nu}_{\rm e} + {\rm D} \to {\rm e}^+ + {\rm n} + {\rm n})_{\rm \delta. p.}|^2} =
g^2_{\rm A}M^6_{\rm N}\frac{144 G^2_{\rm V}Q_{\rm D}}{\pi^2}\,
G^2_{\rm \pi
NN}\,\Bigg( E_{{\rm e}^+}
E_{\bar{\nu}_{\rm e}} - \frac{1}{3}\vec{k}_{{\rm e}^+}\cdot
\vec{k}_{\bar{\nu}_{\rm e}}\Bigg). \eqno ({\rm H}.3)
$$
The integration over the phase volume of the 
(${\rm n n e^+}$)--state we perform in the non--relativistic limit near threshold and in the rest frame
of the deuteron
$$
\frac{1}{2}\int\frac{d^3p_1}{(2\pi)^3 2E_1}\frac{d^3p_2}{(2\pi)^3 2E_2}
\frac{d^3k_{{\rm e}^+}}{(2\pi)^3 2E_{{\rm e}^+}}(2\pi)^4\,\delta^{(4)}(Q +
k_{{\bar{\nu}_{\rm e}}} - p_1 - p_2 - k_{{\rm e}^+})
\Bigg( E_{{\rm e}^+}
E_{\bar{\nu}_{\rm e}} - \frac{1}{3}
\vec{k}_{{\rm e}^+}\cdot
\vec{k}_{\bar{\nu}_{\rm e}}\Bigg)
$$
$$
\hspace{-0.3in}= \frac{E_{\bar{\nu}_{\rm e}}M^3_{\rm
N}}{1024\pi^2}\,\Bigg(\frac{E_{\rm
th}}{M_{\rm N}}
\Bigg)^{\!\!7/2}\Bigg(\frac{2 m_{\rm e}}{E_{\rm
th}}\Bigg)^{\!\!3/2}\frac{8}{\pi E^2_{\rm th}}
\int\!\!\!\int dT_{\rm e^+} dT_{\rm nn}\,
\sqrt{T_{\rm e^+}T_{\rm nn}}\,
\delta\Big(E_{\bar{\nu}_{\rm e}}- E_{\rm th} - T_{\rm e^+}
 - T_{\rm nn}\Big) =
$$
$$
\hspace{-3.3in}= \frac{E_{\bar{\nu}_{\rm e}}M^3_{\rm
N}}{1024\pi^2}\,\Bigg(\frac{E_{\rm
th}}{M_{\rm N}}
\Bigg)^{\!\!7/2}\Bigg(\frac{2 m_{\rm e}}{E_{\rm th}}\Bigg)^{\!\!3/2}\Bigg
(\frac{E_{\bar{\nu}_{\rm e}}}
{E_{\rm th}} - 1\Bigg)^{\!\!2}. \eqno ({\rm H}.4)
$$
The cross section for the process $\bar{\nu}_{\rm e}$ + 
D $\to$ e$^+$ + n + n  calculated near threshold is defined by
$$
\sigma^{\rm \bar{\nu}_{\rm e} D}_{\rm cc}(E_{\bar{\nu}_{\rm e}}) =
\sigma_0\,\Bigg(\frac{E_{\bar{\nu}_{\rm e}}}
{E_{\rm th}} - 1\Bigg)^{\!\!2}, \eqno ({\rm H}.5)
$$
where $\sigma_0$ is defined by Eq.~(8.15). 
The cross section Eq.~({\rm H}.5) agrees with the cross section calculated near threshold by Weneser in the PMA [44,45].

\newpage

\section*{Figure Caption}

\noindent Fig.~1. One--nucleon loop diagrams describing the amplitude of the p + p $\to$ D + W$^+$ transition. The diagram in Fig.~1b can be reduced to the diagram in Fig.~1a by a charge conjugation transformation applied to the virtual nucleons. This yields the factor 2 in Eq.~({\rm C}.13).

\newpage 
\begin{figure}
\centerline{\epsfxsize=9cm \epsfbox{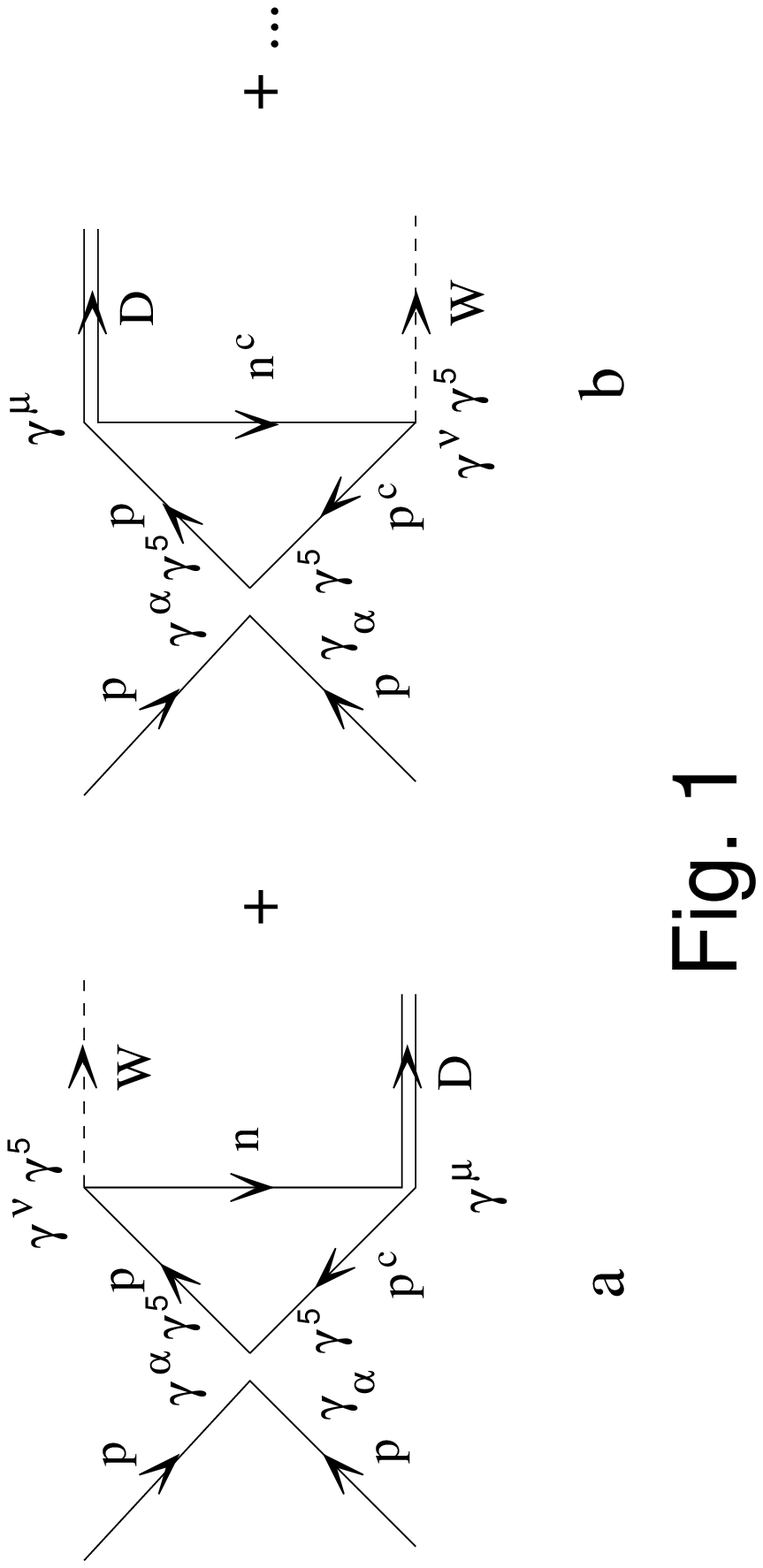}}
\end{figure}

\end{document}